\documentclass[aps,prb,twocolumn,groupedaddress,superscriptaddress,nobibnotes]{revtex4-2}
\usepackage{amsmath}
\usepackage{amssymb}
\usepackage{graphicx}
\usepackage{dcolumn}
\usepackage{bm}
\usepackage{color}
\usepackage{subfig}
\usepackage[normalem]{ulem}
\usepackage[none]{hyphenat}
\usepackage{float}
\usepackage{soul}
\usepackage{array}
\setulcolor{red}
\usepackage{braket}
\usepackage{xr-hyper}
\usepackage{hyperref}
\usepackage{etoolbox}
\usepackage{multirow}
\newcommand{\CXY}{Cr$XY$}
\newcommand{\MXY}{$MXY$}

\newcommand{\MBT}{MnBi$_2$Te$_4$}

\newcommand{\PreserveBackslash}[1]{\let\temp=\\#1\let\\=\temp}
\let\PBS=\PreserveBackslash

\newcounter{lastbib}

\makeatletter
\newenvironment{theSIbibliography}[1]{%
  \section*{\refname}%
  \markboth{\MakeUppercase\refname}{\MakeUppercase\refname}%
  \list{\@biblabel{\@arabic\c@enumiv}}{%
    \settowidth\labelwidth{\@biblabel{#1}}%
    \leftmargin\labelwidth
    \advance\leftmargin\labelsep
    \usecounter{enumiv}%
    \let\p@enumiv\@empty
    \renewcommand\theenumiv{\@arabic\c@enumiv}%
  }%
  \sloppy
  \clubpenalty4000
  \@clubpenalty \clubpenalty
  \widowpenalty4000%
  \sfcode`\.\@m
  \let\old@bibitem\@bibitem
  \def\@bibitem##1{\old@bibitem{SI-##1}}%
  \let\old@lbibitem\@lbibitem
  \def\@lbibitem[##1]##2{\old@lbibitem[##1]{SI-##2}}%
}{%
  \def\@noitemerr{\@latex@warning{Empty `theSIbibliography' environment}}%
  \endlist
}
\makeatother

\begin{document}

\title{Chromium chalcohalide Janus monolayer ferromagnets with perpendicular magnetic anisotropy and high Curie temperature}

\author{M. Bosnar}
\affiliation{Donostia International Physics Center, 20018 Donostia-San Sebasti\'an, Spain}
\affiliation{Present address: Department of Physics, University of Zagreb, Bijenička cesta 32, 10000 Zagreb, Croatia}

\author{J.M. Lendinez}
\email{jm.lendinez@icmm.csic.es}
\affiliation{Instituto de Ciencia de Materiales de Madrid, CSIC, Cantoblanco, 28049 Madrid, Spain}

\author{A.Yu.~Vyazovskaya}
\affiliation{Laboratory of Nanostructured Surfaces and Coatings, Tomsk State University, 634050 Tomsk, Russia}

\author{I.Yu. Sklyadneva}
\affiliation{Donostia International Physics Center, 20018 Donostia-San Sebasti\'an, Spain}

\author{R. Heid}
\affiliation{Institute for Quantum Materials and Technologies, Karlsruhe Institute of Technology, D-76021 Karlsruhe, Germany}

\author{S.\,V. Eremeev}
\affiliation{Institute of Strength Physics and Materials Science, Russian Academy of Sciences, 634021 Tomsk, Russia}
\affiliation{Saint Petersburg State University, Saint Petersburg, Russia, 199034}

\author{U. Atxitia}
\affiliation{Instituto de Ciencia de Materiales de Madrid, CSIC, Cantoblanco, 28049 Madrid, Spain}

\author{S. Gallego}
\affiliation{Instituto de Ciencia de Materiales de Madrid, CSIC, Cantoblanco, 28049 Madrid, Spain}

\author{E.V.~Chulkov}
\affiliation{Donostia International Physics Center, 20018 Donostia-San Sebasti\'an, Spain}
\affiliation{Saint Petersburg State University, Saint Petersburg, Russia, 199034}

\author{A. Arnau}
\affiliation{Centro de Física de Materiales (CFM-MPC), Centro Mixto (CSIC-UPV/EHU), 20018 Donostia-San Sebasti\'an, Spain}
\affiliation{Donostia International Physics Center, 20018 Donostia-San Sebasti\'an, Spain}
\affiliation{Departamento de Pol\'imeros y Materiales Avanzados: F\'isica, Qu\'imica y Tecnolog\'ia, Facultad de Ciencias Qu\'imicas, Universidad del Pa\'is Vasco UPV/EHU, 20018 Donostia-San Sebasti\'an, Spain}

\author{M.M.~Otrokov}
\email{mikhail.otrokov@unizar.es}
\affiliation{Instituto de Nanociencia y Materiales de Aragón (INMA), CSIC-Universidad de Zaragoza, Zaragoza 50009, Spain}

\begin{abstract}

Using density functional theory, we revisit the magnetic properties of a recently proposed family of noncentrosymmetric two-dimensional magnetic materials, chromium chalcohalide monolayers, \CXY{} ($X = \mathrm{S}, \mathrm{Se}, \mathrm{Te}$; $Y = \mathrm{Cl}, \mathrm{Br}, \mathrm{I}$). These systems consist of three atomic planes stacked in the $X$–Cr–$Y$ sequence, which breaks inversion symmetry, giving rise to their designation as "Janus" monolayers. We consider both 1T and 1H structural polymorphs of \CXY{}, beginning with an assessment of their dynamical stability via phonon spectra calculations. All systems, except for the 1H-CrTe$Y$ variants, are found to be dynamically stable. Among the two polymorphs, the 1T phase is consistently more favorable, with energy gains exceeding 0.55~eV per formula unit. Our total-energy calculations reveal that all dynamically stable \CXY{} monolayers exhibit ferromagnetic coupling. However, robust out-of-plane magnetic anisotropy is observed only in the CrSI and CrSeI compositions, for both 1T and 1H structures. The perpendicular magnetic anisotropy results from a constructive interplay between single-ion and anisotropic exchange contributions that overcome the dipole–dipole interaction. We further quantify the Dzyaloshinskii-Moriya interaction (DMI) in CrSI and CrSeI for both polymorphs, and reveal a weak-to-moderate DMI strength as compared to the isotropic exchange interaction term. Finally, for systems exhibiting ferromagnetic coupling and perpendicular magnetic anisotropy, the exchange and anisotropy parameters derived from density functional theory calculations are employed as inputs for large-scale atomistic spin dynamics simulations to probe the temperature evolution of real-space magnetic structures. The calculated Curie temperatures are at least 210 K for 1T-CrSI, 235–260 K for 1H-CrSeI, and 370–410 K for 1H-CrSI. In contrast, the sizable DMI in 1T-CrSeI results in a worm-like domain ground state at zero external field and enables the stabilization of skyrmions under a perpendicular magnetic field. We then discuss and illustrate how \CXY\, monolayers can be used to tune the Bychkov-Rashba split surface state of BiTeI or to enhance magnetism of thin films of the intrinsic antiferromagnetic topological insulator \MBT.

\end{abstract}

\maketitle

\section{Introduction}

The observation of ferromagnetism at finite temperatures in atomically thin van der Waals layers \cite{Huang.nat2017, Gong.nat2017} has sparked intense interest in two-dimensional (2D) layered magnetic materials. From a theoretical standpoint, the Mermin-Wagner theorem \cite{Mermin.prl1966} prohibits magnetic ordering in the 2D isotropic Heisenberg model at finite temperatures. However, the presence of easy-axis magnetic anisotropy lifts this restriction. The recent experimental observation \cite{Huang.nat2017, Gong.nat2017} of this fundamental phenomenon has motivated extensive theoretical and experimental investigations. One particularly appealing feature of 2D magnetic layered materials is their compatibility with van der Waals heterostructure engineering, enabling precise control over strain, composition, as well as optical and electronic properties. As a result, these materials offer a rich platform for both applied and fundamental studies \cite{Wang.acsn2022}.

While most of the currently studied 2D van der Waals magnets are intrinsically centrosymmetric, breaking inversion symmetry can give rise to new functionalities when spin-orbit coupling (SOC) is present. In noncentrosymmetric systems, SOC leads to momentum-dependent splitting of otherwise spin-degenerate bands, a phenomenon known as the Bychkov-Rashba effect \cite{BychkovRashba}. It is believed that electronic states combining a strong Bychkov-Rashba effect with significant exchange splitting play a crucial role in current-induced magnetization switching in heterostructures containing heavy elements \cite{Miron.nat2011}, which is promising for magnetic data storage technologies.

Another motivation for studying such systems stems from the proposal that a ferromagnetic (FM) material with Bychkov-Rashba splitting (e.g., V-doped BiTeI \cite{Klimovskikh.srep2017}), when proximitized to an $s$-wave superconductor, can host a topologically non-trivial superconducting phase supporting zero-energy Majorana modes \cite{Sau.prl2010}. Such Majorana modes are of great interest for quantum technologies, including topological qubits for quantum computation, simulation, and sensing \cite{DasSarma.npjqi2015}.

Moreover, the combination of broken inversion symmetry, magnetism, and strong SOC gives rise to another fundamental phenomenon: the Dzyaloshinskii-Moriya interaction (DMI) \cite{Dzyaloshinskii1957, Moriya.prl1960, Moriya.pr1960}. DMI is an antisymmetric exchange interaction that arises from a first-order spin-orbit correction to the conventional Heisenberg exchange. At the microscopic level, it favors chiral non-collinear spin textures and plays a crucial role in stabilizing chiral domain walls \cite{Ryu.natn2013} and magnetic skyrmions \cite{Yu.natm2011, Fert.natn2013}. The latter are particularly appealing due to their topological protection, nanoscale dimensions, and extremely low driving current requirements, making them strong candidates for development of future spintronic devices and integrated circuits.

Finally, noncentrosymmetric 2D materials can also exhibit intrinsic piezoelectricity and ferroelasticity \cite{Dong.acsn2017, Zhang.nl2019}. These properties make them attractive candidates for applications in electromechanical sensors, actuators, transducers, and energy harvesters, particularly within the fields of nanorobotics, piezotronics, and nanoelectromechanical systems \cite{Lin.mtn2018, Dong.acsn2017}.

To realize the functionalities described above in 2D magnetic van der Waals materials, inversion symmetry can, in principle, be broken through various external means, such as combining different compounds, applying a bias voltage, or introducing strain \cite{Yao.prb2008, Liu.prb2018, Deng.nat2018, Bobkov.prm2024}. However, the most straightforward and desirable approach is to employ a 2D van der Waals magnet that intrinsically lacks inversion symmetry. 

Recently, non-magnetic, noncentrosymmetric 2D transition metal dichalcogenides (TMDs), known as Janus monolayers (MLs), have garnered increasing attention \cite{Yagmurcukardes.apr2020, Zhang.jmca2020}. These systems consist of three atomic layers stacked in the sequence $X_1$–$M$–$X_2$ ($M = \mathrm{Nb},\ \mathrm{Ti},\ \mathrm{Ta},\ \mathrm{Mo},\ \ldots$; $X_1,\ X_2 = \mathrm{S},\ \mathrm{Se},\ \mathrm{Te}$, with $X_1 \neq X_2$), where the top and bottom faces are terminated by different atoms, thereby breaking inversion symmetry. Experimental studies have recently shown that Janus TMD MLs such as MoSSe can be synthesized via a two-step process involving chemical vapor deposition of MoS$_2$, followed by sulfur removal and thermal selenization \cite{Lu.nnano2017, Zhang.acsn2017}.

The experimental realization of MoSSe Janus MLs, combined with the possibility of tuning their electronic properties through appropriate chalcogen pair selection, has motivated theoretical investigations of magnetic analogues of these systems, such as Mn$X_1X_2$ ($X_1,\ X_2 = \mathrm{S},\ \mathrm{Se},\ \mathrm{Te}$;\ $X_1 \neq X_2$) \cite{LiangJ.prb2020}. Notably, this study reports that the DMI amplitudes in MnSeTe and MnSTe Janus MLs are comparable to those found in ferromagnet/heavy-metal heterostructures \cite{Yang.prl2015, Blanco-rey.prb2022}, suggesting that these materials are promising candidates for the realization of magnetic skyrmions.

More recently, transition metal chalco-\emph{halide} Janus MLs Cr$XY$ ($X = \mathrm{S},\ \mathrm{Se},\ \mathrm{Te}$; $Y = \mathrm{Cl},\ \mathrm{Br},\ \mathrm{I}$) have been theoretically proposed \cite{Xiao.pccp2020, Hou.npjcm2022, Guo.apl2022}. These systems may offer advantages over purely chalcogen-based Janus MLs, as the presence of more electronegative halogen atoms is expected to induce stronger electric dipoles, thereby enhancing the Bychkov-Rashba effect, DMI, and piezoelectric response. Additionally, in systems such as $M$TeI, the SOC is stronger than in $M$TeSe due to the higher atomic number of iodine.

Several Janus Cr$XY$ MLs were initially predicted to be FM semiconductors with high Curie temperatures ($T_\text{C}$)~\cite{Xiao.pccp2020, Hou.npjcm2022, Guo.apl2022}. Reported $T_\text{C}$ values reached 161--275~K for CrS$Y$ and 550~K for CrSeBr, climbing as high as 956~K for CrTeI. Some of these findings were refined by subsequent studies~\cite{Li.prb2023, Guan.pccp2023}, which revealed that compositions featuring heavier chalcogens and halogens ($X = \text{Se, Te}$ and $Y = \text{Br, I}$) are not simple ferromagnets but are instead prone to pronounced non-collinearities. Such states are driven by sizable DMI and frustrated isotropic exchange. Predicted examples include the coexistence of worm-like domains and skyrmions at zero field for CrTeI~\cite{Li.prb2023}, while CrSeCl and CrSeBr have been proposed to host bimerons~\cite{Li.prb2023, Guan.pccp2023}. Some of these materials are also expected to exhibit strong vertical piezoelectric responses, with coefficients exceeding those of most known 2D materials~\cite{Xiao.pccp2020, Guo.apl2022}.

In Refs.~\cite{Xiao.pccp2020, Hou.npjcm2022, Guo.apl2022, Li.prb2023, Guan.pccp2023}, the distorted octahedral (1T) structure of \CXY{} MLs has been considered, corresponding to the structure of CrSe$_2$ and CrTe$_2$
\cite{Lasek.ssr2021}. It is well established that the trigonal prismatic (1H) phase is frequently realized within the TMD family \cite{Lasek.ssr2021}. While this structural variant had not been explored for $MXY$ Janus MLs in the aforementioned studies, more recent works have started to address this possibility.

In recent years, the family of Janus $MXY$ MLs has grown considerably, now encompassing a wide range of 3$d$ transition metals ($M = \mathrm{Ti},\ \mathrm{V},\ \mathrm{Cr},\ \mathrm{Mn},\ \mathrm{Fe},\ \mathrm{Co},\ \mathrm{Ni}$) \cite{Mahmoodabadi.cms2023, Rahman.jem2023, Caglayan.prb2024, Su.pccp2024, Chang.mat2024, Huang.apl2024, Dai.cms2025}, although not all combinations of chalcogen and halogen elements ($X = \mathrm{S},\ \mathrm{Se}, \mathrm{Te}$; $Y = \mathrm{Cl}, \mathrm{Br}, \mathrm{I}$) have yet been explored. While the majority of these systems have been investigated in the 1T structure, the 1H phase has also been considered in several cases \cite{Caglayan.prb2024, Su.pccp2024, Chang.mat2024, Rahman.jem2023}, primarily for V- and Ti-based MLs.

\begin{figure*}
    \centering
    \includegraphics[width=1\textwidth]{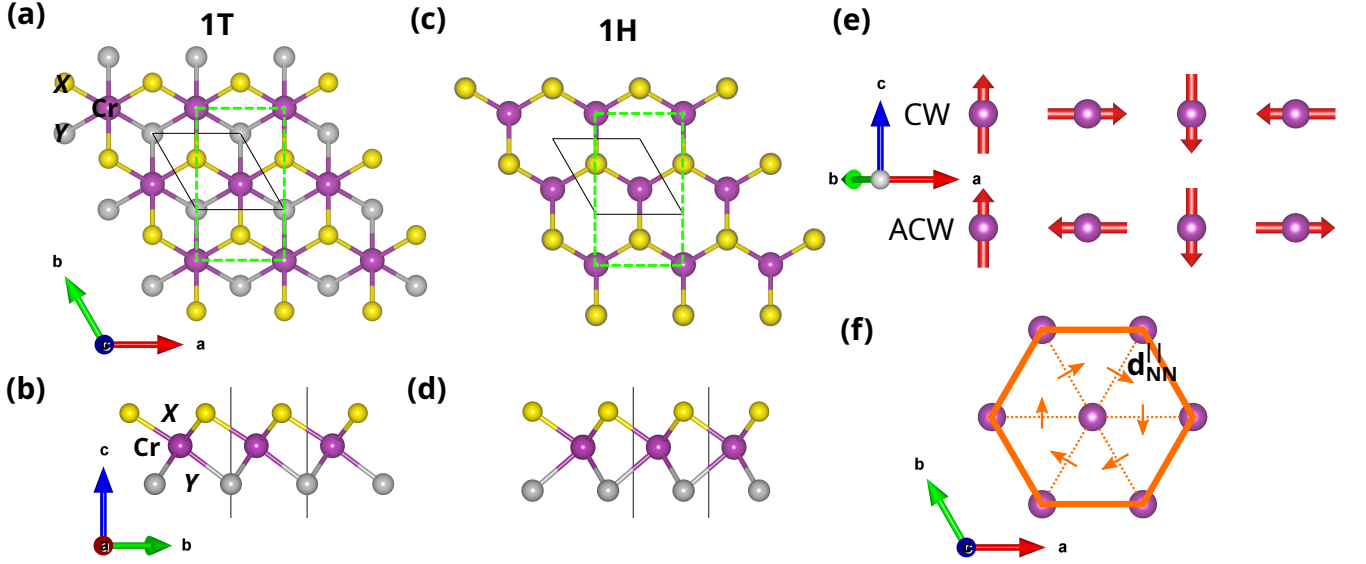}
	\caption{Top (a,c) and side (b,d) views of the crystal structures of the \CXY{} Janus MLs ($X=$ S, Se, Te and $Y=$Cl, Br, I) with the 1T (a,b) and 1H (c,d) structures. Apart from the $(1\times 1)$ cell with one atom per plane, panels (a,c) also show the rectangular $(1\times \sqrt{3})$ cell (green dashes) with two atoms per plane, used for the calculation of the isotropic and anisotropic exchange coupling parameters. (e) Schematic illustration of the clock-wise (CW) and anti clock-wise (ACW) spin spirals, with the Cr local moments rotating within the $xz$ plane. (f) DMI vectors for the nearest neighbors coupling.}
    \label{fig:struc}
\end{figure*}

Despite significant attention received by the Cr-based \MXY{} Janus MLs \cite{Xiao.pccp2020, Hou.npjcm2022, Guo.apl2022, Guan.pccp2023, Jain.jpcm2023, Li.prb2023, Mahmoodabadi.cms2023, Rahman.jem2023, Wang.nl2024, Wu.jpcs2024, Yang.mathor2024}, key aspects of their magnetic behavior still remain to be clarified. Namely, 
concerning the 1T-\CXY{} systems, the shape anisotropy contribution arising from dipole-dipole interactions between local moments has been considered neither in Refs.~\cite{Xiao.pccp2020, Hou.npjcm2022, Guo.apl2022, Guan.pccp2023}, nor in the majority of subsequent studies \cite{Mahmoodabadi.cms2023, Rahman.jem2023, Caglayan.prb2024, Su.pccp2024, Chang.mat2024, Huang.apl2024, Dai.cms2025}. Neglecting this contribution may result in an inaccurate description of magnetic anisotropy, both quantitatively and qualitatively. The recent study by Li et al.~\cite{Li.prb2023} explicitly incorporated the dipole-dipole interaction into the spin Hamiltonian and accounted for it in atomistic simulations. However, this study does not address $X=$ S case. Also, the critical temperatures of the FM systems were not estimated. More importantly, as far as the \CXY\, MLs in the 1H structure are considered, until now, their magnetic properties had only been addressed using machine learning approach \cite{Mahmoodabadi.cms2023}. In the latter work, the exchange coupling parameters were predicted by using trained models, identifying 1H CrSI, CrTeBr and CrTeI MLs as ferromagnets, while the rest of the 1H-\CXY\, MLs were claimed to be antiferromagnetic (AFM). However, an independent verification of these machine learning predictions using the state-of-the-art density functional theory (DFT) calculations in combination with Monte Carlo or atomistic spin dynamics simulations is still lacking for 1H-\CXY. Moreover, upon estimating the critical temperatures within the random-phase approximation, an assumption of a universal, positive magnetocrystalline anisotropy energy of 1 meV was made for all 1H-\CXY\, MLs \cite{Mahmoodabadi.cms2023}, corresponding to an out-of-plane easy axis. The validity of this assumption remains untested. 

Therefore, in the present work, we revisit the magnetic properties of the \CXY\, Janus MLs ($X = \mathrm{S},\ \mathrm{Se},\ \mathrm{Te}$;\ $Y = \mathrm{Cl},\ \mathrm{Br},\ \mathrm{I}$) in both 1T- and 1H structures using DFT calculations in combination with atomistic spin dynamics (ASD) simulations. Our goal is reliably identifying among these systems stable ferromagnets with robust perpendicular magnetic anisotropy (PMA) and elevated Curie temperatures. Such van der Waals materials are of great interest for magnetic storage \cite{Hu.apx2018, Wu.nl2022}, spin-orbit torque \cite{Wang.sciadv2019, Wang.ncomms2023}, as well as the functional components of heterostructures for observation of topological superconductivity \cite{Sau.prl2010, DasSarma.npjqi2015}, or quantum anomalous Hall effect \cite{Hou.npjcm2022, Chang.rmp2023}.

Our analysis includes determination of the isotropic and anisotropic exchange coupling, singe-ion and shape anisotropy, as well as the DMI from first principles. We pay particular attention to the influence of the Hubbard $U_\text{eff}$ parameter applied to the Cr 3$d$ states, assessing the robustness of magnetic properties with respect to its variation. For the systems showing a stable tendency towards the FM order with PMA, the exchange and anisotropy parameters, obtained from DFT, are then used in the large-scale ASD simulations to identify the real space magnetic structures both at zero Kelvin and finite temperatures, up to the transition point to the paramagnetic state. In this way, we identify among all 1T- and 1H-\CXY\, systems the following three: 1T-CrSI and 1H-Cr$X$I ($X = \mathrm{S},\ \mathrm{Se}$) that possess the sought-for FM behavior with PMA. The resulting $T_\text{C}$ is predicted to be at least 210~K for 1T-CrSI, while for the 1H polymorphs $T_\text{C}$ values are equal to 235–260 K for 1H-CrSeI and well above room temperature for 1H-CrSI (370-410 K). Finally, we propose functional heterostructures, in which 1T-CrSI MLs are interfaced with (i) BiTeI \cite{Ishizaka.nmat2011, Eremeev.prl2012, Eremeev.jetpl2012,Maass2016}, whose surfaces host 2D states with a giant Bychkov-Rashba splitting, and (ii) thin films of the intrinsic AFM topological insulator \MBT\ \cite{Otrokov.nat2019, Otrokov.prl2019, Li.sciadv2019, Zhang.prl2019}.

\section{Computational details}

All DFT calculations, except for those of the phonon spectra, were performed using the projector augmented-wave (PAW) method~\cite{Blochl.prb1994}, as implemented in the VASP code~\cite{vasp1,vasp2,vasp3}. The generalized gradient approximation (GGA) was employed to describe the exchange-correlation potential~\cite{Perdew.prl1996}. The plane wave cutoff energy was set to $500$ eV. 

To account for the strongly localized nature of the Cr $3d$ states, we used the GGA+$U$ approach~\cite{Anisimov1991}, with the effective Hubbard parameter defined as $U_{\mathrm{eff}} = U - J$~\cite{Dudarev.prb1998}. The values of $U_{\mathrm{eff}} = 2.1$~eV (as in Ref.~\cite{Hou.npjcm2022}), 3.0~eV, and 4.0~eV were considered for the Cr $3d$ states. For each chosen $U_{\mathrm{eff}}$ value, the crystal structure was fully relaxed and the magnetic properties were calculated for that geometry. All reported quantities at a given $U_{\mathrm{eff}}$ correspond to the structure optimized with the same $U_{\mathrm{eff}}$. This ensures internal consistency between structure and magnetism and allows us to assess the robustness of the results with respect to the choice of $U_{\mathrm{eff}}$, whose exact value remains unknown for the \CXY{} MLs studied here. In other words, we do not attempt to determine an unambiguous value of $U_{\mathrm{eff}}$, since even specialized schemes for its evaluation (e.g., the linear-response approach \cite{Cococcioni.prb2005}) cannot guarantee agreement with future experiments. Our strategy is therefore to examine the stability of the calculated properties against variations of $U_{\mathrm{eff}}$. If a property (e.g., the magnetic anisotropy energy) remains qualitatively unchanged across the tested range, it can be regarded as a robust theoretical prediction. Conversely, when the behavior depends qualitatively on $U_{\mathrm{eff}}$, no definitive prediction is made.

The calculations were performed in a model of repeating films separated by a vacuum gap of about 13~\AA. In-plane cells of different periodicities were employed depending on the specific task, as detailed below.

For each \CXY{} ML (and for each $U_{\mathrm{eff}}$), both the lattice parameters and atomic positions were optimized, with a force convergence criterion of 0.01~eV/{\AA} applied to the latter. A hexagonal $(1 \times 1)$ cell containing one Cr atom was used for structural optimization, which implies that all systems were optimized in the FM state. This magnetic configuration was subsequently verified through the calculation of the nearest neighbor (NN) isotropic exchange coupling constant $J$. For these $J$ calculations, a rectangular $(1 \times \sqrt{3})$ supercell containing two Cr atoms was employed (shown in Figs. \ref{fig:struc}a,c), enabling the modeling of both FM and AFM configurations. The Brillouin zone was sampled using a $11 \times 11 \times 1$ $k$-point mesh for the $(1 \times 1)$ cell, and a $11 \times 7 \times 1$ mesh for the $(1 \times \sqrt{3})$ cell. For selected systems, we have also calculated the isotropic exchange coupling constants up to the third-NNs (Suppl.~Table \ref{stab:j_third} of the Supplementary Material~\cite{supplementary_material}) using OstravaJ code \cite{OstravaJ}. In these calculations, we used a hexagonal $(4 \times 4)$ cell and a $3 \times 3 \times 1$ $k$-point mesh. All calculations described in this paragraph were carried out within the scalar-relativistic approximation.

Phonon spectra were obtained for the FM state using scalar relativistic spin-polarized calculations employing the linear response technique in combination with the mixed-basis pseudopotential method \cite{Heid:99,Meyer}. The plane-wave basis with a cut-off energy of 24 Ry was supplemented with local $s$- and $p$-type functions for all atoms and local $d$-type functions for Cr at each atomic site. To assess the effect of local correlation on the electronic band structure, a Hubbard correction ($U_{\mathrm{eff}} = 2.1$~eV) for the localized strongly correlated $3d$ electrons of Cr was used. Some systems were also tested with $U_{\mathrm{eff}} = 3$~eV, and no qualitative dependence was found. The Brillouin zone integrations were performed on a uniform $12\times12\times1$ k-point mesh using a Gaussian energy smearing scheme.

The $(1 \times \sqrt{3})$ cell was also used for the magnetic anisotropy calculations. These were performed including SOC, using a non-self-consistent approach. The choice of a non-self-consistent calculation was motivated by convergence issues encountered in self-consistent SOC calculations for the AFM configurations in various \CXY{} systems, which hindered the reliable extraction of single-ion anisotropy and anisotropic exchange parameters. However, since the FM phase converged well in self-consistent runs, we verified the accuracy of the non-self-consistent results against fully self-consistent calculations and found good agreement. A $k$-point mesh of $33 \times 21 \times 1$ was used in these calculations.

To compute the DMI, we constructed a $(4 \times 1)$ supercell based on the $(1 \times 1)$ hexagonal unit cell (see Suppl. Fig.~\ref{sfig:dmi}). The Brillouin zone was sampled using a $k$-mesh of $11 \times 45 \times 1$. A total energy convergence threshold of $10^{-8}$~eV was applied in all DMI calculations.

To study the temperature dependent magnetic properties, we employed ASD simulations using a classical Heisenberg spin Hamiltonian with parameters extracted from the DFT calculations. The temporal evolution of each atomic spin was described using the stochastic Landau-Lifshitz-Gilbert (LLG) equation. See Suppl. Sec. \ref{ssec:asd} for more details. 
The LLG-ASD simulations were also tested against the Monte Carlo simulations, performed using the VAMPIRE code \cite{MainVampire}, and a good agreement was found (Suppl. Sec. \ref{sssec:asd_vs_mc}).

Interfaces of the 1T-CrSI Janus ML with a giant Bychkov-Rashba semiconductor are considered on the example of 1T-CrSI and BiTeI, the pair of materials which has perfect matching in the in-plain parameters of their ($2\times2$) and ($\sqrt{3}\times\sqrt{3}$) superlattices. Two halogen--chalcogen interfaces were considered: Te-I and I-S interfaces at the Te- and I-terminated BiTeI surfaces, respectively. The BiTeI substrate was modeled by eight trilayers with either H- or F-passivated opposite side of the slab for Te or I terminated surface, respectively \cite{Eremeev.prl2012, Eremeev.jetpl2012}. The CrSI/BiTeI-($2/\sqrt{3}\times2/\sqrt{3}$) superstructures were optimized using the GGA+$U$ approach with the $k$-point mesh of $6 \times 6 \times 1$. The atoms of the topmost BiTeI trilayer and Janus adlayer were allowed to relax in three directions.

The 1T-CrSI/[\MBT]$_{n\mathrm{SL}}$ heterostructures were constructed by expanding the lattice parameter of the $(2\times2)$ supercell of CrSI by $0.2\,\%$ to match that of the $(\sqrt{3}\times\sqrt{3})$ structure of [\MBT]$_{n\mathrm{SL}}$ (we used the same lattice parameter for $n=1$ and 2). To determine the optimal adsorption registry, the same stacking arrangements as those considered in Suppl. Fig. S4 of Ref.~\cite{Hou.npjcm2022} for CrSBr/Bi$_2$Se$_3$ were constructed for the CrSI/[\MBT]$_{\mathrm{1SL}}$ heterostructure. These structures were fully relaxed, and their total energies were calculated for both FM and AFM interlayer spin configurations. 
In these calculations, the Hubbard $U_\mathrm{eff}$ correction and SOC were included, along with van der Waals interactions treated using the Grimme DFT-D3 model with Becke–Johnson damping~\cite{Grimme.jcc2011}. Only the case of $U_{\mathrm{eff}} = 3\,\mathrm{eV}$ applied to the Cr $3d$ orbitals was considered, while $U_{\mathrm{eff}} = 5.34\,\mathrm{eV}$ was used for the Mn $3d$ orbitals, following previous works~\cite{Eremeev.jac2017, Otrokov.nat2019, Otrokov.prl2019}. For both structural relaxations and static self-consistent total energy calculations, a $9\times9\times1$ Monkhorst–Pack sampling of the Brillouin zone was used. All other computational parameters and convergence thresholds were identical to those employed in the calculations for individual \CXY\ Janus MLs. For evaluation of the SOC contribution to the magnetic anisotropy energy, a denser $21\times21\times1$ Monkhorst–Pack grid was adopted, and the electronic convergence threshold was tightened to $10^{-8}\,\mathrm{eV}$. For both CrSI/BiTeI and CrSI/\MBT, the plane wave cutoff energy was set to $250$ eV to facilitate the treatment of the large cells.

To calculate the Chern number, Wannier functions and the corresponding tight-binding Hamiltonian were constructed using the Wannier90 package~\cite{Mostofi2014CPC, Pizzi2020JOPCM}, and the Wannier charge center summation was performed using the WannierTools package~\cite{Wu2018:CPC}.

\section{Results and discussion}

\begin{figure}
    \centering
    \includegraphics[width=1\columnwidth]{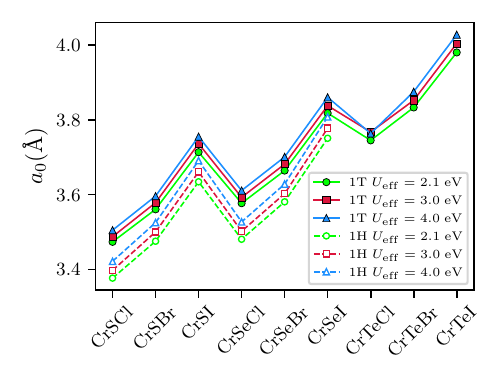}
    \caption{Optimized lattice constants $a_0$ (\AA) of the \CXY{} Janus MLs ($X = \mathrm{S},\ \mathrm{Se},\ \mathrm{Te}$;\ $Y = \mathrm{Cl},\ \mathrm{Br},\ \mathrm{I}$) for the 1T and 1H structures in the FM ordering case, as calculated for different Hubbard parameters $U_{\mathrm{eff}} = 2.1$~eV (as in Ref.~\cite{Hou.npjcm2022}), 3.0~eV, and 4.0~eV applied on the Cr $3d$ states.  Data for the 1H-CrTe$Y$ systems are not shown due to their dynamical instability (see Suppl. Fig.~\ref{sfig:phonons_1h}). All $a_0$ values presented in the graph are also listed in Suppl. Table~\ref{stab:struc}.}
    \label{fig:latt_const}
\end{figure}

\subsection{Crystal structure}

Figs.~\ref{fig:struc}a-d display the crystal structures of the 1T and 1H phases of the \CXY{} MLs ($X = \mathrm{S},\ \mathrm{Se},\ \mathrm{Te}$;\ $Y = \mathrm{Cl},\ \mathrm{Br},\ \mathrm{I}$), which are characterized by ABC and ABA stackings, respectively.  Both the 1T and 1H phases belong to the same $P3m1$ (No. 156) space group despite their different local coordinations \cite{Guo.cms2019, Yin.matadv2021, Oreshonkov.mats2022}. In the 1T (octahedral) polytype each Cr sits in a slightly distorted octahedron, while in the 1H (trigonal‐prismatic) polytype the Cr occupies a similarly distorted prism.

To assess the dynamical stability of these systems, we performed phonon spectra calculations, with the results presented in Suppl. Figs.~\ref{sfig:phonons_1t} and ~\ref{sfig:phonons_1h}. The 1T-\CXY{} MLs, as well as the 1H structures with $X = \mathrm{S}$ and $\mathrm{Se}$, have neither imaginary modes nor phonon dispersion anomalies, 
indicating their dynamical stability. In contrast, the 1H phase of the tellurides is dynamically unstable, 
showing a strong softening of the lowest vibrational mode, characterized by displacements of all atoms along the stacking direction ($z$).
Consequently, the 1H-CrTe$Y$ systems are excluded from further analysis.

The calculated lattice constants $a_0$ of \CXY{} MLs are shown in Fig.~\ref{fig:latt_const}. A systematic trend is observed for both structural polymorphs: $a_0$ increases with the atomic number of either the chalcogen ($X$) or halogen ($Y$) atom, which is consistent with the increase of the corresponding ionic radii. Additionally, increasing $U_{\mathrm{eff}}$ results in a larger $a_0$, which is attributed to the stronger localization of the Cr-$3d$ electrons, reducing their hybridization with the $p$ orbitals of $X$ and $Y$. An exception to this monotonic trend occurs for 1T-CrTeCl, where $a_0$ slightly decreases when increasing $U_{\mathrm{eff}}$ from 3.0 to 4.0~eV, a behavior that correlates with anomalies also found in its exchange coupling and magnetic anisotropy energy (see below). This irregularity may be linked to the system becoming gapless at $U_{\mathrm{eff}} = 4.0$~eV (Suppl. Fig. \ref{sfig:bands_1t-crtey}). Overall, for the 1T polymorphs, $a_0$ ranges from approximately 3.47~\AA{} (CrSCl) to 4.03~\AA{} (CrTeI), while the 1H structures exhibit lattice constants that are systematically $\sim$2.5\% smaller.

\begin{figure}
    \centering
    \includegraphics[width=1.0\columnwidth]{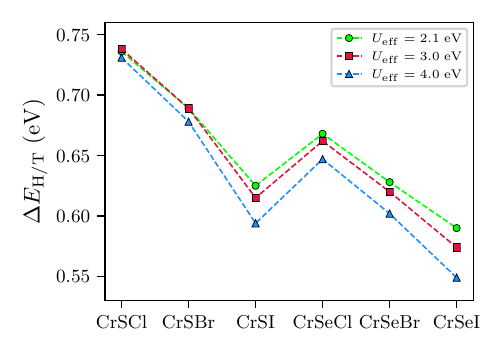}
    \caption{Total energy difference of the 1H and 1T phases, $\Delta E_\text{H/T}=E_\text{1H} - E_\text{1T}$ (in eV). The tellurides are not shown because their 1H phase is dynamically unstable.}
    \label{fig:1h_vs_1t_struc}
\end{figure}

\begin{figure*}
    \centering
    \includegraphics[scale=1.0]{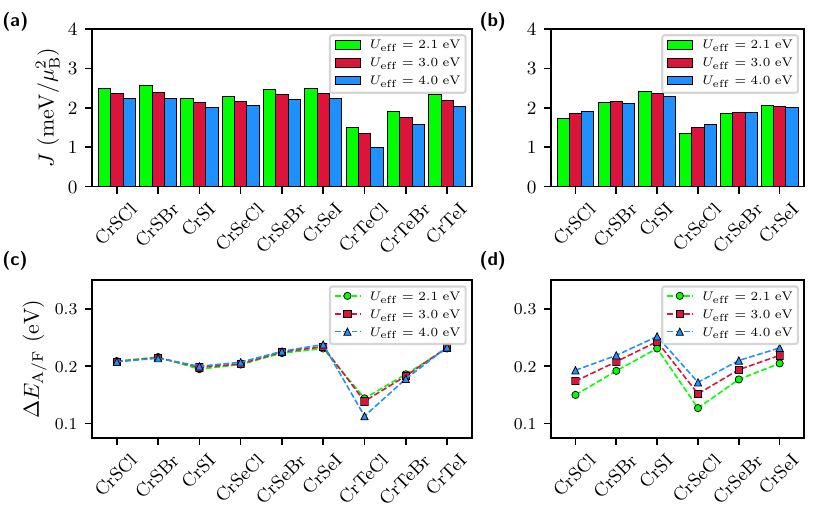}
    \caption{(a,b) Nearest-neighbor isotropic Heisenberg exchange coupling parameters $J$ (meV/$\mu_\mathrm{B}^2$) of the \CXY{} Janus MLs ($X = \mathrm{S},\ \mathrm{Se},\ \mathrm{Te}$;\ $Y = \mathrm{Cl},\ \mathrm{Br},\ \mathrm{I}$) for the 1T (a) and 1H (b) structures. Panels (c) and (d) show the corresponding total energy differences, $\Delta E_{\mathrm{A/F}} = E_{\mathrm{AFM}} - E_{\mathrm{FM}}$, for the 1T (c) and 1H (d) polymorphs, respectively. All $J$ values are listed in Suppl. Table~\ref{stab:j}, and $\Delta E_{\mathrm{A/F}}$ in Suppl. Table~\ref{stab:e_mag}. The supercells used in these calculations are shown by green dashes in Figs.~\ref{fig:struc}a and \ref{fig:struc}c.}
    \label{fig:j}
\end{figure*}

We now address the relative energetic stability of the \CXY's two structural polymorphs, 1T and 1H. Figure~\ref{fig:1h_vs_1t_struc} presents the total energy difference, defined as $\Delta E_\text{H/T} = E_\text{1H} - E_\text{1T}$, evaluated for the FM state (the choice of the latter state is justified below). Across all compositions considered, the 1T structure is consistently more favorable, exhibiting an energy gain of at least 0.55~eV per formula unit (f.u.). This preference is particularly pronounced in MLs incorporating lighter chalcogen and halogen elements.

These findings suggest that the 1T phase is more likely to be realized in experiment, given its clear energetic favorability. Nonetheless, the formation of the 1H polymorph cannot be excluded. Such a scenario has been observed in TMD MLs, where metastable polymorphs with significantly higher formation energies have been successfully synthesized under appropriate conditions \cite{Sokolikova.csr2020}. For instance, although the 1T phases of MoS$_2$ and WS$_2$ are typically 0.8–0.9~eV/f.u. higher in energy than their 1H counterparts \cite{Singh.2dmat2015}, they have been obtained via hydrothermal synthesis under a magnetic field of 9~T, which selectively stabilized the 1T phase due to its higher magnetic susceptibility \cite{Ding.acsn2019}. Other non-equilibrium strategies have also enabled the growth of metastable TMD polymorphs \cite{Sokolikova.ncomms2019, Sokolikova.csr2020}. By analogy, we propose that the 1H polymorph of \CXY{} MLs, while not being energetically most favorable, may likewise be realized under suitable growth conditions. Motivated by this possibility, we proceed to investigate the magnetic properties of both 1T and 1H structures in the following sections. Crucially, Raman spectroscopy provides a direct structural fingerprint that can distinguish these two polymorphs experimentally, so we have calculated their characteristic $A_1$ and $E$ modes frequencies, details of which are presented in Suppl. Sec. \ref{ssec:raman}.

\subsection{Magnetic characterization}
\label{sec:mag}

\subsubsection{Magnetic ordering}

To investigate the magnetic coupling, we compute the total energy differences $\Delta E_{\mathrm{A/F}} = E_{\mathrm{AFM}} - E_{\mathrm{FM}}$ within the scalar-relativistic approximation (i.e., neglecting SOC), and map the results to the isotropic exchange term of the Heisenberg Hamiltonian (Suppl. Sec.~\ref{ssec:hh}),
\begin{align}
H_{\mathrm{iso}} &= -\frac{J}{2} \sum_{i,j} \mathbf{M}_i \cdot \mathbf{M}_j \, ,
\label{eq:h_iso}
\end{align}
 to extract the nearest-neighbor exchange constant $J$ for each \CXY{} system. Here, $J > 0$ ($J < 0$) indicates FM (AFM) coupling, and the indices $i$ and $j$ run over nearest-neighbor (NN) Cr atoms within the layer. Hereinafter, in the Hamiltonian terms, we employ the magnetic moment vector $\mathbf{M}_i$ (in units of $\mu_B$) rather than the spin vector $\mathbf{S}_i$, since the local Cr moments of \CXY\ span an intermediate range of 3.2-3.8~$\mu_B$ (Suppl. Table~\ref{stab:magmom}), preventing a clear assignment to either $S = 3/2$ or $S = 2$.

Total energy differences between AFM and FM configurations, $\Delta E_\text{A/F}$, indicate that all dynamically stable \CXY{} MLs show FM NN interaction. In every case, $\Delta E_\text{A/F}>0$ (Fig.~\ref{fig:j}c,d) and, hence, the corresponding exchange constants $J$ (Fig.~\ref{fig:j}a,b) are positive, confirming FM coupling between nearest neighbors, in agreement with previous studies \cite{Hou.npjcm2022, Guan.pccp2023}. Regarding the microscopic mechanism underlying this FM interaction, our  crystal orbital Hamilton population method calculations \cite{Dronskowski1993,Deringer2011,LOBSTER-2016} reveal a negligible overlap of the Cr-$d$ states from the adjacent sites even for CrSCl, which features the smallest lattice constants ($a^\text{1T}_0 \simeq 3.49$~\AA{} and $a^\text{1H}_0 \simeq 3.40$~\AA\ in average). This indicates that the direct Cr-Cr exchange is marginal in all \CXY\, and that the Cr–$X$($Y$)–Cr superexchange pathways dominate. For the gapped systems (all \CXY{} except for 1H-Cr$X$I with $X=$ S, Se and 1T-CrTeCl at $U_\text{eff} = 4$~eV, see Suppl. Figs. \ref{sfig:bands_1t-crsy}-\ref{sfig:bands_1t-crtey}), the superexchange interaction appears to follow the Goodenough–Kanamori rules \cite{Anderson.pr1959, Kanamori1959, Goodenough1963}: the NN Cr atoms are connected via $X$ and $Y$ atoms forming Cr–$X(Y)$–Cr bonds with angles close to 90$^\circ$, which favors FM coupling. 
For the half-metallic 1H-Cr$X$I (\,$X=\mathrm{S,\,Se}$; see Suppl. Figs. \ref{sfig:bands_1h-crsi}-\ref{sfig:bands_1h-crsei}) the same Cr-$X(Y)$-Cr bridges now operate in the double-exchange regime \cite{Zener1951-II, Vergniory.prb2014, Liao.prb2023}, giving the enhanced FM $J$ (as compared to 1H-\CXY\ with $Y=$ Cl or Br that are gapped),  thanks to the free carriers.

Previous studies have shown that the FM NN exchange is the dominant interaction in 1T-\CXY{} Janus MLs for $X=$ S, Se~\cite{Hou.npjcm2022, Li.prb2023}. However, the $X=$ Te systems are prone to magnetic frustration due to a sizable third-neighbor interaction ~\cite{Hou.npjcm2022, Li.prb2023}, while in the 1T selenides and sulfides this effect is found to be weak. However, we do not investigate frustration effects in CrTe$Y$, as they do not exhibit a clear tendency toward perpendicular magnetic anisotropy (see Sec. \ref{sec:mae}): since identifying systems with an out-of-plane easy axis, and therefore potentially high Curie temperatures, is the primary objective of this study, we focus exclusively on candidates that meet this criterion.

The strong drop of $J$ found for 1T-CrTeCl (Fig.~\ref{fig:j}a) is in agreement with previous reports in Refs. \cite{Hou.npjcm2022, Li.prb2023}. Our analysis of the projected densities of states reveals that this drop correlates to the decrease of the strength of hybridization between Cr-$3d$ and ($X,Y$)-$p$-states, as shown in the Suppl. Figs. \ref{sfig:dos-overlap} and \ref{sfig:overlaps}. It should also be said that in our calculations, $J$ for 1T-CrTeCl may be also to some extent artificially suppressed because of the NN only description. Indeed, in this case, the further‐neighbor couplings are inadvertently absorbed into the effective NN term -- an effect that could be non-negligible in 1T-CrTeCl, whose $J_3$ is negative and significant in magnitude \cite{Hou.npjcm2022, Li.prb2023}.

Regarding the 1H-Cr$X$I MLs ($X$=S, Se), a recent machine learning study \cite{Mahmoodabadi.cms2023} predicted them to mostly be AFM, with 1H-CrSI as the only FM case. This stands in contrast to our direct DFT results, which consistently yield an FM ground state for all these 1H systems. It is noteworthy, however, that the machine learning model's predictions for the 1T-phase align well with our findings, as both studies identify a predominantly FM ground state (with 1T-CrSCl being the only AFM exception in the machine learning work). The observed discrepancy for the 1H-phase likely stems from inherent limitations of the data-driven approach, whose predictive accuracy is critically dependent on the scope and inherent biases of its training dataset. In contrast, the \emph{ab initio} methodology employed here self-consistently treats the charge density of each specific structural and magnetic configuration, thereby offering a more reliable determination of the magnetic coupling.

It is instructive to compare our results with the previously reported values of $J$ in other 2D van der Waals magnets. In monolayer CrI$_3$ \cite{Huang.nat2017}, the NN isotropic exchange constant $J$, obtained using the same approach as we use here, has been reported to be equal to 0.5 meV/$\mu_\mathrm{B}^2$ 
\cite{Delgado.jpm2024}. For a single septuple layer of \MBT{} \cite{Otrokov.nat2019}, the corresponding value is much lower, $J = 0.07$~meV/$\mu_\mathrm{B}^2$ \cite{Delgado.jpm2024}. In comparison, the NN exchange couplings in the \CXY{} Janus MLs are significantly larger, typically ranging from about 1.5 to 2.5~meV/$\mu_\mathrm{B}^2$, depending on the composition, structure, and $U_{\mathrm{eff}}$. These values indicate that FM interactions in \CXY\ MLs are generally stronger than those in CrI$_3$, in many cases by up to 3–5 times, and exceed those of \MBT{} by more than an order of magnitude in all cases.

The local magnetic moments on Cr atoms, $m_\text{Cr}$, range between 3.2 and 3.8~$\mu_\mathrm{B}$ (Suppl. Table~\ref{stab:magmom}), with the lower end tending to the $3d^3$ configuration and the upper end approaching $3d^4$. This reflects a crossover from Cr$^{3+}$-like to Cr$^{2+}$-like character, depending on the chemical environment ($X$, $Y$) and the degree of electronic localization, which increases with $U_{\mathrm{eff}}$. The value of $m_\text{Cr}$ shows a weak upward trend with the atomic number $Z$ of both the chalcogen and halogen atoms, as well as with increasing $U_{\mathrm{eff}}$. Also, the 1H phase tends to exhibit slightly larger local moments than the 1T phase.

It is also worth noting that sizable magnetic moments (antiparallel to the Cr moment) are induced on the chalcogen atoms $X$, as shown in Suppl. Table~\ref{stab:magmom}. Depending on the type of $X$ and the value of $U_{\mathrm{eff}}$, the induced moments $m_X$ can reach up to 0.5~$\mu_\mathrm{B}$. In contrast, the moments induced on the halogen atoms $Y$ are significantly smaller, generally remaining below 0.2~$\mu_\mathrm{B}$. The overall trends are as follows: increasing $U_{\mathrm{eff}}$ leads to larger induced moments; heavier chalcogen or halogen elements result in larger values of $m_X$ and $m_Y$, which correlates with an increase in $m_\text{Cr}$; the 1H phase typically exhibits larger induced moments than the 1T phase.

\subsubsection{Magnetic anisotropy}
\label{sec:mae}

\begin{figure*}
    \centering
    \includegraphics[scale=1.0]{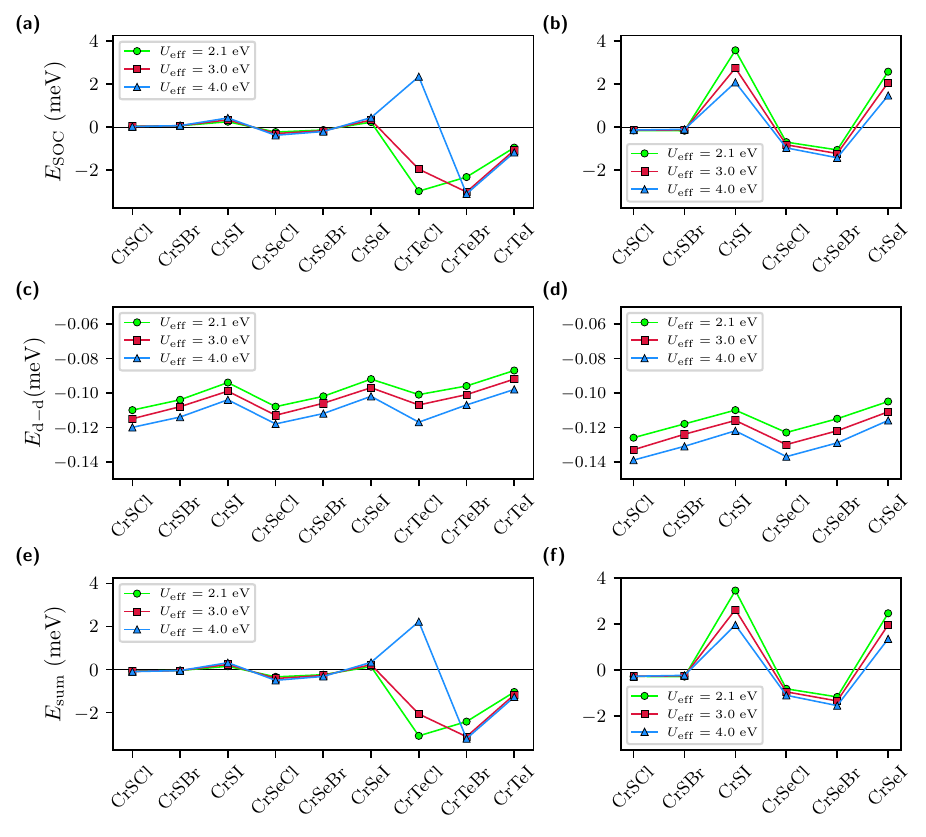}
    \caption{(a,b) Total energy difference between the states with the in-plane (ip) and out-of-plane (oop) directions of the Cr magnetic moment, $E_{\mathrm{SOC}}=E_{\mathrm{ip}} - E_{\mathrm{oop}}$ (meV per Cr atom), and (c,d) the energy of the classical dipole-dipole interaction, $E_{\mathrm{d-d}}$ (meV per Cr atom), and (e,f) $E_{\mathrm{sum}}=E_{\mathrm{SOC}}+E_\mathrm{d-d}$ (meV per Cr atom), of the \CXY\ Janus MLs ($X=$S, Se, Te and $Y=$Cl, Br, I) with the 1T (a,c,e) and 1H (b,d,f) structures, as calculated for the FM ordering. To obtain $E_{\mathrm{SOC}}$, the relativistic approximation is adopted and a non-self-consistent calculation has been performed (see Methods section for justification of the latter). The positive (negative) $E_{\mathrm{sum}}$ corresponds to the oop (ip) magnetic moment direction. All values shown on the graphs are also given in Suppl. Table \ref{stab:mae}.}
    \label{fig:mae}
\end{figure*}

We compute the magnetic anisotropy energy by accounting for both the total energy difference between in-plane and out-of-plane orientations of the Cr magnetic moment, defined as $E_{\mathrm{SOC}} = E_{\mathrm{ip}} - E_{\mathrm{oop}}$, and the contribution from the classical dipole-dipole interaction, $E_{\mathrm{d\text{-}d}}$, which inherently favors aligning magnetic moments within the film plane.

Starting with the $E_{\mathrm{SOC}}$ term, we find that regardless of the $U_{\mathrm{eff}}$ value it is positive for all 1T sulfides as well as for 1T-CrSeI (Fig.~\ref{fig:mae}a and Suppl. Table~\ref{stab:mae}). For CrSCl and CrSBr, $E_{\mathrm{SOC}}$ is about several tens of~$\mu$eV, while for CrSI and CrSeI it reaches several hundreds of~$\mu$eV. At $U_{\mathrm{eff}} = 4$~eV, 1T-CrTeCl also exhibits a positive and sizable $E_{\mathrm{SOC}}$; however, its sign reverses at lower $U_{\mathrm{eff}}$ values, which we attribute to the aforementioned metal-to-insulator transition, as the magnetocrystalline anisotropy is known to depend on the bands occupation \cite{Blanco-rey.njp2019}. Within the 1H polymorph (Fig.~\ref{fig:mae}b), only the iodides display positive $E_{\mathrm{SOC}}$ values, ranging between approximately 1.5 and 3.5~meV depending on $U_{\mathrm{eff}}$, with CrSI exhibiting stronger anisotropy than CrSeI. In contrast, all remaining 1T and 1H MLs exhibit $E_{\mathrm{SOC}} < 0$, indicating a preference for an in-plane orientation of the magnetic moments (regardless of whether the dipole-dipole contribution is included).

We now evaluate precisely the $E_{\text{d-d}}$ term for all dynamically stable systems (Fig.~\ref{fig:mae}c,d, Suppl. Sec.~\ref{ssec:hh} and Suppl. Table~\ref{stab:mae}). Its absolute value is found to lie in the range of approximately 0.09–0.14~meV across all cases and tends to be slightly larger in the 1H polymorphs of \CXY{} than in their 1T counterparts for the same composition, owing to the somewhat larger local magnetic moments $m_\text{Cr}$ and smaller lattice constants $a_0$ in the 1H phase. 

Including this dipolar contribution leads to the magnetic anisotropy energy \cite{Szunyogh.prb1995}, $E_{\text{sum}} = E_{\text{SOC}} + E_{\text{d-d}}$ (Fig.~\ref{fig:mae}e,f and Suppl. Table~\ref{stab:mae}), whose sign indicates whether the magnetic moment prefers to point out-of-plane ($E_{\text{sum}} > 0$) or becomes confined to the plane ($E_{\text{sum}} < 0$). In the 1T-CrSCl and 1T-CrSBr cases, this in-plane scenario is realized, which is precisely due to the inclusion of $E_{\text{d-d}}$, outweighing the positive contribution from $E_{\mathrm{SOC}}$. Missing the dipolar contribution has led to categorizing these systems as having an out-of-plane easy axis \cite{Hou.npjcm2022, Guo.apl2022}.

Ultimately, accounting for the dipole–dipole term results in only CrSI and CrSeI (both in the 1T and 1H phases) exhibiting $E_{\text{sum}} > 0$, indicative of an out-of-plane easy axis. This anisotropy is particularly robust in the 1H polymorph, where its magnitude is barely affected by the dipolar contribution. These four systems thus emerge as promising candidates for sustaining stable FM order at finite temperatures, as they exhibit strong FM coupling and robust PMA across the entire physically relevant range of $U_{\mathrm{eff}}$ (2.1–4.0~eV) for Cr. In contrast, the magnetic anisotropy in 1T-CrTeCl shows strong sensitivity to the choice of $U_{\mathrm{eff}}$, whose exact value is unknown, preventing a definitive prediction in this case. 
This "negative" result is inherently informative, as it underscores the critical sensitivity of magnetic anisotropy to electronic correlation and localization effects, while defining the predictive boundaries of the DFT+$U$ framework in this regime. Should the experimental realization of the 1T-CrTeCl system be achieved, magnetization measurements would be instrumental in definitively resolving the magnetic moment orientation and validating these theoretical insights.

Concerning the systems with $E_{\text{sum}} < 0$, the total energy difference between the cases where $\mathbf{m}_\text{Cr}$ points along two representative in-plane directions $x$ (towards NN) and $y$ (towards NNN) is, in most of these \CXY{} MLs, of the order of a few~$\mu$eV per Cr atom according to our calculations. This makes these two in-plane directions hardly distinguishable within the accuracy of the DFT calculations. This is also consistent with the detailed angle-dependent calculations performed for 1T-CrSeBr and 1T-CrTeI in Ref.~\cite{Xiao.pccp2020}, which indicate an \emph{isotropic} behavior within the $xy$ plane. Therefore, these systems can be considered to lie on the verge of exhibiting easy-\emph{plane}-type magnetic anisotropy (if we neglect the DMI, which, in fact, is significant in these systems \cite{Li.prb2023}). 

To gain deeper insight into the origin of magnetic anisotropy in the 1T and 1H phases of CrSI and CrSeI, we decompose the spin-orbit contribution $E_{\mathrm{SOC}}$ into single-ion ($A$) and exchange anisotropy ($B$) components \cite{Lado.2dmat2017, Lobo.ncomms2024} (see Suppl. Sec.~\ref{ssec:hh})

\begin{align}
H_\text{anis} &= - A \sum_i (M_i^z)^2 - \frac{B}{2} \sum_{i,j} M_i^z M_j^z \, .
\label{eq:h_anis}
\end{align}

\begin{figure}
    \centering
    \includegraphics[width=1.0\columnwidth]{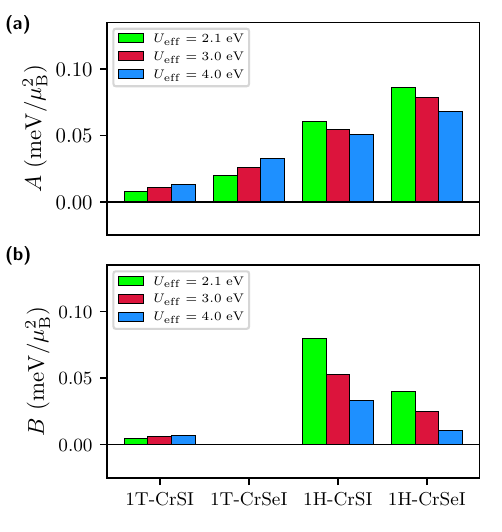}
    \caption{ (a) Single-ion $A$ (meV/$\mu_\mathrm{B}^2$)  and (b)  exchange $B$ (meV/$\mu_\mathrm{B}^2$) anisotropy parameters.}
    \label{fig:sia_ai}
\end{figure}

The results are presented in Fig.~\ref{fig:sia_ai}a,b (as well as Suppl. Table \ref{stab:sia_ai}). For 1T-CrSI, both $A$ and $B$ are positive and comparable in magnitude, with values depending on $U_{\mathrm{eff}}$ in the ranges $0.008 < A < 0.013$~meV/$\mu_\mathrm{B}^2$ and $0.005 < B < 0.007$~meV/$\mu_\mathrm{B}^2$, respectively. In contrast, for 1T-CrSeI, the anisotropic exchange contribution is vanishingly small ($B \simeq 0$~meV/$\mu_\mathrm{B}^2$), and the anisotropy is dominated by the single-ion term, with a positive $A$ ranging from approximately 0.020 to 0.033~meV/$\mu_\mathrm{B}^2$.

The 1H polymorph, in contrast, displays a pronounced enhancement of both $A$ and $B$ relative to the 1T phase. 
For 1H-CrSI, both terms increase dramatically, by factors ranging from several times to more than an order of magnitude depending on $U_{\mathrm{eff}}$. In 1H-CrSeI, $A$ also increases substantially (by a factor of 2–5), and $B$ becomes finite and non-negligible, further contributing to the total anisotropy. As a result, both $A$ and $B$ greatly support the out-of-plane easy axis in the 1H polymorph. 

The larger single-ion anisotropy constants $A$ obtained for the 1H-CrSI and 1H-CrSeI MLs are consistent with their smaller crystal-field splittings: the 1H phase averages $\Delta_{\text{CF}}\simeq 3.15$ eV, whereas the 1T phase is closer to 3.6 eV (see Suppl.\ Sec.~\ref{ssec:cf}).  
According to the simple scaling $A \propto \lambda^{2}/\Delta_{\text{CF}}$\,\cite{Lado.2dmat2017} ($\lambda$ denotes the effective SOC of Cr atom), a reduced $\Delta_{\text{CF}}$ naturally yields a larger single-ion contribution, in qualitative agreement with our DFT results.  
The same reasoning applies to the symmetric anisotropic exchange, which follows $B \propto \lambda^{2}J/\Delta_{\text{CF}}^{\,2}$\,\cite{Yosida1996}: the trends predicted by this expression match those found \emph{ab initio}.  
However, the actual enhancements of $A$ and $B$ in 1H-Cr$X$I relative to 1T are considerably larger than these formulae alone would imply.  
A detailed microscopic analysis of the additional factors required for quantitative agreement lies beyond the scope of the present study.

To further contextualize our findings, we compare the magnetic anisotropy of \CXY{} Janus MLs with that of monolayer CrI$_3$ and a septuple layer of \MBT{}. For CrI$_3$, the single-ion and exchange anisotropy constants have been reported as $A \simeq 0.001$~meV/$\mu_\mathrm{B}^2$ and $B = 0.04$~meV/$\mu_\mathrm{B}^2$, respectively, at $U_{\mathrm{eff}} = 2.1$~eV \cite{Delgado.jpm2024}. For \MBT{}, the corresponding values are $A = 0.017$~meV/$\mu_\mathrm{B}^2$ and $B = -0.002$~meV/$\mu_\mathrm{B}^2$ \cite{Delgado.jpm2024}. In comparison, while the 1T phases of CrSI and CrSeI exhibit similar or slightly lower anisotropy constants, the 1H polymorphs show significantly enhanced values of both $A$ and $B$, surpassing those of CrI$_3$ and \MBT{} by a substantial margin. Combined with the strong exchange couplings, these large anisotropies in 1H Cr$X$I MLs position them as promising candidates for stable 2D ferromagnets with high Curie temperatures.

\subsubsection{Dzyaloshinskii–Moriya interaction}

The DMI term of the spin Hamiltonian reads as
\begin{align}
H_\text{DMI} &=  - \frac{1}{2}\sum_{i,j} \mathbf{D}_{ij} \cdot (\mathbf{M}_i \times \mathbf{M}_j) \, ,
\end{align}
where $\mathbf{D}_{ij}$ is the DMI vector. According to Moriya’s symmetry rules \cite{Moriya.pr1960}, mirror planes passing through the midpoints of the lines connecting neighboring Cr atoms in \CXY{} MLs impose strict constraints on the direction of the DMI vector $\mathbf{D}_{ij}$, requiring it to be perpendicular to those lines (Fig. \ref{fig:struc}f). The DMI vector between a pair of sites $i$ and $j$ can be expressed as
\begin{align}
\mathbf{D}_{ij} = d^{\parallel} (\hat{\mathbf{z}} \times \hat{\mathbf{u}}_{ij}) + d_{ij,z} \hat{\mathbf{z}},
\end{align}
where $d^{\parallel}$ and $d_{ij,z}$ represent the in-plane and out-of-plane components, respectively. Here, $\hat{\mathbf{u}}_{ij}$ is the unit vector pointing from site $i$ to $j$, and $\hat{\mathbf{z}}$ is the unit vector along the out-of-plane direction. In the case of \CXY{} Janus MLs, only the in-plane component $d^{\parallel}$ contributes effectively to the DMI, while the out-of-plane component $d_{ij,z}$ cancels out, as discussed in Refs.~\cite{Yang.prl2015, LiangJ.prb2020, Caglayan.prb2024} and shown in Suppl. Sec.~\ref{ssec:dmi}.

To quantify the DMI strength, we employ the chirality-resolved energy difference method~\cite{Yang.prl2015, LiangJ.prb2020, Caglayan.prb2024}, which enables the extraction of the in-plane DMI component $d^{\parallel}_\text{NN}$ corresponding to nearest-neighbor interactions (next-nearest-neighbor component $d^{\parallel}_\text{NNN}$ is an order of magnitude smaller in 1T-CrSeI ML~\cite{Li.prb2023}):
\begin{align}
d^{\parallel}_\text{NN} &= \frac{\Delta E^{\text{DMI}}}{12 m_\text{Cr}^2},
\end{align}
where $\Delta E^{\text{DMI}} = E^{\text{CW}} - E^{\text{ACW}}$ denotes the total energy difference between clockwise (CW) and anticlockwise (ACW) spin configurations (Fig. \ref{fig:struc}e), as obtained from DFT calculations. The derivation of the above expression is provided in Suppl. Sec.~\ref{ssec:dmi}. A negative (positive) value of $d^{\parallel}_\text{NN}$ corresponds to a preferred CW (ACW) spin chirality.

\begin{figure}
    \centering
    \includegraphics[width=1.0\columnwidth]{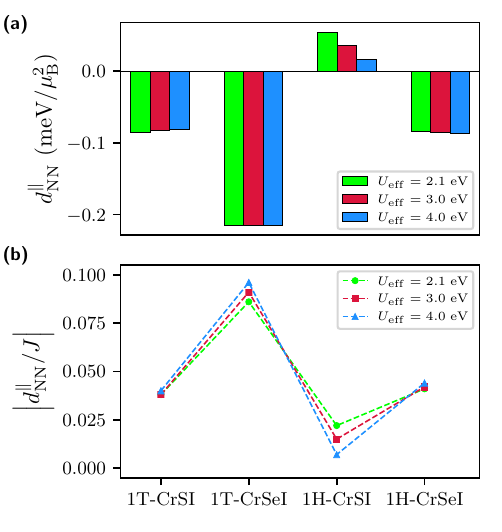}
	\caption{(a) The calculated in-plane DMI parameters $d^{\parallel}_\text{NN}$ (meV/$\mu_\mathrm{B}^2$) for the Cr$X$I MLs. (b) The $|d^{\parallel}_\text{NN}/J|$ ratio. The values shown on this graph are also given in Suppl. Table. \ref{stab:dmi_mCr}.}
    \label{fig:dmi}
\end{figure}

We apply this approach to both 1T and 1H Cr$X$I MLs ($X = \mathrm{S}, \mathrm{Se}$). As shown in Fig.~\ref{fig:dmi}a, this procedure results in relatively robust (with respect to the $U_\mathrm{eff}$ variation) values of $d^{\parallel}_\text{NN}$ in 1T-CrSI, 1H-CrSeI and 1T-CrSeI, with the former two being aproximately equal to 0.085 ~meV/$\mu_\mathrm{B}^2$, while the latter reaching much greater values around 0.22~meV/$\mu_\mathrm{B}^2$. In 1H-CrSI, on the other hand, the values of $d^{\parallel}_\text{NN}$ are much smaller and more sensitive to $U_{\mathrm{eff}}$, falling under 0.02~meV/$\mu_\mathrm{B}^2$ for $U_{\mathrm{eff}} = 4$ eV. 

Compared to the isotropic exchange coupling $J$ in these materials their DMI is rather weak, as the  $|d^{\parallel}_\text{NN} / J|$ ratios mostly lie below 0.045 (Fig.~\ref{fig:dmi}b). This finding is consistent with the observed dominance of exchange splitting over spin–orbit splitting in the electronic structure of the Cr$X$I MLs, shown in Suppl. Figs.~\ref{sfig:bands_1t-crsi}–\ref{sfig:bands_1h-crsei}. Only 1T-CrSeI reaches a DMI-to-exchange ratio of approximately 0.09–0.1, which is just above the threshold where skyrmion formation is typically considered favorable~\cite{Fert.natn2013, Hou.npjcm2022, Li.prb2023, Caglayan.prb2024}.

\begin{figure*}
    \centering
    \includegraphics[width=1.0\textwidth]{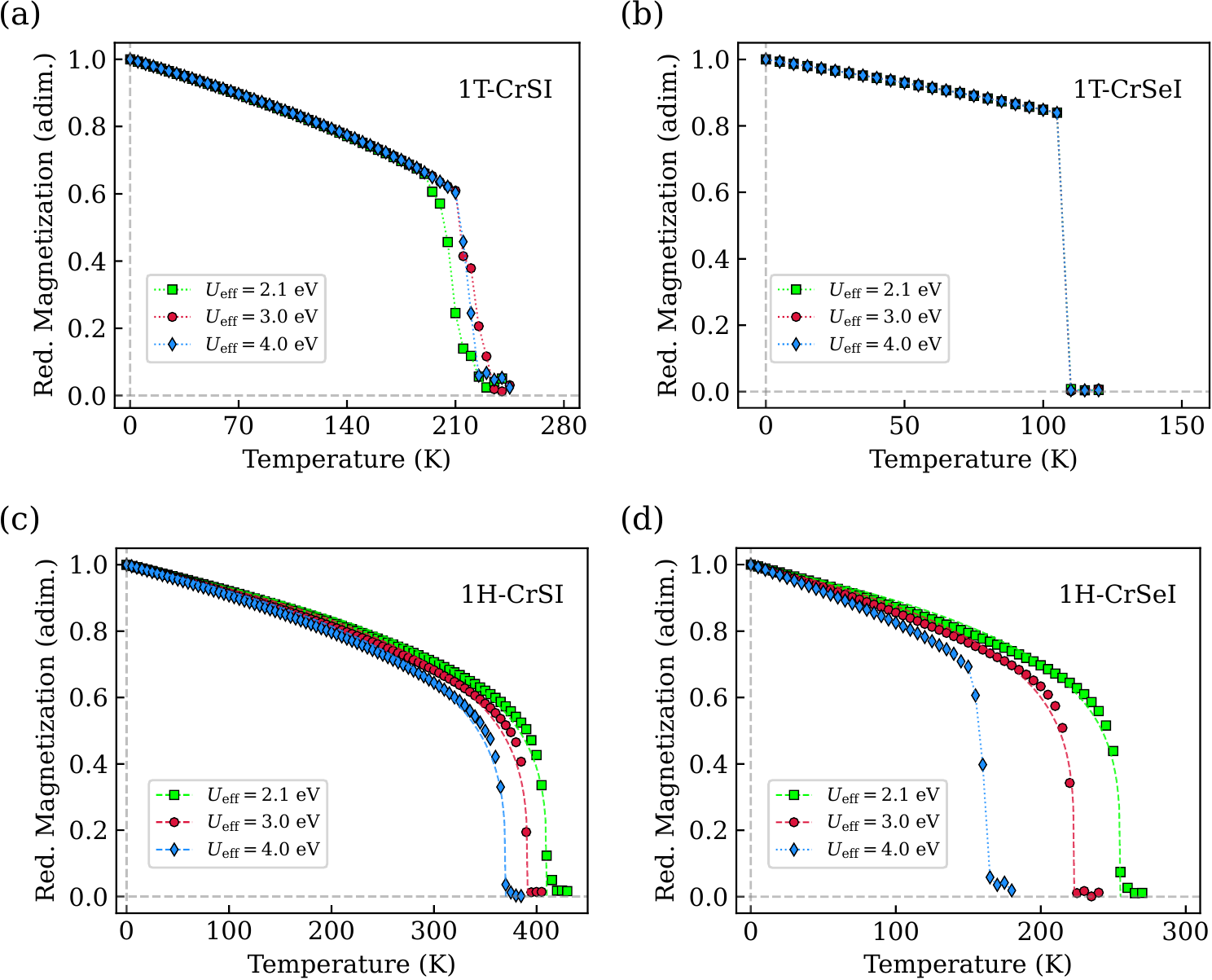}
    \caption{Temperature dependence of the reduced magnetization for (a) 1T-CrSI, (b) 1T-CrSeI, (c) 1H-CrSI, and (d) 1H-CrSeI. Symbols correspond to LLG simulation data-points, dotted lines serve as visual guides, and dashed lines represent fits to the Curie-Bloch equation where applicable.}
    \label{fig:tc}
\end{figure*}

\subsubsection{Temperature dependent magnetic properties}
\label{sec:asd}

To estimate the Curie temperature, $T_{\text{C}}$, of monolayer 1T- and 1H-Cr$X$I ($X = \mathrm{S}, \mathrm{Se}$), we employed atomistic spin dynamics (ASD) simulations to compute the temperature-dependent equilibrium magnetization, presented in Fig.~\ref{fig:tc}. 
As the isotropic exchange coupling beyond NNs could potentially play a non-negligible role \cite{Xie.cm2023, Li.prb2023}, we included $J_1$, $J_2$ and $J_3$ parameters for all these systems; their values are provided in Suppl. Table \ref{stab:j_third}. The model further includes nearest-neighbor anisotropic exchange $B$, single-ion anisotropy $A$, and NN DMI $d^{\parallel}_\text{NN}$. While crucial for the detrmination of magnetic anisotropy in 2D magnets, the dipolar contribution appears to have a negligible effect on the Curie temperature estimation \cite{jenkins_breaking_2022}. The results of our test simulations corroborate those \cite{jenkins_breaking_2022} and also reveal no qualitative changes in the simulated magnetic textures upon the inclusion of the dipole-dipole interaction (Suppl. Sec. \ref{sssec:asd_dd}). The full set of parameters used for the ASD simulations is listed in Suppl. Table~\ref{stab:asd_parameters}. For each temperature, the simulations were initialized from a fully ordered FM state with all spins aligned along the $z$-axis, consistent with the positive magnetic anisotropy energy identified in these MLs. Further methodological details are given in Suppl. Sec.~\ref{sssec:asd_mT}. 

The investigated MLs exhibit markedly distinct magnetic behavior. In the case of 1T-CrSI (Fig. \ref{fig:tc}a), the global FM alignment deteriorates near 210 K, a transition that is independent of the $U_{\mathrm{eff}}$ parameter and is characterized by the formation of FM domains (Supp. Fig. \ref{sfig:stextures_ht}a). This contrasts with 1T-CrSeI (Fig. \ref{fig:tc}b), where the out-of-plane FM configuration collapses even more abruptly at approximately 120 K, indicating that the DMI may stabilize a non-FM ground state (Fig. \ref{sfig:stextures_ht}b). Conversely, the 1H-CrSI monolayer sustains a uniform FM alignment up to its Curie temperature (Fig. \ref{fig:tc}c). We determine $T_{\text{C}}$ for this system to lie within the 370–410 K range, a value which depends on the chosen $U_{\mathrm{eff}}$ parameter. Notably, this places its magnetic ordering well above room temperature and surpasses the $T_{\text{C}}$ of any 2D magnet verified experimentally to date. The robustness of this ordering is further underscored by the negligible impact of the relatively weak DMI on the calculated $T_{\text{C}}$ (see Suppl. Fig.~\ref{sfig:asd_mT_noDMI}c). Finally, for 1H-CrSeI, the calculated $T_{\text{C}}$ values are 260 K and 235 K for $U_{\mathrm{eff}}$ values of 2.1 eV and 3.0 eV, respectively (Fig. \ref{fig:tc}d). However, at $U_{\mathrm{eff}} = 4.0$ eV, the magnetization along the $z$-axis collapses early, coinciding with FM domain formation (Suppl. Fig. \ref{sfig:stextures_ht}c) and precluding a reliable $T_{\text{C}}$ extraction via a Bloch fit. 

As noted previously, the collapse of magnetization in 1T-Cr$X$I at elevated temperatures (Figs. \ref{fig:tc}a,b) may indicate a complex, non-FM ground state. To investigate this possibility, we next examine the ground state magnetic configurations of the 1T and 1H structures in detail. We conducted ASD simulations starting from fully random spin orientations in the paramagnetic state and temperatures above $T_\text{C}$, and gradually cooled the system down to 0 K. Further information is provided in Suppl. Sec.~\ref{sssec:asd_textures}.

For 1T-CrSeI, simulations at all considered values of $U_{\mathrm{eff}}$ consistently yield worm-like domain (WD) magnetic configurations, as shown in Suppl. Fig. \ref{sfig:textures}b. In contrast, the 1H polymorph of CrSeI and both polymorphs of CrSI exhibit a competition between two outcomes: the formation of FM domains and a uniform FM state (Suppl. Figs. \ref{sfig:textures}c,d), with the final result depending on the initial conditions of the simulation and on the system size (Suppl. Figs. \ref{sfig:stextures_ss})).

The 0 K ground state was identified by computing the energy difference between the different magnetic textures obtained from the simulations and the fully ordered FM state,  $\Delta E_0 = E_0^{\mathrm{Sim.}} - E_0^{\mathrm{FM}}$. The results, compiled in Suppl. Table \ref{stab:gs}, confirm that the WD texture is the ground state for 1T-CrSeI. For 1T-CrSI and 1H-CrSeI, the near-zero $\Delta E_0$ values indicate a competition between a uniform FM state and FM domain formation. In 1T-CrSI (1H-CrSeI), the energy difference decreases with decreasing (increasing) $U_\mathrm{eff}$, which reflect intensification of this competition and explains the premature drop in magnetization observed for $U_\mathrm{eff} = 2.1$~eV ($U_\mathrm{eff} = 4.0$~eV). Finally, 1H-CrSI exhibits a stable preference for the uniform FM ground state, consistent with the almost negligible influence of DMI on its magnetic structure.

As previously noted, the substantial $|D_{ij}^\parallel|/J_1$ ratio in 1T-CrSeI suggests a capacity for skyrmion stabilization~\cite{Fert.natn2013, Hou.npjcm2022, Li.prb2023, Caglayan.prb2024}. Consistent with this expectation, Suppl.~Fig.~\ref{sfig:skyrmions} demonstrates the emergence of a skyrmion lattice during field-cooling under a constant magnetic field of $B_z = 0.5$~T.

\begin{figure*}
    \centering
    \includegraphics[width=\textwidth]{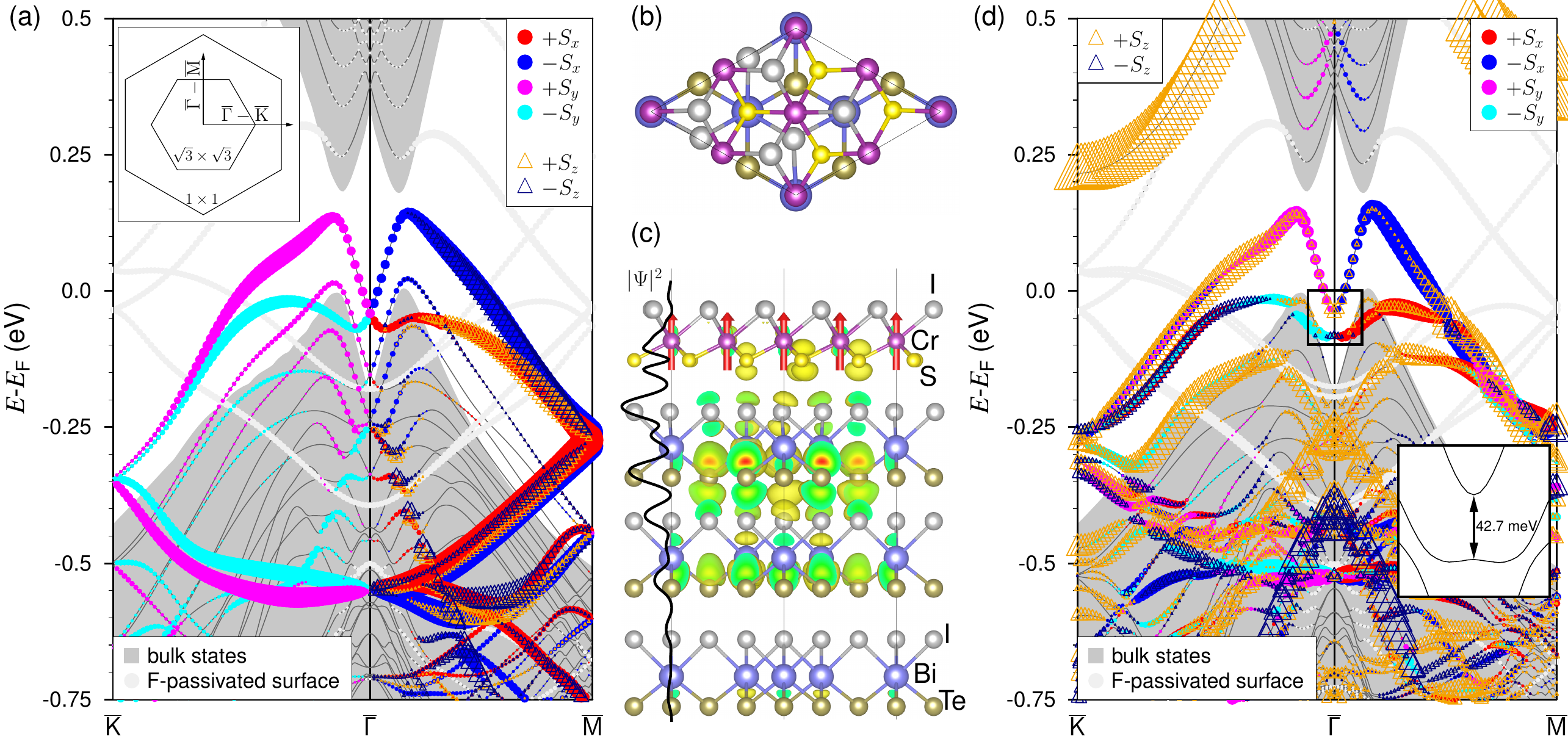}
	\caption{(a) Band structure of the I-terminated BiTeI(0001) surface calculated within the surface Brillouin zone (BZ) of the $(\sqrt{3}\times\sqrt{3})$ plane cell. Red/magenta and blue/cyan circles respectively mark positive and negative $S_x$/$S_y$ projections of spin of the surface-localized states, while orange and dark blue triangles mark positive and negative $S_z$. The inset to panel (a) shows $\bar\Gamma-\bar{\mathrm M}$ and $\bar\Gamma-\bar{\mathrm K}$ directions of the $\sqrt{3}\times\sqrt{3}$ BZ as well as the original $1\times1$ BZ of BiTeI(0001). Darker and lighter gray colors mark the bulk continuum states and surface states localized on the opposite, fluorine-passivated side of the slab, respectively. (b,c) Top (b) and side (c) views of the CrSI/BiTeI-($2/\sqrt{3}\times2/\sqrt{3}$)  structure with the I-S interface. On panel (c), spin magnetic moments of Cr atoms and spatial localization of the gapped Bychkov-Rashba state are also shown. On the left side of panel (c), the black curve shows the Bychkov-Rashba state charge density ($|\Psi|^2(z)$), integrated over $xy$ plane.  (d) Surface band structure of 1T-CrSI/BiTeI-($2/\sqrt{3}\times2/\sqrt{3}$) for the interface structure shown in (b,c). Like in panel (a), red/magenta and blue/cyan circles show $\pm S_\|$ of the surface states; orange and dark blue triangles mark positive and negative $S_z$ spin components, respectively. The inset to panel (d) demonstrates a magnified view of the gapped Bychkov-Rashba split surface state.}
    \label{fig:interface}
\end{figure*}

Let us now compare the $T_\text{C}$s estimated here to those of other 2D magnetic systems. An exfoliated monolayer of CrI$_3$ exhibits an experimentally measured $T_\text{C}$ of 45~K \cite{Huang.nat2017}, while a single septuple layer of \MBT{}, grown by molecular beam epitaxy, shows a $T_\text{C}$ of approximately 15~K \cite{Yang.prx2021, Kagerer.prr2023}. 
As compared to Cr$X$I, these striking differences originate from the combination of strong exchange interactions and robust anisotropy in Cr$X$I Janus MLs. In both 1T and 1H phases of Cr$X$I, the exchange coupling $J$ is consistently 4–5 times stronger than in CrI$_3$. Regarding the anisotropy, the 1T (1H) phases feature single-ion constants $A$ approximately one (nearly two) orders of magnitude larger than those of CrI$_3$, with anisotropic exchange values $B$ that are negligible or smaller than (comparable to or exceeding) those of CrI$_3$, resulting in Curie temperatures {of 4.5-9 times higher. In comparison to \MBT{}, both $J$ and $A,B$ parameters in Cr$X$I MLs are significanlty larger (except for $A$ in the 1T cases, which is somewhat lower than that in \MBT), fully accounting for their substantially enhanced Curie temperatures.

\subsection{Functional interfaces}

\subsubsection{$\mathrm{CrSI/BiTeI}$}

To further substantiate the potential of the here proposed FM Janus MLs with high Curie temperature, we consider the 1T-CrSI/BiTeI interfaces. While in the 1T-CrSI the exchange splitting clearly dominates over the Bychkov-Rashba splitting, the polar semiconductor BiTeI features surface states with gigantic Bychkov-Rashba effect \cite{Ishizaka.nmat2011, Eremeev.prl2012, Eremeev.jetpl2012,Maass2016}. Bringing the two in contact should break the time-reversal symmetry at the BiTeI surface whereby an exchange gap may be expected to open at the crossing point of the two Bychkov-Rashba-split bands \cite{Klimovskikh.srep2017}, which in combination with an $s$-wave superconductor might create a promising platform for pursuing Majorana fermions \cite{Sau.prl2010}. The 1T-CrSI/BiTeI pair benefits from the fact that the in-plane parameters of their ($2\times2$) and ($\sqrt{3}\times\sqrt{3}$) superlattices, respectively, match well, ensuring such an interface can be efficiently treated within DFT.

Surface of BiTeI can be either I or Te terminated or can possess a mixed stacking-fault-induced domains of Te and I terminations \cite{Fiedler2015}. The Te termination supports an electron-like 2D Bychkov-Rashba-split band originating from trapping of conduction band states by the negative surface potential while the hole-like Bychkov-Rashba state resides on the I-terminated surface near the bulk valence band \cite{Ishizaka.nmat2011, Eremeev.prl2012, Eremeev.jetpl2012,Maass2016}. 

When considering possible 1T-CrSI/BiTeI interfaces, both terminations of the BiTeI substrate as well as the orientation of the Janus material -- by iodine or sulfur layer towards the substrate -- should be taken into account. Among the possible variants of interfaces, those that have a halogen--chalcogen contact on the interface plane look realistic, since in this case the alternation of layers is similar to that in the substrate. Thus, we have considered the Te-I interface at the Te-terminated BiTeI(0001) and the I-S interface at the iodine termination of the surface. 

The electronic band structure of the I-terminated BiTeI(0001)-($\sqrt{3}\times\sqrt{3}$) (Fig.~\ref{fig:interface}a) demonstrates the  hole-like Bychkov-Rashba-split surface state as well as additional quantum well subbands, well-known for the whole BiTe$Y$ series ($Y=$ Cl, Br, I) and caused by a deep-penetrating step-like surface potential \cite{Eremeev.prl2012,Eremeev2013}. The folding of the surface state due to the imposed ($\sqrt{3}\times\sqrt{3}$) periodicity can be seen. The spin texture of the surface state demonstrates a Bychkov-Rashba behavior with the in-plane spins $S_\|$ strictly perpendicular to the wave vector: at the $\bar\Gamma-\bar{\mathrm K}$ direction ($k_x$) the spin aligns along $y$ while it points along $x$ for the $\bar\Gamma-\bar{\mathrm M}$ ($k_y$) direction. Note that the $30^\circ$ rotation of the $\sqrt{3}\times\sqrt{3}$ BZ with respect to the $1\times 1$ BZ (see Fig.~\ref{fig:interface}a, inset) leads to the fact the zero $S_z$ (nonzero $S_z$) appears along $\bar\Gamma-\bar{\mathrm K}$ ($\bar\Gamma-\bar{\mathrm M}$) (see Suppl. Fig. ~\ref{sfig:spin_texture} as well), while the opposite holds for the $1\times 1$ BZ \cite{Takayama_PRL2011}.

When brought into contact, 1T-CrSI and BiTeI form an equilibrium interface of the van der Waals type with a distance between the Janus ML and the substrate topmost layer of $\approx3$~\AA, and only a small corrugation of the iodine and sulfur interface layers arising due to the difference in the periods of the superlattices of the adlayer and the substrate (Fig.~\ref{fig:interface}b,c).

Figure~\ref{fig:interface}d shows the band structure of the S–I interface. The band gap of the adlayer turns out to be well aligned with the bulk gap of the substrate, which can be seen by referencing the lowest conduction band of CrSI, which disperses downward from the $\overline{\Gamma}$ to the $\overline{\mathrm{K}}$ point and carries a purely $+S_z$ spin component (cf. Suppl. Fig.~\ref{sfig:bands_1t-crsi}). At the same time, the topmost occupied state of the adlayer strongly hybridizes with the Bychkov–Rashba-split surface state of the substrate. This hybridization causes the latter state, originally localized predominantly within the top trilayer of the pristine substrate, to penetrate into the magnetic adlayer and exhibit significant weight near the sulfur layer, which has sizable induced magnetization (Suppl. Table \ref{stab:magmom}), as well as on the Cr layer (see the spatial distribution of the state and $|\Psi|^2(z)$ in Fig.~\ref{fig:interface}c).

Overall, the weight of this hybridized state within the magnetic trilayer is comparable to its localization in the second trilayer of the substrate. As a result of the hybridization between the Bychkov–Rashba-split surface state of BiTeI and the orbitals of the magnetic 1T-CrSI adlayer at the S–I interface, the degeneracy of the surface state at the BZ center is lifted, leading to the opening of a significant exchange gap of approximately $43$ meV (Fig.~\ref{fig:interface}c, inset). In addition to the gap opening, the hybridization with Janus ML orbitals also induces a noticeable modification of the spin texture of the Bychkov–Rashba-split surface state. This manifests itself not only in the emergence of a finite $S_z$ spin component around the $\overline{\Gamma}$ point and at larger momenta, but also in the breakdown of the conventional in-plane spin–momentum locking: the $S_x$ and $S_y$ spin components are no longer strictly perpendicular to the momentum (see Suppl. Fig. ~\ref{sfig:spin_texture} as well). We tentatively attribute this complex spin texture to the superposition of two different structural periodicities, which modulate the spin orientation in a non-trivial way.

In contrast, at the Te–I interface, where the electron-like Bychkov–Rashba-split surface state of the BiTeI conduction band lies far from the 1T-CrSI ML states (both in energy and momentum), no appreciable hybridization is observed, and the resulting exchange gap is only $\sim 1$ meV (not shown).

\subsubsection{$\mathrm{CrSI}/\mathrm{MnBi}_2\mathrm{Te}_4$}

\begin{figure*}
    \centering
    \includegraphics[scale=1.0]{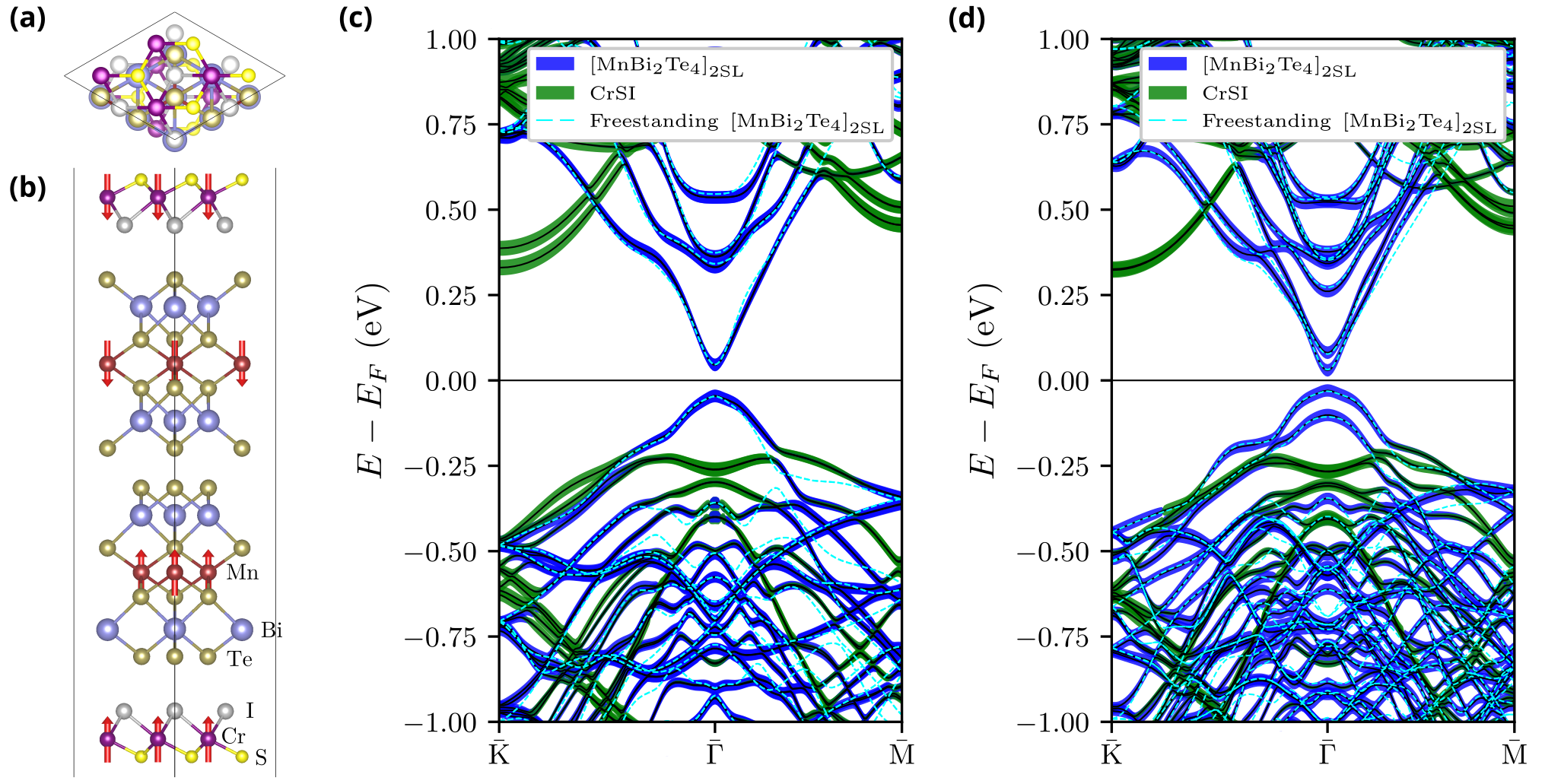}
    \caption{(a) Top and (b) side views of the CrSI/[\MBT]${}_{2\mathrm{SL}}$/CrSI sandwich heterostructure, with the Te-I interfaces. Red arrows indicate the atomic spin directions in the magnetic ground state. (c) Band structure of the CrSI/[\MBT]${}_{2\mathrm{SL}}$/CrSI sandwich, calculated for the crystal and magnetic structures shown in panel (b). Fat bands highlight the [\MBT]$_{2\mathrm{SL}}$ (blue) and CrSI (green) contributions to the band structure. Cyan dashed lines show the band structure of the free-standing [\MBT]${}_{2\mathrm{SL}}$ film. (d) Same as in panel (c), but for the forced FM configuration, where the magnetizations of all four (two) magnetic layers of CrSI/[\MBT]${}_{2\mathrm{SL}}$/CrSI (free-standing [\MBT]${}_{2\mathrm{SL}}$) are aligned parallel.}
    \label{fig:mbt_crsi_interface}
\end{figure*}

Another class of systems whose properties could be enhanced by interfacing with the Janus \CXY\ MLs studied here are thin films of intrinsic magnetic topological insulators of the \MBT\ family~\cite{Otrokov.2dmat2017, Otrokov.nat2019, Otrokov.prl2019}. These materials exhibit a wide range of exotic phenomena~\cite{Liu.nmat2020, Ge.nsr2019, Deng.sci2020, Gao.nat2021, Gao.sci2023, Bosnar.npj2d2023, Qiu.nmat2023, Qiu.nat2025, Lian.nat2025} and have recently attracted significant attention~\cite{Vyazovskaya.commsmat2025}. Their magnetic critical temperatures, around 25 K for \MBT, are already quite remarkable for magnetic topological insulators, yet further enhancement remains an important goal. Here, using the 1T-CrSI/\MBT\ heterostructure as a representative example, we illustrate how such enhancement could potentially be achieved in \MBT\ thin films in a "non-invasive" manner -- that is, without compromising the remarkable properties of \MBT. Favorably, the lattice parameter of the $(\sqrt{3} \times \sqrt{3})$ supercell of \MBT\ closely matches the one of the $(2 \times 2)$ supercell of 1T-CrSI.

The starting point of our reasoning is the behavior of \MBT's critical temperature as a function of film thickness when transitioning from the 2D to the 3D regime. Upon increasing the thickness from a single to a double septuple layer (SL), the critical temperature of $\sim$10 K roughly doubles, according to Monte Carlo simulations~\cite{Otrokov.prl2019}, and then steadily increases with the addition of further SLs, approaching the bulk value of 25 K~\cite{Deng.sci2020, Yang.prx2021, Zhao.nl2021}. Comparing this behavior to that of a hypothetical CrSI/[\MBT]$_{\mathrm{1SL}}$ heterostructure, one may expect the critical temperature of the latter to rise above that of the [\MBT]$_{\mathrm{2SL}}$ film due to two main factors. First, the 1T-CrSI ML exhibits a much higher predicted $T_\mathrm{c}$ than \MBT\ and, second, a stronger \emph{interlayer} exchange coupling may also be expected, since the Cr and Mn layers are closer than the adjacent Mn layers in [\MBT]$_{\mathrm{2SL}}$.

First, we have considered 1T-CrSI/[\MBT]$_{\mathrm{1SL}}$ heterostructures with both S and I atoms located at the interface. In both cases, the bonding between CrSI and \MBT\ is of van der Waals type, with interlayer separations of approximately $3\ \mathrm{\AA}$ and $3.5\ \mathrm{\AA}$ for the Te–S and Te–I interfaces, respectively. The total energy calculations indicate that the Te–S interface is energetically favored over its Te–I counterpart. However, the Te–S interface exhibits significant charge transfer, resulting in a globally metallic electronic structure. In contrast, the Te–I interface shows no noticeable charge transfer, owing to the larger interlayer separation, and the corresponding 1T-CrSI/[\MBT]$_{\mathrm{1SL}}$ band structure remains globally gapped. Therefore, we focus on the Te–I interface in what follows, assuming that the fabrication of such structures could be feasible with suitable technology, which is plausible given the van der Waals nature of both constituents.

The interlayer exchange coupling between CrSI and \MBT\ turns out to be FM and, despite the relatively large van der Waals gap, is quite sizable: the FM state is favored by  $\sim 2.04$ meV per pair of magnetic atoms compared to the AFM state. This value is about twice as large as the corresponding energy difference in [\MBT]$_{\mathrm{2SL}}$~\cite{Otrokov.prl2019}. We have also confirmed that the heterostructure exhibits a positive magnetic anisotropy energy $E_\mathrm{sum}$, which amounts to $0.16$ meV per magnetic atom. Thus, one may indeed expect a significant enhancement of the magnetic critical temperature in CrSI/[\MBT]$_\mathrm{1SL}$ compared to [\MBT]$_\mathrm{2SL}$.

Next, we explore the effect of the CrSI Janus ML on the electronic structure and, consequently, on the topology of 2D \MBT. A single-SL-thick film of \MBT\ is known to be topologically trivial~\cite{Otrokov.prl2019, Li.sciadv2019}, and interfacing it with a 1T-CrSI ML does not induce non-trivial topology either. We therefore consider a double-SL-thick film, for which the zero-plateau quantum anomalous Hall (ZPQAH) state in the AFM ground state and the quantum Hall (QH) state in the forced FM phase have been predicted \emph{ab initio}~\cite{Otrokov.prl2019, Li.sciadv2019, Vyazovskaya.commsmat2025}.

To maximize the enhancement of \MBT\ magnetism, we study a sandwich heterostructure, CrSI/[\MBT]$_{2\mathrm{SL}}$/CrSI (where both interfaces are Te-I), in which the upper and lower SLs are each coupled to a CrSI ML (Fig.~\ref{fig:mbt_crsi_interface}a,b). This configuration is expected to be magnetically more robust than having a single CrSI ML on one side only. Figure~\ref{fig:mbt_crsi_interface}b shows the magnetic ground state of this sandwich structure, determined via total energy calculations. It naturally respects the AFM (Mn–Mn) and FM (Mn–Cr) interlayer exchange couplings and is approximately 0.28 meV per magnetic atom pair lower in energy than the fully FM configuration.

The band structure corresponding to this magnetic ground state is shown in Fig.~\ref{fig:mbt_crsi_interface}c. As in the pristine [\MBT]$_{2\mathrm{SL}}$ case, the forced FM configuration is also physically meaningful and can be realized by applying a moderate external magnetic field to overcome the weak AFM interlayer coupling of \MBT. The corresponding electronic structure is shown in Fig.~\ref{fig:mbt_crsi_interface}d. In both magnetic configurations, the \MBT\ band gap lies within the band gap of 1T-CrSI and the bands of the sandwich heterostructure can be clearly broken down into the [\MBT]$_{2\mathrm{SL}}$ and CrSI contributions, although signatures of hybridization between the two subsystems are also seen.
Crucially, the \MBT-projected part of the band structure follows rather closely the dispersion curves of the free-standing [\MBT]$_{2\mathrm{SL}}$ film (cyan dashed lines in Fig.~\ref{fig:mbt_crsi_interface}c,d), especially as far as the valence band maximum and conduction band minimum are concerned. This suggests that the essential electronic properties of \MBT\, are not affected by CrSI. Accordingly, Wannier charge center calculations of the Chern number $C$ confirm that the system is in the $C = 1$ Chern insulator phase in the forced FM state, and in a ZPQAH state ($C = 0$) in the ground state. Thus, both the QH and ZPQAH states, characteristic of the pristine [\MBT]$_{2\mathrm{SL}}$ film, are preserved in the CrSI/[\MBT]$_{2\mathrm{SL}}$/CrSI sandwich heterostructure. However, unlike the original \MBT\ film, the magnetic ordering in the sandwich is expected to persist up to significantly higher temperatures.

Given the van der Waals nature of the bonding between CrSI and [\MBT]$_\mathrm{2SL}$, we propose that the topologically non-trivial states of thicker \MBT\ films (e.g., intrinsic QAHE in [\MBT]$_\mathrm{3SL}$) sandwiched between CrSI layers are also very likely to be preserved. However, it should be kept in mind that the most pronounced enhancement of the critical temperature is expected at low number of SLs. Thus, the proposed sandwich heterostructures offer a promising route for increasing the operational temperatures of the QAHE~\cite{Deng.sci2020, Lian.nat2025, Wang.ncomms2025, Zhang.ncomms2025}, the QH effect without Landau levels~\cite{Deng.sci2020, Liu.nmat2020, Ge.nsr2019}, axion quasiparticles~\cite{Liu.nmat2020, Qiu.nat2025} in \MBT, and a variety of other exotic phenomena~\cite{Vyazovskaya.commsmat2025}.

\section{Conclusions}

In this work, we have performed a systematic search of stable ferromagnets with robust perpendicular magnetic anisotropy and elevated Curie temperatures among the Cr$XY$ chalcohalide Janus monolayers ($X = \mathrm{S}, \mathrm{Se}, \mathrm{Te}$; $Y = \mathrm{Cl}, \mathrm{Br}, \mathrm{I}$) in both 1T and 1H polymorphs using density functional theory in combination with atomistic spin dynamics simulations. 

Our calculations reveal that the 1T structure is energetically more favorable than the 1H phase across all compositions, with an energy gain exceeding 0.55~eV per formula unit. However, since the 1H structure is dynamically stable for the sulfides and selenides, it may potentially be experimentally accessible under suitable growth conditions. We further find that all dynamically stable systems display ferromagnetic coupling, with robust exchange constants $J$ ranging from 1.5 to 2.5~meV/$\mu_\mathrm{B}^2$, substantially larger than those of such intensely studied 2D magnets as CrI$_3$ and \MBT. The magnetic anisotropy analysis, including both spin–orbit and dipole–dipole contributions, identifies CrSI and CrSeI as the only systems exhibiting out-of-plane easy axes in both 1T and 1H structures. Notably, the 1H polymorph displays particularly strong perpendicular magnetic anisotropy, with single-ion and anisotropic exchange constants generally exceeding those in CrI$_3$ and \MBT.

We next quantify the Dzyaloshinskii–Moriya interaction in Cr$X$I ($X=$ S, Se), finding it to be weak relative to the isotropic exchange in all systems except 1T-CrSeI, where the $D/J$ ratio falls within the regime favorable for skyrmion formation. Using the extracted anisotropy and exchange parameters as inputs for large-scale atomistic spin dynamics simulations, we investigate the evolution of real-space magnetic textures with temperature. Among the studied systems, 1T-CrSI and 1H-Cr$X$I ($X$ = S, Se) exhibit the desired ferromagnetic behavior with perpendicular magnetic anisotropy, yielding Curie temperatures of at least 210 K for 1T-CrSI, 235–260 K for 1H-CrSeI, and 370–410 K (well above room temperature) for 1H-CrSI. In contrast, the sizable Dzyaloshinskii–Moriya interaction in 1T-CrSeI leads to a worm-like domain ground state at zero external field and stabilizes skyrmions under a perpendicular magnetic field.

The remarkable predicted magnetic properties of 1T-CrSI and 1H-Cr$X$I ($X$ = S, Se) make them promising candidates for applications. Specifically, the semiconducting 1T-CrSI monolayer, which exhibits elevated $T_\text{C}$, emerges as compelling candidate for integration into van der Waals heterostructures with thin films of the BiTe$Y$ family of compounds ($Y = \mathrm{Cl},\ \mathrm{Br},\ \mathrm{I}$)~\cite{Ishizaka.nmat2011, Eremeev.prl2012}, or with the \MBT\ family of intrinsic topological insulators~\cite{Otrokov.nat2019, Otrokov.prl2019, Vyazovskaya.commsmat2025}. Our calculations show that, in the former case, the ferromagnetic ordering with perpendicular magnetic anisotropy in 1T-CrSI enables breaking of the time-reversal symmetry at the surface of BiTe$Y$, thereby opening an exchange gap at the crossing point of the Bychkov–Rashba-split surface state~\cite{Klimovskikh.srep2017}. When combined with an $s$-wave superconductor, such a system may provide a promising platform for realizing Majorana fermions~\cite{Sau.prl2010}. Besides, 1T-CrSI ML is also attractive candidate for forming sandwich van der Waals heterostructures with \MBT\ thin films, potentially enabling the realization of quantum anomalous Hall~\cite{Deng.sci2020} or axion insulator~\cite{Liu.nmat2020} phases with enhanced critical temperatures. As for 1H-Cr$X$I ($X = \mathrm{S},\ \mathrm{Se}$), their half-metallic band structure, dominated by spin-up states at the Fermi level, makes them promising ferromagnetic spin-injector materials for spin injection into semiconductors~\cite{Zutic.rmp2004} as well as suitable for the spin-orbit torque observation when interfaced with a heavy metal \cite{Wang.sciadv2019} or topological insulator \cite{Wang.ncomms2023}.

\section*{ACKNOWLEDGMENTS}
M.B., I.Yu.S., A.A., and M.M.O. acknowledge the support by MCIN/AEI/10.13039/501100011033/ (Grant PID2022-138210NB-I00) and "ERDF A way of making Europe". 
M.M.O. acknowledges support  by the Grant CEX2023-001286-S funded by MICIU/AEI/10.13039/501100011033, as well as MCIN with funding from European Union NextGenerationEU (PRTR-C17.I1) promoted by the Government of Aragon. M.B. acknowledges support of the Croatian Science Foundation under the project numbers HRZZ-IP-2022-10-6321 and HRZZ-MOBDOL-2023-12-6938. J.M.L., U.A., and S.G. acknowledge funding from PIE grant from CSIC (20226AT018), and the Spanish Ministerio de Ciencia, Innovación y Universidades grants (PID2021-122980OB-C55, CNS2023-144681, PID2024-157112OB-C52, CEX2024-001445-S) funded by MCIN/AEI/10.13039/501100011033 and by ERDF A way of making Europe and ESF Investing in your future. J.M.L. acknowledges financial support from a PhD grant from Comunidad de Madrid (PIFP-2023/TEC-29997). A.A. acknowledges the grant IT-1527-22 from the Department of Education, Universities and Research of the Basque Government as well as the support by MCIN/AEI/10.13039/501100011033/ (Grant PID2022-137685NB-I00) and "ERDF A way of making Europe". A.Yu.V. gratefully acknowledges the support by the Ministry of Science and Higher Education of the Russian Federation (State Task No. FSWM-2025-0009). S.V.E. and E.V.C. acknowledge Saint-Petersburg State University for a research project No 125022702939-2. S.V.E. acknowledges the support by the Government research assignment for ISPMS SB RAS (Project FWRW-2026-0008).

\clearpage 
\onecolumngrid

\renewcommand{\thefigure}{S\arabic{figure}}
\renewcommand{\thetable}{S\arabic{table}}
\renewcommand{\theequation}{S\arabic{equation}}
\setcounter{section}{0}

\begin{center}
    \textbf{\Large SUPPLEMENTARY MATERIAL}
\end{center}

\section{Phonon calculations}
\label{ssec:phonons}

\begin{figure}[h!]
    \centering    
    \includegraphics[width=\textwidth]{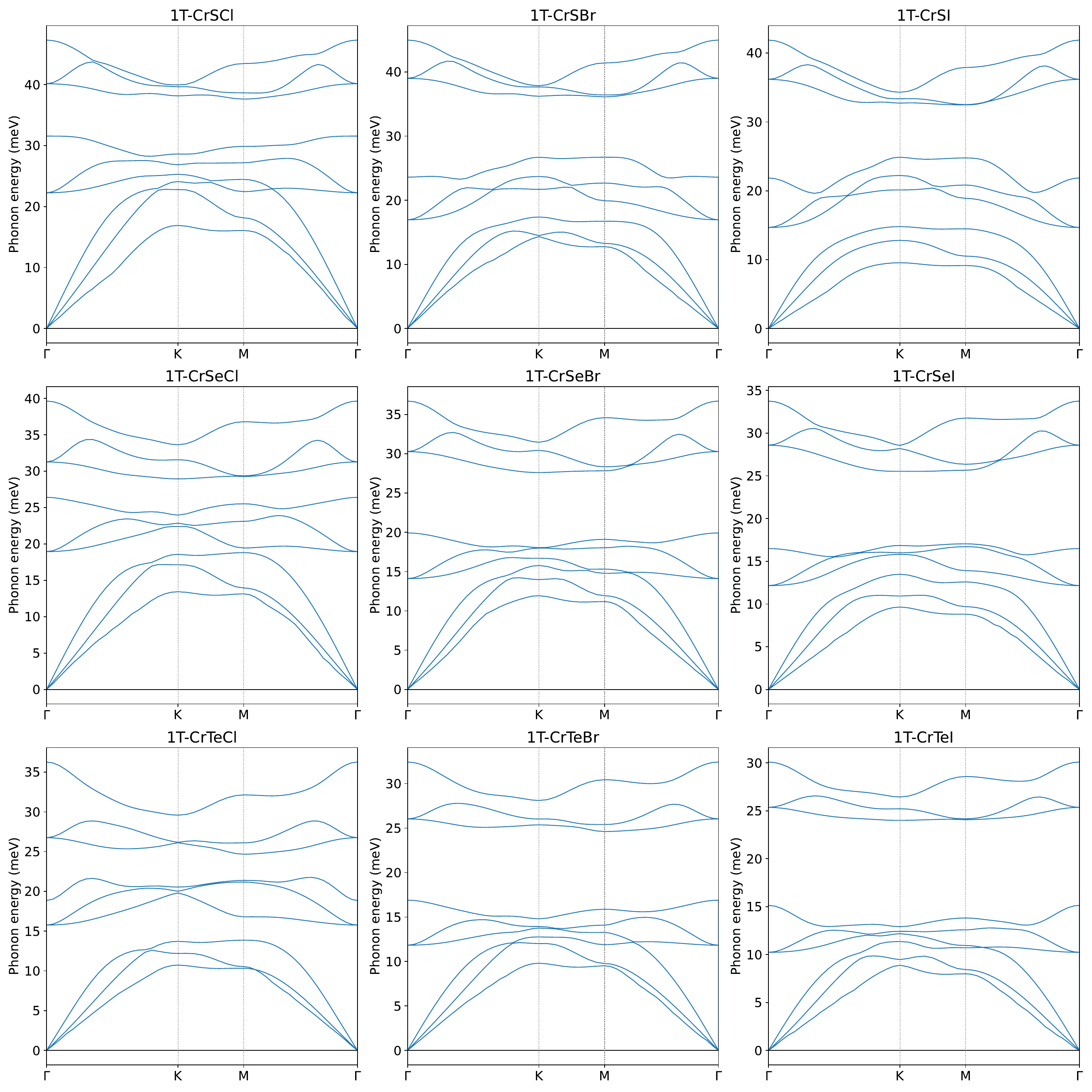}
	\caption{Phonon dispersions, calculated for the 1T-\CXY{} MLs ($X=$S, Se, Te and $Y=$Cl, Br, I) for $U_{\mathrm{eff}}=2.1$ eV.}
    \label{sfig:phonons_1t}
\end{figure}

\begin{figure}[h!]
    \centering    
    \includegraphics[width=\textwidth]{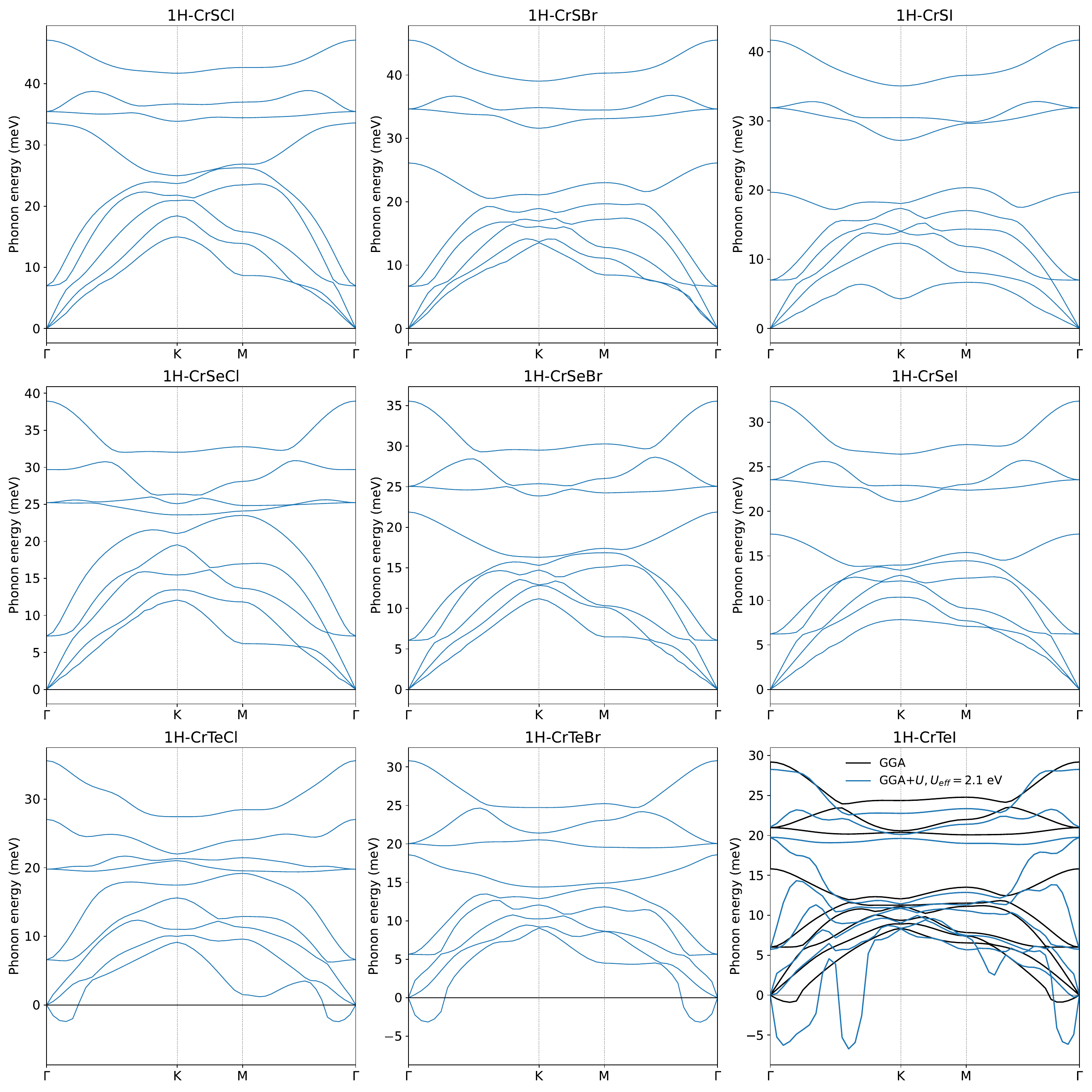}
	\caption{The same as Fig. \ref{sfig:phonons_1t}, but for the 1H-\CXY{} MLs ($X=$S, Se, Te and $Y=$Cl, Br, I). $U_{\mathrm{eff}}=2.1$ eV. For CrTeI (bottom right panel), the GGA spectrum is also shown, featuring imaginary modes, indicating dynamical instability both within GGA and GGA+$U$.}
    \label{sfig:phonons_1h}
\end{figure}

\clearpage

{\renewcommand{\arraystretch}{1.00}%
\begin{table}[!bth]
\caption{Optimized lattice constants $a_0$ (\AA) of the \CXY{} Janus MLs ($X = \mathrm{S},\ \mathrm{Se},\ \mathrm{Te}$;\ $Y = \mathrm{Cl},\ \mathrm{Br},\ \mathrm{I}$) for the 1T and 1H structures in the FM ordering case. Calculations are performed within the scalar-relativistic approximation. Data for the 1H-CrTe$Y$ systems are not shown due to their dynamical instability (see Fig.~\ref{sfig:phonons_1h}).} \label{stab:struc}
\begin{center}
\begin{tabular}%
{>{\centering\arraybackslash}p{0.50cm}%
 >{\centering\arraybackslash}p{2.00cm}%
 >{\centering\arraybackslash}p{1.50cm}
 >{\centering\arraybackslash}p{1.50cm}
 >{\centering\arraybackslash}p{1.50cm}
 >{\centering\arraybackslash}p{1.50cm}
 >{\centering\arraybackslash}p{1.50cm}
 >{\centering\arraybackslash}p{1.50cm}
 >{\centering\arraybackslash}p{1.50cm}
 >{\centering\arraybackslash}p{1.50cm}
 >{\centering\arraybackslash}p{1.50cm}}
\hline \hline
& $U_{\mathrm{eff}}$ (eV) & CrSCl & CrSBr & CrSI & CrSeCl & CrSeBr & CrSeI & CrTeCl & CrTeBr & CrTeI  \\
\hline \hline
   &2.1 &3.4738 &3.5607 &3.7135 &3.5770 &3.6642 &3.8178 &3.7449 &3.8328 &3.9798 \\
1T &3   &3.4877 &3.5784 &3.7361 &3.5928 &3.6814 &3.8377 &3.7673 &3.8519 &4.0036 \\
   &4   &3.5050 &3.5961 &3.7549 &3.6110 &3.7007 &3.8600 &3.7635 &3.8748 &4.0273 \\
\hline \hline
  &2.1&3.3770 &3.4755 &3.6343 &3.4808 &3.5807 &3.7506 &- &- &- \\
1H&3  &3.3977 &3.5003 &3.6607 &3.5022 &3.6026 &3.7772 &- &- &- \\
  &4  &3.4219 &3.5252 &3.6908 &3.5272 &3.6281 &3.8076 &- &- &- \\
\hline \hline
\end{tabular}
\noindent{{Note: For each $U_{\mathrm{eff}}$ value, the corresponding relaxed lattice constants listed here were used consistently in all subsequent calculations of magnetic and electronic properties.}}
\end{center}
\end{table}
}

\section{Raman-active modes}
\label{ssec:raman}

We have calculated the Raman-active modes for the 1T and 1H phases of CrSI and CrSeI, which are the only systems showing robust perpendicular magnetic anisotropy, as shown in the maintext Section \ref{sec:mag}. All optical phonons are  both infrared (IR) and Raman (R) active. The Raman and IR active modes are not mutually exclusive because of lack of inversion symmetry in the \CXY\ MLs. In these structures, there are four Raman active modes: two optical modes belong to $A_1$ and two to the twofold degenerate symmetry class $E$.

The results are shown in Table \ref{stab:raman}.  
For CrSeI, the lowest $E$ modes (at 6 meV for 1H and 12 meV for 1T) are characterized by the in‐phase motion of Cr and Se atoms in the $xy$ plane of the Se–Cr–I trilayer accompanied by displacements of I atoms in the opposite direction. The highest $E$ modes (23.5 meV for 1H and 28.5 meV for 1T) involve mutually perpendicular motion of only Cr and Se atoms. The two $A_1$ modes are stretching vibrations along the stacking ($z$) axis: in the lower‐energy mode the Cr–Se pair oscillates in phase against the I sublattice, whereas in the higher‐energy mode Cr and Se vibrate out of phase with each other while the I atoms remain essentially stationary.

For CrSI, we find a similar pattern: the lowest $E$ modes occur at 7 meV for 1H and 14.7 meV for 1T, while the higher $E$ modes are at 31.8 meV (1H) and 36 meV (1T). The two $A_1$ stretching modes appear at 19.7 meV and 41.6 meV in the 1H phase, and at 21.8 meV and 41.7 meV in the 1T phase.

{\renewcommand{\arraystretch}{1.00}%
\begin{table}[h]
  \centering
  \caption{Raman‐active $A_1$ and $E$ modes for CrSI and CrSeI in the 1H and 1T structures. $U_{\mathrm{eff}}=2.1$\,eV.}
  \label{stab:raman}
  \begin{tabular}%
{>{\centering\arraybackslash}p{2.5cm}
 >{\centering\arraybackslash}p{2.5cm}
 >{\centering\arraybackslash}p{2.5cm}
 >{\centering\arraybackslash}p{3.5cm}
 >{\centering\arraybackslash}p{3.5cm}}
\hline\hline
    Material & Structure & Mode & Energy (meV) & Energy (meV) \\
    \hline
    \multirow{4}{*}{CrSI}
      & \multirow{2}{*}{1T} & $A_1$ & $21.8$  & $41.7$  \\
      &                    & $E$   & $14.7$  & $36.0$  \\
      & \multirow{2}{*}{1H} & $A_1$ & $19.7$  & $41.6$  \\
      &                    & $E$   & $7.0$   & $31.8$  \\
    \hline
    \multirow{4}{*}{CrSeI}
      & \multirow{2}{*}{1T} & $A_1$ & $16.5$  & $33.5$   \\
      &                    & $E$   & $12.0$  & $28.5$   \\
      & \multirow{2}{*}{1H} & $A_1$ & $17.5$  & $32.0$   \\
      &                    & $E$   & $6.0$   & $23.5$   \\
    \hline\hline
  \end{tabular}
\end{table}
}

\clearpage

\newpage

\section{Determination of the isotropic exchange coupling parameter, single ion and exchange anisotropies for the $\mathbf{Cr}XY$ monolayers ($X=\mathbf{S,\ Se,\ Te}$ and $Y=\mathbf{Cl,\ Br,\ I}$)}
\label{ssec:hh}

{\renewcommand{\arraystretch}{1.00}%
\begin{table}[!bth]
\caption{Total energy difference of the AFM and FM configurations inside the Cr layer, $\Delta E_{\mathrm{A/F}}=E_{\mathrm{AFM}} - E_{\mathrm{FM}}$ (eV per Cr pair) of the \CXY\ Janus MLs ($X=$S, Se, Te and $Y=$Cl, Br, I) with the 1T and 1H structures. The scalar-relativistic approximation is adopted.} \label{stab:e_mag}
\begin{center}
\begin{tabular}%
{>{\centering\arraybackslash}p{0.50cm}%
 >{\centering\arraybackslash}p{1.30cm}%
 >{\centering\arraybackslash}p{1.70cm}
 >{\centering\arraybackslash}p{1.40cm}
 >{\centering\arraybackslash}p{1.40cm}
 >{\centering\arraybackslash}p{1.40cm}
 >{\centering\arraybackslash}p{1.40cm}
 >{\centering\arraybackslash}p{1.40cm}
 >{\centering\arraybackslash}p{1.40cm}
 >{\centering\arraybackslash}p{1.40cm}
 >{\centering\arraybackslash}p{1.40cm}
 >{\centering\arraybackslash}p{1.40cm}}
\hline \hline
  & $U_{\mathrm{eff}}$ (eV) & CrSCl & CrSBr & CrSI & CrSeCl & CrSeBr & CrSeI & CrTeCl & CrTeBr & CrTeI  \\
\hline \hline
  &2.1 &0.209 &0.216 &0.195 &0.203 &0.223 &0.231 &0.144 &0.187 &0.231 \\
1T&3   &0.208 &0.215 &0.198 &0.205 &0.225 &0.234 &0.139 &0.183 &0.233 \\
  &4   &0.208 &0.214 &0.200 &0.207 &0.226 &0.238 &0.113 &0.178 &0.234 \\      
  
  \hline \hline
  
  &2.1 &0.150 &0.192 &0.231 &0.127 &0.177 &0.205 &- &- &- \\
1H&3   &0.174 &0.208 &0.243 &0.152 &0.194 &0.219 &- &- &- \\
  &4   &0.193 &0.219 &0.252 &0.172 &0.210 &0.232 &- &- &- \\
\hline \hline
\end{tabular}
\end{center}
\end{table}

{\renewcommand{\arraystretch}{1.00}%
\begin{table}[!bth]
\caption{Local magnetic moment of Cr, $m_\text{Cr}$ ($\mu_\text{B}$), and induced magnetic moments on the $X$ and $Y$ atoms, $m_X$ and $m_Y$ ($\mu_\text{B}$), respectively, of the \CXY\ Janus MLs ($X=$S, Se, Te and $Y=$Cl, Br, I) with the 1T and 1H structures and for the FM ordering case. The scalar-relativistic approximation is adopted.}
\label{stab:magmom}
\begin{center}
\begin{tabular}%
{>{\centering\arraybackslash}p{0.50cm}%
 >{\centering\arraybackslash}p{1.30cm}%
 >{\centering\arraybackslash}p{1.70cm}
 >{\centering\arraybackslash}p{1.40cm}
 >{\centering\arraybackslash}p{1.40cm}
 >{\centering\arraybackslash}p{1.40cm}
 >{\centering\arraybackslash}p{1.40cm}
 >{\centering\arraybackslash}p{1.40cm}
 >{\centering\arraybackslash}p{1.40cm}
 >{\centering\arraybackslash}p{1.40cm}
 >{\centering\arraybackslash}p{1.40cm}
 >{\centering\arraybackslash}p{1.40cm}}
\hline \hline
   & & $U_{\mathrm{eff}}$ (eV)  & CrSCl  & CrSBr  & CrSI & CrSeCl & CrSeBr & CrSeI & CrTeCl & CrTeBr & CrTeI  \\
\hline \hline
   &              &2.1 &3.229  &3.250  &3.297  &3.332  &3.355  &3.400  &3.460  &3.485  &3.522  \\
   &$m_\text{Cr}$ &3   &3.315  &3.345  &3.400  &3.430  &3.459  &3.516  &3.583  &3.608  &3.654  \\
   &              &4   &3.412  &3.448  &3.523  &3.541  &3.575  &3.642  &3.745  &3.739  &3.788  \\
\hline
   &              &2.1 &-0.229 &-0.227 &-0.225 &-0.294 &-0.292 &-0.290 &-0.351 &-0.349 &-0.344 \\
1T & $m_X$        &3   &-0.263 &-0.263 &-0.263 &-0.337 &-0.336 &-0.336 &-0.403 &-0.398 &-0.393 \\
   &              &4   &-0.303 &-0.303 &-0.306 &-0.387 &-0.385 &-0.386 &-0.435 &-0.450 &-0.443 \\
\hline
   &              &2.1 &-0.034 &-0.052 &-0.088 &-0.038 &-0.056 &-0.090 &-0.040 &-0.060 &-0.090 \\
   & $m_Y$        &3   &-0.048 &-0.070 &-0.112 &-0.052 &-0.073 &-0.111 &-0.054 &-0.076 &-0.111 \\
   &              &4   &-0.064 &-0.090 &-0.136 &-0.067 &-0.092 &-0.134 &-0.064 &-0.094 &-0.131 \\
\hline \hline
 &              &2.1 &3.304  &3.337  &3.443  &3.421  &3.455  &3.527  &- &- &- \\
 &$m_\text{Cr}$ &3   &3.421  &3.464  &3.581  &3.548  &3.586  &3.666  &- &- &- \\
 &              &4   &3.547  &3.595  &3.717  &3.681  &3.720  &3.804  &- &- &- \\ 
\hline
 &              &2.1&-0.311 &-0.310 &-0.317 &-0.384 &-0.383 &-0.388 &- &- &- \\
1H&$m_\text{X }$&3  &-0.358 &-0.359 &-0.365 &-0.440 &-0.438 &-0.442 &- &- &-  \\
 &              &4  &-0.410 &-0.410 &-0.414 &-0.499 &-0.494 &-0.494 &- &- &- \\
\hline
 &              &2.1&-0.024 &-0.051 &-0.130 &-0.031 &-0.057 &-0.103 &- &- &- \\
 &$m_\text{Y }$ &3  &-0.046 &-0.079 &-0.163 &-0.052 &-0.082 &-0.133 &- &- &- \\
 &              &4  &-0.069 &-0.106 &-0.190 &-0.073 &-0.108 &-0.162 &- &- &- \\
\hline\hline
\end{tabular}
\end{center}
\end{table}
}

We consider the 2D Heisenberg model with easy-axis anisotropy.

\begin{equation}
\label{eq:h} 
        H = -\frac{J}{2}\sum_{i,j}\mathbf{M}_i\cdot\mathbf{M}_j  - \frac{B}{2}\sum_{i,j}M_i^z M_j^z - A\sum_{i}(M_i^z)^2,   
\end{equation}
where $J>0$ ($J<0$) for a ferromagnet (antiferromagnet), $i$ and $j$ denote nearest neighbors within the layer, $\mathbf M_i$ is the magnetic moment at site $i$ (in $\mu_B$), and $B,A$ are the parameters of the exchange anisotropy and single-site anisotropy, respectively.

To determine $B$ and $A$, let us write the energies of the following 4 magnetic states: ferromagnetic out-of-plane ($E_{FM}^z$), antiferromagnetic out-of-plane ($E_{AFM}^z$), ferromagnetic in-plane ($E_{FM}^x$) and antiferromagnetic in-plane ($E_{AFM}^x$).
According to Eq. \ref{eq:h} and keeping in mind the hexagonal Cr lattice as well as $|\mathbf M_i|=m_\text{Cr}$ we get

\begin{equation}
        E_{\mathrm{FM}}^z  =  -6m_\text{Cr}^2 (J + B) - 2 m_\text{Cr}^2 A \,
\end{equation}
\begin{equation}
        E_{\mathrm{AFM}}^z =  +2m_\text{Cr}^2 (J + B) - 2 m_\text{Cr}^2 A \,
\end{equation}
\begin{equation}
        E_{\mathrm{FM}}^x  =  -6J m_\text{Cr}^2 \,
\end{equation}
\begin{equation}
        E_{\mathrm{AFM}}^x =  +2J m_\text{Cr}^2 \,
\end{equation}

Therefore

\begin{equation}
        \Delta E_{\mathrm{A/F}}^z = E_{\mathrm{AFM}}^z - E_{\mathrm{FM}}^z =  8 m_\text{Cr}^2 (J+B) \,
\end{equation}
\begin{equation}
        \Delta E_{\mathrm{A/F}}^x = E_{\mathrm{AFM}}^x - E_{\mathrm{FM}}^x =  8 J m_\text{Cr}^2          \,
\end{equation}

From the difference of the above two, i.e., $\Delta E_{\mathrm{A/F}}^x - \Delta E_{\mathrm{A/F}}^z = -8 m_\text{Cr}^2 B$ we can determine the exchange anisotropy.  

Therefore, $B  = -(\Delta E_{\mathrm{A/F}}^x -\Delta E_{\mathrm{A/F}}^z)/8 m_\text{Cr}^2$. 
$\Delta E_{\mathrm{A/F}}^x$ and $\Delta E_{\mathrm{A/F}}^z$ are known from the DFT calculations. The $m_\text{Cr}$ values are given in Supplementary Table \ref{stab:magmom}.

Now, we can find $A$ using the above expressions for $E_{\mathrm{FM}}^x$ and $E_{\mathrm{FM}}^z$ on one hand and the DFT result on the other. We therefore have $A = ((E_{\mathrm{FM}}^x - E_{\mathrm{FM}}^z) - 6B m_\text{Cr}^2) / 2 m_\text{Cr}^2$. The $A$ and $B$ values evaluated in this way can be found in Supplementary Table \ref{stab:sia_ai}.

Finally, from the scalar relativistic DFT calculation and Hamiltonian we have the isotropic exchange coupling constant $J=(E_{\mathrm{AFM}}-E_{\mathrm{FM}})/8 m_\text{Cr}^2$, whose values are shown in Supplementary Table \ref{stab:j}.

\makeatletter
\let\oldcite\cite
\renewcommand{\cite}[1]{\oldcite{SI-#1}}
\makeatother

The dipole–dipole interaction energy $E_{d-d}$ was evaluated as
\begin{equation} \label{eq:anisotropy}
        E_{d-d} = E(\hat{m}) - E_{\perp} = \frac{\alpha^2}{4} \sum_{i \neq j} M_i M_j \frac{3 \left( z_{ij}^2 - (\hat{m} \cdot \mathbf{r}_{ij})^2 \right)}{|\mathbf{r}_{ij}|^5},
\end{equation}
i.e., the difference between the energy when all magnetic moments align with an in-plane direction, $E(\hat{m})$, and the reference energy when the moments point perpendicular to the plane, $E_{\perp}$ \cite{Szunyogh.prb1995}. Here, $M_i$ and $M_j$ are the magnitudes of the interacting magnetic moments (in $\mu_B$), $\mathbf{r}_{ij}$ is the vector between them and $z_{ij}$ is its $z$-component, $\hat{m}$ is the unit vector for the in-plane magnetization, and $\alpha$ is the fine-structure constant.

{\renewcommand{\arraystretch}{1.00}%
\begin{table}[!bth]
\caption{Isotropic nearest neighbour exchange coupling constant $J = \Delta E / 8m_\text{Cr}^2$ (meV/$\mu_\mathrm{B}^2$) of the \CXY\ Janus MLs for both 1T and 1H structures.} 
\label{stab:j}
\begin{center}
\begin{tabular}%
{>{\centering\arraybackslash}p{0.50cm} >{\centering\arraybackslash}p{1.7cm}
 >{\centering\arraybackslash}p{1.5cm} >{\centering\arraybackslash}p{1.5cm}
 >{\centering\arraybackslash}p{1.5cm} >{\centering\arraybackslash}p{1.5cm}
 >{\centering\arraybackslash}p{1.5cm} >{\centering\arraybackslash}p{1.5cm}
 >{\centering\arraybackslash}p{1.5cm} >{\centering\arraybackslash}p{1.5cm}
 >{\centering\arraybackslash}p{1.5cm}}
\hline \hline
   & $U_{\mathrm{eff}}$ (eV) & CrSCl & CrSBr & CrSI & CrSeCl & CrSeBr & CrSeI & CrTeCl & CrTeBr & CrTeI \\
\hline \hline
  1T &  2.1 &   2.50 &   2.56 &   2.25 &   2.28 &   2.48 &   2.49 &   1.50 &   1.92 &   2.33 \\
  1T &  3.0 &   2.37 &   2.40 &   2.14 &   2.17 &   2.35 &   2.36 &   1.35 &   1.75 &   2.18 \\
  1T &  4.0 &   2.24 &   2.25 &   2.02 &   2.07 &   2.21 &   2.24 &   1.00 &   1.59 &   2.03 \\
  \hline
  1H &  2.1 &   1.72 &   2.15 &   2.43 &   1.35 &   1.85 &   2.06 &   -&   -&   -\\
  1H &  3.0 &   1.86 &   2.16 &   2.37 &   1.51 &   1.89 &   2.04 &   -&   -&   -\\
  1H &  4.0 &   1.92 &   2.12 &   2.28 &   1.58 &   1.89 &   2.01 &   -&   -&   -\\
  \hline
  \hline
\end{tabular}
\end{center}
\end{table}
}

\renewcommand{\arraystretch}{1.2}
\begin{table}[!bth]
\caption{Isotropic exchange coupling constants calculated up to the third neighbors, $J_1$, $J_2$, and $J_3$ ($\mathrm{meV}/\mu_\mathrm{B}^2$), for the CrSI and CrSeI monolayers in both 1T and 1H phases.}
\label{stab:j_third}
\begin{center}
\begin{tabular}{
    >{\centering\arraybackslash}p{1.5cm} 
    >{\centering\arraybackslash}p{2.5cm} 
    >{\centering\arraybackslash}p{2.5cm} 
    >{\centering\arraybackslash}p{2.5cm} 
    >{\centering\arraybackslash}p{2.5cm} 
}
\hline \hline
\text{Phase} & \text{$U_{\mathrm{eff}}$ (eV)} & \text{$J_1$ } & \text{$J_2$ } & \text{$J_3$ } \\
\hline \hline
\multicolumn{5}{c}{\text{CrSI}} \\
\hline
1T & 2.1 & 2.179 & 0.068 & -0.078 \\
1T & 3.0 & 2.074 & 0.063 & -0.104 \\
1T & 4.0 & 1.956 & 0.061 & -0.132 \\
\hline
1H & 2.1 & 2.259 & 0.163 & -0.018 \\
1H & 3.0 & 2.176 & 0.180 & -0.105 \\
1H & 4.0 & 2.094 & 0.183 & -0.181 \\
\hline
\multicolumn{5}{c}{\text{CrSeI}} \\
\hline
1T & 2.1 & 2.401 & 0.093 & -0.176 \\
1T & 3.0 & 2.275 & 0.089 & -0.210 \\
1T & 4.0 & 2.156 & 0.088 & -0.247 \\
\hline
1H & 2.1 & 2.041 & 0.025 & -0.257 \\
1H & 3.0 & 1.998 & 0.043 & -0.340 \\
1H & 4.0 & 1.952 & 0.054 & -0.404 \\
\hline \hline
\end{tabular}
\end{center}
\end{table}

\clearpage

\newpage

\begin{figure}[h!]
    \centering    
\includegraphics[width=0.99\textwidth]{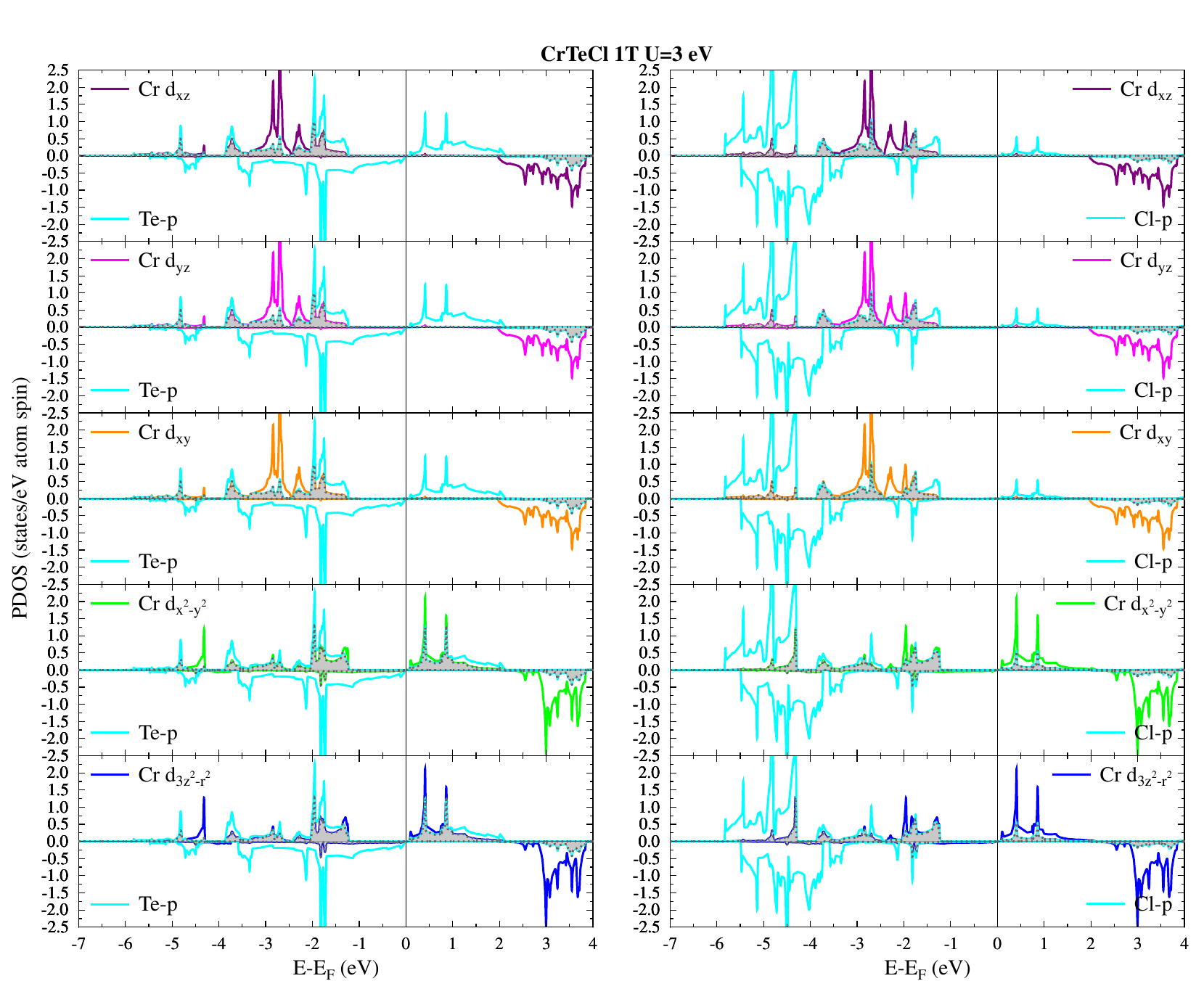}
\caption{1T CrTeCl projected densities of states (PDOS) onto Cr-$3d$-orbitals as well as Te-$p$ and Cl-$p$, shown by color \emph{solid} lines, as calculated for $U_{\mathrm{eff}}=3$ eV. Majority PDOS are plotted as positive values, while minority as negative ones. The dashed grey lines and grey shading represent the overlap parameter defined as $O_{d_i\text{-}p}^{{X/Y}}(E,\sigma) = \min\left[ \text{PDOS}_{\text{Cr}-d_i}(E,\sigma), \text{PDOS}_{X/Y-p}(E,\sigma) \right]$, where $i=d_{xy},\ d_{yz},\ d_{z^2},\ d_{xz},\ d_{x^2 - y^2}$, $p=p_x+p_y+p_z$ (these orbitals are to a high accuracy degenerate) and $\sigma = \{ \uparrow, \downarrow \}$. We introduce this parameter as a measure of the $p-d$ hybridization: its integral reflects the shared area under the two PDOS curves ($d$ and $p$) and hence is related to the hybridization of the $X/Y$-$p$ and Cr-$3d$ states. }
\label{sfig:dos-overlap}
\end{figure}

\begin{figure}[h!]
    \centering    
\includegraphics[width=0.85\textwidth]{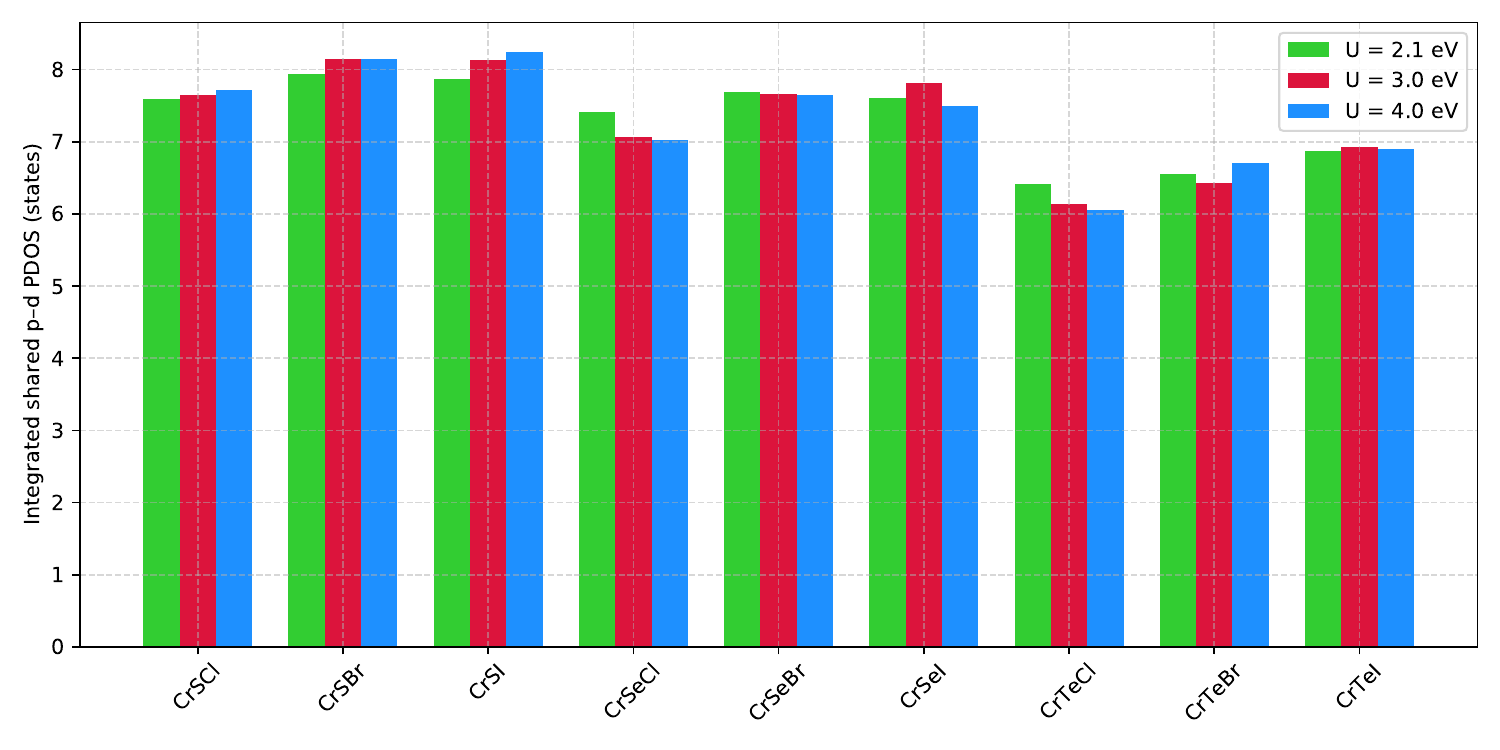}
\caption{Integrated shared PDOS,
$I_{d_i\text{-}p}^{{X/Y}} = \sum_{\sigma=\uparrow,\downarrow}         \int_{-\infty}^{E_F}  O_{d_i\text{-}p}^{{X/Y}}(E,\sigma)\,dE$,
where $O_{d_i\text{-}p}^{{X/Y}}(E,\sigma)$ is defined in the caption of Fig. \ref{sfig:dos-overlap}) for 1T-\CXY. In essence, it is the shared area under the $p$ and $d$ PDOS curves. The larger (smaller) $I_{d_i\text{-}p}^{{X/Y}}$ reflects the stronger (weaker) overlap of the the $p$ and $d$ PDOS curves, indicating stronger (weaker) $p-d$ hybridization. Note a significant drop for 1T-CrTeCl, which correlates with the drop of $J$ seen in the maintext 
	Fig.~\ref{fig:j}.}
\label{sfig:overlaps}
\end{figure}

\clearpage

\newpage

{\renewcommand{\arraystretch}{0.90}%
\begin{table}[!bth]
\caption{Total energy difference between the states with in-plane (ip) and out-of-plane (oop) directions of the Cr magnetic moment, $E_{\mathrm{SOC}} = E_{\mathrm{ip}} - E_{\mathrm{oop}}$ (meV per Cr atom); the energy of the classical dipole–dipole interaction, $E_{\text{d-d}}$ (meV per Cr atom); and their sum, $E_{\mathrm{sum}} = E_{\mathrm{SOC}} + E_{\text{d-d}}$ (meV per Cr atom), for the \CXY{} Janus MLs ($X = \mathrm{S},\ \mathrm{Se},\ \mathrm{Te}$;\ $Y = \mathrm{Cl},\ \mathrm{Br},\ \mathrm{I}$) in both the 1T and 1H structures under FM ordering. The relativistic approximation is adopted, and all values are obtained from non-self-consistent calculations. Positive (negative) values of $E_{\mathrm{sum}}$ correspond to an out-of-plane (in-plane) orientation of the magnetic moment. }
\label{stab:mae}
\begin{center}
\begin{tabular}%
{>{\PBS\centering\hspace{0pt}} p{0.60cm}%
 >{\PBS\centering\hspace{0pt}} p{1.40cm}%
 >{\PBS\centering\hspace{0pt}} p{1.70cm}
 >{\PBS\centering\hspace{0pt}} p{1.40cm}
 >{\PBS\centering\hspace{0pt}} p{1.40cm}
 >{\PBS\centering\hspace{0pt}} p{1.40cm}
 >{\PBS\centering\hspace{0pt}} p{1.40cm}
 >{\PBS\centering\hspace{0pt}} p{1.40cm}
 >{\PBS\centering\hspace{0pt}} p{1.40cm}
 >{\PBS\centering\hspace{0pt}} p{1.40cm}
 >{\PBS\centering\hspace{0pt}} p{1.40cm}
 >{\PBS\centering\hspace{0pt}} p{1.40cm}}
\hline \hline
& & $U_{\mathrm{eff}}$ (eV)& CrSCl & CrSBr & CrSI & CrSeCl & CrSeBr & CrSeI & CrTeCl & CrTeBr & CrTeI  \\
\hline \hline
   &              &2.1 & 0.036 & 0.059 & 0.246 &-0.234 &-0.138 & 0.227 &-2.978 &-2.322 &-0.956 \\
1T &$E_{\mathrm{SOC}}$     &3   & 0.029 & 0.062 & 0.334 &-0.303 &-0.171 & 0.323 &-1.957 &-3.017 &-1.078 \\
   &              &4   & 0.023 & 0.065 & 0.420 &-0.383 &-0.208 & 0.435 & 2.339 &-3.101 &-1.176 \\
\hline
   &              &2.1 &-0.110 &-0.104 &-0.094 &-0.108 &-0.102 &-0.092 &-0.101 &-0.096 &-0.087 \\
1T &$E_{\mathrm{d-d}}$     &3   &-0.115 &-0.108 &-0.099 &-0.113 &-0.106 &-0.097 &-0.107 &-0.101 &-0.092 \\
   &              &4   &-0.120 &-0.114 &-0.104 &-0.118 &-0.112 &-0.102 &-0.117 &-0.107 &-0.098 \\
\hline
   &              &2.1 &-0.074 &-0.045 & 0.152 &-0.342 &-0.240 & 0.135 &-3.079 &-2.418 &-1.043 \\
1T &$E_{\mathrm{sum}}$     &3   &-0.086 &-0.046 & 0.235 &-0.416 &-0.277 & 0.226 &-2.064 &-3.118 &-1.170 \\
   &              &4   &-0.097 &-0.049 & 0.316 &-0.501 &-0.320 & 0.333 & 2.223 &-3.208 &-1.274 \\     
   
\hline \hline

   &              &2.1 &-0.159 &-0.173 & 3.563 &-0.701 &-1.062 & 2.569 &- &- &- \\
1H &$E_{\mathrm{SOC}}$     &3   &-0.148 &-0.133 & 2.740 &-0.826 &-1.230 & 2.066 &- &- &- \\
   &              &4   &-0.139 &-0.107 & 2.077 &-0.969 &-1.426 & 1.465 &- &- &- \\
\hline
   &              &2.1 &-0.126 &-0.118 &-0.110 &-0.123 &-0.115 &-0.105 &- &- &- \\
1H &$E_{\mathrm{d-d}}$     &3   &-0.133 &-0.124 &-0.116 &-0.130 &-0.122 &-0.111 &- &- &- \\
   &              &4   &-0.139 &-0.131 &-0.122 &-0.137 &-0.129 &-0.116 &- &- &- \\
\hline
   &              &2.1 &-0.285 &-0.291 & 3.453 &-0.824 &-1.177 & 2.464 &- &- &- \\
1H &$E_{\mathrm{sum}}$     &3   &-0.281 &-0.257 & 2.624 &-0.956 &-1.352 & 1.955 &- &- &- \\
   &              &4   &-0.278 &-0.238 & 1.955 &-1.106 &-1.555 & 1.349 &- &- &- \\     
\hline \hline
\end{tabular}
\end{center}
\end{table}

\begin{table}[h]
\centering
\caption{Single-ion ($A$) and exchange ($B$) anisotropies (in meV/$\mu_\mathrm{B}^2$) for CrSI and CrSeI in both 1T and 1H polymorphs.}
\label{stab:sia_ai}
\begin{tabular}{ccccc}
\hline\hline
       & Phase & $U_{\text{eff}}$ (eV) & $A$   & $B$    \\
\hline
CrSI   & 1T    & 2.1                   & 0.008 & 0.005 \\
CrSI   & 1T    & 3.0                   & 0.011 & 0.006 \\
CrSI   & 1T    & 4.0                   & 0.013 & 0.007 \\
CrSeI  & 1T    & 2.1                   & 0.020 & 0.000 \\
CrSeI  & 1T    & 3.0                   & 0.026 & 0.000 \\
CrSeI  & 1T    & 4.0                   & 0.033 & 0.000 \\
CrSI   & 1H    & 2.1                   & 0.061 & 0.080 \\
CrSI   & 1H    & 3.0                   & 0.055 & 0.053 \\
CrSI   & 1H    & 4.0                   & 0.051 & 0.033 \\
CrSeI  & 1H    & 2.1                   & 0.086 & 0.040 \\
CrSeI  & 1H    & 3.0                   & 0.079 & 0.025 \\
CrSeI  & 1H    & 4.0                   & 0.068 & 0.011 \\
\hline\hline
\end{tabular}
\end{table}

\section{Crystal field splittings in 1T and 1H $\mathbf{Cr}X\mathbf{I}$ ($X=\mathbf{S,\ Se}$) monolayers}
\label{ssec:cf}

\begin{figure}[h!]
    \centering    
\includegraphics[width=0.99\textwidth]{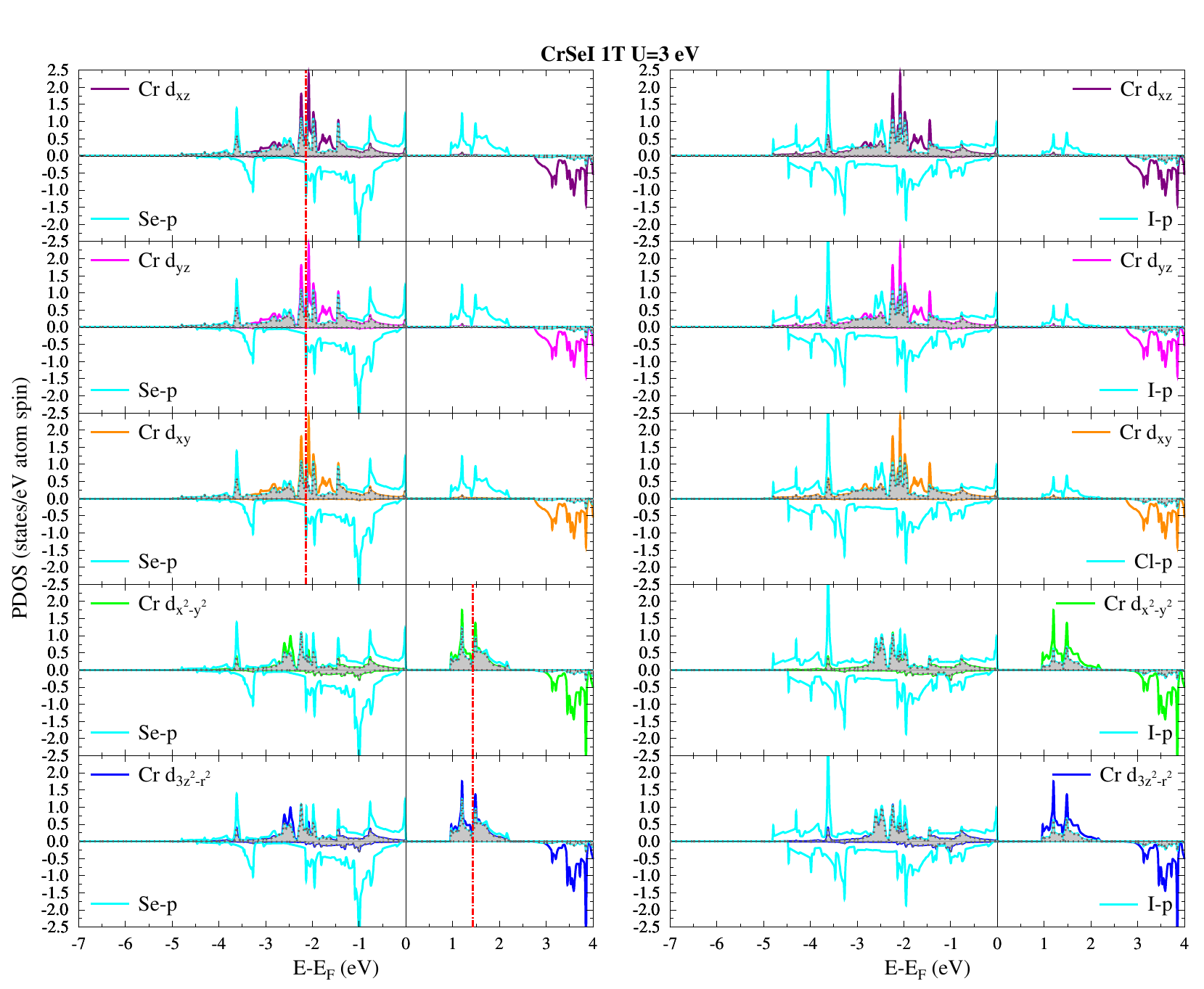}
\caption{1T CrSeI projected densities of states (PDOS) onto Cr-$d$-orbitals as well as Se-$p$ and I-$p$, shown by color \emph{solid} lines, as calculated for $U_{eff}=3$ eV. Majority PDOS are plotted as positive values, while minority as negative ones. As in Fig. \ref{sfig:dos-overlap}, the dashed grey lines and grey shading represent the overlap parameter defined as $O_{d_i\text{-}p}^{{X/Y}}(E,\sigma) = \min\left[ \text{PDOS}_{\text{Cr}-d_i}(E,\sigma), \text{PDOS}_{X/Y-p}(E,\sigma) \right]$, where $i=d_{xy},\ d_{yz},\ d_{z^2},\ d_{xz},\ d_{x^2 - y^2}$, $p=p_x+p_y+p_z$ and $\sigma = \{ \uparrow,\ \downarrow \}$. The vertical red dash-and-dot lines denote the centers of gravity of the relevant Cr-$3d$ states for the determination of the crystal field splitting.}
\label{sfig:dos_1t-crsei}
\end{figure}

Both 1T and 1H polymorphs of Cr\textit{XY} Janus MLs possess the same space group P3m1 (No. 156), associated with the \(C_{3v}\) point group. However, they differ in the local coordination of the Cr atoms: in the 1T polytype, each Cr sits in a distorted octahedral environment, whereas in the 1H structure, the Cr atom is coordinated by a distorted trigonal prism. In both polytypes, the reduced symmetry lifts the degeneracy of the Cr \(3d\) orbitals, and the crystal-field (CF) splitting must be described using the irreducible representations of \(C_{3v}\): \(a_1\) and \(e\) levels. In the 1T case, although the exact degeneracies of the \(t_{2g}\)-like and \(e_g\)-like subsets are broken due to distortion, the intra-group splittings are small (on the order of 1 meV), see Fig. \ref{sfig:dos_1t-crsei} where the projected densities of states of the Cr-$d_{xy},\ d_{xz},\ d_{yz}$ orbitals hardly display any noticeable difference. We therefore retain the approximate labels:
\[
t_{2g}\text{-like} \approx \{ d_{xy},\ d_{xz},\ d_{yz} \}, \quad
e_g\text{-like} \approx \{ d_{x^2 - y^2},\ d_{z^2} \},
\]
and define the CF splitting as the difference between the centers of gravity (CGs) of these subsets in the spin-up channel (their CG's are marked by the vertical dash-and-dot lines in Fig. \ref{sfig:dos_1t-crsei}):
\[
\Delta_\mathrm{CF}^{\mathrm{1T}} = \mathrm{CG}_{e_g} - \mathrm{CG}_{t_{2g}}.
\]

In the 1H case (Fig. \ref{sfig:dos_1h-crsei}), the orbitals transform as:
\[
a_1 \equiv d_{z^2}, \quad
e' \equiv \{ d_{xy},\ d_{x^2 - y^2} \}, \quad
e'' \equiv \{ d_{xz},\ d_{yz} \}.
\]
Here, the appropriate crystal-field splitting is defined as the energy separation between the occupied \(e'\) and the unoccupied \(e''\) doublets  in the spin-up channel (see the vertical dash-and-dot lines in Fig. \ref{sfig:dos_1h-crsei}):
\[
\Delta_\mathrm{CF}^{1\mathrm{H}} = \mathrm{CG}_{e''} - \mathrm{CG}_{e'}.
\]

\begin{figure}[h!]
    \centering    
\includegraphics[width=0.99\textwidth]{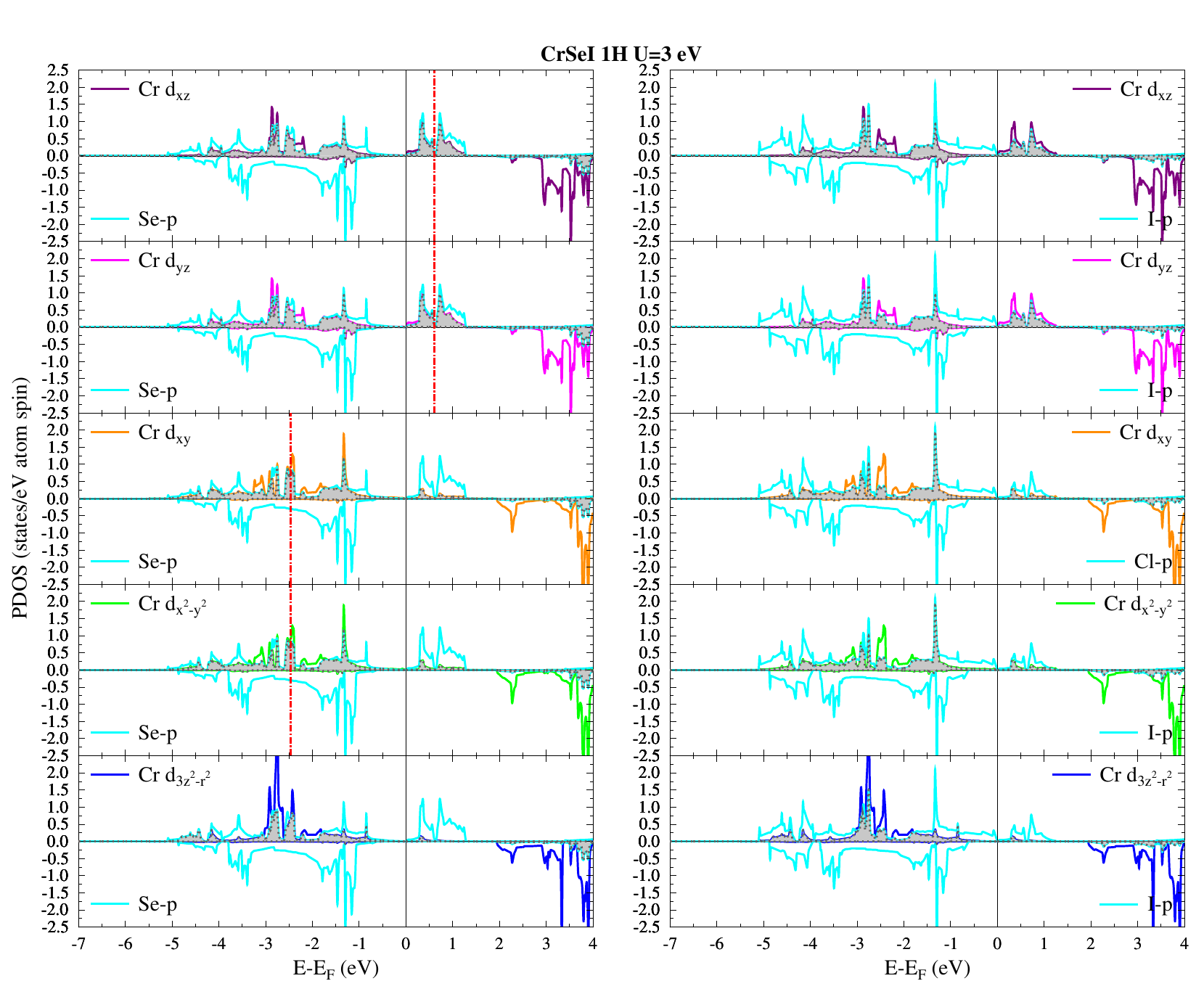}
\caption{1H CrSeI projected densities of states (PDOS) onto Cr-$d$-orbitals as well as Se-$p$ and I-$p$, shown by color \emph{solid} lines, as calculated for $U_{\mathrm{eff}}=3$ eV. Majority PDOS are plotted as positive values, while minority as negative ones. As in Fig. \ref{sfig:dos-overlap}, the dashed grey lines and grey shading represent the overlap parameter defined as $O_{d_i\text{-}p}^{{X/Y}}(E,\sigma) = \min\left[ \text{PDOS}_{\text{Cr}-d_i}(E,\sigma), \text{PDOS}_{X/Y-p}(E,\sigma) \right]$, where $i=d_{xy},\ d_{yz},\ d_{z^2},\ d_{xz},\ d_{x^2 - y^2}$, $p=p_x+p_y+p_z$ and $\sigma = \{ \uparrow,\ \downarrow \}$. The vertical red dash-and-dot lines denote the centers of gravity of the relevant Cr-$3d$ states for the determination of the crystal field splitting.}
\label{sfig:dos_1h-crsei}
\end{figure}

The numerical values of \(\Delta_\mathrm{CF}\), computed from the spin-up PDOS, are summarized in Table \ref{stab:cf}:

\begin{table}[H]
\centering
\caption{Crystal-field splittings \(\Delta_\mathrm{CF}^\mathrm{1T}\) and \(\Delta_\mathrm{CF}^\mathrm{1H}\) extracted from spin-up PDOS for CrSI and CrSeI MLs.}
\label{stab:cf}
\begin{tabular}{cccc}
\hline\hline
Material & Phase & \(U_\mathrm{eff}\) (eV) & $\Delta_\mathrm{CF}$ (eV) \\
\hline
CrSI     & 1T    & 2.1                     & 3.320 \\
CrSI     & 1T    & 3.0                     & 3.659 \\
CrSI     & 1T    & 4.0                     & 4.043 \\
CrSeI    & 1T    & 2.1                     & 3.243 \\
CrSeI    & 1T    & 3.0                     & 3.563 \\
CrSeI    & 1T    & 4.0                     & 3.908 \\
CrSI     & 1H    & 2.1                     & 2.985 \\
CrSI     & 1H    & 3.0                     & 3.199 \\
CrSI     & 1H    & 4.0                     & 3.440 \\
CrSeI    & 1H    & 2.1                     & 2.851 \\
CrSeI    & 1H    & 3.0                     & 3.086 \\
CrSeI    & 1H    & 4.0                     & 3.351 \\
\hline\hline
\end{tabular}
\end{table}

According to second-order perturbation theory, the single-ion anisotropy constant \(A\) scales as:
\[
A \propto \frac{\lambda^2}{\Delta_\mathrm{CF}},
\]
where \(\lambda\) is the spin-orbit coupling constant and \(\Delta_\mathrm{CF}\) \cite{Lado.2dmat2017}. Thus, the enhancement of \(A\) in the 1H polymorphs correlates with the smaller $\Delta_\mathrm{CF}$ values they display.

\clearpage
\newpage

\begin{figure}[h!]
    \centering    
\includegraphics[width=0.6\textwidth]{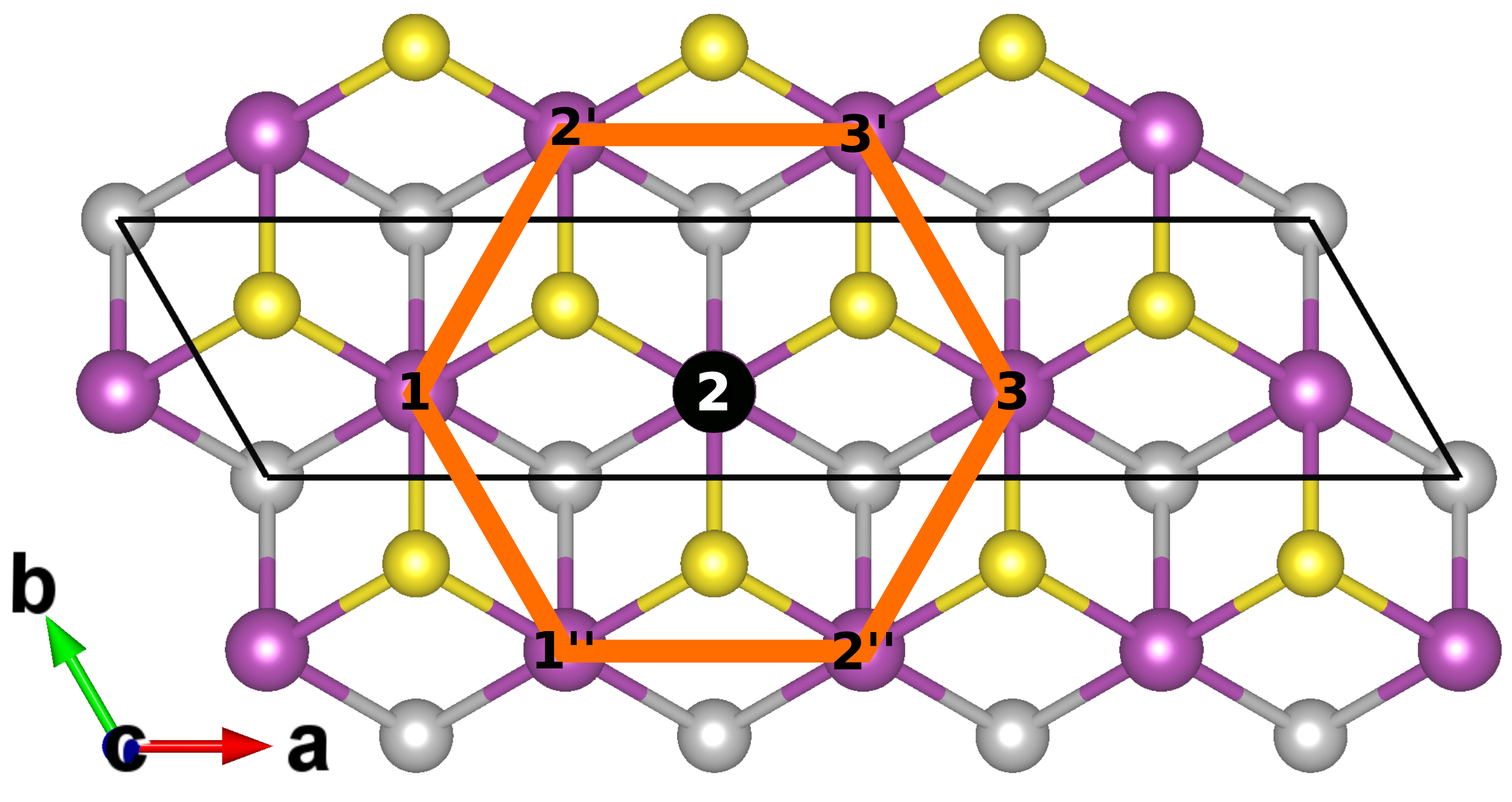}
\caption{Top view of the $(4 \times 1)$ cell used for the spin-spiral calculations. The hexagon with labels denotes the NNs of the atom 2.}
\label{sfig:dmi}
\end{figure}

\section{Dzyaloshinskii–Moriya interaction}
\label{ssec:dmi}

To derive the in-plane Dzyaloshinskii–Moriya interaction (DMI)} component $d^{\parallel}_\text{NN}$ corresponding to nearest-neighbor (NN) interactions, we start from the DMI term of the spin Hamiltonian:
\[
H_{\text{DMI}} = \frac{1}{2}\sum_{i,j} \mathbf{D}_{ij} \cdot \left( \mathbf
{M}_i \times \mathbf{M}_j \right),
\]
where $\mathbf{D}_{ij}$ is the DMI vector for the $i \to j$ bond, $\mathbf{M}_i$ is the magnetic moment at site $i$, and the sum runs over all pairs of nearest neighbors $i,j$.

In Fig. \ref{sfig:dmi}, the site $i=2$ has six nearest neighbors: $j=$ 3, 3$'$, 2$'$, 1, 1$''$, 2$''$. The DMI energy for this site, $E_2^{\text{DMI}}$, therefore is
\[
\begin{aligned}
E_2^{\text{DMI}} =\; &\frac{1}{2} \left[ \mathbf{D}_{23} \cdot (\mathbf{M}_2 \times \mathbf{M}_3)
+ \mathbf{D}_{2 2''} \cdot (\mathbf{M}_2 \times \mathbf{M}_{2''}) \right. \\
&\left. + \mathbf{D}_{2 1''} \cdot (\mathbf{M}_2 \times \mathbf{M}_{1''}) 
+ \mathbf{D}_{21} \cdot (\mathbf{M}_2 \times \mathbf{M}_1) \right. \\
&\left. + \mathbf{D}_{2 2'} \cdot (\mathbf{M}_2 \times \mathbf{M}_{2'}) 
+ \mathbf{D}_{2 3'} \cdot (\mathbf{M}_2 \times \mathbf{M}_{3'}) \right]
\end{aligned}
\]

For the case of the CW and ACW spirals in the $xz$ plane (maintext Fig. \ref{fig:struc}e) the latter equation results in
\[
\begin{aligned}
E_2^{\text{CW}} &= \frac{m_\text{Cr}^2}{2} \left( D_{23}^y - D_{2 1''}^y - D_{21}^y + D_{2 3'}^y \right), \\
E_2^{\text{ACW}} &= -\frac{m_\text{Cr}^2}{2} \left( D_{23}^y - D_{2 1''}^y - D_{21}^y + D_{2 3'}^y \right).
\end{aligned}
\]

Taking the difference between the two leads to
\[
\Delta E_2^{\text{DMI}} = E_2^{\text{CW}} - E_2^{\text{ACW}} = m_\text{Cr}^2 \left( D_{23}^y - D_{2 1''}^y - D_{21}^y + D_{2 3'}^y \right)
\]

The Moriya’s symmetry rule \cite{Moriya.pr1960} takes the form of $\mathbf{D}_{ij} = d^{\parallel} (\hat{\mathbf{z}} \times \hat{\mathbf{u}}_{ij})$ (the $d_{ij,z}$ term is absent, only the $y$ components enter the $\Delta E_2^{\text{DMI}}$ expression above), the DMI vectors are perpendicular to the vectors $\hat{\mathbf{u}}_{ij}$ connecting the sites $i$ to $j$, as shown in the maintext Fig. \ref{fig:struc}f for the NN coupling. Taking into account the $y$ components of the DMI vectors and their absolute values $d^{\parallel}_\text{NN}$, we have:

\[
D_{23}^y - D_{2 1''}^y - D_{21}^y + D_{2 3'}^y = 3 d^{\parallel}_\text{NN}.
\]

Hence, 
\[
\Delta E_2^{\text{DMI}} = 3 m_\text{Cr}^2 d^{\parallel}_\text{NN}.
\]

Since we have four sites in the cell 
\[
\Delta E^{\text{DMI}} = 4 \Delta E_2^{\text{DMI}} = 12  
m_\text{Cr}^2 d^{\parallel}_\text{NN}. \]

If we now consider the spirals in the $xy$ plane, it leads to
\[
\begin{aligned}
E_2^{\text{CW}} &= \frac{m_\text{Cr}^2}{2} \left( -D_{23}^z + D_{2 1''}^z + D_{21}^z - D_{2 3'}^z \right), \\
E_2^{\text{ACW}} &= -\frac{m_\text{Cr}^2}{2} \left( D_{23}^z - D_{2 1''}^z - D_{21}^z + D_{2 3'}^z \right).
\end{aligned}
\]

In this case, the Moriya’s symmetry rule \cite{Moriya.pr1960} takes the form of $\mathbf{D}_{ij} = d_{ij,z} \hat{\mathbf{z}}$, i.e. there are only $z$-components present. Since in the expressions above $D_{23}^z = D_{2 1''}^z = D_{21}^z = D_{2 3'}^z =d_{ij,z}$, both $E_2^{\text{CW}}=0$ and $E_2^{\text{ACW}}=0$

\begin{table}[h!]
\centering
\caption{In-plane DMI component $d^{\parallel}_{NN}$ (meV/$\mu_\mathrm{B}^2$), NN isotropic exchange coupling parameter $J$ (meV/$\mu_\mathrm{B}^2$), and their ratio.}
\label{stab:dmi_mCr}
\begin{tabular}{lcccccc}
\hline\hline
Material & Phase & $U_{\text{eff}}$ (eV) & $d^{\parallel}_{NN}$ (meV/$\mu_\mathrm{B}^2$) & $J$ (meV/$\mu_\mathrm{B}^2$) & $|d^{\parallel}_{NN}/J|$ \\
\hline
CrSI  & 1T & 2.1 & -0.086 & 2.25 & 0.038 \\
CrSI  & 1T & 3.0 & -0.082 & 2.14 & 0.038 \\
CrSI  & 1T & 4.0 & -0.081 & 2.02 & 0.040 \\
CrSeI & 1T & 2.1 & -0.215 & 2.49 & 0.086 \\
CrSeI & 1T & 3.0 & -0.215 & 2.36 & 0.091 \\
CrSeI & 1T & 4.0 & -0.215 & 2.24 & 0.096 \\
CrSI  & 1H & 2.1 &  0.054 & 2.43 & 0.022 \\
CrSI  & 1H & 3.0 &  0.035 & 2.37 & 0.015 \\
CrSI  & 1H & 4.0 &  0.017 & 2.28 & 0.007 \\
CrSeI & 1H & 2.1 & -0.084 & 2.06 & 0.041 \\
CrSeI & 1H & 3.0 & -0.086 & 2.04 & 0.042 \\
CrSeI & 1H & 4.0 & -0.087 & 2.01 & 0.044 \\
\hline\hline
\end{tabular}
\end{table}

\clearpage
\newpage

\section{Atomistic spin dynamics simulations}
\label{ssec:asd}

We conducted atomistic spin dynamic (ASD) computational simulations based on a classical Heisenberg Hamiltonian,

\begin{multline}
\label{eq:ham_asd_corr}
    \mathcal{H}=-\frac{J_1}{2}\sum\limits_{\langle i,j \rangle} \mathbf{S}_i \cdot \mathbf{S}_j - \frac{J_2}{2}\sum\limits_{\langle\langle i,j \rangle\rangle} \mathbf{S}_i\cdot \mathbf{S}_j  - \frac{J_3}{2}\sum\limits_{\langle\langle\langle i,j \rangle\rangle\rangle} \mathbf{S}_i \cdot \mathbf{S}_j - \\ - \dfrac{1}{2}\sum\limits_{\langle i,j \rangle} \mathbf{D}_{ij}\cdot(\mathbf{S}_i \times \mathbf{S}_j) - \dfrac{B}{2}\sum\limits_{\langle i,j \rangle} S_i^z \, S_j^z  - \sum\limits_{i} A (S_i^z)^2.
\end{multline}

Here, $|\mathbf{S}_i|=1$ denotes the normalized classical spin vector at site $i$; $J_1$, $J_2$ and $J_3$ are nearest neighbor (NN), second-NN and third-NN isotropic Heisenberg exchange parameters respectively; $\mathbf{D}_{ij}$ corresponds to the NN Dzyaloshinskii-Moriya interaction vector; $B$ describes the anisotropic NN exchange interaction and $A$ accounts for the single-ion anisotropy.

The stochastic Landau-Lifshitz-Gilbert equation governs the temporal evolution of each atomic spin  \cite{nowak_classical_2007}:
\begin{equation}
\label{eq:llg_corr}
    \frac{\partial \mathbf{S}_i}{\partial t}=-\frac{|\gamma_i|}{(1+\lambda^2_i)\mu_i} \left[ (\mathbf{S}_i \times \mathbf{H}_i) -\lambda_i (\mathbf{S}_i \times (\mathbf{S}_i \times \mathbf{H}_i)) \right],
\end{equation}
where  $\mathbf{H}_i=\partial\mathcal{H}/\partial\mathbf{S}_i+\mathbf{\zeta}_i$ represents the effective field at site $i$, with thermal fluctuations described by the stochastic field $\mathbf{\zeta}_i$ in the form of Gaussian white noise. The gyromagnetic ratio and the atomic damping are $\gamma_i=1.76\times10^{11}\, \mathrm{T^{-1}}\,\mathrm{s^{-1}}$ and $\lambda_i=0.1$,  respectively. The atomistic magnetic moments $\mu_i$ were extracted from the local magnetic moment of Cr atoms listed in Table~\ref{stab:magmom}.

The large 
ASD simulations were conducted using a previously developed GPU-accelerated code based on the \textsc{NVIDIA CUDA C-API}.

\subsection{Equilibrium magnetization dependence with temperature}
\label{sssec:asd_mT}

We computed the reduced magnetization, $\tilde{m}=\langle \frac{1}{N} \, \sum_i S_i^z\rangle$, in thermal equilibrium at different temperatures, where $N$ is the total number of spins in the simulation box. We considered large spin grids consisting of $512\times256\times1$ unit cells with periodic boundary conditions applied within the  $xy$ plane. The resulting magnetization curves as a function of temperature were fitted using a Curie-Bloch equation $\tilde{m}(T) = (1-T/T_\text{C})^\beta$, where $\beta$ and $T_{\text{C}}$ were extracted as fitting parameters.

\subsection{Magnetic ground state ($0$~K).}
\label{sssec:asd_textures}

Fig.~\ref{sfig:textures}a illustrates the cooling simulations carried out to investigate the magnetic ground state of Cr$X$I MLs ($X=$ Se, S). To thoroughly explore the resulting spin textures, each simulation was repeated six times per system, starting from different random configurations and using distinct initialization seeds for the thermal field term in the stochastic Landau-Lifshitz-Gilbert equation.

The magnetic ground state was identified by comparing the magnetic energy of the spin textures obtained in the ASD simulations with the fully ordered FM state, $E_0^{\text{FM}}$, as calculated from the ASD Hamiltonian:
\begin{equation}
    \label{eq:delta_U} 
    \Delta E_0=E_0^{\rm{Sim.}}-E_0^{\mathrm{FM}}= E_0^{\rm{Sim.}}- \left\{-\left(\dfrac{z_{\mathrm{1}}(J_1+B)+z_{\mathrm{2}}J_2+z_{\mathrm{3}}J_3}{2}\right)-A\right\},
\end{equation}
where $z_1=z_2=z_3=6$ denote the number of NN, second-NN and third-NN. Clearly, $E_0^{\text{Sim.}\,  \text{FM}}=E_0^{\text{FM}}$.

\newpage

\begin{table}[h]
\centering

\caption{\label{stab:asd_parameters} 
Atomistic parameters for the CrSI and CrSeI monolayers in both 1T and 1H phases.}
\footnotesize
\begin{tabular}{>{\PBS\centering\hspace{0pt}} p{1.80cm}%
 >{\PBS\centering\hspace{0pt}} p{1.80cm}%
 >{\PBS\centering\hspace{0pt}} p{1.80cm}
 >{\PBS\centering\hspace{0pt}} p{1.80cm}
 >{\PBS\centering\hspace{0pt}} p{1.80cm}
 >{\PBS\centering\hspace{0pt}} p{1.80cm}
 >{\PBS\centering\hspace{0pt}} p{1.80cm}
 >{\PBS\centering\hspace{0pt}} p{2.00cm}}
 
\hline\hline
$U_\mathrm{eff}$ (eV)  & $\mu_{\rm{Cr}}$ ($\mu_{B}$) & $J_1$ (meV) & $J_2$ (meV) & $J_3$ (meV)& $A$ (meV)& $B$ (meV)& $d_{NN}^\parallel$ (meV) \\ \hline \hline
\multicolumn{8}{c}{1T-CrSI} \\ \hline
2.1 & 3.297 & 23.684 & 0.738  & -0.843 & 0.087 & 0.054  & -0.935 \\
3.0 & 3.400 & 23.981 &  0.724 & -1.204 & 0.127 &  0.069 & -0.948 \\
4.0 & 3.523 & 24.282 & 0.753  & -1.643 & 0.161 & 0.087  & -1.005 \\
\hline
\multicolumn{8}{c}{1T-CrSeI} \\ \hline
2.1 & 3.400 & 27.754 & 1.074  & -2.040 & 0.231 & 0.000  & -2.485 \\
3.0 & 3.516 & 28.122 & 1.099  & -2.597 & 0.321 &  0.000 & -2.658 \\
4.0 & 3.642 & 28.591 & 1.168  & -3.280 & 0.437 & 0.000  & -2.852 \\
\hline
\multicolumn{8}{c}{1H-CrSI} \\ \hline
2.1 & 3.443 & 26.781 & 1.937  & -0.216 & 0.723 & 0.948  & 0.640 \\
3.0 & 3.581 & 27.901 &  2.311 & -1.348 & 0.705 &  0.680 & 0.449 \\
4.0 & 3.717 & 28.924 & 2.527  & -2.503 & 0.705 & 0.456  & 0.235 \\  \hline
\multicolumn{8}{c}{1H-CrSeI} \\ \hline
2.1 & 3.527 & 25.388 & 0.309  & -3.193 & 1.070 & 0.498  & -1.045 \\
3.0 & 3.666 & 26.858 &  0.578 & -4.568 & 1.062 &  0.336 & -1.156 \\
4.0 & 3.804 & 28.253 & 0.783  & -5.843 & 0.984 & 0.159  & -1.259 \\
\hline\hline
\end{tabular}
\label{tab:scaled_params}
\end{table}

\begin{figure}[h!]
\begin{center}
\includegraphics[width=1.0 \textwidth]{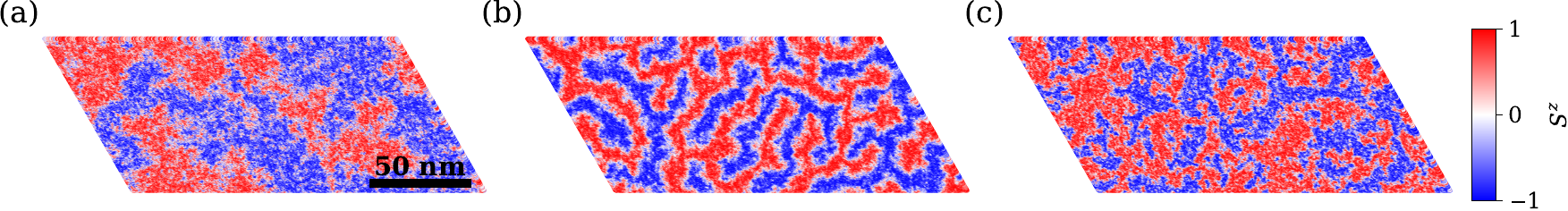}
\caption{\label{sfig:stextures_ht} Spin textures at different temperatures obtained by initializing all spins in a fully ordered FM state for (a) 1T-CrSI ($U_{\rm{eff}}=3.0$~eV) at $T=240$~K, (b) 1T-CrSeI ($U_{\rm{eff}}=3.0$~eV) at $150$~K and (c) 1H-CrSI ($U_{\rm{eff}}=4.0$~eV) at $T=180$~K.}
\vspace{-0.5cm}
\end{center}
\end{figure}

\begin{figure}[h!]
    \centering    
\includegraphics[width=1.0\textwidth]{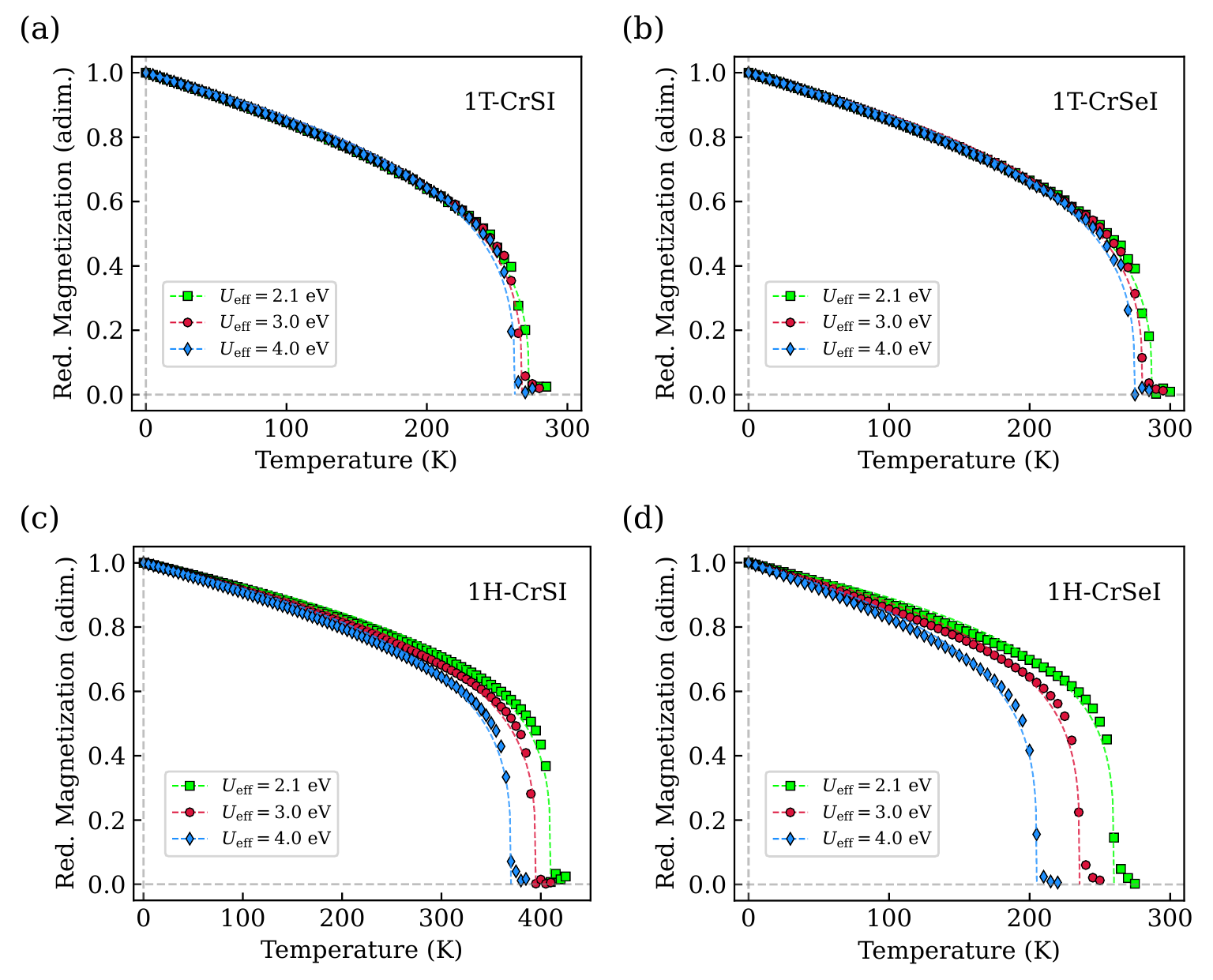}
\caption{Temperature dependence of the reduced magnetization excluding DMI from the calculations for (a) 1T-CrSI, (b) 1T-CrSeI, (c) 1H-CrSI, and (d) 1H-CrSeI. Symbols correspond to ASD simulations data-points. Dashed lines represent fits to the Curie-Bloch equation.}
\label{sfig:asd_mT_noDMI}
\end{figure}

\begin{figure*}[h!]
    \centering
    \includegraphics[width=1.0\textwidth]{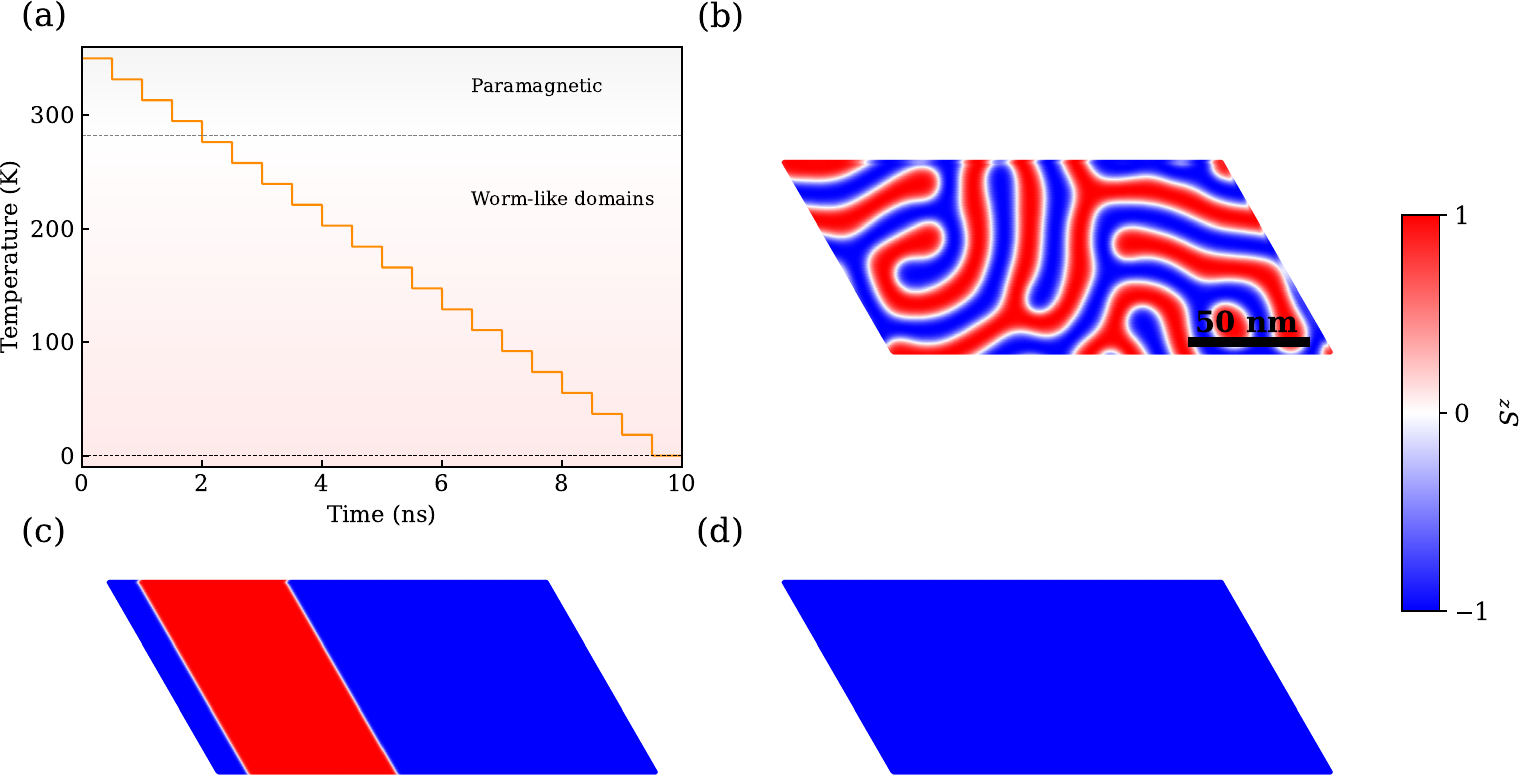}
    \caption{(a) Temperature evolution over time during the gradual cooling simulations for 1T-CrSeI. (b) Worm-like domain (WD) spin textures stabilized in 1T-CrSeI at 0~K after the cooling process. (c) Ferromagnetic domain (FMD) texture and (d) a fully FM configuration obtained in the 1H-CrSeI system for different initial configurations. All structures shown here correspond to $U_{\rm eff} = 2.1$~eV.}
    \label{sfig:textures}
\end{figure*}

\begin{table}[h!]
\centering
\caption{Magnetic textures obtained from ASD cooling simulations for CrSI and CrSeI monolayers in the 1T and 1H phases: ferromagnetic (FM), ferromagnetic domains (FMD), and worm-like domains (WD). $\Delta E_0$ denotes the per-spin energy difference, defined in Eq.~\ref{eq:delta_U}, between each magnetic texture and the fully FM state.}
\label{stab:gs}
\small
\begin{tabular}{>{\PBS\centering\hspace{0pt}} p{2.00cm}%
 >{\PBS\centering\hspace{0pt}} p{2.40cm}%
 >{\PBS\centering\hspace{0pt}} p{3.20cm} |
 >{\PBS\centering\hspace{0pt}} p{2.00cm}
 >{\PBS\centering\hspace{0pt}} p{2.40cm}
 >{\PBS\centering\hspace{0pt}} p{3.20cm}}
    \hline\hline
     $U_{\mathrm{eff}}$ (eV) & Sim. Textures & $\Delta E_0$ (meV/spin) & $U_{\mathrm{eff}}$ (eV) & Sim. Configs. & $\Delta E_0$ (meV/spin) \\ \hline\hline
       \multicolumn{3}{c|}{1T-CrSI $\qquad \; \;$  } & \multicolumn{3}{c}{1T-CrSeI  $\qquad \; \;$ } \\ \hline
       2.1 & FM/FMD & 0.017 & 2.1 & WD & -0.083 \\ 
       3.0 & FM/FMD & 0.022 & 3.0 & WD & -0.088\\ 
       4.0 & FM/FMD & 0.024 & 4.0 & WD & -0.105\\ \hline
       \multicolumn{3}{c|}{1H-CrSI $\qquad \; \;$  } & \multicolumn{3}{c}{1H-CrSeI  $\qquad \; \;$ } \\ \hline
       2.1 & FM/FMD & 0.165 & 2.1 & FM/FMD & 0.072 \\ 
       3.0 & FM/FMD & 0.138 & 3.0 & FM/FMD & 0.051\\ 
       4.0 & FM/FMD & 0.117 & 4.0 & FM/FMD & 0.027\\ \hline
    \end{tabular}
\vspace{-0.5cm}
\end{table}

\begin{figure}[h!]
\begin{center}
\includegraphics[width=1.0 \textwidth]{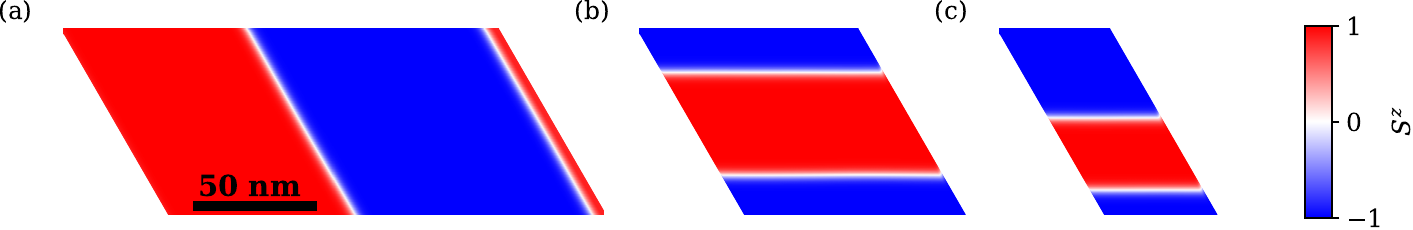}
\caption{\label{sfig:stextures_ss} Ferromagnetic domains stabilized at $0$~K in 1T-CrSI ($U_{\mathrm{eff}} = 3.0$~eV) for different system sizes: (a) $512 \times 256 \times 1$ unit cells, (b) $256 \times 256 \times 1$ unit cells, and (c) $128 \times 256 \times 1$ unit cells.}
\vspace{-0.5cm}
\end{center}
\end{figure}

\begin{figure}[h!]
\begin{center}
\includegraphics[width=1.0 \textwidth]{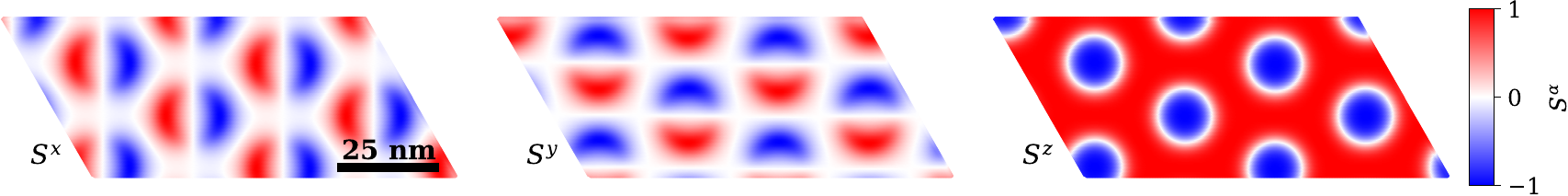}
\caption{\label{sfig:skyrmions} Skyrmion lattice formation in 1T-CrSeI after cooling in the presence of an applied magnetic field $B_z=0.5$ T. $U_{\rm{eff}}=2.1$ eV.}
\vspace{-0.5cm}
\end{center}
\end{figure}

\clearpage

\subsection{Effect of the dipolar contribution in the ASD simulations.}
\label{sssec:asd_dd}

Inclusion of the dipole-dipole interaction in the ASD simulations is computationally demanding due to its long-range nature, which significantly increases the numerical burden. We have, nevertheless, evaluated its influence for representative cases in both the $1\text{T}$ and $1\text{H}$ phases of Cr$X$I ($X = \text{S, Se}$). Our analysis confirms that the primary conclusions derived from simulations omitting this contribution remain valid. This is exemplified in Fig.~\ref{sfig:asd_dd}, which demonstrates that the estimated Curie temperature for $1\text{H}$-CrSI is insensitive to the inclusion of dipole-dipole term. 

This finding is consistent with previous studies~\cite{jenkins_breaking_2022}, which similarly concluded that the dipolar contribution to the $T_{\text{C}}$ of 2D ferromagnets is negligible, even in the limit of vanishing magnetocrystalline anisotropy. Consequently, dipole-dipole interaction was omitted in the majority of our simulations to enhance computational efficiency and facilitate the use of larger supercells, thereby minimizing finite-size effects in the determination of $T_{\text{C}}$~\cite{jenkins_breaking_2022}. Furthermore, we observed no qualitative changes in the simulated magnetic textures upon the inclusion of the dipole-dipole interaction (Fig.~\ref{sfig:asd_dd_textures}). We reiterate, however, that taking this interaction into account is crucial for the correct evaluation of the magnetic anisotropy energy.

\begin{figure}[h!]
\begin{center}
\includegraphics[width=0.5 \textwidth]{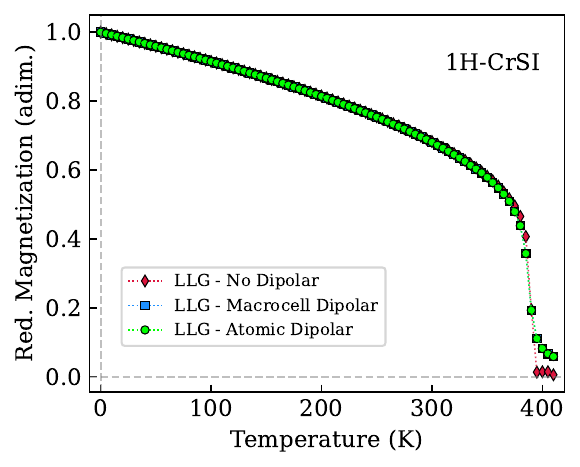}
\caption{\label{sfig:asd_dd} Temperature dependence of the reduced magnetization for $1\text{H}$-CrSI, simulated via the ASD-LLG method with and without the inclusion of the dipole-dipole interaction. Two distinct numerical treatments implemented in the VAMPIRE code~\cite{MainVampire} were evaluated: the \textit{Macrocell} approach, in which the magnetic moments of large macrocells are averaged and treated as a single unit for the calculation of the dipolar interaction, and the \textit{Atomic dipolar} approach, in which the individual dipole-dipole interactions between all spins in the lattice are explicitly considered. The simulations demonstrate that dipolar contribution has no appreciable effect on either the equilibrium magnetization or the Curie temperature.}
\vspace{-0.5cm}
\end{center}
\end{figure}

\begin{figure}[h!]
\begin{center}
\includegraphics[width=0.99\textwidth]{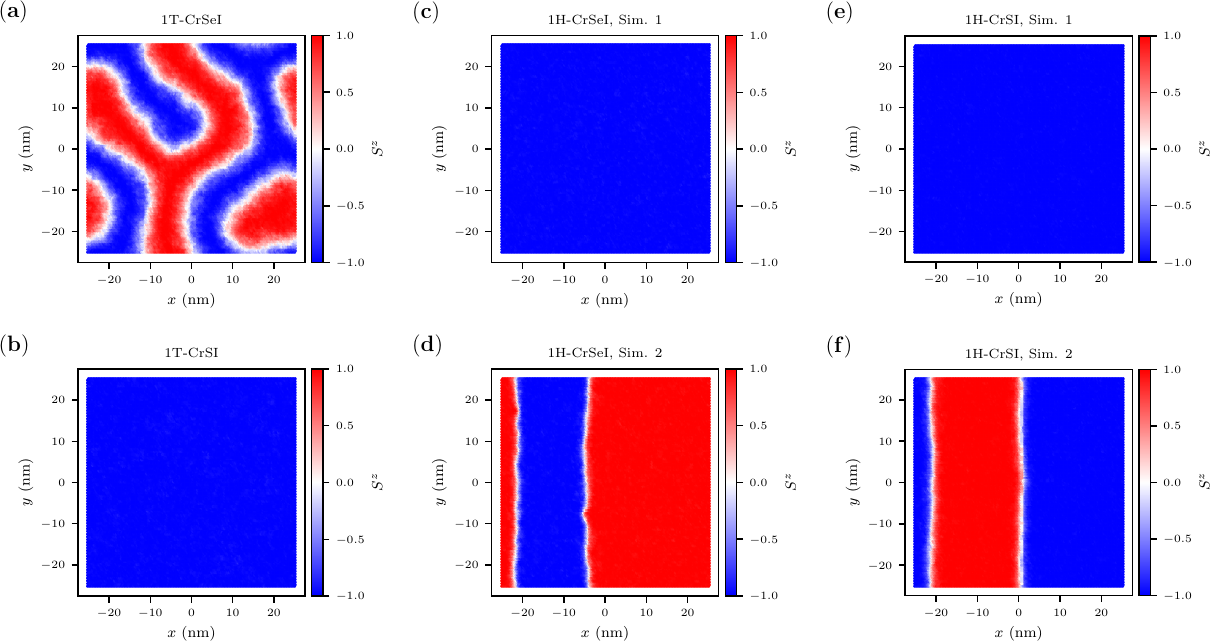}
\caption{\label{sfig:asd_dd_textures} 
Spin textures obtained by initializing all spins in random orientations and including dipole–dipole interactions for (a) 1T-CrSeI, (b) 1T-CrSI, (c,d) 1H-CrSeI, and (e,f) 1H-CrSI. The systems were allowed to relax from the initial configuration for 1 nanosecond at a constant temperature of 10 K. The dipolar interactions were computed using the atomic dipolar approach as implemented in the VAMPIRE code \cite{MainVampire}. The simulations show that all systems exhibit the same qualitative behavior as in calculations where dipolar interactions are neglected. For 1H-CrSeI and 1H-CrSI, different initialization seeds yielded distinct magnetic textures.}
\vspace{-0.5cm}
\end{center}
\end{figure}

\subsection{Comparing the LLG and Monte Carlo approaches}
\label{sssec:asd_vs_mc}

Fig.~\ref{sfig:asd_vs_mc} shows the comparison of the ASD-LLG simulations with those made using Monte Carlo method, as implemented in the open-source VAMPIRE code \cite{MainVampire}, using the same spin Hamiltonian as in the ASD simulations (Eq. \ref{eq:ham_asd_corr}), but neglecting the DMI. As seen in Fig.~\ref{sfig:asd_vs_mc}, the results of the MC simulations agree well with those of the ASD simulations, consistently to the results of such a comparison in the review article \cite{MainVampire} (see Fig. 5 of the latter reference).

\begin{figure}[h!]
\begin{center}
\includegraphics[width=0.5 \textwidth]{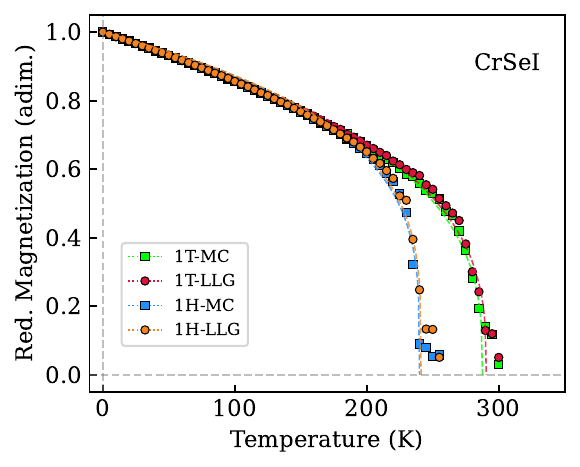}
\caption{\label{sfig:asd_vs_mc} Reduced magnetization curves as a function of temperature for 1T- and 1H-CrSeI with $U_{\rm{eff}} = 3.0$ eV, obtained from Monte Carlo (MC) and LLG based ASD simulations (LLG). Note that DMI is neglected in these simulations.}
\vspace{-0.5cm}
\end{center}
\end{figure}

\clearpage

\section{Electronic band structures of $\mathbf{Cr}X\mathbf{I}$ MLs ($X= \mathbf{S},\ \mathbf{Se}$)}

Figures~\ref{sfig:bands_1t-crsi}–\ref{sfig:bands_1h-crsei} show the calculated band structures of the 1T and 1H Cr$X$I MLs ($X = \mathrm{S}, \mathrm{Se}$). The 1T-CrSI and 1T-CrSeI systems are semiconductors with indirect band gaps of several tenths of an eV, whereas their 1H counterparts exhibit half-metallic behavior. 

In all cases shown, the exchange splittings dominate over the spin–orbit splittings, as evidenced by the spin textures displayed in panels (a) of each figure. This effect is particularly pronounced in the 1H polymorphs, whose low-energy bands exhibit stronger Cr character (see panels (b)), and which also show a higher induced magnetic moment on the $X$ and I atoms (Table~\ref{stab:magmom}) compared to the 1T phases. On the other hand, the more pronounced in-plane spin components observed in the 1T cases correlate with their stronger DMI.

\begin{figure}[h!]
    \centering    
    \includegraphics[width=0.9\textwidth]{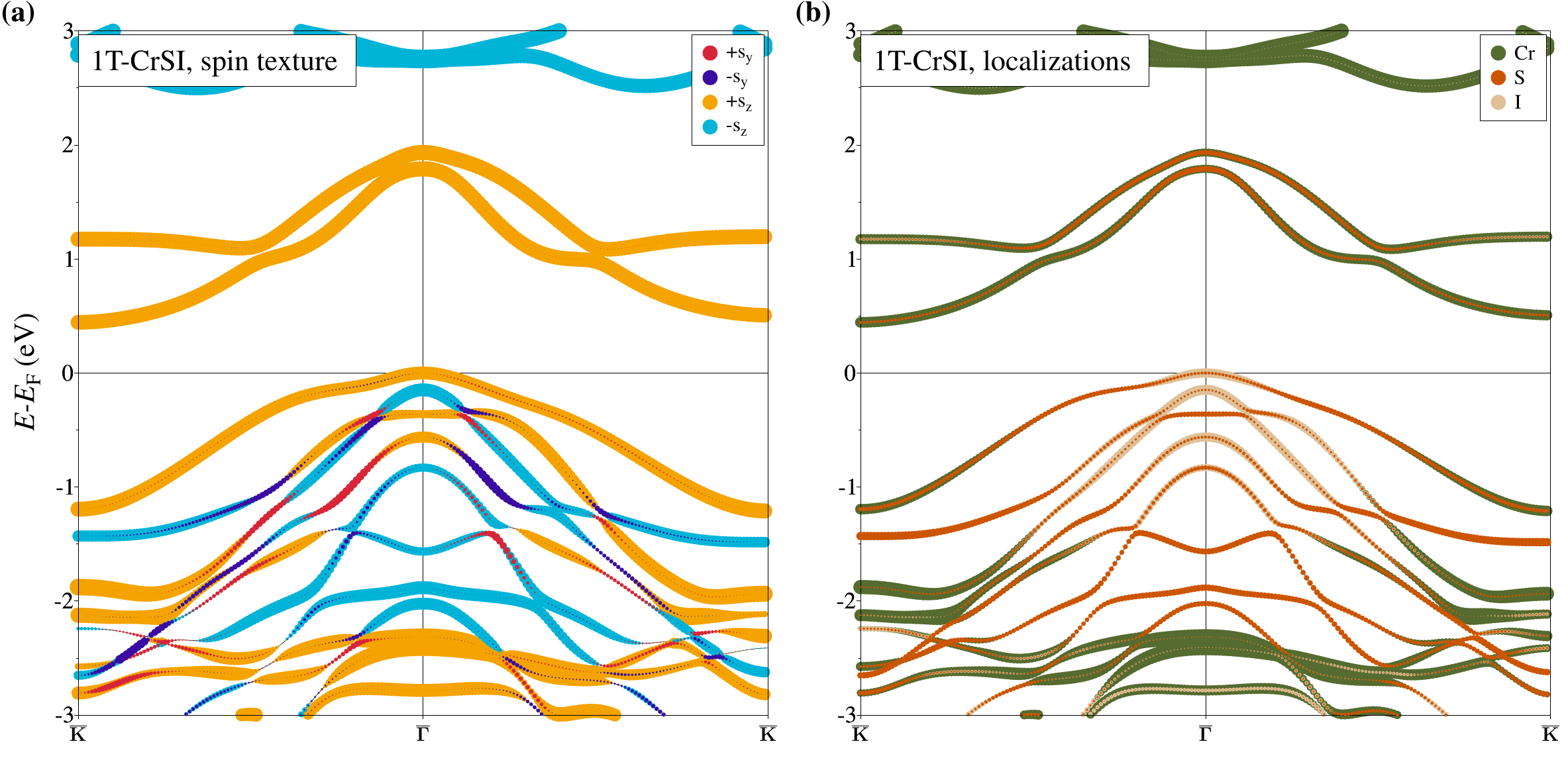}
    \caption{Calculated low-energy electronic band structure of the FM 1T-CrSI ML, resolved by spin (a) and atomic character (b). $U_\text{eff} = 3$~eV.}
    \label{sfig:bands_1t-crsi}
\end{figure}

\begin{figure}[h!]
    \centering    
    \includegraphics[width=0.9\textwidth]{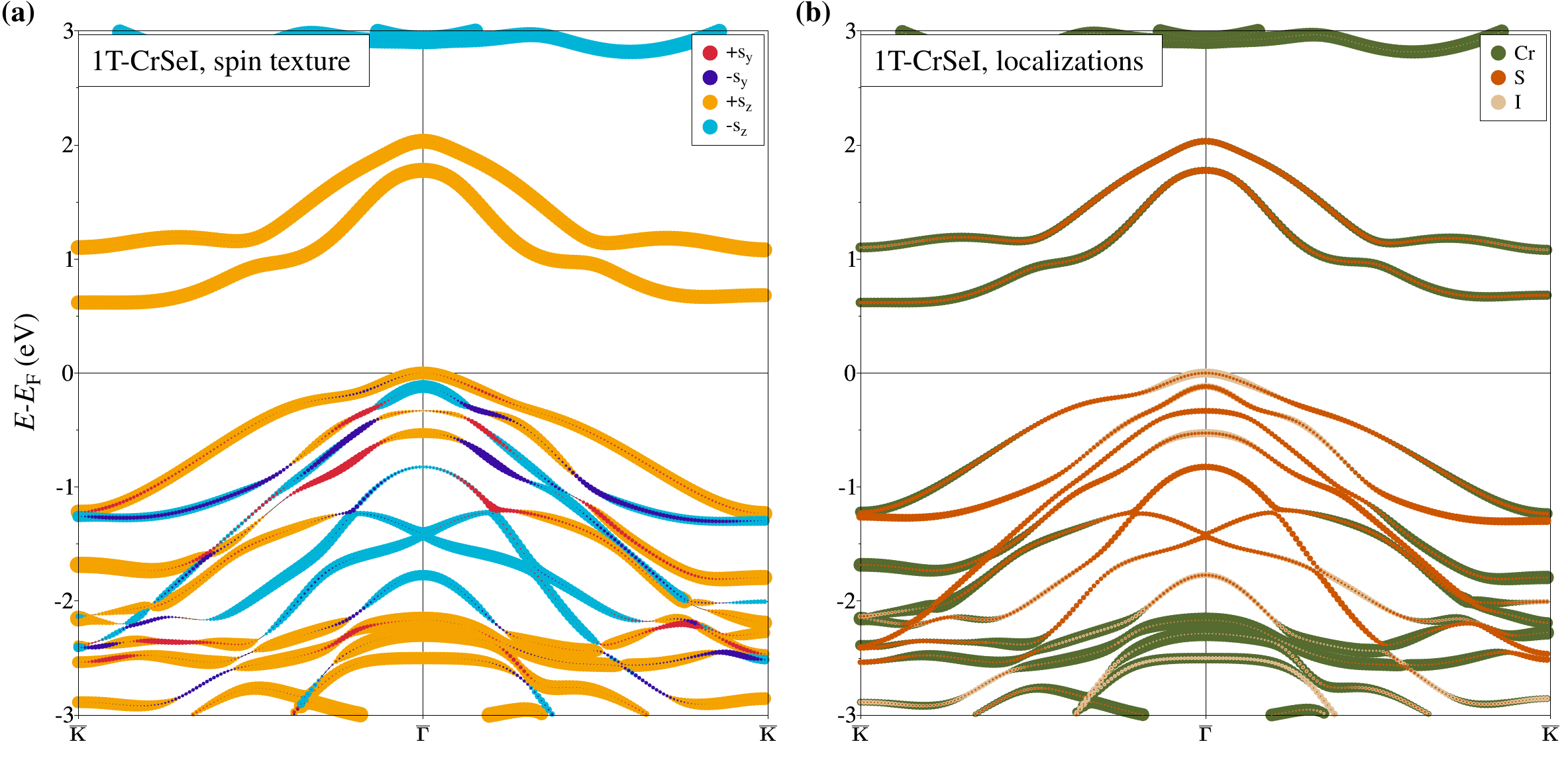}
    \caption{Calculated low-energy electronic band structure of the FM 1T-CrSeI ML, resolved by spin (a) and atomic character (b). $U_\text{eff} = 3$~eV.}
    \label{sfig:bands_1t-crsei}
\end{figure}

\begin{figure}[h!]
    \centering    
    \includegraphics[width=0.9\textwidth]{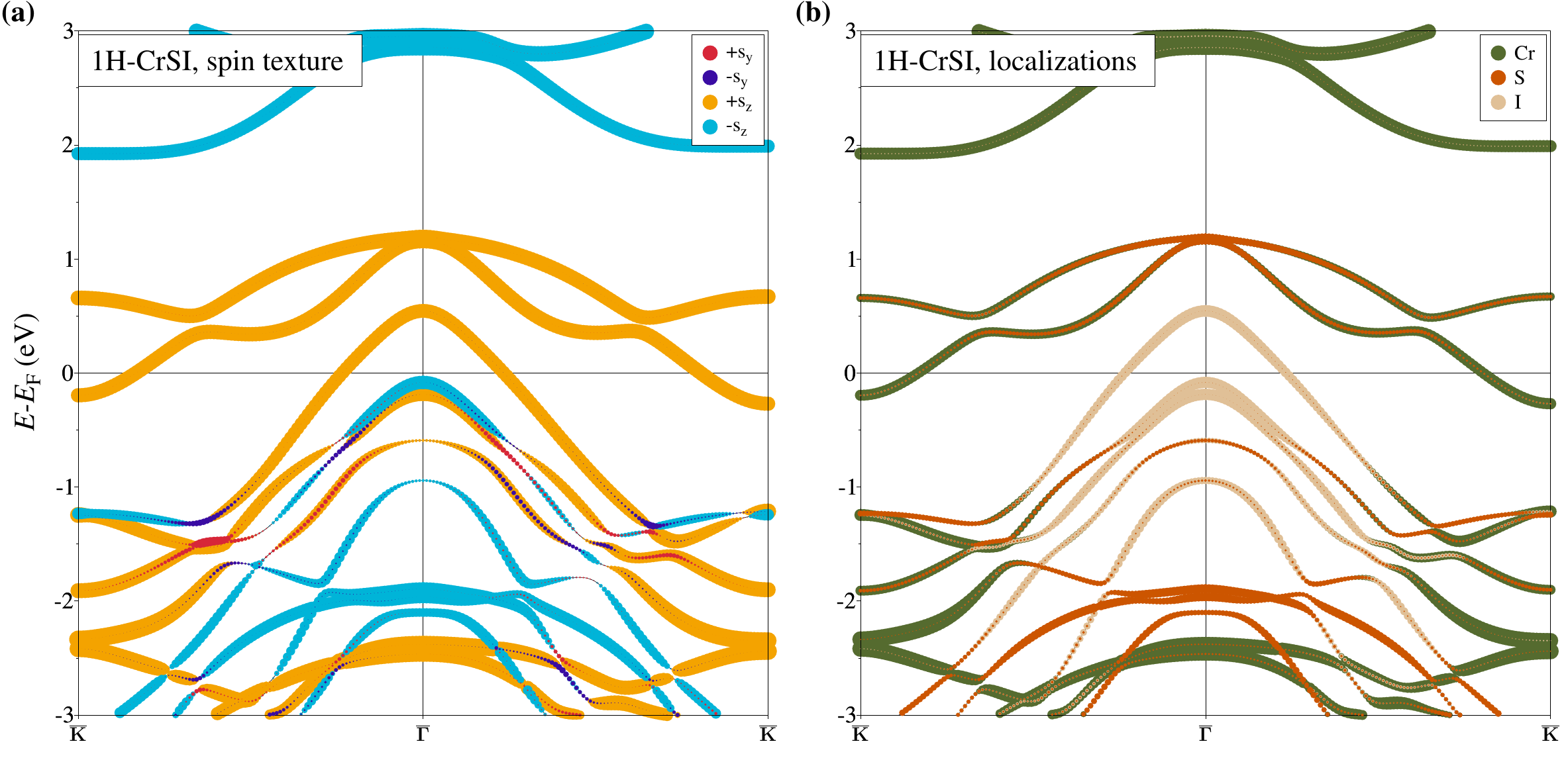}
    \caption{Calculated low-energy electronic band structure of the FM 1H-CrSI ML, resolved by spin (a) and atomic character (b). $U_\text{eff} = 3$~eV.}
    \label{sfig:bands_1h-crsi}
\end{figure}

\begin{figure}[h!]
    \centering    
    \includegraphics[width=0.9\textwidth]{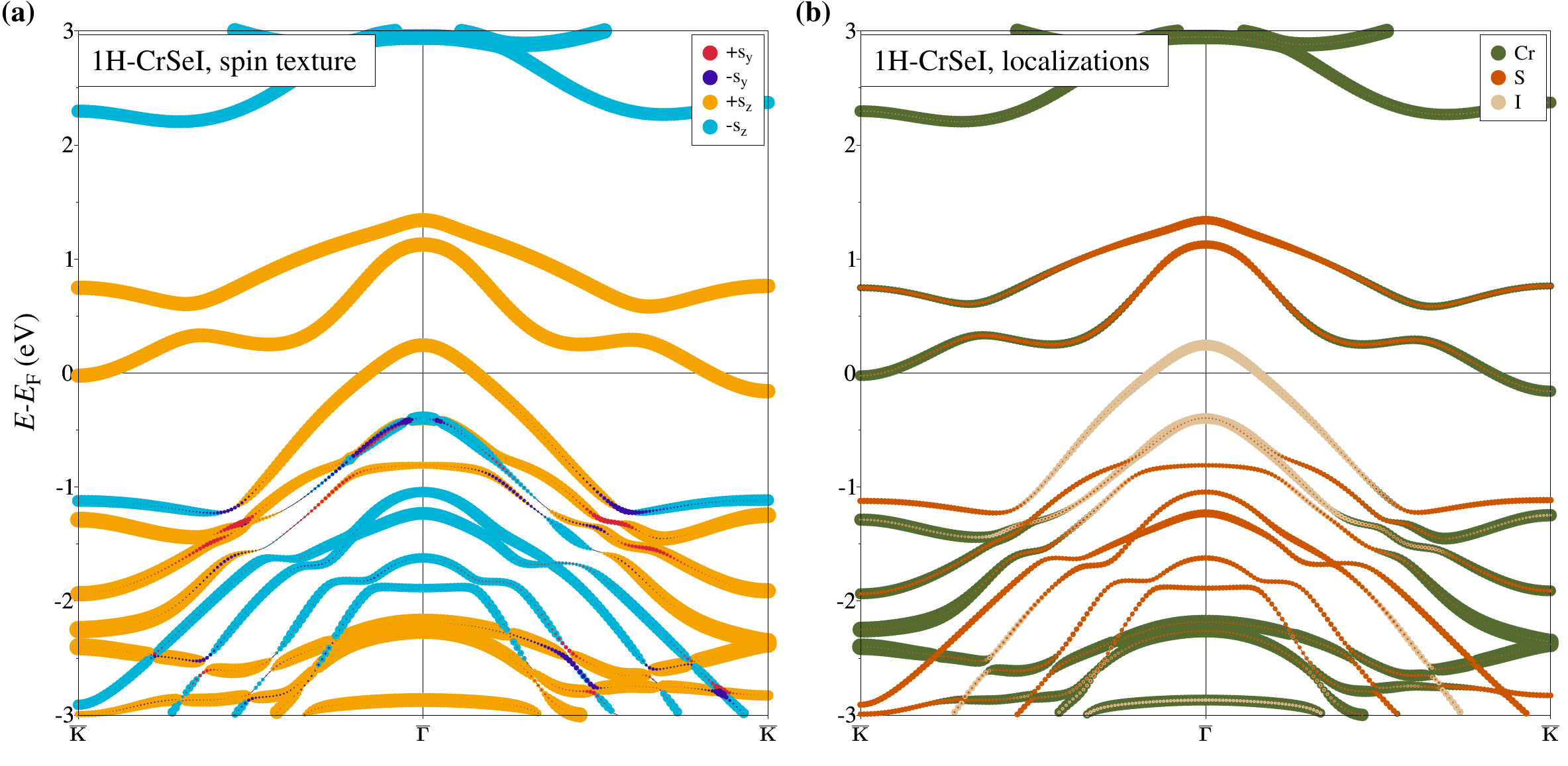}
    \caption{Calculated low-energy electronic band structure of the FM 1H-CrSeI ML, resolved by spin (a) and atomic character (b). $U_\text{eff} = 3$~eV.}
    \label{sfig:bands_1h-crsei}
\end{figure}

\clearpage

The band structures of all \CXY\ MLs studied here (except 1H tellurides), calculated as a function of the magnetization direction and $U_\text{eff}$ value, are shown in Figs. \ref{sfig:bands_1t-crsy}-\ref{sfig:bands_1t-crtey} below.

\begin{figure}[h!]
    \centering    
    \includegraphics[width=0.99\textwidth]{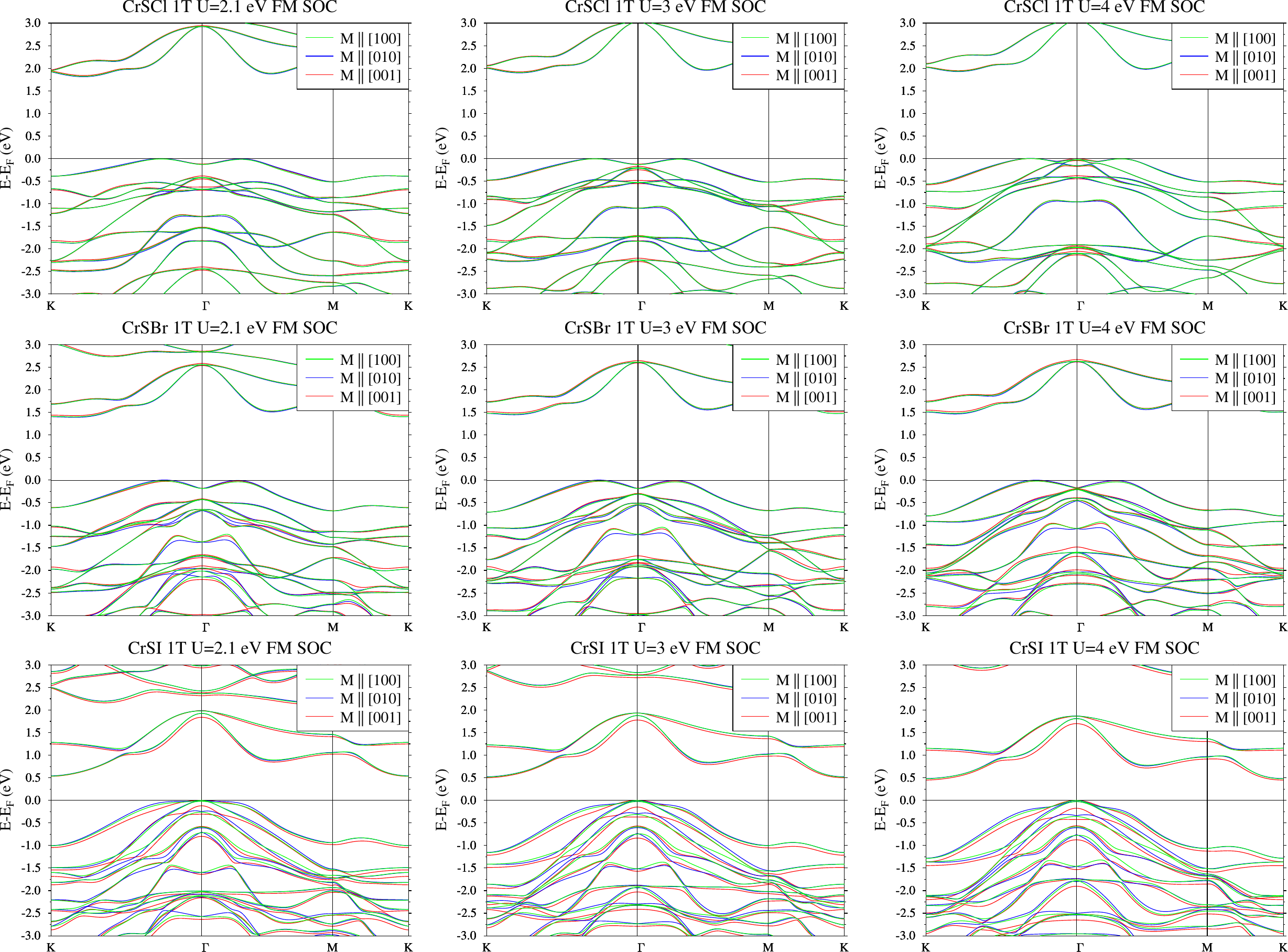}
    \caption{Calculated low-energy electronic band structure of the 1T-CrS$Y$ MLs for the FM state with the magnetization parallel to  $x$ (green), $y$ (blue), and $z$ (red) directions for different $U_\text{eff}$.}
    \label{sfig:bands_1t-crsy}
\end{figure}

\begin{figure}[h!]
    \centering    
    \includegraphics[width=0.99\textwidth]{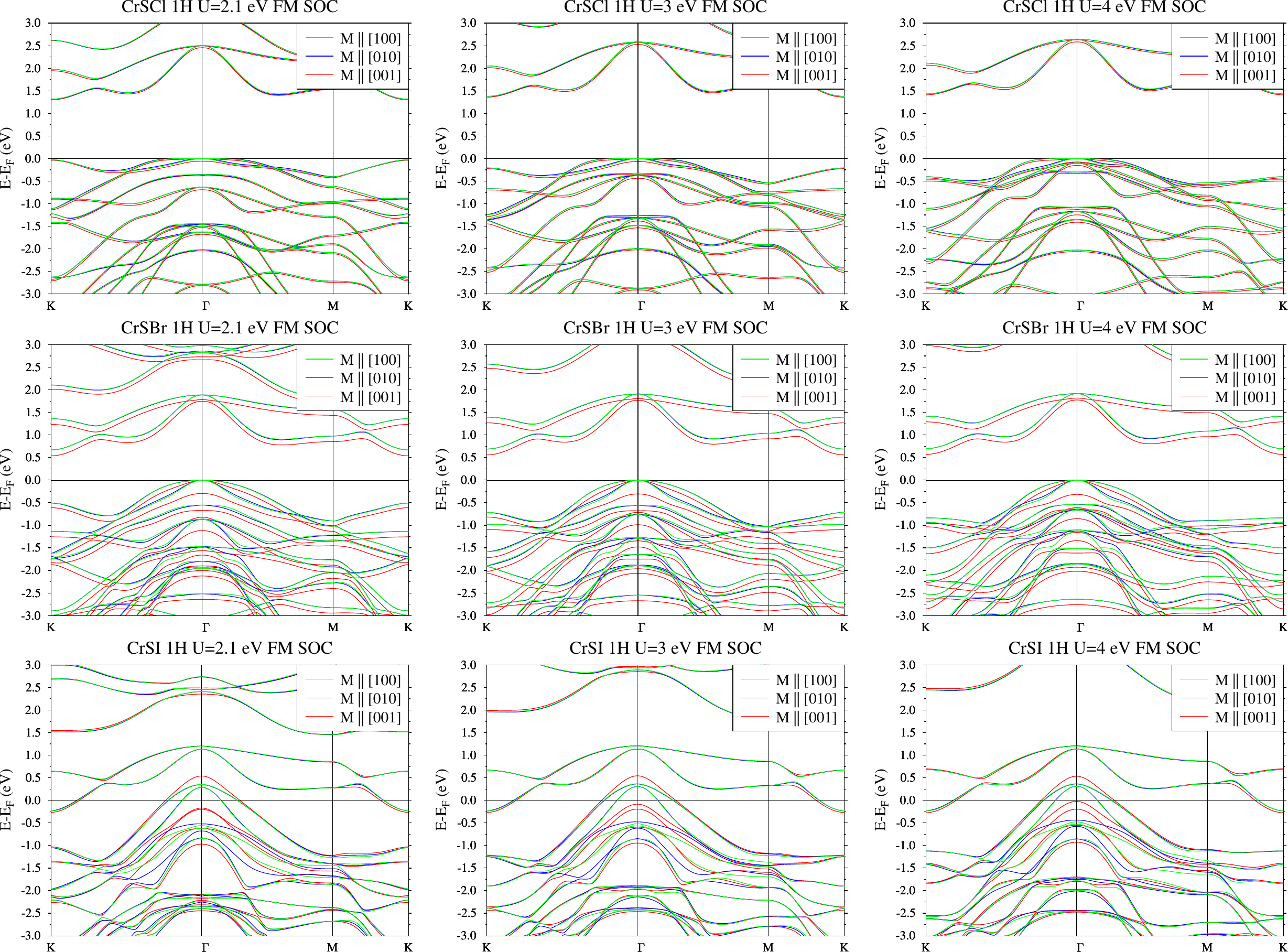}
        \caption{The same as Fig. \ref{sfig:bands_1t-crsy}, but for the 1H-CrS$Y$ MLs}
    \label{sfig:bands_1h-crsy}
\end{figure}

\begin{figure}[h!]
    \centering    
    \includegraphics[width=0.99\textwidth]{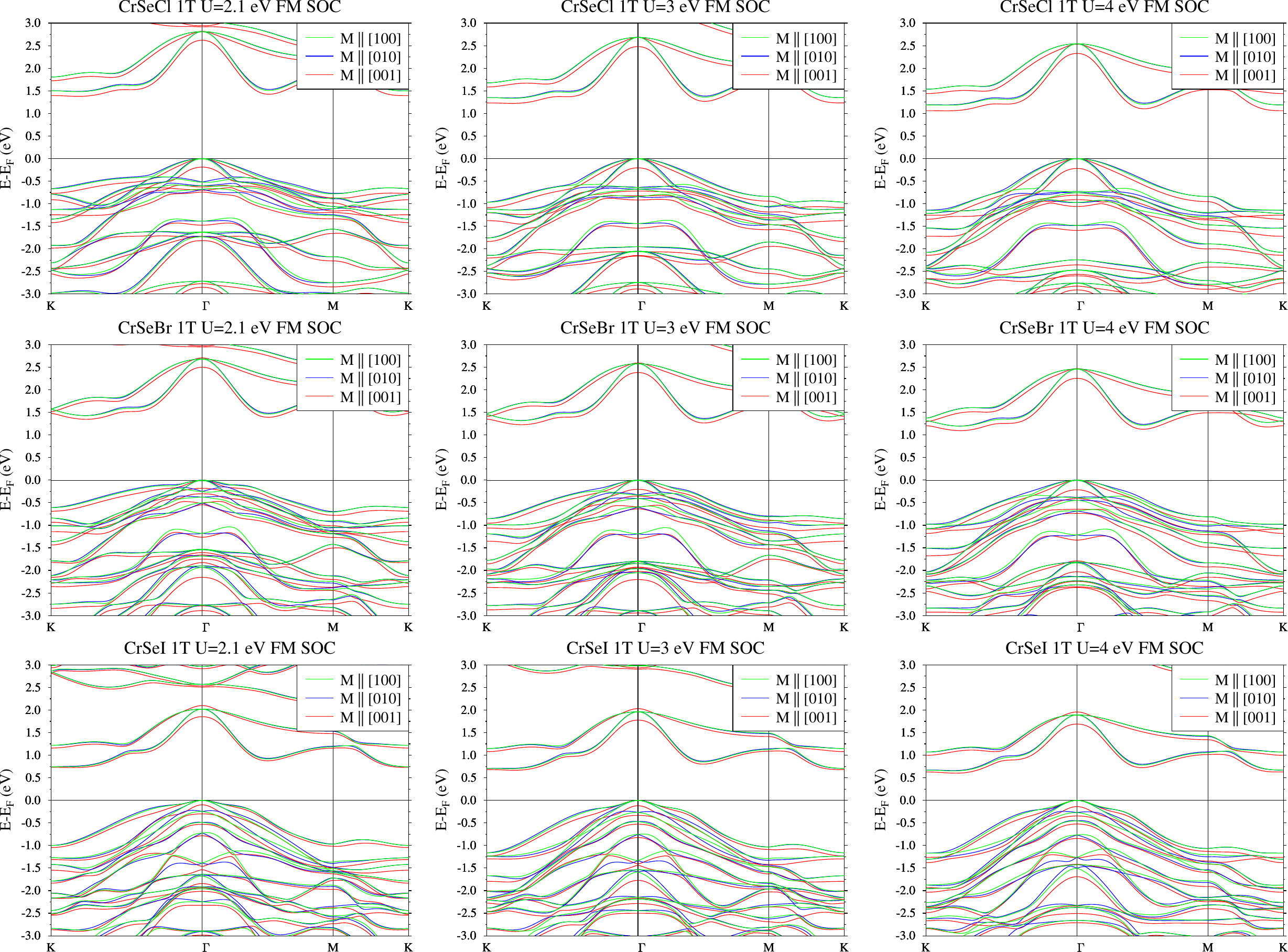}
    \caption{Calculated low-energy electronic band structure of the 1T-CrSe$Y$ MLs for the FM state with the magnetization parallel to  $x$ (green), $y$ (blue), and $z$ (red) directions for different $U_\text{eff}$.}
    \label{sfig:bands_1t-crsey}
\end{figure}

\begin{figure}[h!]
    \centering    
    \includegraphics[width=0.99\textwidth]{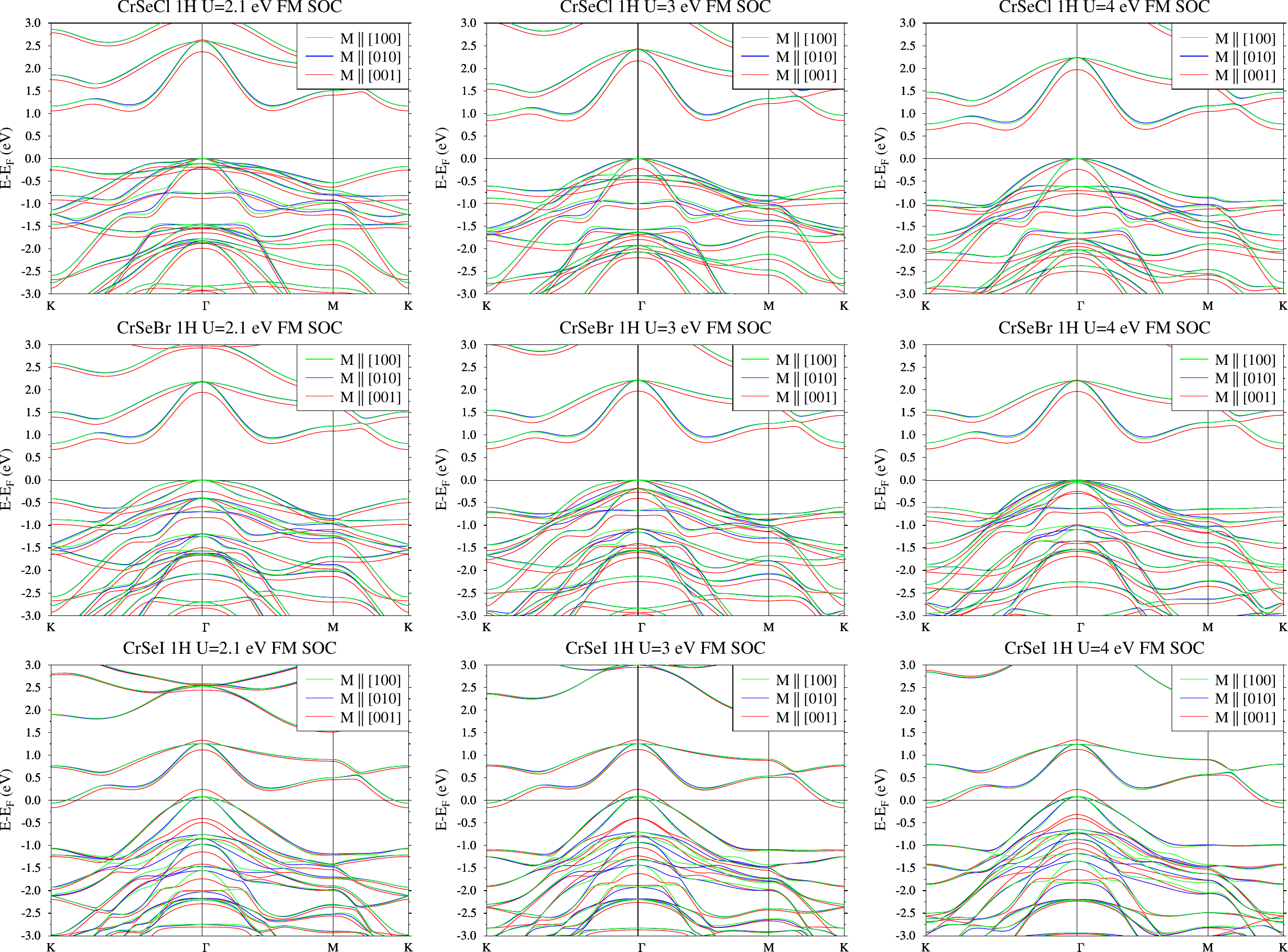}
        \caption{The same as Fig. \ref{sfig:bands_1t-crsy}, but for the 1H-CrSe$Y$ MLs}
    \label{sfig:bands_1h-crsey}
\end{figure}

\begin{figure}[h!]
    \centering    
    \includegraphics[width=0.99\textwidth]{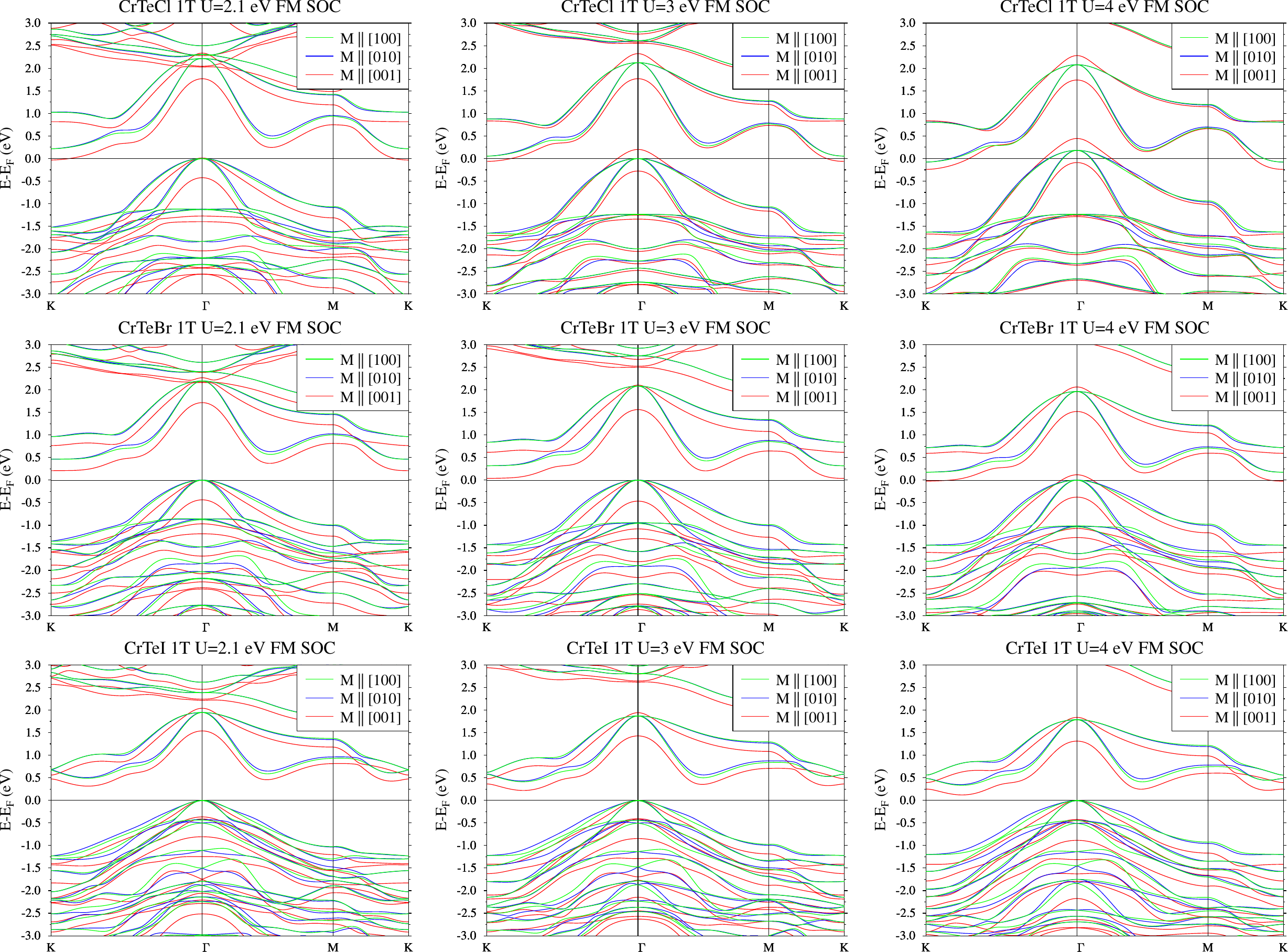}
    \caption{Calculated low-energy electronic band structure of the 1T-CrTe$Y$ MLs for the FM state with the magnetization parallel to  $x$ (green), $y$ (blue), and $z$ (red) directions for different $U_\text{eff}$. Note that for $U_\text{eff}=2.1$ and 3 eV, 1T-CrTeCl has Cr moments locked in the $xy$ plane, while for $U_\text{eff}=4$ eV they point along $z$, see Table \ref{stab:mae}. Therefore, at $U_\text{eff}=2.1, 3$ eV the spectrum of 1T-CrTeCl is gapped (green or blue lines), while at $U_\text{eff}=4$ eV it is gapless (red lines).}
    \label{sfig:bands_1t-crtey}
\end{figure}

\clearpage\newpage


\begin{figure}[h!]
    \centering    
    \includegraphics[width=0.75\textwidth]{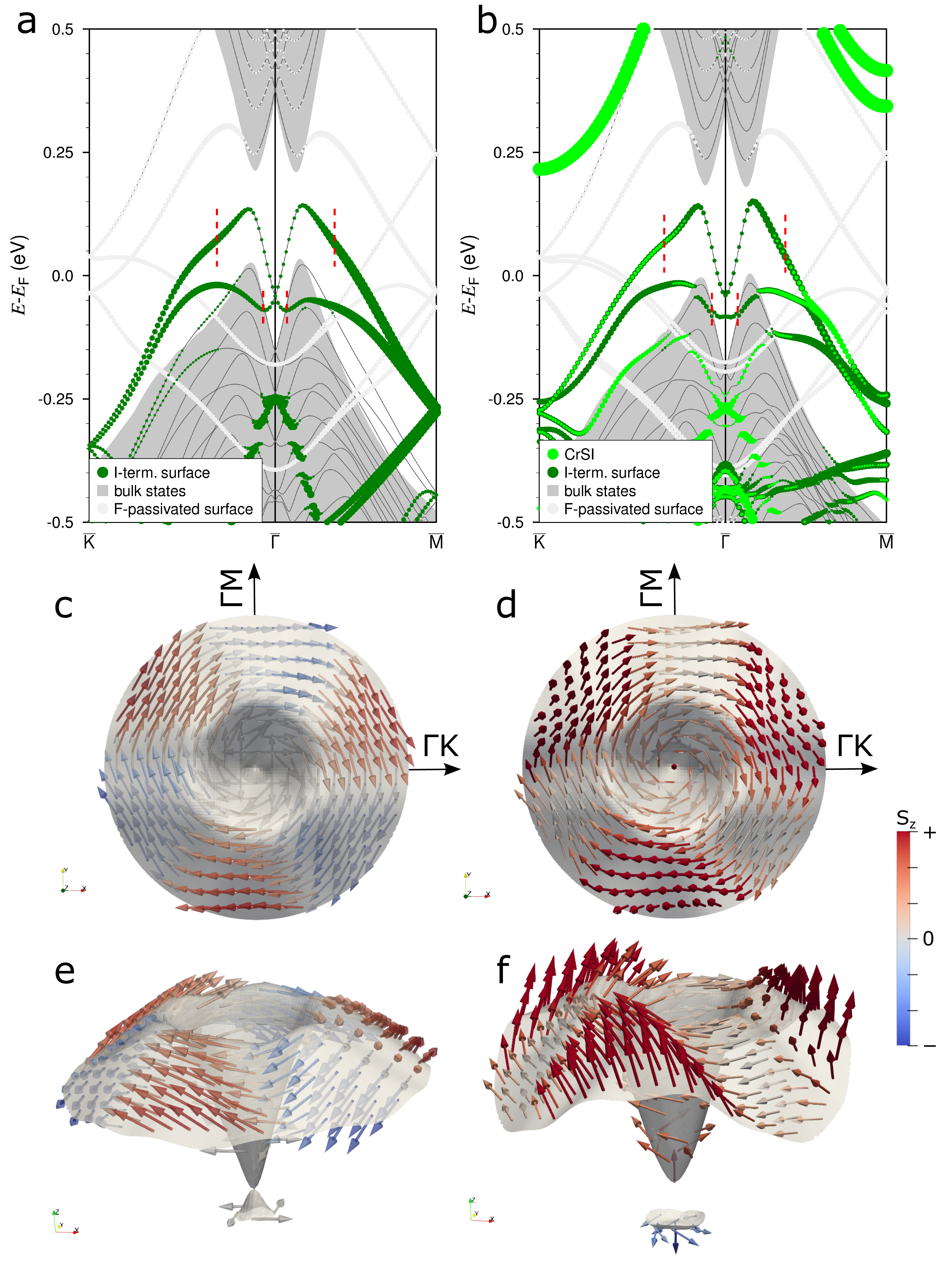}
        \caption{Spin texture of the Bychkov-Rashba surface state on the clean $\mathrm{BiTeI}(0001)$ surface and at the $\mathrm{1T-CrSI/BiTeI}-(2/\sqrt{3}\times2/\sqrt{3})$ interface. Panels (a) and (b) show surface band structures of $\mathrm{BiTeI}(001)$ and $\mathrm{CrSI/BiTeI}-(2/\sqrt{3}\times2/\sqrt{3})$. Dark green circles in both panels mark the states localized within topmost trilayer of the I-terminated surface while light green circles in (b) show the states localized within the CrSI Janus film. Darker and lighter gray colors mark the bulk continuum states and surface states localized on the opposite, fluorine-passivated side of the slab, respectively. Vertical red dashed line segments delimit the Bychkov-Rashba state areas for which the spin texture is shown in top (c,d) and side (e,f) views for the clean surface (c,e) and the interface (d,f).}
    \label{sfig:spin_texture}
\end{figure}

\clearpage\newpage

\begin{theSIbibliography}{99}
  \setcounter{enumiv}{\value{lastbib}}

\expandafter\ifx\csname natexlab\endcsname\relax\def\natexlab#1{#1}\fi
\expandafter\ifx\csname bibnamefont\endcsname\relax
  \def\bibnamefont#1{#1}\fi
\expandafter\ifx\csname bibfnamefont\endcsname\relax
  \def\bibfnamefont#1{#1}\fi
\expandafter\ifx\csname citenamefont\endcsname\relax
  \def\citenamefont#1{#1}\fi
\expandafter\ifx\csname url\endcsname\relax
  \def\url#1{\texttt{#1}}\fi
\expandafter\ifx\csname urlprefix\endcsname\relax\def\urlprefix{URL }\fi
\providecommand{\bibinfo}[2]{#2}
\providecommand{\eprint}[2][]{\url{#2}}

\bibitem[{\citenamefont{Szunyogh et~al.}(1995)\citenamefont{Szunyogh,
  \'Ujfalussy, and Weinberger}}]{Szunyogh.prb1995}
\bibinfo{author}{\bibfnamefont{L.}~\bibnamefont{Szunyogh}},
  \bibinfo{author}{\bibfnamefont{B.}~\bibnamefont{\'Ujfalussy}},
  \bibnamefont{and}
  \bibinfo{author}{\bibfnamefont{P.}~\bibnamefont{Weinberger}},
  \bibinfo{journal}{Phys. Rev. B} \textbf{\bibinfo{volume}{51}},
  \bibinfo{pages}{9552} (\bibinfo{year}{1995}).

\bibitem[{\citenamefont{Lado and Fern{\'a}ndez-Rossier}(2017)}]{Lado.2dmat2017}
\bibinfo{author}{\bibfnamefont{J.~L.} \bibnamefont{Lado}} \bibnamefont{and}
  \bibinfo{author}{\bibfnamefont{J.}~\bibnamefont{Fern{\'a}ndez-Rossier}},
  \bibinfo{journal}{2D Mater.} \textbf{\bibinfo{volume}{4}},
  \bibinfo{pages}{035002} (\bibinfo{year}{2017}).

\bibitem[{\citenamefont{Moriya}(1960)}]{Moriya.pr1960}
\bibinfo{author}{\bibfnamefont{T.}~\bibnamefont{Moriya}},
  \bibinfo{journal}{Phys. Rev.} \textbf{\bibinfo{volume}{120}},
  \bibinfo{pages}{91} (\bibinfo{year}{1960}).

\bibitem[{\citenamefont{Nowak}(2007)}]{nowak_classical_2007}
\bibinfo{author}{\bibfnamefont{U.}~\bibnamefont{Nowak}},
  \emph{\bibinfo{title}{Classical {Spin} {Models}, in Handbook of {Magnetism}
  and {Advanced} {Magnetic} {Materials}}} (\bibinfo{publisher}{John Wiley \&
  Sons, New York}, \bibinfo{year}{2007}).

\bibitem[{\citenamefont{Jenkins et~al.}(2022)\citenamefont{Jenkins, Rózsa,
  Atxitia, Evans, Novoselov, and Santos}}]{jenkins_breaking_2022}
\bibinfo{author}{\bibfnamefont{S.}~\bibnamefont{Jenkins}},
  \bibinfo{author}{\bibfnamefont{L.}~\bibnamefont{Rózsa}},
  \bibinfo{author}{\bibfnamefont{U.}~\bibnamefont{Atxitia}},
  \bibinfo{author}{\bibfnamefont{R.~F.~L.} \bibnamefont{Evans}},
  \bibinfo{author}{\bibfnamefont{K.~S.} \bibnamefont{Novoselov}},
  \bibnamefont{and} \bibinfo{author}{\bibfnamefont{E.~J.~G.}
  \bibnamefont{Santos}}, \bibinfo{journal}{Nat. Commun.}
  \textbf{\bibinfo{volume}{13}}, \bibinfo{pages}{6917} (\bibinfo{year}{2022}).

\bibitem[{\citenamefont{Evans et~al.}(2014)\citenamefont{Evans, Fan,
  Chureemart, Ostler, Ellis, and Chantrell}}]{MainVampire}
\bibinfo{author}{\bibfnamefont{R.~F.~L.} \bibnamefont{Evans}},
  \bibinfo{author}{\bibfnamefont{W.~J.} \bibnamefont{Fan}},
  \bibinfo{author}{\bibfnamefont{P.}~\bibnamefont{Chureemart}},
  \bibinfo{author}{\bibfnamefont{T.~A.} \bibnamefont{Ostler}},
  \bibinfo{author}{\bibfnamefont{M.~O.~A.} \bibnamefont{Ellis}},
  \bibnamefont{and} \bibinfo{author}{\bibfnamefont{R.~W.}
  \bibnamefont{Chantrell}}, \bibinfo{journal}{Journal of Physics: Condensed
  Matter} \textbf{\bibinfo{volume}{26}}, \bibinfo{pages}{103202}
  (\bibinfo{year}{2014}).

\end{theSIbibliography}


\begin{thebibliography}{120}%
\makeatletter
\providecommand \@ifxundefined [1]{%
 \@ifx{#1\undefined}
}%
\providecommand \@ifnum [1]{%
 \ifnum #1\expandafter \@firstoftwo
 \else \expandafter \@secondoftwo
 \fi
}%
\providecommand \@ifx [1]{%
 \ifx #1\expandafter \@firstoftwo
 \else \expandafter \@secondoftwo
 \fi
}%
\providecommand \natexlab [1]{#1}%
\providecommand \enquote  [1]{``#1''}%
\providecommand \bibnamefont  [1]{#1}%
\providecommand \bibfnamefont [1]{#1}%
\providecommand \citenamefont [1]{#1}%
\providecommand \href@noop [0]{\@secondoftwo}%
\providecommand \href [0]{\begingroup \@sanitize@url \@href}%
\providecommand \@href[1]{\@@startlink{#1}\@@href}%
\providecommand \@@href[1]{\endgroup#1\@@endlink}%
\providecommand \@sanitize@url [0]{\catcode `\\12\catcode `\$12\catcode
  `\&12\catcode `\#12\catcode `\^12\catcode `\_12\catcode `\%12\relax}%
\providecommand \@@startlink[1]{}%
\providecommand \@@endlink[0]{}%
\providecommand \url  [0]{\begingroup\@sanitize@url \@url }%
\providecommand \@url [1]{\endgroup\@href {#1}{\urlprefix }}%
\providecommand \urlprefix  [0]{URL }%
\providecommand \Eprint [0]{\href }%
\providecommand \doibase [0]{https://doi.org/}%
\providecommand \selectlanguage [0]{\@gobble}%
\providecommand \bibinfo  [0]{\@secondoftwo}%
\providecommand \bibfield  [0]{\@secondoftwo}%
\providecommand \translation [1]{[#1]}%
\providecommand \BibitemOpen [0]{}%
\providecommand \bibitemStop [0]{}%
\providecommand \bibitemNoStop [0]{.\EOS\space}%
\providecommand \EOS [0]{\spacefactor3000\relax}%
\providecommand \BibitemShut  [1]{\csname bibitem#1\endcsname}%
\let\auto@bib@innerbib\@empty
\bibitem [{\citenamefont {Huang}\ \emph {et~al.}(2017)\citenamefont {Huang},
  \citenamefont {Clark}, \citenamefont {Navarro-Moratalla}, \citenamefont
  {Klein}, \citenamefont {Cheng}, \citenamefont {Seyler}, \citenamefont
  {Zhong}, \citenamefont {Schmidgall}, \citenamefont {McGuire}, \citenamefont
  {Cobden} \emph {et~al.}}]{Huang.nat2017}%
  \BibitemOpen
  \bibfield  {author} {\bibinfo {author} {\bibfnamefont {B.}~\bibnamefont
  {Huang}}, \bibinfo {author} {\bibfnamefont {G.}~\bibnamefont {Clark}},
  \bibinfo {author} {\bibfnamefont {E.}~\bibnamefont {Navarro-Moratalla}},
  \bibinfo {author} {\bibfnamefont {D.~R.}\ \bibnamefont {Klein}}, \bibinfo
  {author} {\bibfnamefont {R.}~\bibnamefont {Cheng}}, \bibinfo {author}
  {\bibfnamefont {K.~L.}\ \bibnamefont {Seyler}}, \bibinfo {author}
  {\bibfnamefont {D.}~\bibnamefont {Zhong}}, \bibinfo {author} {\bibfnamefont
  {E.}~\bibnamefont {Schmidgall}}, \bibinfo {author} {\bibfnamefont {M.~A.}\
  \bibnamefont {McGuire}}, \bibinfo {author} {\bibfnamefont {D.~H.}\
  \bibnamefont {Cobden}}, \emph {et~al.},\ }\bibfield  {title} {\bibinfo
  {title} {Layer-dependent ferromagnetism in a van der {{Waals}} crystal down
  to the monolayer limit},\ }\href
  {https://doi.org/https://doi.org/10.1038/nature22391} {\bibfield  {journal}
  {\bibinfo  {journal} {Nature}\ }\textbf {\bibinfo {volume} {546}},\ \bibinfo
  {pages} {270} (\bibinfo {year} {2017})}\BibitemShut {NoStop}%
\bibitem [{\citenamefont {Gong}\ \emph {et~al.}(2017)\citenamefont {Gong},
  \citenamefont {Li}, \citenamefont {Li}, \citenamefont {Ji}, \citenamefont
  {Stern}, \citenamefont {Xia}, \citenamefont {Cao}, \citenamefont {Bao},
  \citenamefont {Wang}, \citenamefont {Wang} \emph {et~al.}}]{Gong.nat2017}%
  \BibitemOpen
  \bibfield  {author} {\bibinfo {author} {\bibfnamefont {C.}~\bibnamefont
  {Gong}}, \bibinfo {author} {\bibfnamefont {L.}~\bibnamefont {Li}}, \bibinfo
  {author} {\bibfnamefont {Z.}~\bibnamefont {Li}}, \bibinfo {author}
  {\bibfnamefont {H.}~\bibnamefont {Ji}}, \bibinfo {author} {\bibfnamefont
  {A.}~\bibnamefont {Stern}}, \bibinfo {author} {\bibfnamefont
  {Y.}~\bibnamefont {Xia}}, \bibinfo {author} {\bibfnamefont {T.}~\bibnamefont
  {Cao}}, \bibinfo {author} {\bibfnamefont {W.}~\bibnamefont {Bao}}, \bibinfo
  {author} {\bibfnamefont {C.}~\bibnamefont {Wang}}, \bibinfo {author}
  {\bibfnamefont {Y.}~\bibnamefont {Wang}}, \emph {et~al.},\ }\bibfield
  {title} {\bibinfo {title} {Discovery of intrinsic ferromagnetism in
  two-dimensional van der {{Waals}} crystals},\ }\href@noop {} {\bibfield
  {journal} {\bibinfo  {journal} {Nature}\ }\textbf {\bibinfo {volume} {546}},\
  \bibinfo {pages} {265} (\bibinfo {year} {2017})}\BibitemShut {NoStop}%
\bibitem [{\citenamefont {Mermin}\ and\ \citenamefont
  {Wagner}(1966)}]{Mermin.prl1966}%
  \BibitemOpen
  \bibfield  {author} {\bibinfo {author} {\bibfnamefont {N.~D.}\ \bibnamefont
  {Mermin}}\ and\ \bibinfo {author} {\bibfnamefont {H.}~\bibnamefont
  {Wagner}},\ }\bibfield  {title} {\bibinfo {title} {Absence of ferromagnetism
  or antiferromagnetism in one- or two-dimensional isotropic {Heisenberg}
  models},\ }\href {https://doi.org/10.1103/PhysRevLett.17.1133} {\bibfield
  {journal} {\bibinfo  {journal} {Phys. Rev. Lett.}\ }\textbf {\bibinfo
  {volume} {17}},\ \bibinfo {pages} {1133} (\bibinfo {year}
  {1966})}\BibitemShut {NoStop}%
\bibitem [{\citenamefont {Wang}\ \emph {et~al.}(2022)\citenamefont {Wang},
  \citenamefont {Bedoya-Pinto}, \citenamefont {Blei}, \citenamefont {Dismukes},
  \citenamefont {Hamo}, \citenamefont {Jenkins}, \citenamefont {Koperski},
  \citenamefont {Liu}, \citenamefont {Sun}, \citenamefont {Telford} \emph
  {et~al.}}]{Wang.acsn2022}%
  \BibitemOpen
  \bibfield  {author} {\bibinfo {author} {\bibfnamefont {Q.~H.}\ \bibnamefont
  {Wang}}, \bibinfo {author} {\bibfnamefont {A.}~\bibnamefont {Bedoya-Pinto}},
  \bibinfo {author} {\bibfnamefont {M.}~\bibnamefont {Blei}}, \bibinfo {author}
  {\bibfnamefont {A.~H.}\ \bibnamefont {Dismukes}}, \bibinfo {author}
  {\bibfnamefont {A.}~\bibnamefont {Hamo}}, \bibinfo {author} {\bibfnamefont
  {S.}~\bibnamefont {Jenkins}}, \bibinfo {author} {\bibfnamefont
  {M.}~\bibnamefont {Koperski}}, \bibinfo {author} {\bibfnamefont
  {Y.}~\bibnamefont {Liu}}, \bibinfo {author} {\bibfnamefont {Q.-C.}\
  \bibnamefont {Sun}}, \bibinfo {author} {\bibfnamefont {E.~J.}\ \bibnamefont
  {Telford}}, \emph {et~al.},\ }\bibfield  {title} {\bibinfo {title} {The
  magnetic genome of two-dimensional van der {Waals} materials},\ }\href
  {https://doi.org/10.1021/acsnano.1c09150} {\bibfield  {journal} {\bibinfo
  {journal} {ACS Nano}\ }\textbf {\bibinfo {volume} {16}},\ \bibinfo {pages}
  {6960} (\bibinfo {year} {2022})}\BibitemShut {NoStop}%
\bibitem [{\citenamefont {Bychkov}\ and\ \citenamefont
  {Rashba}(1984)}]{BychkovRashba}%
  \BibitemOpen
  \bibfield  {author} {\bibinfo {author} {\bibfnamefont {Y.~A.}\ \bibnamefont
  {Bychkov}}\ and\ \bibinfo {author} {\bibfnamefont {E.~I.}\ \bibnamefont
  {Rashba}},\ }\bibfield  {title} {\bibinfo {title} {Properties of a {2D}
  electron gas with lifted spectral degeneracy},\ }\href@noop {} {\bibfield
  {journal} {\bibinfo  {journal} {JETP Lett}\ }\textbf {\bibinfo {volume}
  {39}},\ \bibinfo {pages} {78} (\bibinfo {year} {1984})}\BibitemShut {NoStop}%
\bibitem [{\citenamefont {Miron}\ \emph {et~al.}(2011)\citenamefont {Miron},
  \citenamefont {Garello}, \citenamefont {Gaudin}, \citenamefont {Zermatten},
  \citenamefont {Costache}, \citenamefont {Auffret}, \citenamefont {Bandiera},
  \citenamefont {Rodmacq}, \citenamefont {Schuhl},\ and\ \citenamefont
  {Gambardella}}]{Miron.nat2011}%
  \BibitemOpen
  \bibfield  {author} {\bibinfo {author} {\bibfnamefont {I.~M.}\ \bibnamefont
  {Miron}}, \bibinfo {author} {\bibfnamefont {K.}~\bibnamefont {Garello}},
  \bibinfo {author} {\bibfnamefont {G.}~\bibnamefont {Gaudin}}, \bibinfo
  {author} {\bibfnamefont {P.-J.}\ \bibnamefont {Zermatten}}, \bibinfo {author}
  {\bibfnamefont {M.~V.}\ \bibnamefont {Costache}}, \bibinfo {author}
  {\bibfnamefont {S.}~\bibnamefont {Auffret}}, \bibinfo {author} {\bibfnamefont
  {S.}~\bibnamefont {Bandiera}}, \bibinfo {author} {\bibfnamefont
  {B.}~\bibnamefont {Rodmacq}}, \bibinfo {author} {\bibfnamefont
  {A.}~\bibnamefont {Schuhl}},\ and\ \bibinfo {author} {\bibfnamefont
  {P.}~\bibnamefont {Gambardella}},\ }\bibfield  {title} {\bibinfo {title}
  {Perpendicular switching of a single ferromagnetic layer induced by in-plane
  current injection},\ }\href {https://doi.org/10.1038/s41598-017-03507-0}
  {\bibfield  {journal} {\bibinfo  {journal} {Nature}\ }\textbf {\bibinfo
  {volume} {476}},\ \bibinfo {pages} {189} (\bibinfo {year}
  {2011})}\BibitemShut {NoStop}%
\bibitem [{\citenamefont {Klimovskikh}\ \emph {et~al.}(2017)\citenamefont
  {Klimovskikh}, \citenamefont {Shikin}, \citenamefont {Otrokov}, \citenamefont
  {Ernst}, \citenamefont {Rusinov}, \citenamefont {Tereshchenko}, \citenamefont
  {Golyashov}, \citenamefont {S{\'a}nchez-Barriga}, \citenamefont {Varykhalov},
  \citenamefont {Rader} \emph {et~al.}}]{Klimovskikh.srep2017}%
  \BibitemOpen
  \bibfield  {author} {\bibinfo {author} {\bibfnamefont {I.}~\bibnamefont
  {Klimovskikh}}, \bibinfo {author} {\bibfnamefont {A.}~\bibnamefont {Shikin}},
  \bibinfo {author} {\bibfnamefont {M.}~\bibnamefont {Otrokov}}, \bibinfo
  {author} {\bibfnamefont {A.}~\bibnamefont {Ernst}}, \bibinfo {author}
  {\bibfnamefont {I.}~\bibnamefont {Rusinov}}, \bibinfo {author} {\bibfnamefont
  {O.}~\bibnamefont {Tereshchenko}}, \bibinfo {author} {\bibfnamefont
  {V.}~\bibnamefont {Golyashov}}, \bibinfo {author} {\bibfnamefont
  {J.}~\bibnamefont {S{\'a}nchez-Barriga}}, \bibinfo {author} {\bibfnamefont
  {A.~Y.}\ \bibnamefont {Varykhalov}}, \bibinfo {author} {\bibfnamefont
  {O.}~\bibnamefont {Rader}}, \emph {et~al.},\ }\bibfield  {title} {\bibinfo
  {title} {Giant magnetic band gap in the {Rashba}-split surface state of
  vanadium-doped {BiTeI}: A combined photoemission and ab initio study},\
  }\href {https://doi.org/https://doi.org/10.1038/nature10309} {\bibfield
  {journal} {\bibinfo  {journal} {Sci. Rep.}\ }\textbf {\bibinfo {volume}
  {7}},\ \bibinfo {pages} {3353} (\bibinfo {year} {2017})}\BibitemShut
  {NoStop}%
\bibitem [{\citenamefont {Sau}\ \emph {et~al.}(2010)\citenamefont {Sau},
  \citenamefont {Lutchyn}, \citenamefont {Tewari},\ and\ \citenamefont
  {Das~Sarma}}]{Sau.prl2010}%
  \BibitemOpen
  \bibfield  {author} {\bibinfo {author} {\bibfnamefont {J.~D.}\ \bibnamefont
  {Sau}}, \bibinfo {author} {\bibfnamefont {R.~M.}\ \bibnamefont {Lutchyn}},
  \bibinfo {author} {\bibfnamefont {S.}~\bibnamefont {Tewari}},\ and\ \bibinfo
  {author} {\bibfnamefont {S.}~\bibnamefont {Das~Sarma}},\ }\bibfield  {title}
  {\bibinfo {title} {Generic new platform for topological quantum computation
  using semiconductor heterostructures},\ }\href
  {https://doi.org/10.1103/PhysRevLett.104.040502} {\bibfield  {journal}
  {\bibinfo  {journal} {Phys. Rev. Lett.}\ }\textbf {\bibinfo {volume} {104}},\
  \bibinfo {pages} {040502} (\bibinfo {year} {2010})}\BibitemShut {NoStop}%
\bibitem [{\citenamefont {Sarma}\ \emph {et~al.}(2015)\citenamefont {Sarma},
  \citenamefont {Freedman},\ and\ \citenamefont {Nayak}}]{DasSarma.npjqi2015}%
  \BibitemOpen
  \bibfield  {author} {\bibinfo {author} {\bibfnamefont {S.~D.}\ \bibnamefont
  {Sarma}}, \bibinfo {author} {\bibfnamefont {M.}~\bibnamefont {Freedman}},\
  and\ \bibinfo {author} {\bibfnamefont {C.}~\bibnamefont {Nayak}},\ }\bibfield
   {title} {\bibinfo {title} {Majorana zero modes and topological quantum
  computation},\ }\href {https://doi.org/https://doi.org/10.1038/npjqi.2015.1}
  {\bibfield  {journal} {\bibinfo  {journal} {npj Quantum Inf.}\ }\textbf
  {\bibinfo {volume} {1}},\ \bibinfo {pages} {1} (\bibinfo {year}
  {2015})}\BibitemShut {NoStop}%
\bibitem [{\citenamefont {Dzialoshinskii}(1957)}]{Dzyaloshinskii1957}%
  \BibitemOpen
  \bibfield  {author} {\bibinfo {author} {\bibfnamefont {I.}~\bibnamefont
  {Dzialoshinskii}},\ }\bibfield  {title} {\bibinfo {title} {Thermodynamic
  theory of weak ferromagnetism in antiferromagnetic substances},\ }\href@noop
  {} {\bibfield  {journal} {\bibinfo  {journal} {Soviet Physics JETP-USSR}\
  }\textbf {\bibinfo {volume} {5}},\ \bibinfo {pages} {1259} (\bibinfo {year}
  {1957})}\BibitemShut {NoStop}%
\bibitem [{\citenamefont {Moriya}(1960{\natexlab{a}})}]{Moriya.prl1960}%
  \BibitemOpen
  \bibfield  {author} {\bibinfo {author} {\bibfnamefont {T.}~\bibnamefont
  {Moriya}},\ }\bibfield  {title} {\bibinfo {title} {New mechanism of
  anisotropic superexchange interaction},\ }\href
  {https://doi.org/10.1103/PhysRevLett.4.228} {\bibfield  {journal} {\bibinfo
  {journal} {Phys. Rev. Lett.}\ }\textbf {\bibinfo {volume} {4}},\ \bibinfo
  {pages} {228} (\bibinfo {year} {1960}{\natexlab{a}})}\BibitemShut {NoStop}%
\bibitem [{\citenamefont {Moriya}(1960{\natexlab{b}})}]{Moriya.pr1960}%
  \BibitemOpen
  \bibfield  {author} {\bibinfo {author} {\bibfnamefont {T.}~\bibnamefont
  {Moriya}},\ }\bibfield  {title} {\bibinfo {title} {Anisotropic superexchange
  interaction and weak ferromagnetism},\ }\href
  {https://doi.org/10.1103/PhysRev.120.91} {\bibfield  {journal} {\bibinfo
  {journal} {Phys. Rev.}\ }\textbf {\bibinfo {volume} {120}},\ \bibinfo {pages}
  {91} (\bibinfo {year} {1960}{\natexlab{b}})}\BibitemShut {NoStop}%
\bibitem [{\citenamefont {Ryu}\ \emph {et~al.}(2013)\citenamefont {Ryu},
  \citenamefont {Thomas}, \citenamefont {Yang},\ and\ \citenamefont
  {Parkin}}]{Ryu.natn2013}%
  \BibitemOpen
  \bibfield  {author} {\bibinfo {author} {\bibfnamefont {K.-S.}\ \bibnamefont
  {Ryu}}, \bibinfo {author} {\bibfnamefont {L.}~\bibnamefont {Thomas}},
  \bibinfo {author} {\bibfnamefont {S.-H.}\ \bibnamefont {Yang}},\ and\
  \bibinfo {author} {\bibfnamefont {S.}~\bibnamefont {Parkin}},\ }\bibfield
  {title} {\bibinfo {title} {Chiral spin torque at magnetic domain walls},\
  }\href {https://doi.org/https://doi.org/10.1038/nnano.2013.102} {\bibfield
  {journal} {\bibinfo  {journal} {Nature Nanotech}\ }\textbf {\bibinfo {volume}
  {8}},\ \bibinfo {pages} {527} (\bibinfo {year} {2013})}\BibitemShut {NoStop}%
\bibitem [{\citenamefont {Yu}\ \emph {et~al.}(2011)\citenamefont {Yu},
  \citenamefont {Kanazawa}, \citenamefont {Onose}, \citenamefont {Kimoto},
  \citenamefont {Zhang}, \citenamefont {Ishiwata}, \citenamefont {Matsui},\
  and\ \citenamefont {Tokura}}]{Yu.natm2011}%
  \BibitemOpen
  \bibfield  {author} {\bibinfo {author} {\bibfnamefont {X.}~\bibnamefont
  {Yu}}, \bibinfo {author} {\bibfnamefont {N.}~\bibnamefont {Kanazawa}},
  \bibinfo {author} {\bibfnamefont {Y.}~\bibnamefont {Onose}}, \bibinfo
  {author} {\bibfnamefont {K.}~\bibnamefont {Kimoto}}, \bibinfo {author}
  {\bibfnamefont {W.}~\bibnamefont {Zhang}}, \bibinfo {author} {\bibfnamefont
  {S.}~\bibnamefont {Ishiwata}}, \bibinfo {author} {\bibfnamefont
  {Y.}~\bibnamefont {Matsui}},\ and\ \bibinfo {author} {\bibfnamefont
  {Y.}~\bibnamefont {Tokura}},\ }\bibfield  {title} {\bibinfo {title} {Near
  room-temperature formation of a skyrmion crystal in thin-films of the
  helimagnet {FeGe}},\ }\href
  {https://doi.org/https://doi.org/10.1038/nmat2916} {\bibfield  {journal}
  {\bibinfo  {journal} {Nature Mater.}\ }\textbf {\bibinfo {volume} {10}},\
  \bibinfo {pages} {106} (\bibinfo {year} {2011})}\BibitemShut {NoStop}%
\bibitem [{\citenamefont {Fert}\ \emph {et~al.}(2013)\citenamefont {Fert},
  \citenamefont {Cros},\ and\ \citenamefont {Sampaio}}]{Fert.natn2013}%
  \BibitemOpen
  \bibfield  {author} {\bibinfo {author} {\bibfnamefont {A.}~\bibnamefont
  {Fert}}, \bibinfo {author} {\bibfnamefont {V.}~\bibnamefont {Cros}},\ and\
  \bibinfo {author} {\bibfnamefont {J.}~\bibnamefont {Sampaio}},\ }\bibfield
  {title} {\bibinfo {title} {Skyrmions on the track},\ }\href
  {https://doi.org/https://doi.org/10.1038/nnano.2013.29} {\bibfield  {journal}
  {\bibinfo  {journal} {Nature Nanotech}\ }\textbf {\bibinfo {volume} {8}},\
  \bibinfo {pages} {152} (\bibinfo {year} {2013})}\BibitemShut {NoStop}%
\bibitem [{\citenamefont {Dong}\ \emph {et~al.}(2017)\citenamefont {Dong},
  \citenamefont {Lou},\ and\ \citenamefont {Shenoy}}]{Dong.acsn2017}%
  \BibitemOpen
  \bibfield  {author} {\bibinfo {author} {\bibfnamefont {L.}~\bibnamefont
  {Dong}}, \bibinfo {author} {\bibfnamefont {J.}~\bibnamefont {Lou}},\ and\
  \bibinfo {author} {\bibfnamefont {V.~B.}\ \bibnamefont {Shenoy}},\ }\bibfield
   {title} {\bibinfo {title} {Large in-plane and vertical piezoelectricity in
  {Janus} transition metal dichalchogenides},\ }\href
  {https://doi.org/10.1021/acsnano.7b03313} {\bibfield  {journal} {\bibinfo
  {journal} {ACS Nano}\ }\textbf {\bibinfo {volume} {11}},\ \bibinfo {pages}
  {8242} (\bibinfo {year} {2017})}\BibitemShut {NoStop}%
\bibitem [{\citenamefont {Zhang}\ \emph
  {et~al.}(2019{\natexlab{a}})\citenamefont {Zhang}, \citenamefont {Nie},
  \citenamefont {Sanvito},\ and\ \citenamefont {Du}}]{Zhang.nl2019}%
  \BibitemOpen
  \bibfield  {author} {\bibinfo {author} {\bibfnamefont {C.}~\bibnamefont
  {Zhang}}, \bibinfo {author} {\bibfnamefont {Y.}~\bibnamefont {Nie}}, \bibinfo
  {author} {\bibfnamefont {S.}~\bibnamefont {Sanvito}},\ and\ \bibinfo {author}
  {\bibfnamefont {A.}~\bibnamefont {Du}},\ }\bibfield  {title} {\bibinfo
  {title} {First-principles prediction of a room-temperature ferromagnetic
  {Janus} {VSSe} monolayer with piezoelectricity, ferroelasticity, and large
  valley polarization},\ }\href {https://doi.org/10.1021/acs.nanolett.8b05050}
  {\bibfield  {journal} {\bibinfo  {journal} {Nano Lett.}\ }\textbf {\bibinfo
  {volume} {19}},\ \bibinfo {pages} {1366} (\bibinfo {year}
  {2019}{\natexlab{a}})}\BibitemShut {NoStop}%
\bibitem [{\citenamefont {Lin}\ \emph {et~al.}(2018)\citenamefont {Lin},
  \citenamefont {Pan},\ and\ \citenamefont {Wang}}]{Lin.mtn2018}%
  \BibitemOpen
  \bibfield  {author} {\bibinfo {author} {\bibfnamefont {P.}~\bibnamefont
  {Lin}}, \bibinfo {author} {\bibfnamefont {C.}~\bibnamefont {Pan}},\ and\
  \bibinfo {author} {\bibfnamefont {Z.~L.}\ \bibnamefont {Wang}},\ }\bibfield
  {title} {\bibinfo {title} {Two-dimensional nanomaterials for novel
  piezotronics and piezophototronics},\ }\href
  {https://doi.org/https://doi.org/10.1016/j.mtnano.2018.11.006} {\bibfield
  {journal} {\bibinfo  {journal} {Materials Today Nano}\ }\textbf {\bibinfo
  {volume} {4}},\ \bibinfo {pages} {17} (\bibinfo {year} {2018})}\BibitemShut
  {NoStop}%
\bibitem [{\citenamefont {Yao}\ \emph {et~al.}(2008)\citenamefont {Yao},
  \citenamefont {Xiao},\ and\ \citenamefont {Niu}}]{Yao.prb2008}%
  \BibitemOpen
  \bibfield  {author} {\bibinfo {author} {\bibfnamefont {W.}~\bibnamefont
  {Yao}}, \bibinfo {author} {\bibfnamefont {D.}~\bibnamefont {Xiao}},\ and\
  \bibinfo {author} {\bibfnamefont {Q.}~\bibnamefont {Niu}},\ }\bibfield
  {title} {\bibinfo {title} {Valley-dependent optoelectronics from inversion
  symmetry breaking},\ }\href {https://doi.org/10.1103/PhysRevB.77.235406}
  {\bibfield  {journal} {\bibinfo  {journal} {Phys. Rev. B}\ }\textbf {\bibinfo
  {volume} {77}},\ \bibinfo {pages} {235406} (\bibinfo {year}
  {2008})}\BibitemShut {NoStop}%
\bibitem [{\citenamefont {Liu}\ \emph {et~al.}(2018)\citenamefont {Liu},
  \citenamefont {Shi}, \citenamefont {Lu},\ and\ \citenamefont
  {Anantram}}]{Liu.prb2018}%
  \BibitemOpen
  \bibfield  {author} {\bibinfo {author} {\bibfnamefont {J.}~\bibnamefont
  {Liu}}, \bibinfo {author} {\bibfnamefont {M.}~\bibnamefont {Shi}}, \bibinfo
  {author} {\bibfnamefont {J.}~\bibnamefont {Lu}},\ and\ \bibinfo {author}
  {\bibfnamefont {M.~P.}\ \bibnamefont {Anantram}},\ }\bibfield  {title}
  {\bibinfo {title} {Analysis of electrical-field-dependent
  {Dzyaloshinskii-Moriya} interaction and magnetocrystalline anisotropy in a
  two-dimensional ferromagnetic monolayer},\ }\href
  {https://doi.org/10.1103/PhysRevB.97.054416} {\bibfield  {journal} {\bibinfo
  {journal} {Phys. Rev. B}\ }\textbf {\bibinfo {volume} {97}},\ \bibinfo
  {pages} {054416} (\bibinfo {year} {2018})}\BibitemShut {NoStop}%
\bibitem [{\citenamefont {Deng}\ \emph {et~al.}(2018)\citenamefont {Deng},
  \citenamefont {Yu}, \citenamefont {Song}, \citenamefont {Zhang},
  \citenamefont {Wang}, \citenamefont {Sun}, \citenamefont {Yi}, \citenamefont
  {Wu}, \citenamefont {Wu}, \citenamefont {Zhu} \emph {et~al.}}]{Deng.nat2018}%
  \BibitemOpen
  \bibfield  {author} {\bibinfo {author} {\bibfnamefont {Y.}~\bibnamefont
  {Deng}}, \bibinfo {author} {\bibfnamefont {Y.}~\bibnamefont {Yu}}, \bibinfo
  {author} {\bibfnamefont {Y.}~\bibnamefont {Song}}, \bibinfo {author}
  {\bibfnamefont {J.}~\bibnamefont {Zhang}}, \bibinfo {author} {\bibfnamefont
  {N.~Z.}\ \bibnamefont {Wang}}, \bibinfo {author} {\bibfnamefont
  {Z.}~\bibnamefont {Sun}}, \bibinfo {author} {\bibfnamefont {Y.}~\bibnamefont
  {Yi}}, \bibinfo {author} {\bibfnamefont {Y.~Z.}\ \bibnamefont {Wu}}, \bibinfo
  {author} {\bibfnamefont {S.}~\bibnamefont {Wu}}, \bibinfo {author}
  {\bibfnamefont {J.}~\bibnamefont {Zhu}}, \emph {et~al.},\ }\bibfield  {title}
  {\bibinfo {title} {Gate-tunable room-temperature ferromagnetism in
  two-dimensional {Fe}${}_3${GeTe}${}_2$},\ }\href
  {https://doi.org/https://doi.org/10.1038/s41586-018-0626-9} {\bibfield
  {journal} {\bibinfo  {journal} {Nature}\ }\textbf {\bibinfo {volume} {563}},\
  \bibinfo {pages} {94} (\bibinfo {year} {2018})}\BibitemShut {NoStop}%
\bibitem [{\citenamefont {Bobkov}\ \emph {et~al.}(2024)\citenamefont {Bobkov},
  \citenamefont {Bokai}, \citenamefont {Otrokov}, \citenamefont {Bobkov},\ and\
  \citenamefont {Bobkova}}]{Bobkov.prm2024}%
  \BibitemOpen
  \bibfield  {author} {\bibinfo {author} {\bibfnamefont {G.~A.}\ \bibnamefont
  {Bobkov}}, \bibinfo {author} {\bibfnamefont {K.~A.}\ \bibnamefont {Bokai}},
  \bibinfo {author} {\bibfnamefont {M.~M.}\ \bibnamefont {Otrokov}}, \bibinfo
  {author} {\bibfnamefont {A.~M.}\ \bibnamefont {Bobkov}},\ and\ \bibinfo
  {author} {\bibfnamefont {I.~V.}\ \bibnamefont {Bobkova}},\ }\bibfield
  {title} {\bibinfo {title} {Gate-controlled proximity effect in
  superconductor/ferromagnet van der waals heterostructures},\ }\href
  {https://doi.org/10.1103/PhysRevMaterials.8.104801} {\bibfield  {journal}
  {\bibinfo  {journal} {Phys. Rev. Mater.}\ }\textbf {\bibinfo {volume} {8}},\
  \bibinfo {pages} {104801} (\bibinfo {year} {2024})}\BibitemShut {NoStop}%
\bibitem [{\citenamefont {Yagmurcukardes}\ \emph {et~al.}(2020)\citenamefont
  {Yagmurcukardes}, \citenamefont {Qin}, \citenamefont {Ozen}, \citenamefont
  {Sayyad}, \citenamefont {Peeters}, \citenamefont {Tongay},\ and\
  \citenamefont {Sahin}}]{Yagmurcukardes.apr2020}%
  \BibitemOpen
  \bibfield  {author} {\bibinfo {author} {\bibfnamefont {M.}~\bibnamefont
  {Yagmurcukardes}}, \bibinfo {author} {\bibfnamefont {Y.}~\bibnamefont {Qin}},
  \bibinfo {author} {\bibfnamefont {S.}~\bibnamefont {Ozen}}, \bibinfo {author}
  {\bibfnamefont {M.}~\bibnamefont {Sayyad}}, \bibinfo {author} {\bibfnamefont
  {F.~M.}\ \bibnamefont {Peeters}}, \bibinfo {author} {\bibfnamefont
  {S.}~\bibnamefont {Tongay}},\ and\ \bibinfo {author} {\bibfnamefont
  {H.}~\bibnamefont {Sahin}},\ }\bibfield  {title} {\bibinfo {title} {Quantum
  properties and applications of {2D} {Janus} crystals and their
  superlattices},\ }\href {https://doi.org/10.1063/1.5135306} {\bibfield
  {journal} {\bibinfo  {journal} {Applied Physics Reviews}\ }\textbf {\bibinfo
  {volume} {7}},\ \bibinfo {pages} {011311} (\bibinfo {year}
  {2020})}\BibitemShut {NoStop}%
\bibitem [{\citenamefont {Zhang}\ \emph {et~al.}(2020)\citenamefont {Zhang},
  \citenamefont {Yang}, \citenamefont {Gong}, \citenamefont {Pan},
  \citenamefont {Wang}, \citenamefont {Guo}, \citenamefont {Zhang},\ and\
  \citenamefont {Fu}}]{Zhang.jmca2020}%
  \BibitemOpen
  \bibfield  {author} {\bibinfo {author} {\bibfnamefont {L.}~\bibnamefont
  {Zhang}}, \bibinfo {author} {\bibfnamefont {Z.}~\bibnamefont {Yang}},
  \bibinfo {author} {\bibfnamefont {T.}~\bibnamefont {Gong}}, \bibinfo {author}
  {\bibfnamefont {R.}~\bibnamefont {Pan}}, \bibinfo {author} {\bibfnamefont
  {H.}~\bibnamefont {Wang}}, \bibinfo {author} {\bibfnamefont {Z.}~\bibnamefont
  {Guo}}, \bibinfo {author} {\bibfnamefont {H.}~\bibnamefont {Zhang}},\ and\
  \bibinfo {author} {\bibfnamefont {X.}~\bibnamefont {Fu}},\ }\bibfield
  {title} {\bibinfo {title} {Recent advances in emerging {Janus}
  two-dimensional materials: from fundamental physics to device applications},\
  }\href {https://doi.org/10.1039/D0TA01999B} {\bibfield  {journal} {\bibinfo
  {journal} {J. Mater. Chem. A}\ }\textbf {\bibinfo {volume} {8}},\ \bibinfo
  {pages} {8813} (\bibinfo {year} {2020})}\BibitemShut {NoStop}%
\bibitem [{\citenamefont {Lu}\ \emph {et~al.}(2017)\citenamefont {Lu},
  \citenamefont {Zhu}, \citenamefont {Xiao}, \citenamefont {Chuu},
  \citenamefont {Han}, \citenamefont {Chiu}, \citenamefont {Cheng},
  \citenamefont {Yang}, \citenamefont {Wei}, \citenamefont {Yang} \emph
  {et~al.}}]{Lu.nnano2017}%
  \BibitemOpen
  \bibfield  {author} {\bibinfo {author} {\bibfnamefont {A.-Y.}\ \bibnamefont
  {Lu}}, \bibinfo {author} {\bibfnamefont {H.}~\bibnamefont {Zhu}}, \bibinfo
  {author} {\bibfnamefont {J.}~\bibnamefont {Xiao}}, \bibinfo {author}
  {\bibfnamefont {C.-P.}\ \bibnamefont {Chuu}}, \bibinfo {author}
  {\bibfnamefont {Y.}~\bibnamefont {Han}}, \bibinfo {author} {\bibfnamefont
  {M.-H.}\ \bibnamefont {Chiu}}, \bibinfo {author} {\bibfnamefont {C.-C.}\
  \bibnamefont {Cheng}}, \bibinfo {author} {\bibfnamefont {C.-W.}\ \bibnamefont
  {Yang}}, \bibinfo {author} {\bibfnamefont {K.-H.}\ \bibnamefont {Wei}},
  \bibinfo {author} {\bibfnamefont {Y.}~\bibnamefont {Yang}}, \emph {et~al.},\
  }\bibfield  {title} {\bibinfo {title} {Janus monolayers of transition metal
  dichalcogenides},\ }\href@noop {} {\bibfield  {journal} {\bibinfo  {journal}
  {Nature Nanotech.}\ }\textbf {\bibinfo {volume} {12}},\ \bibinfo {pages}
  {744} (\bibinfo {year} {2017})}\BibitemShut {NoStop}%
\bibitem [{\citenamefont {Zhang}\ \emph {et~al.}(2017)\citenamefont {Zhang},
  \citenamefont {Jia}, \citenamefont {Kholmanov}, \citenamefont {Dong},
  \citenamefont {Er}, \citenamefont {Chen}, \citenamefont {Guo}, \citenamefont
  {Jin}, \citenamefont {Shenoy}, \citenamefont {Shi} \emph
  {et~al.}}]{Zhang.acsn2017}%
  \BibitemOpen
  \bibfield  {author} {\bibinfo {author} {\bibfnamefont {J.}~\bibnamefont
  {Zhang}}, \bibinfo {author} {\bibfnamefont {S.}~\bibnamefont {Jia}}, \bibinfo
  {author} {\bibfnamefont {I.}~\bibnamefont {Kholmanov}}, \bibinfo {author}
  {\bibfnamefont {L.}~\bibnamefont {Dong}}, \bibinfo {author} {\bibfnamefont
  {D.}~\bibnamefont {Er}}, \bibinfo {author} {\bibfnamefont {W.}~\bibnamefont
  {Chen}}, \bibinfo {author} {\bibfnamefont {H.}~\bibnamefont {Guo}}, \bibinfo
  {author} {\bibfnamefont {Z.}~\bibnamefont {Jin}}, \bibinfo {author}
  {\bibfnamefont {V.~B.}\ \bibnamefont {Shenoy}}, \bibinfo {author}
  {\bibfnamefont {L.}~\bibnamefont {Shi}}, \emph {et~al.},\ }\bibfield  {title}
  {\bibinfo {title} {{Janus} monolayer transition-metal dichalcogenides},\
  }\href {https://doi.org/10.1021/acsnano.7b03186} {\bibfield  {journal}
  {\bibinfo  {journal} {ACS Nano}\ }\textbf {\bibinfo {volume} {11}},\ \bibinfo
  {pages} {8192} (\bibinfo {year} {2017})}\BibitemShut {NoStop}%
\bibitem [{\citenamefont {Liang}\ \emph {et~al.}(2020)\citenamefont {Liang},
  \citenamefont {Wang}, \citenamefont {Du}, \citenamefont {Hallal},
  \citenamefont {Garcia}, \citenamefont {Chshiev}, \citenamefont {Fert},\ and\
  \citenamefont {Yang}}]{LiangJ.prb2020}%
  \BibitemOpen
  \bibfield  {author} {\bibinfo {author} {\bibfnamefont {J.}~\bibnamefont
  {Liang}}, \bibinfo {author} {\bibfnamefont {W.}~\bibnamefont {Wang}},
  \bibinfo {author} {\bibfnamefont {H.}~\bibnamefont {Du}}, \bibinfo {author}
  {\bibfnamefont {A.}~\bibnamefont {Hallal}}, \bibinfo {author} {\bibfnamefont
  {K.}~\bibnamefont {Garcia}}, \bibinfo {author} {\bibfnamefont
  {M.}~\bibnamefont {Chshiev}}, \bibinfo {author} {\bibfnamefont
  {A.}~\bibnamefont {Fert}},\ and\ \bibinfo {author} {\bibfnamefont
  {H.}~\bibnamefont {Yang}},\ }\bibfield  {title} {\bibinfo {title} {Very large
  {Dzyaloshinskii-Moriya} interaction in two-dimensional {Janus} manganese
  dichalcogenides and its application to realize skyrmion states},\ }\href
  {https://doi.org/10.1103/PhysRevB.101.184401} {\bibfield  {journal} {\bibinfo
   {journal} {Phys. Rev. B}\ }\textbf {\bibinfo {volume} {101}},\ \bibinfo
  {pages} {184401} (\bibinfo {year} {2020})}\BibitemShut {NoStop}%
\bibitem [{\citenamefont {Yang}\ \emph {et~al.}(2015)\citenamefont {Yang},
  \citenamefont {Thiaville}, \citenamefont {Rohart}, \citenamefont {Fert},\
  and\ \citenamefont {Chshiev}}]{Yang.prl2015}%
  \BibitemOpen
  \bibfield  {author} {\bibinfo {author} {\bibfnamefont {H.}~\bibnamefont
  {Yang}}, \bibinfo {author} {\bibfnamefont {A.}~\bibnamefont {Thiaville}},
  \bibinfo {author} {\bibfnamefont {S.}~\bibnamefont {Rohart}}, \bibinfo
  {author} {\bibfnamefont {A.}~\bibnamefont {Fert}},\ and\ \bibinfo {author}
  {\bibfnamefont {M.}~\bibnamefont {Chshiev}},\ }\bibfield  {title} {\bibinfo
  {title} {Anatomy of {Dzyaloshinskii-Moriya} interaction at
  $\mathrm{Co}/\mathrm{Pt}$ interfaces},\ }\href
  {https://doi.org/10.1103/PhysRevLett.115.267210} {\bibfield  {journal}
  {\bibinfo  {journal} {Phys. Rev. Lett.}\ }\textbf {\bibinfo {volume} {115}},\
  \bibinfo {pages} {267210} (\bibinfo {year} {2015})}\BibitemShut {NoStop}%
\bibitem [{\citenamefont {Blanco-Rey}\ \emph {et~al.}(2022)\citenamefont
  {Blanco-Rey}, \citenamefont {Bihlmayer}, \citenamefont {Arnau},\ and\
  \citenamefont {Cerd\'a}}]{Blanco-rey.prb2022}%
  \BibitemOpen
  \bibfield  {author} {\bibinfo {author} {\bibfnamefont {M.}~\bibnamefont
  {Blanco-Rey}}, \bibinfo {author} {\bibfnamefont {G.}~\bibnamefont
  {Bihlmayer}}, \bibinfo {author} {\bibfnamefont {A.}~\bibnamefont {Arnau}},\
  and\ \bibinfo {author} {\bibfnamefont {J.~I.}\ \bibnamefont {Cerd\'a}},\
  }\bibfield  {title} {\bibinfo {title} {Nature of interfacial
  {Dzyaloshinskii-Moriya} interactions in graphene/{Co}/{Pt}(111) multilayer
  heterostructures},\ } \href {https://doi.org/10.1103/PhysRevB.106.064426}
  {\bibfield  {journal} {\bibinfo  {journal} {Phys. Rev. B}\ }\textbf {\bibinfo
  {volume} {106}},\ \bibinfo {pages} {064426} (\bibinfo {year}
  {2022})}\BibitemShut {NoStop}%
\bibitem [{\citenamefont {Xiao}\ \emph {et~al.}(2020)\citenamefont {Xiao},
  \citenamefont {Xu}, \citenamefont {Xiao}, \citenamefont {Wang},\ and\
  \citenamefont {Dai}}]{Xiao.pccp2020}%
  \BibitemOpen
  \bibfield  {author} {\bibinfo {author} {\bibfnamefont {W.-Z.}\ \bibnamefont
  {Xiao}}, \bibinfo {author} {\bibfnamefont {L.}~\bibnamefont {Xu}}, \bibinfo
  {author} {\bibfnamefont {G.}~\bibnamefont {Xiao}}, \bibinfo {author}
  {\bibfnamefont {L.-L.}\ \bibnamefont {Wang}},\ and\ \bibinfo {author}
  {\bibfnamefont {X.-Y.}\ \bibnamefont {Dai}},\ }\bibfield  {title} {\bibinfo
  {title} {Two-dimensional hexagonal chromium chalco-halides with large
  vertical piezoelectricity{,} high-temperature ferromagnetism{,} and high
  magnetic anisotropy},\ }\href {https://doi.org/10.1039/D0CP02293D} {\bibfield
   {journal} {\bibinfo  {journal} {Phys. Chem. Chem. Phys.}\ }\textbf {\bibinfo
  {volume} {22}},\ \bibinfo {pages} {14503} (\bibinfo {year}
  {2020})}\BibitemShut {NoStop}%
\bibitem [{\citenamefont {Hou}\ \emph {et~al.}(2022)\citenamefont {Hou},
  \citenamefont {Xue}, \citenamefont {Qiu}, \citenamefont {Wang},\ and\
  \citenamefont {Wu}}]{Hou.npjcm2022}%
  \BibitemOpen
  \bibfield  {author} {\bibinfo {author} {\bibfnamefont {Y.}~\bibnamefont
  {Hou}}, \bibinfo {author} {\bibfnamefont {F.}~\bibnamefont {Xue}}, \bibinfo
  {author} {\bibfnamefont {L.}~\bibnamefont {Qiu}}, \bibinfo {author}
  {\bibfnamefont {Z.}~\bibnamefont {Wang}},\ and\ \bibinfo {author}
  {\bibfnamefont {R.}~\bibnamefont {Wu}},\ }\bibfield  {title} {\bibinfo
  {title} {Multifunctional two-dimensional van der {Waals} {Janus} magnet
  {Cr}-based dichalcogenide halides},\ }\href
  {https://doi.org/https://doi.org/10.1038/s41524-022-00802-x} {\bibfield
  {journal} {\bibinfo  {journal} {npj Comput. Mater.}\ }\textbf {\bibinfo
  {volume} {8}},\ \bibinfo {pages} {120} (\bibinfo {year} {2022})}\BibitemShut
  {NoStop}%
\bibitem [{\citenamefont {Guo}\ \emph {et~al.}(2022)\citenamefont {Guo},
  \citenamefont {Guo}, \citenamefont {Zhu},\ and\ \citenamefont
  {Ang}}]{Guo.apl2022}%
  \BibitemOpen
  \bibfield  {author} {\bibinfo {author} {\bibfnamefont {S.-D.}\ \bibnamefont
  {Guo}}, \bibinfo {author} {\bibfnamefont {X.-S.}\ \bibnamefont {Guo}},
  \bibinfo {author} {\bibfnamefont {Y.-T.}\ \bibnamefont {Zhu}},\ and\ \bibinfo
  {author} {\bibfnamefont {Y.-S.}\ \bibnamefont {Ang}},\ }\bibfield  {title}
  {\bibinfo {title} {Predicted ferromagnetic monolayer {CrSCl} with large
  vertical piezoelectric response: {A} first-principles study},\ }\href
  {https://doi.org/https://doi.org/10.1063/5.0109033} {\bibfield  {journal}
  {\bibinfo  {journal} {Appl. Phys. Lett.}\ }\textbf {\bibinfo {volume}
  {121}},\ \bibinfo {pages} {062403} (\bibinfo {year} {2022})}\BibitemShut
  {NoStop}%
\bibitem [{\citenamefont {Li}\ \emph {et~al.}(2023)\citenamefont {Li},
  \citenamefont {Yu}, \citenamefont {Liang}, \citenamefont {Ga},\ and\
  \citenamefont {Yang}}]{Li.prb2023}%
  \BibitemOpen
  \bibfield  {author} {\bibinfo {author} {\bibfnamefont {P.}~\bibnamefont
  {Li}}, \bibinfo {author} {\bibfnamefont {D.}~\bibnamefont {Yu}}, \bibinfo
  {author} {\bibfnamefont {J.}~\bibnamefont {Liang}}, \bibinfo {author}
  {\bibfnamefont {Y.}~\bibnamefont {Ga}},\ and\ \bibinfo {author}
  {\bibfnamefont {H.}~\bibnamefont {Yang}},\ }\bibfield  {title} {\bibinfo
  {title} {Topological spin textures in {1T}-phase janus magnets: Interplay
  between {Dzyaloshinskii-Moriya} interaction, magnetic frustration, and
  isotropic higher-order interactions},\ }\href
  {https://doi.org/10.1103/PhysRevB.107.054408} {\bibfield  {journal} {\bibinfo
   {journal} {Phys. Rev. B}\ }\textbf {\bibinfo {volume} {107}},\ \bibinfo
  {pages} {054408} (\bibinfo {year} {2023})}\BibitemShut {NoStop}%
\bibitem [{\citenamefont {Guan}\ \emph {et~al.}(2023)\citenamefont {Guan},
  \citenamefont {Shen}, \citenamefont {Xue}, \citenamefont {Zhong},
  \citenamefont {Wu},\ and\ \citenamefont {Song}}]{Guan.pccp2023}%
  \BibitemOpen
  \bibfield  {author} {\bibinfo {author} {\bibfnamefont {Z.}~\bibnamefont
  {Guan}}, \bibinfo {author} {\bibfnamefont {Z.}~\bibnamefont {Shen}}, \bibinfo
  {author} {\bibfnamefont {Y.}~\bibnamefont {Xue}}, \bibinfo {author}
  {\bibfnamefont {T.}~\bibnamefont {Zhong}}, \bibinfo {author} {\bibfnamefont
  {X.}~\bibnamefont {Wu}},\ and\ \bibinfo {author} {\bibfnamefont
  {C.}~\bibnamefont {Song}},\ }\bibfield  {title} {\bibinfo {title} {Electronic
  properties{,} skyrmions and bimerons in janus {CrXY} ({X}{,} {Y} = {S}{,}
  {Se}{,} {Te}{,} {Cl}{,} {Br}{,} {I}{,} and {X} $\neq$ {Y}) monolayers},\
  }\href {https://doi.org/10.1039/D3CP02470A} {\bibfield  {journal} {\bibinfo
  {journal} {Phys. Chem. Chem. Phys.}\ }\textbf {\bibinfo {volume} {25}},\
  \bibinfo {pages} {24968} (\bibinfo {year} {2023})}\BibitemShut {NoStop}%
\bibitem [{\citenamefont {Lasek}\ \emph {et~al.}(2021)\citenamefont {Lasek},
  \citenamefont {Li}, \citenamefont {Kolekar}, \citenamefont {Coelho},
  \citenamefont {Guo}, \citenamefont {Zhang}, \citenamefont {Wang},\ and\
  \citenamefont {Batzill}}]{Lasek.ssr2021}%
  \BibitemOpen
  \bibfield  {author} {\bibinfo {author} {\bibfnamefont {K.}~\bibnamefont
  {Lasek}}, \bibinfo {author} {\bibfnamefont {J.}~\bibnamefont {Li}}, \bibinfo
  {author} {\bibfnamefont {S.}~\bibnamefont {Kolekar}}, \bibinfo {author}
  {\bibfnamefont {P.~M.}\ \bibnamefont {Coelho}}, \bibinfo {author}
  {\bibfnamefont {L.}~\bibnamefont {Guo}}, \bibinfo {author} {\bibfnamefont
  {M.}~\bibnamefont {Zhang}}, \bibinfo {author} {\bibfnamefont
  {Z.}~\bibnamefont {Wang}},\ and\ \bibinfo {author} {\bibfnamefont
  {M.}~\bibnamefont {Batzill}},\ }\bibfield  {title} {\bibinfo {title}
  {{Synthesis and characterization of 2D transition metal dichalcogenides:
  Recent progress from a vacuum surface science perspective}},\ }\href
  {https://doi.org/https://doi.org/10.1016/j.surfrep.2021.100523} {\bibfield
  {journal} {\bibinfo  {journal} {Surf. Sci. Rep.}\ }\textbf {\bibinfo {volume}
  {76}},\ \bibinfo {pages} {100523} (\bibinfo {year} {2021})}\BibitemShut
  {NoStop}%
\bibitem [{\citenamefont {Mahmoodabadi}\ \emph {et~al.}(2023)\citenamefont
  {Mahmoodabadi}, \citenamefont {Modarresi}, \citenamefont {Roknabadi},\ and\
  \citenamefont {Mogulkoc}}]{Mahmoodabadi.cms2023}%
  \BibitemOpen
  \bibfield  {author} {\bibinfo {author} {\bibfnamefont {A.~N.}\ \bibnamefont
  {Mahmoodabadi}}, \bibinfo {author} {\bibfnamefont {M.}~\bibnamefont
  {Modarresi}}, \bibinfo {author} {\bibfnamefont {M.~R.}\ \bibnamefont
  {Roknabadi}},\ and\ \bibinfo {author} {\bibfnamefont {A.}~\bibnamefont
  {Mogulkoc}},\ }\bibfield  {title} {\bibinfo {title} {Efficient discovery of
  room temperature magnetic transition metal monolayers assisted by artificial
  neural network},\ }\href
  {https://doi.org/https://doi.org/10.1016/j.commatsci.2023.112166} {\bibfield
  {journal} {\bibinfo  {journal} {Comput. Mater. Sci.}\ }\textbf {\bibinfo
  {volume} {224}},\ \bibinfo {pages} {112166} (\bibinfo {year}
  {2023})}\BibitemShut {NoStop}%
\bibitem [{\citenamefont {Rahman}(2023)}]{Rahman.jem2023}%
  \BibitemOpen
  \bibfield  {author} {\bibinfo {author} {\bibfnamefont {A.~U.}\ \bibnamefont
  {Rahman}},\ }\bibfield  {title} {\bibinfo {title} {Strain induces
  ferromagnetism in a {Janus} transition metal dichalcogenides: {CrSTe}-{1H}
  monolayer},\ }\href {https://doi.org/10.1007/s11664-022-10075-1} {\bibfield
  {journal} {\bibinfo  {journal} {J. Electron. Mater.}\ }\textbf {\bibinfo
  {volume} {52}},\ \bibinfo {pages} {1036} (\bibinfo {year}
  {2023})}\BibitemShut {NoStop}%
\bibitem [{\citenamefont {Caglayan}\ \emph {et~al.}(2024)\citenamefont
  {Caglayan}, \citenamefont {Mogulkoc}, \citenamefont {Mogulkoc}, \citenamefont
  {Modarresi},\ and\ \citenamefont {Rudenko}}]{Caglayan.prb2024}%
  \BibitemOpen
  \bibfield  {author} {\bibinfo {author} {\bibfnamefont {R.}~\bibnamefont
  {Caglayan}}, \bibinfo {author} {\bibfnamefont {A.}~\bibnamefont {Mogulkoc}},
  \bibinfo {author} {\bibfnamefont {Y.}~\bibnamefont {Mogulkoc}}, \bibinfo
  {author} {\bibfnamefont {M.}~\bibnamefont {Modarresi}},\ and\ \bibinfo
  {author} {\bibfnamefont {A.~N.}\ \bibnamefont {Rudenko}},\ }\bibfield
  {title} {\bibinfo {title} {{Dzyaloshinskii-Moriya} interaction and nontrivial
  spin textures in the {Janus} semiconductor monolayers $\mathrm{VXY}$
  $({X}=\mathrm{Cl}, \mathrm{Br}, \mathrm{I}; {Y}=\mathrm{S}, \mathrm{Se},
  \mathrm{Te})$},\ }\href {https://doi.org/10.1103/PhysRevB.110.094440}
  {\bibfield  {journal} {\bibinfo  {journal} {Phys. Rev. B}\ }\textbf {\bibinfo
  {volume} {110}},\ \bibinfo {pages} {094440} (\bibinfo {year}
  {2024})}\BibitemShut {NoStop}%
\bibitem [{\citenamefont {Su}\ \emph {et~al.}(2024)\citenamefont {Su},
  \citenamefont {Peng}, \citenamefont {Yan}, \citenamefont {Lin}, \citenamefont
  {Huang},\ and\ \citenamefont {Liu}}]{Su.pccp2024}%
  \BibitemOpen
  \bibfield  {author} {\bibinfo {author} {\bibfnamefont {B.}~\bibnamefont
  {Su}}, \bibinfo {author} {\bibfnamefont {X.}~\bibnamefont {Peng}}, \bibinfo
  {author} {\bibfnamefont {Z.}~\bibnamefont {Yan}}, \bibinfo {author}
  {\bibfnamefont {L.}~\bibnamefont {Lin}}, \bibinfo {author} {\bibfnamefont
  {X.}~\bibnamefont {Huang}},\ and\ \bibinfo {author} {\bibfnamefont {J.-M.}\
  \bibnamefont {Liu}},\ }\bibfield  {title} {\bibinfo {title} {Large valley
  polarization and the valley-dependent {Hall} effect in a {Janus} {TiTeBr}
  monolayer},\ }\href {https://doi.org/10.1039/D4CP00318G} {\bibfield
  {journal} {\bibinfo  {journal} {Phys. Chem. Chem. Phys.}\ }\textbf {\bibinfo
  {volume} {26}},\ \bibinfo {pages} {11722} (\bibinfo {year}
  {2024})}\BibitemShut {NoStop}%
\bibitem [{\citenamefont {Chang}\ \emph {et~al.}(2024)\citenamefont {Chang},
  \citenamefont {Zhang}, \citenamefont {Deng}, \citenamefont {Wu},\ and\
  \citenamefont {Zhang}}]{Chang.mat2024}%
  \BibitemOpen
  \bibfield  {author} {\bibinfo {author} {\bibfnamefont {Y.}~\bibnamefont
  {Chang}}, \bibinfo {author} {\bibfnamefont {Z.}~\bibnamefont {Zhang}},
  \bibinfo {author} {\bibfnamefont {L.}~\bibnamefont {Deng}}, \bibinfo {author}
  {\bibfnamefont {Y.}~\bibnamefont {Wu}},\ and\ \bibinfo {author}
  {\bibfnamefont {X.}~\bibnamefont {Zhang}},\ }\bibfield  {title} {\bibinfo
  {title} {Ferrovalley and quantum anomalous {Hall} effect in {Janus} {TiTeCl}
  monolayer},\ }\href {https://doi.org/10.3390/ma17133331} {\bibfield
  {journal} {\bibinfo  {journal} {Materials}\ }\textbf {\bibinfo {volume}
  {17}},\ \bibinfo {pages} {13} (\bibinfo {year} {2024})}\BibitemShut {NoStop}%
\bibitem [{\citenamefont {Huang}\ \emph {et~al.}(2024)\citenamefont {Huang},
  \citenamefont {Li}, \citenamefont {Wang}, \citenamefont {Zhao}, \citenamefont
  {Shen}, \citenamefont {Chen},\ and\ \citenamefont {Hou}}]{Huang.apl2024}%
  \BibitemOpen
  \bibfield  {author} {\bibinfo {author} {\bibfnamefont {X.-F.}\ \bibnamefont
  {Huang}}, \bibinfo {author} {\bibfnamefont {K.-J.}\ \bibnamefont {Li}},
  \bibinfo {author} {\bibfnamefont {Z.}~\bibnamefont {Wang}}, \bibinfo {author}
  {\bibfnamefont {S.-B.}\ \bibnamefont {Zhao}}, \bibinfo {author}
  {\bibfnamefont {B.}~\bibnamefont {Shen}}, \bibinfo {author} {\bibfnamefont
  {Z.-X.}\ \bibnamefont {Chen}},\ and\ \bibinfo {author} {\bibfnamefont
  {Y.}~\bibnamefont {Hou}},\ }\bibfield  {title} {\bibinfo {title} {Above
  room-temperature two-dimensional ferromagnetic half-metals in {Mn}-based
  {Janus} magnets},\ }\href {https://doi.org/10.1063/5.0214167} {\bibfield
  {journal} {\bibinfo  {journal} {Appl. Phys. Lett.}\ }\textbf {\bibinfo
  {volume} {124}},\ \bibinfo {pages} {252402} (\bibinfo {year}
  {2024})}\BibitemShut {NoStop}%
\bibitem [{\citenamefont {Dai}\ and\ \citenamefont {Tian}(2025)}]{Dai.cms2025}%
  \BibitemOpen
  \bibfield  {author} {\bibinfo {author} {\bibfnamefont {S.}~\bibnamefont
  {Dai}}\ and\ \bibinfo {author} {\bibfnamefont {Z.}~\bibnamefont {Tian}},\
  }\bibfield  {title} {\bibinfo {title} {Tunable magnetic properties of
  two-dimensional {Janus} {NiClI} monolayer},\ }\href
  {https://doi.org/https://doi.org/10.1016/j.commatsci.2024.113472} {\bibfield
  {journal} {\bibinfo  {journal} {Comput. Mater. Sci.}\ }\textbf {\bibinfo
  {volume} {246}},\ \bibinfo {pages} {113472} (\bibinfo {year}
  {2025})}\BibitemShut {NoStop}%
\bibitem [{\citenamefont {Jain}\ \emph {et~al.}(2023)\citenamefont {Jain},
  \citenamefont {Mandal},\ and\ \citenamefont {Bera}}]{Jain.jpcm2023}%
  \BibitemOpen
  \bibfield  {author} {\bibinfo {author} {\bibfnamefont {A.}~\bibnamefont
  {Jain}}, \bibinfo {author} {\bibfnamefont {D.}~\bibnamefont {Mandal}},\ and\
  \bibinfo {author} {\bibfnamefont {C.}~\bibnamefont {Bera}},\ }\bibfield
  {title} {\bibinfo {title} {Quasi-harmonic approach to evaluate pyroelectric
  properties in {Janus} {CrSeBr} monolayer},\ }\href
  {https://doi.org/10.1088/1361-648X/ace411} {\bibfield  {journal} {\bibinfo
  {journal} {J. Phys.: Condens. Matter}\ }\textbf {\bibinfo {volume} {35}},\
  \bibinfo {pages} {415401} (\bibinfo {year} {2023})}\BibitemShut {NoStop}%
\bibitem [{\citenamefont {Wang}\ \emph {et~al.}(2024)\citenamefont {Wang},
  \citenamefont {Xue}, \citenamefont {Qiu}, \citenamefont {Wang}, \citenamefont
  {Wu},\ and\ \citenamefont {Hou}}]{Wang.nl2024}%
  \BibitemOpen
  \bibfield  {author} {\bibinfo {author} {\bibfnamefont {Z.-Q.}\ \bibnamefont
  {Wang}}, \bibinfo {author} {\bibfnamefont {F.}~\bibnamefont {Xue}}, \bibinfo
  {author} {\bibfnamefont {L.}~\bibnamefont {Qiu}}, \bibinfo {author}
  {\bibfnamefont {Z.}~\bibnamefont {Wang}}, \bibinfo {author} {\bibfnamefont
  {R.}~\bibnamefont {Wu}},\ and\ \bibinfo {author} {\bibfnamefont
  {Y.}~\bibnamefont {Hou}},\ }\bibfield  {title} {\bibinfo {title} {Switching
  intrinsic magnetic skyrmions with controllable magnetic anisotropy in van der
  {Waals} multiferroic heterostructures},\ }\href
  {https://doi.org/10.1021/acs.nanolett.3c05024} {\bibfield  {journal}
  {\bibinfo  {journal} {Nano Lett.}\ }\textbf {\bibinfo {volume} {24}},\
  \bibinfo {pages} {4117} (\bibinfo {year} {2024})}\BibitemShut {NoStop}%
\bibitem [{\citenamefont {Wu}\ \emph {et~al.}(2024)\citenamefont {Wu},
  \citenamefont {Liao}, \citenamefont {Nie}, \citenamefont {Wang},
  \citenamefont {Xia}, \citenamefont {Xiong},\ and\ \citenamefont
  {Guo}}]{Wu.jpcs2024}%
  \BibitemOpen
  \bibfield  {author} {\bibinfo {author} {\bibfnamefont {Y.-F.}\ \bibnamefont
  {Wu}}, \bibinfo {author} {\bibfnamefont {J.-J.}\ \bibnamefont {Liao}},
  \bibinfo {author} {\bibfnamefont {Y.-Z.}\ \bibnamefont {Nie}}, \bibinfo
  {author} {\bibfnamefont {X.-G.}\ \bibnamefont {Wang}}, \bibinfo {author}
  {\bibfnamefont {Q.-L.}\ \bibnamefont {Xia}}, \bibinfo {author} {\bibfnamefont
  {R.}~\bibnamefont {Xiong}},\ and\ \bibinfo {author} {\bibfnamefont {G.-H.}\
  \bibnamefont {Guo}},\ }\bibfield  {title} {\bibinfo {title} {Ferroelectric
  control of ferromagnetism in {CrTeI}/{In}$_2${Se}$_3$ heterostructure: {A}
  first-principles study},\ }\href
  {https://doi.org/https://doi.org/10.1016/j.jpcs.2024.112000} {\bibfield
  {journal} {\bibinfo  {journal} {J. Phys. Chem. Solids}\ }\textbf {\bibinfo
  {volume} {190}},\ \bibinfo {pages} {112000} (\bibinfo {year}
  {2024})}\BibitemShut {NoStop}%
\bibitem [{\citenamefont {Yang}\ \emph {et~al.}(2024)\citenamefont {Yang},
  \citenamefont {Dou}, \citenamefont {Li}, \citenamefont {Dai}, \citenamefont
  {Huang},\ and\ \citenamefont {Ma}}]{Yang.mathor2024}%
  \BibitemOpen
  \bibfield  {author} {\bibinfo {author} {\bibfnamefont {J.}~\bibnamefont
  {Yang}}, \bibinfo {author} {\bibfnamefont {K.}~\bibnamefont {Dou}}, \bibinfo
  {author} {\bibfnamefont {X.}~\bibnamefont {Li}}, \bibinfo {author}
  {\bibfnamefont {Y.}~\bibnamefont {Dai}}, \bibinfo {author} {\bibfnamefont
  {B.}~\bibnamefont {Huang}},\ and\ \bibinfo {author} {\bibfnamefont
  {Y.}~\bibnamefont {Ma}},\ }\bibfield  {title} {\bibinfo {title}
  {Strain-driven skyrmion–bimeron switching in topological magnetic monolayer
  {CrSeBr}},\ }\href {https://doi.org/10.1039/D4MH00734D} {\bibfield  {journal}
  {\bibinfo  {journal} {Mater. Horiz.}\ }\textbf {\bibinfo {volume} {11}},\
  \bibinfo {pages} {5374} (\bibinfo {year} {2024})}\BibitemShut {NoStop}%
\bibitem [{\citenamefont {Hu}\ \emph {et~al.}(2018)\citenamefont {Hu},
  \citenamefont {Wang}, \citenamefont {Zhao},\ and\ \citenamefont
  {Wu}}]{Hu.apx2018}%
  \BibitemOpen
  \bibfield  {author} {\bibinfo {author} {\bibfnamefont {J.}~\bibnamefont
  {Hu}}, \bibinfo {author} {\bibfnamefont {P.}~\bibnamefont {Wang}}, \bibinfo
  {author} {\bibfnamefont {J.}~\bibnamefont {Zhao}},\ and\ \bibinfo {author}
  {\bibfnamefont {R.}~\bibnamefont {Wu}},\ }\bibfield  {title} {\bibinfo
  {title} {Engineering magnetic anisotropy in two-dimensional magnetic
  materials},\ }\href {https://doi.org/10.1080/23746149.2018.1432415} {\bibfield  {journal} {\bibinfo  {journal}
  {Adv. Phys. X}\ }\textbf {\bibinfo {volume} {3}},\ \bibinfo {pages}
  {1432415} (\bibinfo {year} {2018})}\BibitemShut {NoStop}%
\bibitem [{\citenamefont {Wu}\ \emph {et~al.}(2022)\citenamefont {Wu},
  \citenamefont {Wang}, \citenamefont {Verma}, \citenamefont {Rao},
  \citenamefont {Cheng}, \citenamefont {Guo}, \citenamefont {Cao},
  \citenamefont {Watanabe}, \citenamefont {Taniguchi}, \citenamefont {Lau}
  \emph {et~al.}}]{Wu.nl2022}%
  \BibitemOpen
  \bibfield  {author} {\bibinfo {author} {\bibfnamefont {G.}~\bibnamefont
  {Wu}}, \bibinfo {author} {\bibfnamefont {D.}~\bibnamefont {Wang}}, \bibinfo
  {author} {\bibfnamefont {N.}~\bibnamefont {Verma}}, \bibinfo {author}
  {\bibfnamefont {R.}~\bibnamefont {Rao}}, \bibinfo {author} {\bibfnamefont
  {Y.}~\bibnamefont {Cheng}}, \bibinfo {author} {\bibfnamefont
  {S.}~\bibnamefont {Guo}}, \bibinfo {author} {\bibfnamefont {G.}~\bibnamefont
  {Cao}}, \bibinfo {author} {\bibfnamefont {K.}~\bibnamefont {Watanabe}},
  \bibinfo {author} {\bibfnamefont {T.}~\bibnamefont {Taniguchi}}, \bibinfo
  {author} {\bibfnamefont {C.~N.}\ \bibnamefont {Lau}}, \emph {et~al.},\
  }\bibfield  {title} {\bibinfo {title} {Enhancing perpendicular magnetic
  anisotropy in garnet ferrimagnet by interfacing with few-layer {WTe}${}_2$},\
  }\href {https://doi.org/10.1021/acs.nanolett.1c04237} {\bibfield  {journal} {\bibinfo  {journal} {Nano Letters}\
  }\textbf {\bibinfo {volume} {22}},\ \bibinfo {pages} {1115} (\bibinfo {year}
  {2022})}\BibitemShut {NoStop}%
\bibitem [{\citenamefont {Wang}\ \emph {et~al.}(2019)\citenamefont {Wang},
  \citenamefont {Tang}, \citenamefont {Xia}, \citenamefont {He}, \citenamefont
  {Zhang}, \citenamefont {Liu}, \citenamefont {Wan}, \citenamefont {Fang},
  \citenamefont {Guo}, \citenamefont {Yang} \emph {et~al.}}]{Wang.sciadv2019}%
  \BibitemOpen
  \bibfield  {author} {\bibinfo {author} {\bibfnamefont {X.}~\bibnamefont
  {Wang}}, \bibinfo {author} {\bibfnamefont {J.}~\bibnamefont {Tang}}, \bibinfo
  {author} {\bibfnamefont {X.}~\bibnamefont {Xia}}, \bibinfo {author}
  {\bibfnamefont {C.}~\bibnamefont {He}}, \bibinfo {author} {\bibfnamefont
  {J.}~\bibnamefont {Zhang}}, \bibinfo {author} {\bibfnamefont
  {Y.}~\bibnamefont {Liu}}, \bibinfo {author} {\bibfnamefont {C.}~\bibnamefont
  {Wan}}, \bibinfo {author} {\bibfnamefont {C.}~\bibnamefont {Fang}}, \bibinfo
  {author} {\bibfnamefont {C.}~\bibnamefont {Guo}}, \bibinfo {author}
  {\bibfnamefont {W.}~\bibnamefont {Yang}}, \emph {et~al.},\ }\bibfield
  {title} {\bibinfo {title} {Current-driven magnetization switching in a van
  der {Waals} ferromagnet {Fe$_3$GeTe$_2$}},\ }\href {10.1126/sciadv.aaw8904} {\bibfield
  {journal} {\bibinfo  {journal} {Sci.~Adv.}\ }\textbf {\bibinfo
  {volume} {5}},\ \bibinfo {pages} {eaaw8904} (\bibinfo {year}
  {2019})}\BibitemShut {NoStop}%
\bibitem [{\citenamefont {Wang}\ \emph {et~al.}(2023)\citenamefont {Wang},
  \citenamefont {Wu}, \citenamefont {Zhang}, \citenamefont {Liu}, \citenamefont
  {Chen}, \citenamefont {Pandey}, \citenamefont {Yin}, \citenamefont {Wei},
  \citenamefont {Lei}, \citenamefont {Shi} \emph {et~al.}}]{Wang.ncomms2023}%
  \BibitemOpen
  \bibfield  {author} {\bibinfo {author} {\bibfnamefont {H.}~\bibnamefont
  {Wang}}, \bibinfo {author} {\bibfnamefont {H.}~\bibnamefont {Wu}}, \bibinfo
  {author} {\bibfnamefont {J.}~\bibnamefont {Zhang}}, \bibinfo {author}
  {\bibfnamefont {Y.}~\bibnamefont {Liu}}, \bibinfo {author} {\bibfnamefont
  {D.}~\bibnamefont {Chen}}, \bibinfo {author} {\bibfnamefont {C.}~\bibnamefont
  {Pandey}}, \bibinfo {author} {\bibfnamefont {J.}~\bibnamefont {Yin}},
  \bibinfo {author} {\bibfnamefont {D.}~\bibnamefont {Wei}}, \bibinfo {author}
  {\bibfnamefont {N.}~\bibnamefont {Lei}}, \bibinfo {author} {\bibfnamefont
  {S.}~\bibnamefont {Shi}}, \emph {et~al.},\ }\bibfield  {title} {\bibinfo
  {title} {Room temperature energy-efficient spin-orbit torque switching in
  two-dimensional van der {Waals} {Fe$_3$GeTe$_2$} induced by topological
  insulators},\ }\href {https://doi.org/10.1038/s41467-023-40714-y} {\bibfield  {journal} {\bibinfo  {journal}
  {Nat. Commun.}\ }\textbf {\bibinfo {volume} {14}},\ \bibinfo {pages} {5173}
  (\bibinfo {year} {2023})}\BibitemShut {NoStop}%
\bibitem [{\citenamefont {Chang}\ \emph {et~al.}(2023)\citenamefont {Chang},
  \citenamefont {Liu},\ and\ \citenamefont {MacDonald}}]{Chang.rmp2023}%
  \BibitemOpen
  \bibfield  {author} {\bibinfo {author} {\bibfnamefont {C.-Z.}\ \bibnamefont
  {Chang}}, \bibinfo {author} {\bibfnamefont {C.-X.}\ \bibnamefont {Liu}},\
  and\ \bibinfo {author} {\bibfnamefont {A.~H.}\ \bibnamefont {MacDonald}},\
  }\bibfield  {title} {\bibinfo {title} {Colloquium: Quantum anomalous {Hall}
  effect},\ }\href {https://doi.org/10.1103/RevModPhys.95.011002} {\bibfield
  {journal} {\bibinfo  {journal} {Rev. Mod. Phys.}\ }\textbf {\bibinfo {volume}
  {95}},\ \bibinfo {pages} {011002} (\bibinfo {year} {2023})}\BibitemShut
  {NoStop}%
\bibitem [{\citenamefont {Ishizaka}\ \emph {et~al.}(2011)\citenamefont
  {Ishizaka}, \citenamefont {Bahramy}, \citenamefont {Murakawa}, \citenamefont
  {Sakano}, \citenamefont {Shimojima}, \citenamefont {Sonobe}, \citenamefont
  {Koizumi}, \citenamefont {Shin}, \citenamefont {Miyahara}, \citenamefont
  {Kimura} \emph {et~al.}}]{Ishizaka.nmat2011}%
  \BibitemOpen
  \bibfield  {author} {\bibinfo {author} {\bibfnamefont {K.}~\bibnamefont
  {Ishizaka}}, \bibinfo {author} {\bibfnamefont {M.}~\bibnamefont {Bahramy}},
  \bibinfo {author} {\bibfnamefont {H.}~\bibnamefont {Murakawa}}, \bibinfo
  {author} {\bibfnamefont {M.}~\bibnamefont {Sakano}}, \bibinfo {author}
  {\bibfnamefont {T.}~\bibnamefont {Shimojima}}, \bibinfo {author}
  {\bibfnamefont {T.}~\bibnamefont {Sonobe}}, \bibinfo {author} {\bibfnamefont
  {K.}~\bibnamefont {Koizumi}}, \bibinfo {author} {\bibfnamefont
  {S.}~\bibnamefont {Shin}}, \bibinfo {author} {\bibfnamefont {H.}~\bibnamefont
  {Miyahara}}, \bibinfo {author} {\bibfnamefont {A.}~\bibnamefont {Kimura}},
  \emph {et~al.},\ }\bibfield  {title} {\bibinfo {title} {Giant {Rashba}-type
  spin splitting in bulk {BiTeI}},\ }\href
  {https://doi.org/https://doi.org/10.1038/nmat3051} {\bibfield  {journal}
  {\bibinfo  {journal} {Nat. Mater.}\ }\textbf {\bibinfo {volume} {10}},\
  \bibinfo {pages} {521} (\bibinfo {year} {2011})}\BibitemShut {NoStop}%
\bibitem [{\citenamefont {Eremeev}\ \emph
  {et~al.}(2012{\natexlab{a}})\citenamefont {Eremeev}, \citenamefont {Nechaev},
  \citenamefont {Koroteev}, \citenamefont {Echenique},\ and\ \citenamefont
  {Chulkov}}]{Eremeev.prl2012}%
  \BibitemOpen
  \bibfield  {author} {\bibinfo {author} {\bibfnamefont {S.~V.}\ \bibnamefont
  {Eremeev}}, \bibinfo {author} {\bibfnamefont {I.~A.}\ \bibnamefont
  {Nechaev}}, \bibinfo {author} {\bibfnamefont {Y.~M.}\ \bibnamefont
  {Koroteev}}, \bibinfo {author} {\bibfnamefont {P.~M.}\ \bibnamefont
  {Echenique}},\ and\ \bibinfo {author} {\bibfnamefont {E.~V.}\ \bibnamefont
  {Chulkov}},\ }\bibfield  {title} {\bibinfo {title} {Ideal two-dimensional
  electron systems with a giant {Rashba}-type spin splitting in real materials:
  Surfaces of bismuth tellurohalides},\ }\href
  {https://doi.org/10.1103/PhysRevLett.108.246802} {\bibfield  {journal}
  {\bibinfo  {journal} {Phys. Rev. Lett.}\ }\textbf {\bibinfo {volume} {108}},\
  \bibinfo {pages} {246802} (\bibinfo {year} {2012}{\natexlab{a}})}\BibitemShut
  {NoStop}%
\bibitem [{\citenamefont {Eremeev}\ \emph
  {et~al.}(2012{\natexlab{b}})\citenamefont {Eremeev}, \citenamefont
  {Nechaev},\ and\ \citenamefont {Chulkov}}]{Eremeev.jetpl2012}%
  \BibitemOpen
  \bibfield  {author} {\bibinfo {author} {\bibfnamefont {S.~V.}\ \bibnamefont
  {Eremeev}}, \bibinfo {author} {\bibfnamefont {I.}~\bibnamefont {Nechaev}},\
  and\ \bibinfo {author} {\bibfnamefont {E.~V.}\ \bibnamefont {Chulkov}},\
  }\bibfield  {title} {\bibinfo {title} {Giant {Rashba}-type spin splitting at
  polar surfaces of {BiTeI}},\ }\href
  {https://doi.org/https://doi.org/10.1134/S0021364012190071} {\bibfield
  {journal} {\bibinfo  {journal} {JETP Lett.}\ }\textbf {\bibinfo {volume}
  {96}},\ \bibinfo {pages} {437} (\bibinfo {year}
  {2012}{\natexlab{b}})}\BibitemShut {NoStop}%
\bibitem [{\citenamefont {Maa{\ss}}\ \emph {et~al.}(2016)\citenamefont
  {Maa{\ss}}, \citenamefont {Bentmann}, \citenamefont {Seibel}, \citenamefont
  {Tusche}, \citenamefont {Eremeev}, \citenamefont {Peixoto}, \citenamefont
  {Tereshchenko}, \citenamefont {Kokh}, \citenamefont {Chulkov}, \citenamefont
  {Kirschner},\ and\ \citenamefont {Reinert}}]{Maass2016}%
  \BibitemOpen
  \bibfield  {author} {\bibinfo {author} {\bibfnamefont {H.}~\bibnamefont
  {Maa{\ss}}}, \bibinfo {author} {\bibfnamefont {H.}~\bibnamefont {Bentmann}},
  \bibinfo {author} {\bibfnamefont {C.}~\bibnamefont {Seibel}}, \bibinfo
  {author} {\bibfnamefont {C.}~\bibnamefont {Tusche}}, \bibinfo {author}
  {\bibfnamefont {S.~V.}\ \bibnamefont {Eremeev}}, \bibinfo {author}
  {\bibfnamefont {T.~R.~F.}\ \bibnamefont {Peixoto}}, \bibinfo {author}
  {\bibfnamefont {O.~E.}\ \bibnamefont {Tereshchenko}}, \bibinfo {author}
  {\bibfnamefont {K.~A.}\ \bibnamefont {Kokh}}, \bibinfo {author}
  {\bibfnamefont {E.~V.}\ \bibnamefont {Chulkov}}, \bibinfo {author}
  {\bibfnamefont {J.}~\bibnamefont {Kirschner}},\ and\ \bibinfo {author}
  {\bibfnamefont {F.}~\bibnamefont {Reinert}},\ }\bibfield  {title} {\bibinfo
  {title} {Spin-texture inversion in the giant {Rashba} semiconductor
  {BiTeI}},\ }\href {https://doi.org/10.1038/ncomms11621} {\bibfield  {journal}
  {\bibinfo  {journal} {Nat. Commun.}\ }\textbf {\bibinfo {volume} {7}},\
  \bibinfo {pages} {11621} (\bibinfo {year} {2016})}\BibitemShut {NoStop}%
\bibitem [{\citenamefont {Otrokov}\ \emph
  {et~al.}(2019{\natexlab{a}})\citenamefont {Otrokov}, \citenamefont
  {Klimovskikh}, \citenamefont {Bentmann}, \citenamefont {Estyunin},
  \citenamefont {Zeugner}, \citenamefont {Aliev}, \citenamefont {Ga{\ss}},
  \citenamefont {Wolter}, \citenamefont {Koroleva}, \citenamefont {Shikin}
  \emph {et~al.}}]{Otrokov.nat2019}%
  \BibitemOpen
  \bibfield  {author} {\bibinfo {author} {\bibfnamefont {M.~M.}\ \bibnamefont
  {Otrokov}}, \bibinfo {author} {\bibfnamefont {I.~I.}\ \bibnamefont
  {Klimovskikh}}, \bibinfo {author} {\bibfnamefont {H.}~\bibnamefont
  {Bentmann}}, \bibinfo {author} {\bibfnamefont {D.}~\bibnamefont {Estyunin}},
  \bibinfo {author} {\bibfnamefont {A.}~\bibnamefont {Zeugner}}, \bibinfo
  {author} {\bibfnamefont {Z.~S.}\ \bibnamefont {Aliev}}, \bibinfo {author}
  {\bibfnamefont {S.}~\bibnamefont {Ga{\ss}}}, \bibinfo {author} {\bibfnamefont
  {A.}~\bibnamefont {Wolter}}, \bibinfo {author} {\bibfnamefont
  {A.}~\bibnamefont {Koroleva}}, \bibinfo {author} {\bibfnamefont {A.~M.}\
  \bibnamefont {Shikin}}, \emph {et~al.},\ }\bibfield  {title} {\bibinfo
  {title} {Prediction and observation of an antiferromagnetic topological
  insulator},\ }\href
  {https://doi.org/https://doi.org/10.1038/s41586-019-1840-9} {\bibfield
  {journal} {\bibinfo  {journal} {Nature}\ }\textbf {\bibinfo {volume} {576}},\
  \bibinfo {pages} {416} (\bibinfo {year} {2019}{\natexlab{a}})}\BibitemShut
  {NoStop}%
\bibitem [{\citenamefont {Otrokov}\ \emph
  {et~al.}(2019{\natexlab{b}})\citenamefont {Otrokov}, \citenamefont {Rusinov},
  \citenamefont {Blanco-Rey}, \citenamefont {Hoffmann}, \citenamefont
  {Vyazovskaya}, \citenamefont {Eremeev}, \citenamefont {Ernst}, \citenamefont
  {Echenique}, \citenamefont {Arnau},\ and\ \citenamefont
  {Chulkov}}]{Otrokov.prl2019}%
  \BibitemOpen
  \bibfield  {author} {\bibinfo {author} {\bibfnamefont {M.~M.}\ \bibnamefont
  {Otrokov}}, \bibinfo {author} {\bibfnamefont {I.~P.}\ \bibnamefont
  {Rusinov}}, \bibinfo {author} {\bibfnamefont {M.}~\bibnamefont {Blanco-Rey}},
  \bibinfo {author} {\bibfnamefont {M.}~\bibnamefont {Hoffmann}}, \bibinfo
  {author} {\bibfnamefont {A.~Y.}\ \bibnamefont {Vyazovskaya}}, \bibinfo
  {author} {\bibfnamefont {S.~V.}\ \bibnamefont {Eremeev}}, \bibinfo {author}
  {\bibfnamefont {A.}~\bibnamefont {Ernst}}, \bibinfo {author} {\bibfnamefont
  {P.~M.}\ \bibnamefont {Echenique}}, \bibinfo {author} {\bibfnamefont
  {A.}~\bibnamefont {Arnau}},\ and\ \bibinfo {author} {\bibfnamefont {E.~V.}\
  \bibnamefont {Chulkov}},\ }\bibfield  {title} {\bibinfo {title} {{Unique
  Thickness-Dependent Properties of the van der {Waals} Interlayer
  Antiferromagnet ${\mathrm{MnBi}}_{2}{\mathrm{Te}}_{4}$ Films}},\ }\href
  {https://doi.org/10.1103/PhysRevLett.122.107202} {\bibfield  {journal}
  {\bibinfo  {journal} {Phys. Rev. Lett.}\ }\textbf {\bibinfo {volume} {122}},\
  \bibinfo {pages} {107202} (\bibinfo {year} {2019}{\natexlab{b}})}\BibitemShut
  {NoStop}%
\bibitem [{\citenamefont {Li}\ \emph {et~al.}(2019)\citenamefont {Li},
  \citenamefont {Li}, \citenamefont {Du}, \citenamefont {Wang}, \citenamefont
  {Gu}, \citenamefont {Zhang}, \citenamefont {He}, \citenamefont {Duan},\ and\
  \citenamefont {Xu}}]{Li.sciadv2019}%
  \BibitemOpen
  \bibfield  {author} {\bibinfo {author} {\bibfnamefont {J.}~\bibnamefont
  {Li}}, \bibinfo {author} {\bibfnamefont {Y.}~\bibnamefont {Li}}, \bibinfo
  {author} {\bibfnamefont {S.}~\bibnamefont {Du}}, \bibinfo {author}
  {\bibfnamefont {Z.}~\bibnamefont {Wang}}, \bibinfo {author} {\bibfnamefont
  {B.-L.}\ \bibnamefont {Gu}}, \bibinfo {author} {\bibfnamefont {S.-C.}\
  \bibnamefont {Zhang}}, \bibinfo {author} {\bibfnamefont {K.}~\bibnamefont
  {He}}, \bibinfo {author} {\bibfnamefont {W.}~\bibnamefont {Duan}},\ and\
  \bibinfo {author} {\bibfnamefont {Y.}~\bibnamefont {Xu}},\ }\bibfield
  {title} {\bibinfo {title} {{Intrinsic magnetic topological insulators in van
  der {Waals} layered MnBi$_2$Te$_4$-family materials}},\ }\href
  {https://doi.org/10.1126/sciadv.aaw5685} {\bibfield  {journal} {\bibinfo
  {journal} {Sci. Adv.}\ }\textbf {\bibinfo {volume} {5}},\ \bibinfo {pages}
  {eaaw5685} (\bibinfo {year} {2019})}\BibitemShut {NoStop}%
\bibitem [{\citenamefont {Zhang}\ \emph
  {et~al.}(2019{\natexlab{b}})\citenamefont {Zhang}, \citenamefont {Shi},
  \citenamefont {Zhu}, \citenamefont {Xing}, \citenamefont {Zhang},\ and\
  \citenamefont {Wang}}]{Zhang.prl2019}%
  \BibitemOpen
  \bibfield  {author} {\bibinfo {author} {\bibfnamefont {D.}~\bibnamefont
  {Zhang}}, \bibinfo {author} {\bibfnamefont {M.}~\bibnamefont {Shi}}, \bibinfo
  {author} {\bibfnamefont {T.}~\bibnamefont {Zhu}}, \bibinfo {author}
  {\bibfnamefont {D.}~\bibnamefont {Xing}}, \bibinfo {author} {\bibfnamefont
  {H.}~\bibnamefont {Zhang}},\ and\ \bibinfo {author} {\bibfnamefont
  {J.}~\bibnamefont {Wang}},\ }\bibfield  {title} {\bibinfo {title}
  {Topological axion states in the magnetic insulator {MnBi}$_{2}${Te}$_{4}$
  with the quantized magnetoelectric effect},\ }\href
  {https://doi.org/10.1103/PhysRevLett.122.206401} {\bibfield  {journal}
  {\bibinfo  {journal} {Phys. Rev. Lett.}\ }\textbf {\bibinfo {volume} {122}},\
  \bibinfo {pages} {206401} (\bibinfo {year} {2019}{\natexlab{b}})}\BibitemShut
  {NoStop}%
\bibitem [{\citenamefont {Bl\"ochl}(1994)}]{Blochl.prb1994}%
  \BibitemOpen
  \bibfield  {author} {\bibinfo {author} {\bibfnamefont {P.~E.}\ \bibnamefont
  {Bl\"ochl}},\ }\bibfield  {title} {\bibinfo {title} {Projector augmented-wave
  method},\ }\href {https://doi.org/10.1103/PhysRevB.50.17953} {\bibfield
  {journal} {\bibinfo  {journal} {Phys. Rev. B}\ }\textbf {\bibinfo {volume}
  {50}},\ \bibinfo {pages} {17953} (\bibinfo {year} {1994})}\BibitemShut
  {NoStop}%
\bibitem [{\citenamefont {Kresse}\ and\ \citenamefont {Hafner}(1993)}]{vasp1}%
  \BibitemOpen
  \bibfield  {author} {\bibinfo {author} {\bibfnamefont {G.}~\bibnamefont
  {Kresse}}\ and\ \bibinfo {author} {\bibfnamefont {J.}~\bibnamefont
  {Hafner}},\ }\bibfield  {title} {\bibinfo {title} {Ab initio molecular
  dynamics for liquid metals},\ }\href
  {https://doi.org/10.1103/PhysRevB.47.558} {\bibfield  {journal} {\bibinfo
  {journal} {Phys. Rev. B}\ }\textbf {\bibinfo {volume} {47}},\ \bibinfo
  {pages} {558} (\bibinfo {year} {1993})}\BibitemShut {NoStop}%
\bibitem [{\citenamefont {Kresse}\ and\ \citenamefont {Hafner}(1994)}]{vasp2}%
  \BibitemOpen
  \bibfield  {author} {\bibinfo {author} {\bibfnamefont {G.}~\bibnamefont
  {Kresse}}\ and\ \bibinfo {author} {\bibfnamefont {J.}~\bibnamefont
  {Hafner}},\ }\bibfield  {title} {\bibinfo {title} {Ab initio
  molecular-dynamics simulation of the liquid-metal--amorphous-semiconductor
  transition in germanium},\ }\href {https://doi.org/10.1103/PhysRevB.49.14251}
  {\bibfield  {journal} {\bibinfo  {journal} {Phys. Rev. B}\ }\textbf {\bibinfo
  {volume} {49}},\ \bibinfo {pages} {14251} (\bibinfo {year}
  {1994})}\BibitemShut {NoStop}%
\bibitem [{\citenamefont {Kresse}\ and\ \citenamefont
  {Furthmüller}(1996)}]{vasp3}%
  \BibitemOpen
  \bibfield  {author} {\bibinfo {author} {\bibfnamefont {G.}~\bibnamefont
  {Kresse}}\ and\ \bibinfo {author} {\bibfnamefont {J.}~\bibnamefont
  {Furthmüller}},\ }\bibfield  {title} {\bibinfo {title} {Efficiency of
  ab-initio total energy calculations for metals and semiconductors using a
  plane-wave basis set},\ }\href
  {https://doi.org/https://doi.org/10.1016/0927-0256(96)00008-0} {\bibfield
  {journal} {\bibinfo  {journal} {Comput. Mater. Sci.}\ }\textbf {\bibinfo
  {volume} {6}},\ \bibinfo {pages} {15 } (\bibinfo {year} {1996})}\BibitemShut
  {NoStop}%
\bibitem [{\citenamefont {Perdew}\ \emph {et~al.}(1996)\citenamefont {Perdew},
  \citenamefont {Burke},\ and\ \citenamefont {Ernzerhof}}]{Perdew.prl1996}%
  \BibitemOpen
  \bibfield  {author} {\bibinfo {author} {\bibfnamefont {J.~P.}\ \bibnamefont
  {Perdew}}, \bibinfo {author} {\bibfnamefont {K.}~\bibnamefont {Burke}},\ and\
  \bibinfo {author} {\bibfnamefont {M.}~\bibnamefont {Ernzerhof}},\ }\bibfield
  {title} {\bibinfo {title} {Generalized gradient approximation made simple},\
  }\href {https://doi.org/10.1103/PhysRevLett.77.3865} {\bibfield  {journal}
  {\bibinfo  {journal} {Phys. Rev. Lett.}\ }\textbf {\bibinfo {volume} {77}},\
  \bibinfo {pages} {3865} (\bibinfo {year} {1996})}\BibitemShut {NoStop}%
\bibitem [{\citenamefont {Anisimov}\ \emph {et~al.}(1991)\citenamefont
  {Anisimov}, \citenamefont {Zaanen},\ and\ \citenamefont
  {Andersen}}]{Anisimov1991}%
  \BibitemOpen
  \bibfield  {author} {\bibinfo {author} {\bibfnamefont {V.~I.}\ \bibnamefont
  {Anisimov}}, \bibinfo {author} {\bibfnamefont {J.}~\bibnamefont {Zaanen}},\
  and\ \bibinfo {author} {\bibfnamefont {O.~K.}\ \bibnamefont {Andersen}},\
  }\bibfield  {title} {\bibinfo {title} {Band theory and {Mott} insulators:
  {Hubbard} {U} instead of {Stoner} {I}},\ }\href
  {https://doi.org/10.1103/PhysRevB.44.943} {\bibfield  {journal} {\bibinfo
  {journal} {Phys. Rev. B}\ }\textbf {\bibinfo {volume} {44}},\ \bibinfo
  {pages} {943} (\bibinfo {year} {1991})}\BibitemShut {NoStop}%
\bibitem [{\citenamefont {Dudarev}\ \emph {et~al.}(1998)\citenamefont
  {Dudarev}, \citenamefont {Botton}, \citenamefont {Savrasov}, \citenamefont
  {Humphreys},\ and\ \citenamefont {Sutton}}]{Dudarev.prb1998}%
  \BibitemOpen
  \bibfield  {author} {\bibinfo {author} {\bibfnamefont {S.~L.}\ \bibnamefont
  {Dudarev}}, \bibinfo {author} {\bibfnamefont {G.~A.}\ \bibnamefont {Botton}},
  \bibinfo {author} {\bibfnamefont {S.~Y.}\ \bibnamefont {Savrasov}}, \bibinfo
  {author} {\bibfnamefont {C.~J.}\ \bibnamefont {Humphreys}},\ and\ \bibinfo
  {author} {\bibfnamefont {A.~P.}\ \bibnamefont {Sutton}},\ }\bibfield  {title}
  {\bibinfo {title} {Electron-energy-loss spectra and the structural stability
  of nickel oxide: An {LSDA+U} study},\ }\href
  {https://doi.org/10.1103/PhysRevB.57.1505} {\bibfield  {journal} {\bibinfo
  {journal} {Phys. Rev. B}\ }\textbf {\bibinfo {volume} {57}},\ \bibinfo
  {pages} {1505} (\bibinfo {year} {1998})}\BibitemShut {NoStop}%
\bibitem [{\citenamefont {Cococcioni}\ and\ \citenamefont
  {De~Gironcoli}(2005)}]{Cococcioni.prb2005}%
  \BibitemOpen
  \bibfield  {author} {\bibinfo {author} {\bibfnamefont {M.}~\bibnamefont
  {Cococcioni}}\ and\ \bibinfo {author} {\bibfnamefont {S.}~\bibnamefont
  {De~Gironcoli}},\ }\bibfield  {title} {\bibinfo {title} {Linear response
  approach to the calculation of the effective interaction parameters in the
  {LDA}+{U} method},\ }\href {https://doi.org/10.1103/PhysRevB.71.035105} {\bibfield  {journal} {\bibinfo  {journal}
  {Phys. Rev. B}\ }\textbf
  {\bibinfo {volume} {71}},\ \bibinfo {pages} {035105} (\bibinfo {year}
  {2005})}\BibitemShut {NoStop}%
\bibitem [{\citenamefont {Bosnar}\ \emph {et~al.}(2026)\citenamefont
  {Bosnar}, \citenamefont {Lendinez}, \citenamefont {Vyazovskaya}, \citenamefont {Sklyadneva}, \citenamefont {Heid}, \citenamefont {Eremeev}, \citenamefont {Atxitia}, \citenamefont {Gallego} \citenamefont {Chulkov}, \citenamefont {Arnau},\ and\ \citenamefont{Otrokov}}] {supplementary_material}%
  \BibitemOpen
  \bibfield  {author} {\bibinfo {author} {\bibfnamefont {M.}~\bibnamefont {Bosnar}}, \bibinfo {author} {\bibfnamefont {J.~M.}~\bibfnamefont{Lendinez}}, \bibinfo {author} {\bibfnamefont {A.~Yu.}~\bibfnamefont{Vyazovskaya}}, \bibinfo {author} {\bibfnamefont {I.~Yu.}~\bibfnamefont{Sklyadneva}}, \bibinfo {author} {\bibfnamefont {R.}~\bibfnamefont{Heid}}, \bibinfo {author} {\bibfnamefont {S.~V.}~\bibfnamefont{Eremeev}}, \bibinfo {author} {\bibfnamefont {U.}~\bibfnamefont{Atxitia}}, \bibinfo {author} {\bibfnamefont {S.}~\bibfnamefont{Gallego}}, \bibinfo {author} {\bibfnamefont {E.~V.}~\bibfnamefont{Chulkov}}, \bibinfo {author} {\bibfnamefont {A.}~\bibfnamefont{Arnau}},\ and\ \bibinfo {author} {\bibfnamefont {M.~M.}~\bibfnamefont{Otrokov}},\ }\bibfield {title}{\bibinfo {tile} {{Supplementary} {Material} for ”{Chromium} chalcohalide {Janus}
monolayer ferromagnets with perpendicular magnetic anisotropy
and high {Curie} temperature”}}
  \BibitemShut {Nostop}%
\bibitem [{\citenamefont {Priessnitz}\ and\ \citenamefont
  {Legut}(2025)}]{OstravaJ}%
  \BibitemOpen
  \bibfield  {author} {\bibinfo {author} {\bibfnamefont {J.}~\bibnamefont
  {Priessnitz}}\ and\ \bibinfo {author} {\bibfnamefont {D.}~\bibnamefont
  {Legut}},\ }\bibfield  {title} {\bibinfo {title} {Ostrava{J}: a tool for
  calculating magnetic exchange interactions via {DFT}},\ }\href {https://doi.org/10.48550/arXiv.2501.08251}
  {\bibfield  {journal} {\bibinfo  {journal} {arXiv preprint arXiv:2501.08251}\
  } (\bibinfo {year} {2025})}\BibitemShut {NoStop}%
\bibitem [{\citenamefont {Heid}\ and\ \citenamefont {Bohnen}(1999)}]{Heid:99}%
  \BibitemOpen
  \bibfield  {author} {\bibinfo {author} {\bibfnamefont {R.}~\bibnamefont
  {Heid}}\ and\ \bibinfo {author} {\bibfnamefont {K.~P.}\ \bibnamefont
  {Bohnen}},\ }\bibfield  {title} {\bibinfo {title} {Linear response in a
  density-functional mixed-basis approach},\ }\href
  {https://doi.org/10.1103/PhysRevB.60.R3709} {\bibfield  {journal} {\bibinfo
  {journal} {Phys. Rev. B}\ }\textbf {\bibinfo {volume} {60}},\ \bibinfo
  {pages} {R3709} (\bibinfo {year} {1999})}\BibitemShut {NoStop}%
\bibitem [{\citenamefont {Meyer}\ \emph {et~al.}()\citenamefont {Meyer},
  \citenamefont {Els\"asser}, \citenamefont {Lechermann},\ and\ \citenamefont
  {F\"ahnle}}]{Meyer}%
  \BibitemOpen
  \bibfield  {author} {\bibinfo {author} {\bibfnamefont {B.}~\bibnamefont
  {Meyer}}, \bibinfo {author} {\bibfnamefont {C.}~\bibnamefont {Els\"asser}},
  \bibinfo {author} {\bibfnamefont {F.}~\bibnamefont {Lechermann}},\ and\
  \bibinfo {author} {\bibfnamefont {M.}~\bibnamefont {F\"ahnle}},\ }\href@noop
  {} {\bibinfo  {journal} {FORTRAN90 Program for Mixed-Basis Pseudopotential
  Calculations for Crystals, Max-Planck-Institut f\"ur Metallforschung,
  Stuttgart (unpublished)}\ }\BibitemShut {NoStop}%
\bibitem [{\citenamefont {Evans}\ \emph {et~al.}(2014)\citenamefont {Evans},
  \citenamefont {Fan}, \citenamefont {Chureemart}, \citenamefont {Ostler},
  \citenamefont {Ellis},\ and\ \citenamefont {Chantrell}}]{MainVampire}%
  \BibitemOpen
\bibfield  {journal} {  }\bibfield  {author} {\bibinfo {author} {\bibfnamefont
  {R.~F.~L.}\ \bibnamefont {Evans}}, \bibinfo {author} {\bibfnamefont {W.~J.}\
  \bibnamefont {Fan}}, \bibinfo {author} {\bibfnamefont {P.}~\bibnamefont
  {Chureemart}}, \bibinfo {author} {\bibfnamefont {T.~A.}\ \bibnamefont
  {Ostler}}, \bibinfo {author} {\bibfnamefont {M.~O.~A.}\ \bibnamefont
  {Ellis}},\ and\ \bibinfo {author} {\bibfnamefont {R.~W.}\ \bibnamefont
  {Chantrell}},\ }\bibfield  {title} {\bibinfo {title} {Constrained {Monte Carlo}
  method and calculation of the temperature dependence of magnetic
  anisotropy},\ }\href {https://doi.org/10.1088/0953-8984/26/10/103202}
  {\bibfield  {journal} {\bibinfo  {journal} {Journal of Physics: Condensed
  Matter}\ }\textbf {\bibinfo {volume} {26}},\ \bibinfo {pages} {103202}
  (\bibinfo {year} {2014})}\BibitemShut {NoStop}%
\bibitem [{\citenamefont {Grimme}\ \emph {et~al.}(2011)\citenamefont {Grimme},
  \citenamefont {Ehrlich},\ and\ \citenamefont {Goerigk}}]{Grimme.jcc2011}%
  \BibitemOpen
  \bibfield  {author} {\bibinfo {author} {\bibfnamefont {S.}~\bibnamefont
  {Grimme}}, \bibinfo {author} {\bibfnamefont {S.}~\bibnamefont {Ehrlich}},\
  and\ \bibinfo {author} {\bibfnamefont {L.}~\bibnamefont {Goerigk}},\
  }\bibfield  {title} {\bibinfo {title} {Effect of the damping function in
  dispersion corrected density functional theory},\ }\href
  {https://doi.org/https://doi.org/10.1002/jcc.21759} {\bibfield  {journal}
  {\bibinfo  {journal} {J. Comput. Chem.}\ }\textbf {\bibinfo {volume} {32}},\
  \bibinfo {pages} {1456} (\bibinfo {year} {2011})}\BibitemShut {NoStop}%
\bibitem [{\citenamefont {Eremeev}\ \emph {et~al.}(2017)\citenamefont
  {Eremeev}, \citenamefont {Otrokov},\ and\ \citenamefont
  {Chulkov}}]{Eremeev.jac2017}%
  \BibitemOpen
  \bibfield  {author} {\bibinfo {author} {\bibfnamefont {S.~V.}\ \bibnamefont
  {Eremeev}}, \bibinfo {author} {\bibfnamefont {M.~M.}\ \bibnamefont
  {Otrokov}},\ and\ \bibinfo {author} {\bibfnamefont {E.~V.}\ \bibnamefont
  {Chulkov}},\ }\bibfield  {title} {\bibinfo {title} {{Competing rhombohedral
  and monoclinic crystal structures in {MnPn}${}_2${Ch}${}_4$ compounds: An
  ab-initio study}},\ }\href
  {https://doi.org/https://doi.org/10.1016/j.jallcom.2017.03.121} {\bibfield
  {journal} {\bibinfo  {journal} {J. Alloys Compd.}\ }\textbf {\bibinfo
  {volume} {709}},\ \bibinfo {pages} {172} (\bibinfo {year}
  {2017})}\BibitemShut {NoStop}%
\bibitem [{\citenamefont {Mostofi}\ \emph {et~al.}(2014)\citenamefont
  {Mostofi}, \citenamefont {Yates}, \citenamefont {Pizzi}, \citenamefont {Lee},
  \citenamefont {Souza}, \citenamefont {Vanderbilt},\ and\ \citenamefont
  {Marzari}}]{Mostofi2014CPC}%
  \BibitemOpen
  \bibfield  {author} {\bibinfo {author} {\bibfnamefont {A.~A.}\ \bibnamefont
  {Mostofi}}, \bibinfo {author} {\bibfnamefont {J.~R.}\ \bibnamefont {Yates}},
  \bibinfo {author} {\bibfnamefont {G.}~\bibnamefont {Pizzi}}, \bibinfo
  {author} {\bibfnamefont {Y.-S.}\ \bibnamefont {Lee}}, \bibinfo {author}
  {\bibfnamefont {I.}~\bibnamefont {Souza}}, \bibinfo {author} {\bibfnamefont
  {D.}~\bibnamefont {Vanderbilt}},\ and\ \bibinfo {author} {\bibfnamefont
  {N.}~\bibnamefont {Marzari}},\ }\bibfield  {title} {\bibinfo {title} {An
  updated version of {Wannier}90: A tool for obtaining maximally-localised
  {Wannier} functions},\ }\href
  {https://doi.org/https://doi.org/10.1016/j.cpc.2014.05.003} {\bibfield
  {journal} {\bibinfo  {journal} {Comput. Phys. Commun.}\ }\textbf {\bibinfo
  {volume} {185}},\ \bibinfo {pages} {2309} (\bibinfo {year}
  {2014})}\BibitemShut {NoStop}%
\bibitem [{\citenamefont {Pizzi}\ \emph {et~al.}(2020)\citenamefont {Pizzi},
  \citenamefont {Vitale}, \citenamefont {Arita}, \citenamefont {Blügel},
  \citenamefont {Freimuth}, \citenamefont {G{\'{e}}ranton}, \citenamefont
  {Gibertini}, \citenamefont {Gresch}, \citenamefont {Johnson}, \citenamefont
  {Koretsune}, \citenamefont {Iba{\~{n}}ez-Azpiroz}, \citenamefont {Lee},
  \citenamefont {Lihm}, \citenamefont {Marchand}, \citenamefont {Marrazzo},
  \citenamefont {Mokrousov}, \citenamefont {Mustafa}, \citenamefont {Nohara},
  \citenamefont {Nomura}, \citenamefont {Paulatto}, \citenamefont
  {Ponc{\'{e}}}, \citenamefont {Ponweiser}, \citenamefont {Qiao}, \citenamefont
  {Thöle}, \citenamefont {Tsirkin}, \citenamefont {Wierzbowska}, \citenamefont
  {Marzari}, \citenamefont {Vanderbilt}, \citenamefont {Souza}, \citenamefont
  {Mostofi},\ and\ \citenamefont {Yates}}]{Pizzi2020JOPCM}%
  \BibitemOpen
  \bibfield  {author} {\bibinfo {author} {\bibfnamefont {G.}~\bibnamefont
  {Pizzi}}, \bibinfo {author} {\bibfnamefont {V.}~\bibnamefont {Vitale}},
  \bibinfo {author} {\bibfnamefont {R.}~\bibnamefont {Arita}}, \bibinfo
  {author} {\bibfnamefont {S.}~\bibnamefont {Blügel}}, \bibinfo {author}
  {\bibfnamefont {F.}~\bibnamefont {Freimuth}}, \bibinfo {author}
  {\bibfnamefont {G.}~\bibnamefont {G{\'{e}}ranton}}, \bibinfo {author}
  {\bibfnamefont {M.}~\bibnamefont {Gibertini}}, \bibinfo {author}
  {\bibfnamefont {D.}~\bibnamefont {Gresch}}, \bibinfo {author} {\bibfnamefont
  {C.}~\bibnamefont {Johnson}}, \bibinfo {author} {\bibfnamefont
  {T.}~\bibnamefont {Koretsune}}, \bibinfo {author} {\bibfnamefont
  {J.}~\bibnamefont {Iba{\~{n}}ez-Azpiroz}}, \bibinfo {author} {\bibfnamefont
  {H.}~\bibnamefont {Lee}}, \bibinfo {author} {\bibfnamefont {J.-M.}\
  \bibnamefont {Lihm}}, \bibinfo {author} {\bibfnamefont {D.}~\bibnamefont
  {Marchand}}, \bibinfo {author} {\bibfnamefont {A.}~\bibnamefont {Marrazzo}},
  \bibinfo {author} {\bibfnamefont {Y.}~\bibnamefont {Mokrousov}}, \bibinfo
  {author} {\bibfnamefont {J.~I.}\ \bibnamefont {Mustafa}}, \bibinfo {author}
  {\bibfnamefont {Y.}~\bibnamefont {Nohara}}, \bibinfo {author} {\bibfnamefont
  {Y.}~\bibnamefont {Nomura}}, \bibinfo {author} {\bibfnamefont
  {L.}~\bibnamefont {Paulatto}}, \bibinfo {author} {\bibfnamefont
  {S.}~\bibnamefont {Ponc{\'{e}}}}, \bibinfo {author} {\bibfnamefont
  {T.}~\bibnamefont {Ponweiser}}, \bibinfo {author} {\bibfnamefont
  {J.}~\bibnamefont {Qiao}}, \bibinfo {author} {\bibfnamefont {F.}~\bibnamefont
  {Thöle}}, \bibinfo {author} {\bibfnamefont {S.~S.}\ \bibnamefont {Tsirkin}},
  \bibinfo {author} {\bibfnamefont {M.}~\bibnamefont {Wierzbowska}}, \bibinfo
  {author} {\bibfnamefont {N.}~\bibnamefont {Marzari}}, \bibinfo {author}
  {\bibfnamefont {D.}~\bibnamefont {Vanderbilt}}, \bibinfo {author}
  {\bibfnamefont {I.}~\bibnamefont {Souza}}, \bibinfo {author} {\bibfnamefont
  {A.~A.}\ \bibnamefont {Mostofi}},\ and\ \bibinfo {author} {\bibfnamefont
  {J.~R.}\ \bibnamefont {Yates}},\ }\bibfield  {title} {\bibinfo {title}
  {Wannier90 as a community code: new features and applications},\ }\href
  {https://doi.org/10.1088/1361-648x/ab51ff} {\bibfield  {journal} {\bibinfo
  {journal} {J. Condens. Matter Phys.}\ }\textbf {\bibinfo {volume} {32}},\
  \bibinfo {pages} {165902} (\bibinfo {year} {2020})}\BibitemShut {NoStop}%
\bibitem [{\citenamefont {Wu}\ \emph {et~al.}(2018)\citenamefont {Wu},
  \citenamefont {Zhang}, \citenamefont {Song}, \citenamefont {Troyer},\ and\
  \citenamefont {Soluyanov}}]{Wu2018:CPC}%
  \BibitemOpen
  \bibfield  {author} {\bibinfo {author} {\bibfnamefont {Q.}~\bibnamefont
  {Wu}}, \bibinfo {author} {\bibfnamefont {S.}~\bibnamefont {Zhang}}, \bibinfo
  {author} {\bibfnamefont {H.-F.}\ \bibnamefont {Song}}, \bibinfo {author}
  {\bibfnamefont {M.}~\bibnamefont {Troyer}},\ and\ \bibinfo {author}
  {\bibfnamefont {A.~A.}\ \bibnamefont {Soluyanov}},\ }\bibfield  {title}
  {\bibinfo {title} {{WannierTools} : An open-source software package for novel
  topological materials},\ }\href
  {https://doi.org/https://doi.org/10.1016/j.cpc.2017.09.033} {\bibfield
  {journal} {\bibinfo  {journal} {Comput. Phys. Commun.}\ }\textbf {\bibinfo
  {volume} {224}},\ \bibinfo {pages} {405 } (\bibinfo {year}
  {2018})}\BibitemShut {NoStop}%
\bibitem [{\citenamefont {Guo}\ \emph {et~al.}(2019)\citenamefont {Guo},
  \citenamefont {Li},\ and\ \citenamefont {Guo}}]{Guo.cms2019}%
  \BibitemOpen
  \bibfield  {author} {\bibinfo {author} {\bibfnamefont {S.-D.}\ \bibnamefont
  {Guo}}, \bibinfo {author} {\bibfnamefont {Y.-F.}\ \bibnamefont {Li}},\ and\
  \bibinfo {author} {\bibfnamefont {X.-S.}\ \bibnamefont {Guo}},\ }\bibfield
  {title} {\bibinfo {title} {Predicted {Janus} monolayer {ZrSSe} with enhanced
  n-type thermoelectric properties compared with monolayer {ZrS}${}_2$},\
  }\href {https://doi.org/https://doi.org/10.1016/j.commatsci.2019.01.035}
  {\bibfield  {journal} {\bibinfo  {journal} {Comput. Mater. Sci.}\ }\textbf
  {\bibinfo {volume} {161}},\ \bibinfo {pages} {16} (\bibinfo {year}
  {2019})}\BibitemShut {NoStop}%
\bibitem [{\citenamefont {Yin}\ \emph {et~al.}(2021)\citenamefont {Yin},
  \citenamefont {Tan}, \citenamefont {Ding}, \citenamefont {Wen}, \citenamefont
  {Li}, \citenamefont {Teobaldi},\ and\ \citenamefont {Liu}}]{Yin.matadv2021}%
  \BibitemOpen
  \bibfield  {author} {\bibinfo {author} {\bibfnamefont {W.-J.}\ \bibnamefont
  {Yin}}, \bibinfo {author} {\bibfnamefont {H.-J.}\ \bibnamefont {Tan}},
  \bibinfo {author} {\bibfnamefont {P.-J.}\ \bibnamefont {Ding}}, \bibinfo
  {author} {\bibfnamefont {B.}~\bibnamefont {Wen}}, \bibinfo {author}
  {\bibfnamefont {X.-B.}\ \bibnamefont {Li}}, \bibinfo {author} {\bibfnamefont
  {G.}~\bibnamefont {Teobaldi}},\ and\ \bibinfo {author} {\bibfnamefont
  {L.-M.}\ \bibnamefont {Liu}},\ }\bibfield  {title} {\bibinfo {title} {Recent
  advances in low-dimensional {Janus} materials: theoretical and simulation
  perspectives},\ }\href {https://doi.org/10.1039/D1MA00660F} {\bibfield
  {journal} {\bibinfo  {journal} {Mater. Adv.}\ }\textbf {\bibinfo {volume}
  {2}},\ \bibinfo {pages} {7543} (\bibinfo {year} {2021})}\BibitemShut
  {NoStop}%
\bibitem [{\citenamefont {Oreshonkov}\ \emph {et~al.}(2022)\citenamefont
  {Oreshonkov}, \citenamefont {Sukhanova},\ and\ \citenamefont
  {Popov}}]{Oreshonkov.mats2022}%
  \BibitemOpen
  \bibfield  {author} {\bibinfo {author} {\bibfnamefont {A.~S.}\ \bibnamefont
  {Oreshonkov}}, \bibinfo {author} {\bibfnamefont {E.~V.}\ \bibnamefont
  {Sukhanova}},\ and\ \bibinfo {author} {\bibfnamefont {Z.~I.}\ \bibnamefont
  {Popov}},\ }\bibfield  {title} {\bibinfo {title} {Raman spectroscopy of
  {Janus} {MoSSe} monolayer polymorph modifications using density functional
  theory},\ }\href {https://doi.org/https://doi.org/10.3390/ma15113988}
  {\bibfield  {journal} {\bibinfo  {journal} {Materials}\ }\textbf {\bibinfo
  {volume} {15}},\ \bibinfo {pages} {3988} (\bibinfo {year}
  {2022})}\BibitemShut {NoStop}%
\bibitem [{\citenamefont {Sokolikova}\ and\ \citenamefont
  {Mattevi}(2020)}]{Sokolikova.csr2020}%
  \BibitemOpen
  \bibfield  {author} {\bibinfo {author} {\bibfnamefont {M.~S.}\ \bibnamefont
  {Sokolikova}}\ and\ \bibinfo {author} {\bibfnamefont {C.}~\bibnamefont
  {Mattevi}},\ }\bibfield  {title} {\bibinfo {title} {Direct synthesis of
  metastable phases of {2D} transition metal dichalcogenides},\ }\href
  {https://doi.org/10.1039/D0CS00143K} {\bibfield  {journal} {\bibinfo
  {journal} {Chem. Soc. Rev.}\ }\textbf {\bibinfo {volume} {49}},\ \bibinfo
  {pages} {3952} (\bibinfo {year} {2020})}\BibitemShut {NoStop}%
\bibitem [{\citenamefont {Singh}\ \emph {et~al.}(2015)\citenamefont {Singh},
  \citenamefont {Shirodkar},\ and\ \citenamefont {Waghmare}}]{Singh.2dmat2015}%
  \BibitemOpen
  \bibfield  {author} {\bibinfo {author} {\bibfnamefont {A.}~\bibnamefont
  {Singh}}, \bibinfo {author} {\bibfnamefont {S.~N.}\ \bibnamefont
  {Shirodkar}},\ and\ \bibinfo {author} {\bibfnamefont {U.~V.}\ \bibnamefont
  {Waghmare}},\ }\bibfield  {title} {\bibinfo {title} {{1H} and {1T}
  polymorphs, structural transitions and anomalous properties of {(Mo, W)(S,
  Se)}${}_2$ monolayers: first-principles analysis},\ }\href
  {https://doi.org/10.1088/2053-1583/2/3/035013} {\bibfield  {journal}
  {\bibinfo  {journal} {2D Mater.}\ }\textbf {\bibinfo {volume} {2}},\ \bibinfo
  {pages} {035013} (\bibinfo {year} {2015})}\BibitemShut {NoStop}%
\bibitem [{\citenamefont {Ding}\ \emph {et~al.}(2019)\citenamefont {Ding},
  \citenamefont {Hu}, \citenamefont {Dai}, \citenamefont {Tang}, \citenamefont
  {Wei}, \citenamefont {Sheng}, \citenamefont {Liang}, \citenamefont {Shao},
  \citenamefont {Song}, \citenamefont {Liu} \emph {et~al.}}]{Ding.acsn2019}%
  \BibitemOpen
  \bibfield  {author} {\bibinfo {author} {\bibfnamefont {W.}~\bibnamefont
  {Ding}}, \bibinfo {author} {\bibfnamefont {L.}~\bibnamefont {Hu}}, \bibinfo
  {author} {\bibfnamefont {J.}~\bibnamefont {Dai}}, \bibinfo {author}
  {\bibfnamefont {X.}~\bibnamefont {Tang}}, \bibinfo {author} {\bibfnamefont
  {R.}~\bibnamefont {Wei}}, \bibinfo {author} {\bibfnamefont {Z.}~\bibnamefont
  {Sheng}}, \bibinfo {author} {\bibfnamefont {C.}~\bibnamefont {Liang}},
  \bibinfo {author} {\bibfnamefont {D.}~\bibnamefont {Shao}}, \bibinfo {author}
  {\bibfnamefont {W.}~\bibnamefont {Song}}, \bibinfo {author} {\bibfnamefont
  {Q.}~\bibnamefont {Liu}}, \emph {et~al.},\ }\bibfield  {title} {\bibinfo
  {title} {Highly ambient-stable {1T-MoS}${}_2$ and {1T-WS}${}_2$ by
  hydrothermal synthesis under high magnetic fields},\ }\href
  {https://doi.org/https://doi.org/10.1021/acsnano.8b07744} {\bibfield
  {journal} {\bibinfo  {journal} {ACS Nano}\ }\textbf {\bibinfo {volume}
  {13}},\ \bibinfo {pages} {1694} (\bibinfo {year} {2019})}\BibitemShut
  {NoStop}%
\bibitem [{\citenamefont {Sokolikova}\ \emph {et~al.}(2019)\citenamefont
  {Sokolikova}, \citenamefont {Sherrell}, \citenamefont {Palczynski},
  \citenamefont {Bemmer},\ and\ \citenamefont
  {Mattevi}}]{Sokolikova.ncomms2019}%
  \BibitemOpen
  \bibfield  {author} {\bibinfo {author} {\bibfnamefont {M.~S.}\ \bibnamefont
  {Sokolikova}}, \bibinfo {author} {\bibfnamefont {P.~C.}\ \bibnamefont
  {Sherrell}}, \bibinfo {author} {\bibfnamefont {P.}~\bibnamefont
  {Palczynski}}, \bibinfo {author} {\bibfnamefont {V.~L.}\ \bibnamefont
  {Bemmer}},\ and\ \bibinfo {author} {\bibfnamefont {C.}~\bibnamefont
  {Mattevi}},\ }\bibfield  {title} {\bibinfo {title} {Direct solution-phase
  synthesis of {1T-WSe}${}_2$ nanosheets},\ }\href
  {https://doi.org/https://doi.org/10.1038/s41467-019-08594-3} {\bibfield
  {journal} {\bibinfo  {journal} {Nat. Commun.}\ }\textbf {\bibinfo {volume}
  {10}},\ \bibinfo {pages} {712} (\bibinfo {year} {2019})}\BibitemShut
  {NoStop}%
\bibitem [{\citenamefont {Dronskowski}\ and\ \citenamefont
  {Bloechl}(1993)}]{Dronskowski1993}%
  \BibitemOpen
  \bibfield  {author} {\bibinfo {author} {\bibfnamefont {R.}~\bibnamefont
  {Dronskowski}}\ and\ \bibinfo {author} {\bibfnamefont {P.~E.}\ \bibnamefont
  {Bloechl}},\ }\bibfield  {title} {\bibinfo {title} {{Crystal orbital Hamilton
  populations (COHP): energy-resolved visualization of chemical bonding in
  solids based on density-functional calculations}},\ }\href
  {https://doi.org/10.1021/j100135a014} {\bibfield  {journal} {\bibinfo
  {journal} {J. Phys. Chem.}\ }\textbf {\bibinfo {volume} {97}},\ \bibinfo
  {pages} {8617} (\bibinfo {year} {1993})}\BibitemShut {NoStop}%
\bibitem [{\citenamefont {Deringer}\ \emph {et~al.}(2011)\citenamefont
  {Deringer}, \citenamefont {Tchougr\'eeff},\ and\ \citenamefont
  {Dronskowski}}]{Deringer2011}%
  \BibitemOpen
  \bibfield  {author} {\bibinfo {author} {\bibfnamefont {V.~L.}\ \bibnamefont
  {Deringer}}, \bibinfo {author} {\bibfnamefont {A.~L.}\ \bibnamefont
  {Tchougr\'eeff}},\ and\ \bibinfo {author} {\bibfnamefont {R.}~\bibnamefont
  {Dronskowski}},\ }\bibfield  {title} {\bibinfo {title} {{Crystal Orbital
  Hamilton Population (COHP) Analysis As Projected from Plane-Wave Basis
  Sets}},\ }\href {https://doi.org/10.1021/jp202489s} {\bibfield  {journal}
  {\bibinfo  {journal} {J. Phys. Chem. A}\ }\textbf {\bibinfo {volume} {115}},\
  \bibinfo {pages} {5461} (\bibinfo {year} {2011})}\BibitemShut {NoStop}%
\bibitem [{\citenamefont {Maintz}\ \emph {et~al.}(2016)\citenamefont {Maintz},
  \citenamefont {Deringer}, \citenamefont {Tchougr\'eeff},\ and\ \citenamefont
  {Dronskowski}}]{LOBSTER-2016}%
  \BibitemOpen
  \bibfield  {author} {\bibinfo {author} {\bibfnamefont {S.}~\bibnamefont
  {Maintz}}, \bibinfo {author} {\bibfnamefont {V.~L.}\ \bibnamefont
  {Deringer}}, \bibinfo {author} {\bibfnamefont {A.~L.}\ \bibnamefont
  {Tchougr\'eeff}},\ and\ \bibinfo {author} {\bibfnamefont {R.}~\bibnamefont
  {Dronskowski}},\ }\bibfield  {title} {\bibinfo {title} {{LOBSTER: A tool to
  extract chemical bonding from plane-wave based DFT}},\ }\href
  {https://doi.org/https://doi.org/10.1002/jcc.24300} {\bibfield  {journal}
  {\bibinfo  {journal} {J. Comput. Chem.}\ }\textbf {\bibinfo {volume} {37}},\
  \bibinfo {pages} {1030} (\bibinfo {year} {2016})}\BibitemShut {NoStop}%
\bibitem [{\citenamefont {Anderson}(1959)}]{Anderson.pr1959}%
  \BibitemOpen
  \bibfield  {author} {\bibinfo {author} {\bibfnamefont {P.~W.}\ \bibnamefont
  {Anderson}},\ }\bibfield  {title} {\bibinfo {title} {New approach to the
  theory of superexchange interactions},\ }\href
  {https://doi.org/10.1103/PhysRev.115.2} {\bibfield  {journal} {\bibinfo
  {journal} {Phys. Rev.}\ }\textbf {\bibinfo {volume} {115}},\ \bibinfo {pages}
  {2} (\bibinfo {year} {1959})}\BibitemShut {NoStop}%
\bibitem [{\citenamefont {Kanamori}(1959)}]{Kanamori1959}%
  \BibitemOpen
  \bibfield  {author} {\bibinfo {author} {\bibfnamefont {J.}~\bibnamefont
  {Kanamori}},\ }\bibfield  {title} {\bibinfo {title} {Superexchange
  interaction and symmetry properties of electron orbitals},\ }\href
  {https://doi.org/https://doi.org/10.1016/0022-3697(59)90061-7} {\bibfield
  {journal} {\bibinfo  {journal} {J. Phys. Chem. Solids}\ }\textbf {\bibinfo
  {volume} {10}},\ \bibinfo {pages} {87} (\bibinfo {year} {1959})}\BibitemShut
  {NoStop}%
\bibitem [{\citenamefont {Goodenough}(1963)}]{Goodenough1963}%
  \BibitemOpen
  \bibfield  {author} {\bibinfo {author} {\bibfnamefont {J.~B.}\ \bibnamefont
  {Goodenough}},\ }\href@noop {} {\emph {\bibinfo {title} {Magnetism and the
  chemical bond}}}\ (\bibinfo  {publisher} {J. Wiley \& Sons, New York, United
  States of America},\ \bibinfo {year} {1963})\BibitemShut {NoStop}%
\bibitem [{\citenamefont {Zener}(1951)}]{Zener1951-II}%
  \BibitemOpen
  \bibfield  {author} {\bibinfo {author} {\bibfnamefont {C.}~\bibnamefont
  {Zener}},\ }\bibfield  {title} {\bibinfo {title} {Interaction between the
  $d$-shells in the transition metals. {II}. {Ferromagnetic} compounds of
  manganese with perovskite structure},\ }\href
  {https://doi.org/10.1103/PhysRev.82.403} {\bibfield  {journal} {\bibinfo
  {journal} {Phys. Rev.}\ }\textbf {\bibinfo {volume} {82}},\ \bibinfo {pages}
  {403} (\bibinfo {year} {1951})}\BibitemShut {NoStop}%
\bibitem [{\citenamefont {Vergniory}\ \emph {et~al.}(2014)\citenamefont
  {Vergniory}, \citenamefont {Otrokov}, \citenamefont {Thonig}, \citenamefont
  {Hoffmann}, \citenamefont {Maznichenko}, \citenamefont {Geilhufe},
  \citenamefont {Zubizarreta}, \citenamefont {Ostanin}, \citenamefont
  {Marmodoro}, \citenamefont {Henk}, \citenamefont {Hergert}, \citenamefont
  {Mertig}, \citenamefont {Chulkov},\ and\ \citenamefont
  {Ernst}}]{Vergniory.prb2014}%
  \BibitemOpen
  \bibfield  {author} {\bibinfo {author} {\bibfnamefont {M.~G.}\ \bibnamefont
  {Vergniory}}, \bibinfo {author} {\bibfnamefont {M.~M.}\ \bibnamefont
  {Otrokov}}, \bibinfo {author} {\bibfnamefont {D.}~\bibnamefont {Thonig}},
  \bibinfo {author} {\bibfnamefont {M.}~\bibnamefont {Hoffmann}}, \bibinfo
  {author} {\bibfnamefont {I.~V.}\ \bibnamefont {Maznichenko}}, \bibinfo
  {author} {\bibfnamefont {M.}~\bibnamefont {Geilhufe}}, \bibinfo {author}
  {\bibfnamefont {X.}~\bibnamefont {Zubizarreta}}, \bibinfo {author}
  {\bibfnamefont {S.}~\bibnamefont {Ostanin}}, \bibinfo {author} {\bibfnamefont
  {A.}~\bibnamefont {Marmodoro}}, \bibinfo {author} {\bibfnamefont
  {J.}~\bibnamefont {Henk}}, \bibinfo {author} {\bibfnamefont {W.}~\bibnamefont
  {Hergert}}, \bibinfo {author} {\bibfnamefont {I.}~\bibnamefont {Mertig}},
  \bibinfo {author} {\bibfnamefont {E.~V.}\ \bibnamefont {Chulkov}},\ and\
  \bibinfo {author} {\bibfnamefont {A.}~\bibnamefont {Ernst}},\ }\bibfield
  {title} {\bibinfo {title} {Exchange interaction and its tuning in magnetic
  binary chalcogenides},\ }\href {https://doi.org/10.1103/PhysRevB.89.165202}
  {\bibfield  {journal} {\bibinfo  {journal} {Phys. Rev. B}\ }\textbf {\bibinfo
  {volume} {89}},\ \bibinfo {pages} {165202} (\bibinfo {year}
  {2014})}\BibitemShut {NoStop}%
\bibitem [{\citenamefont {Liao}\ \emph {et~al.}(2023)\citenamefont {Liao},
  \citenamefont {Nie}, \citenamefont {Wang}, \citenamefont {Luo}, \citenamefont
  {Xia}, \citenamefont {Xiong},\ and\ \citenamefont {Guo}}]{Liao.prb2023}%
  \BibitemOpen
  \bibfield  {author} {\bibinfo {author} {\bibfnamefont {J.-J.}\ \bibnamefont
  {Liao}}, \bibinfo {author} {\bibfnamefont {Y.-Z.}\ \bibnamefont {Nie}},
  \bibinfo {author} {\bibfnamefont {X.-g.}\ \bibnamefont {Wang}}, \bibinfo
  {author} {\bibfnamefont {Z.-y.}\ \bibnamefont {Luo}}, \bibinfo {author}
  {\bibfnamefont {Q.-l.}\ \bibnamefont {Xia}}, \bibinfo {author} {\bibfnamefont
  {R.}~\bibnamefont {Xiong}},\ and\ \bibinfo {author} {\bibfnamefont {G.-h.}\
  \bibnamefont {Guo}},\ }\bibfield  {title} {\bibinfo {title} {Spin
  reorientation and {Curie} temperature promotion in {CrI}${}_{3}$-{Bi} van der
  {Waals} heterostructures},\ }\href
  {https://doi.org/10.1103/PhysRevB.107.184403} {\bibfield  {journal} {\bibinfo
   {journal} {Phys. Rev. B}\ }\textbf {\bibinfo {volume} {107}},\ \bibinfo
  {pages} {184403} (\bibinfo {year} {2023})}\BibitemShut {NoStop}%
\bibitem [{\citenamefont {Delgado}\ \emph {et~al.}(2024)\citenamefont
  {Delgado}, \citenamefont {Otrokov},\ and\ \citenamefont
  {Arnau}}]{Delgado.jpm2024}%
  \BibitemOpen
  \bibfield  {author} {\bibinfo {author} {\bibfnamefont {F.}~\bibnamefont
  {Delgado}}, \bibinfo {author} {\bibfnamefont {M.~M.}\ \bibnamefont
  {Otrokov}},\ and\ \bibinfo {author} {\bibfnamefont {A.}~\bibnamefont
  {Arnau}},\ }\bibfield  {title} {\bibinfo {title} {Spin wave excitations in
  low dimensional systems with large magnetic anisotropy},\ }\href
  {https://doi.org/10.1088/2515-7639/ad558b} {\bibfield  {journal} {\bibinfo
  {journal} {J. Phys. Mater.}\ }\textbf {\bibinfo {volume} {7}},\ \bibinfo
  {pages} {035005} (\bibinfo {year} {2024})}\BibitemShut {NoStop}%
\bibitem [{\citenamefont {Blanco-Rey}\ \emph {et~al.}(2019)\citenamefont
  {Blanco-Rey}, \citenamefont {Cerd{\'a}},\ and\ \citenamefont
  {Arnau}}]{Blanco-rey.njp2019}%
  \BibitemOpen
  \bibfield  {author} {\bibinfo {author} {\bibfnamefont {M.}~\bibnamefont
  {Blanco-Rey}}, \bibinfo {author} {\bibfnamefont {J.~I.}\ \bibnamefont
  {Cerd{\'a}}},\ and\ \bibinfo {author} {\bibfnamefont {A.}~\bibnamefont
  {Arnau}},\ }\bibfield  {title} {\bibinfo {title} {Validity of perturbative
  methods to treat the spin--orbit interaction: application to
  magnetocrystalline anisotropy},\ }\href {https://doi.org/10.1088/1367-2630/ab3060} {\bibfield  {journal}
  {\bibinfo  {journal} {New J. Phys.}\ }\textbf {\bibinfo {volume}
  {21}},\ \bibinfo {pages} {073054} (\bibinfo {year} {2019})}\BibitemShut
  {NoStop}%
\bibitem [{\citenamefont {Szunyogh}\ \emph {et~al.}(1995)\citenamefont
  {Szunyogh}, \citenamefont {\'Ujfalussy},\ and\ \citenamefont
  {Weinberger}}]{Szunyogh.prb1995}%
  \BibitemOpen
  \bibfield  {author} {\bibinfo {author} {\bibfnamefont {L.}~\bibnamefont
  {Szunyogh}}, \bibinfo {author} {\bibfnamefont {B.}~\bibnamefont
  {\'Ujfalussy}},\ and\ \bibinfo {author} {\bibfnamefont {P.}~\bibnamefont
  {Weinberger}},\ }\bibfield  {title} {\bibinfo {title} {Magnetic anisotropy of
  iron multilayers on {Au}(001): First-principles calculations in terms of the
  fully relativistic spin-polarized screened {KKR} method},\ }\href
  {https://doi.org/10.1103/PhysRevB.51.9552} {\bibfield  {journal} {\bibinfo
  {journal} {Phys. Rev. B}\ }\textbf {\bibinfo {volume} {51}},\ \bibinfo
  {pages} {9552} (\bibinfo {year} {1995})}\BibitemShut {NoStop}%
\bibitem [{\citenamefont {Lado}\ and\ \citenamefont
  {Fern{\'a}ndez-Rossier}(2017)}]{Lado.2dmat2017}%
  \BibitemOpen
  \bibfield  {author} {\bibinfo {author} {\bibfnamefont {J.~L.}\ \bibnamefont
  {Lado}}\ and\ \bibinfo {author} {\bibfnamefont {J.}~\bibnamefont
  {Fern{\'a}ndez-Rossier}},\ }\bibfield  {title} {\bibinfo {title} {On the
  origin of magnetic anisotropy in two dimensional {CrI}${}_3$},\ }\href
  {https://doi.org/10.1088/2053-1583/aa75ed} {\bibfield  {journal} {\bibinfo
  {journal} {2D Mater.}\ }\textbf {\bibinfo {volume} {4}},\ \bibinfo {pages}
  {035002} (\bibinfo {year} {2017})}\BibitemShut {NoStop}%
\bibitem [{\citenamefont {Lobo-Checa}\ \emph {et~al.}(2024)\citenamefont
  {Lobo-Checa}, \citenamefont {Hern{\'a}ndez-L{\'o}pez}, \citenamefont
  {Otrokov}, \citenamefont {Piquero-Zulaica}, \citenamefont {Candia},
  \citenamefont {Gargiani}, \citenamefont {Serrate}, \citenamefont {Delgado},
  \citenamefont {Valvidares}, \citenamefont {Cerd{\'a}} \emph
  {et~al.}}]{Lobo.ncomms2024}%
  \BibitemOpen
  \bibfield  {author} {\bibinfo {author} {\bibfnamefont {J.}~\bibnamefont
  {Lobo-Checa}}, \bibinfo {author} {\bibfnamefont {L.}~\bibnamefont
  {Hern{\'a}ndez-L{\'o}pez}}, \bibinfo {author} {\bibfnamefont {M.~M.}\
  \bibnamefont {Otrokov}}, \bibinfo {author} {\bibfnamefont {I.}~\bibnamefont
  {Piquero-Zulaica}}, \bibinfo {author} {\bibfnamefont {A.~E.}\ \bibnamefont
  {Candia}}, \bibinfo {author} {\bibfnamefont {P.}~\bibnamefont {Gargiani}},
  \bibinfo {author} {\bibfnamefont {D.}~\bibnamefont {Serrate}}, \bibinfo
  {author} {\bibfnamefont {F.}~\bibnamefont {Delgado}}, \bibinfo {author}
  {\bibfnamefont {M.}~\bibnamefont {Valvidares}}, \bibinfo {author}
  {\bibfnamefont {J.}~\bibnamefont {Cerd{\'a}}}, \emph {et~al.},\ }\bibfield
  {title} {\bibinfo {title} {Ferromagnetism on an atom-thick \& extended {2D}
  metal-organic coordination network},\ }\href
  {https://doi.org/https://doi.org/10.1038/s41467-024-46115-z} {\bibfield
  {journal} {\bibinfo  {journal} {Nat. Commun.}\ }\textbf {\bibinfo {volume}
  {15}},\ \bibinfo {pages} {1858} (\bibinfo {year} {2024})}\BibitemShut
  {NoStop}%
\bibitem [{\citenamefont {Yosida}(1996)}]{Yosida1996}%
  \BibitemOpen
  \bibfield  {author} {\bibinfo {author} {\bibfnamefont {K.}~\bibnamefont
  {Yosida}},\ }\href@noop {} {\emph {\bibinfo {title} {Theory of magnetism}}}\
  (\bibinfo  {publisher} {Springer, Heidelberg, Germany},\ \bibinfo {year}
  {1996})\BibitemShut {NoStop}%
\bibitem [{\citenamefont {Xie}\ \emph {et~al.}(2023)\citenamefont {Xie},
  \citenamefont {Gonzalez}, \citenamefont {Li}, \citenamefont {Michiardi},
  \citenamefont {Gorovikov}, \citenamefont {Ryu}, \citenamefont {Fender},
  \citenamefont {Zonno}, \citenamefont {Jo}, \citenamefont {Zhdanovich} \emph
  {et~al.}}]{Xie.cm2023}%
  \BibitemOpen
  \bibfield  {author} {\bibinfo {author} {\bibfnamefont {L.~S.}\ \bibnamefont
  {Xie}}, \bibinfo {author} {\bibfnamefont {O.}~\bibnamefont {Gonzalez}},
  \bibinfo {author} {\bibfnamefont {K.}~\bibnamefont {Li}}, \bibinfo {author}
  {\bibfnamefont {M.}~\bibnamefont {Michiardi}}, \bibinfo {author}
  {\bibfnamefont {S.}~\bibnamefont {Gorovikov}}, \bibinfo {author}
  {\bibfnamefont {S.~H.}\ \bibnamefont {Ryu}}, \bibinfo {author} {\bibfnamefont
  {S.~S.}\ \bibnamefont {Fender}}, \bibinfo {author} {\bibfnamefont
  {M.}~\bibnamefont {Zonno}}, \bibinfo {author} {\bibfnamefont {N.~H.}\
  \bibnamefont {Jo}}, \bibinfo {author} {\bibfnamefont {S.}~\bibnamefont
  {Zhdanovich}}, \emph {et~al.},\ }\bibfield  {title} {\bibinfo {title}
  {Comparative electronic structures of the chiral helimagnets
  {Cr$_{1/3}$NbS$_2$ and Cr$_{1/3}$TaS$_2$}},\ }\href {https://doi.org/10.1021/acs.chemmater.3c01564} {\bibfield
  {journal} {\bibinfo  {journal} {Chem.~Mater.}\ }\textbf {\bibinfo
  {volume} {35}},\ \bibinfo {pages} {7239} (\bibinfo {year}
  {2023})}\BibitemShut {NoStop}%
\bibitem [{\citenamefont {Jenkins}\ \emph {et~al.}(2022)\citenamefont
  {Jenkins}, \citenamefont {Rózsa}, \citenamefont {Atxitia}, \citenamefont
  {Evans}, \citenamefont {Novoselov},\ and\ \citenamefont
  {Santos}}]{jenkins_breaking_2022}%
  \BibitemOpen
  \bibfield  {author} {\bibinfo {author} {\bibfnamefont {S.}~\bibnamefont
  {Jenkins}}, \bibinfo {author} {\bibfnamefont {L.}~\bibnamefont {Rózsa}},
  \bibinfo {author} {\bibfnamefont {U.}~\bibnamefont {Atxitia}}, \bibinfo
  {author} {\bibfnamefont {R.~F.~L.}\ \bibnamefont {Evans}}, \bibinfo {author}
  {\bibfnamefont {K.~S.}\ \bibnamefont {Novoselov}},\ and\ \bibinfo {author}
  {\bibfnamefont {E.~J.~G.}\ \bibnamefont {Santos}},\ }\bibfield  {title}
  {\bibinfo {title} {Breaking through the {Mermin}-{Wagner} limit in {2D} van
  der {Waals} magnets},\ }\href {https://doi.org/10.1038/s41467-022-34389-0}
  {\bibfield  {journal} {\bibinfo  {journal} {Nat. Commun.}\ }\textbf {\bibinfo
  {volume} {13}},\ \bibinfo {pages} {6917} (\bibinfo {year}
  {2022})}\BibitemShut {NoStop}%
\bibitem [{\citenamefont {Yang}\ \emph {et~al.}(2021)\citenamefont {Yang},
  \citenamefont {Xu}, \citenamefont {Zhu}, \citenamefont {Niu}, \citenamefont
  {Xu}, \citenamefont {Peng}, \citenamefont {Cheng}, \citenamefont {Jia},
  \citenamefont {Huang}, \citenamefont {Xu}, \citenamefont {Lu},\ and\
  \citenamefont {Ye}}]{Yang.prx2021}%
  \BibitemOpen
  \bibfield  {author} {\bibinfo {author} {\bibfnamefont {S.}~\bibnamefont
  {Yang}}, \bibinfo {author} {\bibfnamefont {X.}~\bibnamefont {Xu}}, \bibinfo
  {author} {\bibfnamefont {Y.}~\bibnamefont {Zhu}}, \bibinfo {author}
  {\bibfnamefont {R.}~\bibnamefont {Niu}}, \bibinfo {author} {\bibfnamefont
  {C.}~\bibnamefont {Xu}}, \bibinfo {author} {\bibfnamefont {Y.}~\bibnamefont
  {Peng}}, \bibinfo {author} {\bibfnamefont {X.}~\bibnamefont {Cheng}},
  \bibinfo {author} {\bibfnamefont {X.}~\bibnamefont {Jia}}, \bibinfo {author}
  {\bibfnamefont {Y.}~\bibnamefont {Huang}}, \bibinfo {author} {\bibfnamefont
  {X.}~\bibnamefont {Xu}}, \bibinfo {author} {\bibfnamefont {J.}~\bibnamefont
  {Lu}},\ and\ \bibinfo {author} {\bibfnamefont {Y.}~\bibnamefont {Ye}},\
  }\bibfield  {title} {\bibinfo {title} {Odd-even layer-number effect and
  layer-dependent magnetic phase diagrams in {MnBi}${}_{2}${Te}${}_{4}$},\
  }\href {https://doi.org/10.1103/PhysRevX.11.011003} {\bibfield  {journal}
  {\bibinfo  {journal} {Phys. Rev. X}\ }\textbf {\bibinfo {volume} {11}},\
  \bibinfo {pages} {011003} (\bibinfo {year} {2021})}\BibitemShut {NoStop}%
\bibitem [{\citenamefont {Kagerer}\ \emph {et~al.}(2023)\citenamefont
  {Kagerer}, \citenamefont {Fornari}, \citenamefont {Buchberger}, \citenamefont
  {Tschirner}, \citenamefont {Veyrat}, \citenamefont {Kamp}, \citenamefont
  {Tcakaev}, \citenamefont {Zabolotnyy}, \citenamefont {Morelh\~ao},
  \citenamefont {Geldiyev}, \citenamefont {M\"uller}, \citenamefont {Fedorov},
  \citenamefont {Rienks}, \citenamefont {Gargiani}, \citenamefont {Valvidares},
  \citenamefont {Folkers}, \citenamefont {Isaeva}, \citenamefont {B\"uchner},
  \citenamefont {Hinkov}, \citenamefont {Claessen}, \citenamefont {Bentmann},\
  and\ \citenamefont {Reinert}}]{Kagerer.prr2023}%
  \BibitemOpen
  \bibfield  {author} {\bibinfo {author} {\bibfnamefont {P.}~\bibnamefont
  {Kagerer}}, \bibinfo {author} {\bibfnamefont {C.~I.}\ \bibnamefont
  {Fornari}}, \bibinfo {author} {\bibfnamefont {S.}~\bibnamefont {Buchberger}},
  \bibinfo {author} {\bibfnamefont {T.}~\bibnamefont {Tschirner}}, \bibinfo
  {author} {\bibfnamefont {L.}~\bibnamefont {Veyrat}}, \bibinfo {author}
  {\bibfnamefont {M.}~\bibnamefont {Kamp}}, \bibinfo {author} {\bibfnamefont
  {A.~V.}\ \bibnamefont {Tcakaev}}, \bibinfo {author} {\bibfnamefont
  {V.}~\bibnamefont {Zabolotnyy}}, \bibinfo {author} {\bibfnamefont {S.~L.}\
  \bibnamefont {Morelh\~ao}}, \bibinfo {author} {\bibfnamefont
  {B.}~\bibnamefont {Geldiyev}}, \bibinfo {author} {\bibfnamefont
  {S.}~\bibnamefont {M\"uller}}, \bibinfo {author} {\bibfnamefont
  {A.}~\bibnamefont {Fedorov}}, \bibinfo {author} {\bibfnamefont
  {E.}~\bibnamefont {Rienks}}, \bibinfo {author} {\bibfnamefont
  {P.}~\bibnamefont {Gargiani}}, \bibinfo {author} {\bibfnamefont
  {M.}~\bibnamefont {Valvidares}}, \bibinfo {author} {\bibfnamefont {L.~C.}\
  \bibnamefont {Folkers}}, \bibinfo {author} {\bibfnamefont {A.}~\bibnamefont
  {Isaeva}}, \bibinfo {author} {\bibfnamefont {B.}~\bibnamefont {B\"uchner}},
  \bibinfo {author} {\bibfnamefont {V.}~\bibnamefont {Hinkov}}, \bibinfo
  {author} {\bibfnamefont {R.}~\bibnamefont {Claessen}}, \bibinfo {author}
  {\bibfnamefont {H.}~\bibnamefont {Bentmann}},\ and\ \bibinfo {author}
  {\bibfnamefont {F.}~\bibnamefont {Reinert}},\ }\bibfield  {title} {\bibinfo
  {title} {Two-dimensional ferromagnetic extension of a topological
  insulator},\ }\href {https://doi.org/10.1103/PhysRevResearch.5.L022019}
  {\bibfield  {journal} {\bibinfo  {journal} {Phys. Rev. Res.}\ }\textbf
  {\bibinfo {volume} {5}},\ \bibinfo {pages} {L022019} (\bibinfo {year}
  {2023})}\BibitemShut {NoStop}%
\bibitem [{\citenamefont {Fiedler}\ \emph {et~al.}(2015)\citenamefont
  {Fiedler}, \citenamefont {Bathon}, \citenamefont {Eremeev}, \citenamefont
  {Tereshchenko}, \citenamefont {Kokh}, \citenamefont {Chulkov}, \citenamefont
  {Sessi}, \citenamefont {Bentmann}, \citenamefont {Bode},\ and\ \citenamefont
  {Reinert}}]{Fiedler2015}%
  \BibitemOpen
  \bibfield  {author} {\bibinfo {author} {\bibfnamefont {S.}~\bibnamefont
  {Fiedler}}, \bibinfo {author} {\bibfnamefont {T.}~\bibnamefont {Bathon}},
  \bibinfo {author} {\bibfnamefont {S.~V.}\ \bibnamefont {Eremeev}}, \bibinfo
  {author} {\bibfnamefont {O.~E.}\ \bibnamefont {Tereshchenko}}, \bibinfo
  {author} {\bibfnamefont {K.~A.}\ \bibnamefont {Kokh}}, \bibinfo {author}
  {\bibfnamefont {E.~V.}\ \bibnamefont {Chulkov}}, \bibinfo {author}
  {\bibfnamefont {P.}~\bibnamefont {Sessi}}, \bibinfo {author} {\bibfnamefont
  {H.}~\bibnamefont {Bentmann}}, \bibinfo {author} {\bibfnamefont
  {M.}~\bibnamefont {Bode}},\ and\ \bibinfo {author} {\bibfnamefont
  {F.}~\bibnamefont {Reinert}},\ }\bibfield  {title} {\bibinfo {title}
  {Termination-dependent surface properties in the giant-{Rashba}
  semiconductors {$\text{BiTe}X$ ($X=\text{Cl}$, Br, I)}},\ }\href
  {https://doi.org/10.1103/PhysRevB.92.235430} {\bibfield  {journal} {\bibinfo
  {journal} {Phys. Rev. B}\ }\textbf {\bibinfo {volume} {92}},\ \bibinfo
  {pages} {235430} (\bibinfo {year} {2015})}\BibitemShut {NoStop}%
\bibitem [{\citenamefont {Eremeev}\ \emph {et~al.}(2013)\citenamefont
  {Eremeev}, \citenamefont {Rusinov}, \citenamefont {Nechaev},\ and\
  \citenamefont {Chulkov}}]{Eremeev2013}%
  \BibitemOpen
  \bibfield  {author} {\bibinfo {author} {\bibfnamefont {S.~V.}\ \bibnamefont
  {Eremeev}}, \bibinfo {author} {\bibfnamefont {I.~P.}\ \bibnamefont
  {Rusinov}}, \bibinfo {author} {\bibfnamefont {I.~A.}\ \bibnamefont
  {Nechaev}},\ and\ \bibinfo {author} {\bibfnamefont {E.~V.}\ \bibnamefont
  {Chulkov}},\ }\bibfield  {title} {\bibinfo {title} {Rashba split surface
  states in {BiTeBr}},\ }\href {https://doi.org/10.1088/1367-2630/15/7/075015}
  {\bibfield  {journal} {\bibinfo  {journal} {New J. Phys.}\ }\textbf {\bibinfo
  {volume} {15}},\ \bibinfo {pages} {075015} (\bibinfo {year}
  {2013})}\BibitemShut {NoStop}%
\bibitem [{\citenamefont {Takayama}\ \emph {et~al.}(2011)\citenamefont
  {Takayama}, \citenamefont {Sato}, \citenamefont {Souma},\ and\ \citenamefont
  {Takahashi}}]{Takayama_PRL2011}%
  \BibitemOpen
  \bibfield  {author} {\bibinfo {author} {\bibfnamefont {A.}~\bibnamefont
  {Takayama}}, \bibinfo {author} {\bibfnamefont {T.}~\bibnamefont {Sato}},
  \bibinfo {author} {\bibfnamefont {S.}~\bibnamefont {Souma}},\ and\ \bibinfo
  {author} {\bibfnamefont {T.}~\bibnamefont {Takahashi}},\ }\bibfield  {title}
  {\bibinfo {title} {Giant out-of-plane spin component and the asymmetry of
  spin polarization in surface rashba states of bismuth thin film},\ }\href
  {https://doi.org/10.1103/PhysRevLett.106.166401} {\bibfield  {journal}
  {\bibinfo  {journal} {Phys. Rev. Lett.}\ }\textbf {\bibinfo {volume} {106}},\
  \bibinfo {pages} {166401} (\bibinfo {year} {2011})}\BibitemShut {NoStop}%
\bibitem [{\citenamefont {Otrokov}\ \emph {et~al.}(2017)\citenamefont
  {Otrokov}, \citenamefont {Menshchikova}, \citenamefont {Vergniory},
  \citenamefont {Rusinov}, \citenamefont {Vyazovskaya}, \citenamefont
  {Koroteev}, \citenamefont {Bihlmayer}, \citenamefont {Ernst}, \citenamefont
  {Echenique}, \citenamefont {Arnau},\ and\ \citenamefont
  {Chulkov}}]{Otrokov.2dmat2017}%
  \BibitemOpen
  \bibfield  {author} {\bibinfo {author} {\bibfnamefont {M.~M.}\ \bibnamefont
  {Otrokov}}, \bibinfo {author} {\bibfnamefont {T.~V.}\ \bibnamefont
  {Menshchikova}}, \bibinfo {author} {\bibfnamefont {M.~G.}\ \bibnamefont
  {Vergniory}}, \bibinfo {author} {\bibfnamefont {I.~P.}\ \bibnamefont
  {Rusinov}}, \bibinfo {author} {\bibfnamefont {A.~Y.}\ \bibnamefont
  {Vyazovskaya}}, \bibinfo {author} {\bibfnamefont {Y.~M.}\ \bibnamefont
  {Koroteev}}, \bibinfo {author} {\bibfnamefont {G.}~\bibnamefont {Bihlmayer}},
  \bibinfo {author} {\bibfnamefont {A.}~\bibnamefont {Ernst}}, \bibinfo
  {author} {\bibfnamefont {P.~M.}\ \bibnamefont {Echenique}}, \bibinfo {author}
  {\bibfnamefont {A.}~\bibnamefont {Arnau}},\ and\ \bibinfo {author}
  {\bibfnamefont {E.~V.}\ \bibnamefont {Chulkov}},\ }\bibfield  {title}
  {\bibinfo {title} {Highly-ordered wide bandgap materials for quantized
  anomalous {{Hall}} and magnetoelectric effects},\ }\href
  {https://doi.org/https://doi.org/10.1088/2053-1583/aa6bec} {\bibfield
  {journal} {\bibinfo  {journal} {2D Mater.}\ }\textbf {\bibinfo {volume}
  {4}},\ \bibinfo {pages} {025082} (\bibinfo {year} {2017})}\BibitemShut
  {NoStop}%
\bibitem [{\citenamefont {Liu}\ \emph {et~al.}(2020)\citenamefont {Liu},
  \citenamefont {Wang}, \citenamefont {Li}, \citenamefont {Wu}, \citenamefont
  {Li}, \citenamefont {Li}, \citenamefont {He}, \citenamefont {Xu},
  \citenamefont {Zhang},\ and\ \citenamefont {Wang}}]{Liu.nmat2020}%
  \BibitemOpen
  \bibfield  {author} {\bibinfo {author} {\bibfnamefont {C.}~\bibnamefont
  {Liu}}, \bibinfo {author} {\bibfnamefont {Y.}~\bibnamefont {Wang}}, \bibinfo
  {author} {\bibfnamefont {H.}~\bibnamefont {Li}}, \bibinfo {author}
  {\bibfnamefont {Y.}~\bibnamefont {Wu}}, \bibinfo {author} {\bibfnamefont
  {Y.}~\bibnamefont {Li}}, \bibinfo {author} {\bibfnamefont {J.}~\bibnamefont
  {Li}}, \bibinfo {author} {\bibfnamefont {K.}~\bibnamefont {He}}, \bibinfo
  {author} {\bibfnamefont {Y.}~\bibnamefont {Xu}}, \bibinfo {author}
  {\bibfnamefont {J.}~\bibnamefont {Zhang}},\ and\ \bibinfo {author}
  {\bibfnamefont {Y.}~\bibnamefont {Wang}},\ }\bibfield  {title} {\bibinfo
  {title} {Robust axion insulator and {Chern} insulator phases in a
  two-dimensional antiferromagnetic topological insulator},\ }\href
  {https://doi.org/https://doi.org/10.1038/s41563-019-0573-3} {\bibfield
  {journal} {\bibinfo  {journal} {Nat. Mater.}\ }\textbf {\bibinfo {volume}
  {19}},\ \bibinfo {pages} {522} (\bibinfo {year} {2020})}\BibitemShut
  {NoStop}%
\bibitem [{\citenamefont {Ge}\ \emph {et~al.}(2020)\citenamefont {Ge},
  \citenamefont {Liu}, \citenamefont {Li}, \citenamefont {Li}, \citenamefont
  {Luo}, \citenamefont {Wu}, \citenamefont {Xu},\ and\ \citenamefont
  {Wang}}]{Ge.nsr2019}%
  \BibitemOpen
  \bibfield  {author} {\bibinfo {author} {\bibfnamefont {J.}~\bibnamefont
  {Ge}}, \bibinfo {author} {\bibfnamefont {Y.}~\bibnamefont {Liu}}, \bibinfo
  {author} {\bibfnamefont {J.}~\bibnamefont {Li}}, \bibinfo {author}
  {\bibfnamefont {H.}~\bibnamefont {Li}}, \bibinfo {author} {\bibfnamefont
  {T.}~\bibnamefont {Luo}}, \bibinfo {author} {\bibfnamefont {Y.}~\bibnamefont
  {Wu}}, \bibinfo {author} {\bibfnamefont {Y.}~\bibnamefont {Xu}},\ and\
  \bibinfo {author} {\bibfnamefont {J.}~\bibnamefont {Wang}},\ }\bibfield
  {title} {\bibinfo {title} {{{High}-{Chern}-number and high-temperature
  quantum {{Hall}} effect without {{Landau}} levels}},\ }\href
  {https://doi.org/10.1093/nsr/nwaa089} {\bibfield  {journal} {\bibinfo
  {journal} {Natl. Sci. Rev.}\ }\textbf {\bibinfo {volume} {7}},\ \bibinfo
  {pages} {1280} (\bibinfo {year} {2020})}\BibitemShut {NoStop}%
\bibitem [{\citenamefont {Deng}\ \emph {et~al.}(2020)\citenamefont {Deng},
  \citenamefont {Yu}, \citenamefont {Shi}, \citenamefont {Guo}, \citenamefont
  {Xu}, \citenamefont {Wang}, \citenamefont {Chen},\ and\ \citenamefont
  {Zhang}}]{Deng.sci2020}%
  \BibitemOpen
  \bibfield  {author} {\bibinfo {author} {\bibfnamefont {Y.}~\bibnamefont
  {Deng}}, \bibinfo {author} {\bibfnamefont {Y.}~\bibnamefont {Yu}}, \bibinfo
  {author} {\bibfnamefont {M.~Z.}\ \bibnamefont {Shi}}, \bibinfo {author}
  {\bibfnamefont {Z.}~\bibnamefont {Guo}}, \bibinfo {author} {\bibfnamefont
  {Z.}~\bibnamefont {Xu}}, \bibinfo {author} {\bibfnamefont {J.}~\bibnamefont
  {Wang}}, \bibinfo {author} {\bibfnamefont {X.~H.}\ \bibnamefont {Chen}},\
  and\ \bibinfo {author} {\bibfnamefont {Y.}~\bibnamefont {Zhang}},\ }\bibfield
   {title} {\bibinfo {title} {Quantum anomalous {Hall} effect in intrinsic
  magnetic topological insulator {MnBi}$_2${Te}$_4$},\ }\href
  {https://doi.org/10.1126/science.aax815} {\bibfield  {journal} {\bibinfo
  {journal} {Science}\ }\textbf {\bibinfo {volume} {367}},\ \bibinfo {pages}
  {895} (\bibinfo {year} {2020})}\BibitemShut {NoStop}%
\bibitem [{\citenamefont {Gao}\ \emph {et~al.}(2021)\citenamefont {Gao},
  \citenamefont {Liu}, \citenamefont {Hu}, \citenamefont {Qiu}, \citenamefont
  {Tzschaschel}, \citenamefont {Ghosh}, \citenamefont {Ho}, \citenamefont
  {B{\'e}rub{\'e}}, \citenamefont {Chen}, \citenamefont {Sun}, \citenamefont
  {Zhang}, \citenamefont {Zhang}, \citenamefont {Wang}, \citenamefont {Wang},
  \citenamefont {Huang}, \citenamefont {Felser}, \citenamefont {Agarwal},
  \citenamefont {Ding}, \citenamefont {Tien}, \citenamefont {Akey},
  \citenamefont {Gardener}, \citenamefont {Singh}, \citenamefont {Wataname},
  \citenamefont {Taniguchi}, \citenamefont {Burch}, \citenamefont {Bell},
  \citenamefont {Zhou}, \citenamefont {Gao}, \citenamefont {Lu}, \citenamefont
  {Bansil}, \citenamefont {Lin}, \citenamefont {Chang}, \citenamefont {Fu},
  \citenamefont {Ma}, \citenamefont {Ni},\ and\ \citenamefont
  {Xu}}]{Gao.nat2021}%
  \BibitemOpen
  \bibfield  {author} {\bibinfo {author} {\bibfnamefont {A.}~\bibnamefont
  {Gao}}, \bibinfo {author} {\bibfnamefont {Y.-F.}\ \bibnamefont {Liu}},
  \bibinfo {author} {\bibfnamefont {C.}~\bibnamefont {Hu}}, \bibinfo {author}
  {\bibfnamefont {J.-X.}\ \bibnamefont {Qiu}}, \bibinfo {author} {\bibfnamefont
  {C.}~\bibnamefont {Tzschaschel}}, \bibinfo {author} {\bibfnamefont
  {B.}~\bibnamefont {Ghosh}}, \bibinfo {author} {\bibfnamefont {S.-C.}\
  \bibnamefont {Ho}}, \bibinfo {author} {\bibfnamefont {D.}~\bibnamefont
  {B{\'e}rub{\'e}}}, \bibinfo {author} {\bibfnamefont {R.}~\bibnamefont
  {Chen}}, \bibinfo {author} {\bibfnamefont {H.}~\bibnamefont {Sun}}, \bibinfo
  {author} {\bibfnamefont {Z.}~\bibnamefont {Zhang}}, \bibinfo {author}
  {\bibfnamefont {X.-Y.}\ \bibnamefont {Zhang}}, \bibinfo {author}
  {\bibfnamefont {Y.-X.}\ \bibnamefont {Wang}}, \bibinfo {author}
  {\bibfnamefont {N.}~\bibnamefont {Wang}}, \bibinfo {author} {\bibfnamefont
  {Z.}~\bibnamefont {Huang}}, \bibinfo {author} {\bibfnamefont
  {C.}~\bibnamefont {Felser}}, \bibinfo {author} {\bibfnamefont
  {A.}~\bibnamefont {Agarwal}}, \bibinfo {author} {\bibfnamefont
  {T.}~\bibnamefont {Ding}}, \bibinfo {author} {\bibfnamefont {H.-J.}\
  \bibnamefont {Tien}}, \bibinfo {author} {\bibfnamefont {A.}~\bibnamefont
  {Akey}}, \bibinfo {author} {\bibfnamefont {J.}~\bibnamefont {Gardener}},
  \bibinfo {author} {\bibfnamefont {B.}~\bibnamefont {Singh}}, \bibinfo
  {author} {\bibfnamefont {K.}~\bibnamefont {Wataname}}, \bibinfo {author}
  {\bibfnamefont {T.}~\bibnamefont {Taniguchi}}, \bibinfo {author}
  {\bibfnamefont {K.~S.}\ \bibnamefont {Burch}}, \bibinfo {author}
  {\bibfnamefont {D.~C.}\ \bibnamefont {Bell}}, \bibinfo {author}
  {\bibfnamefont {B.~B.}\ \bibnamefont {Zhou}}, \bibinfo {author}
  {\bibfnamefont {W.}~\bibnamefont {Gao}}, \bibinfo {author} {\bibfnamefont
  {H.-Z.}\ \bibnamefont {Lu}}, \bibinfo {author} {\bibfnamefont
  {A.}~\bibnamefont {Bansil}}, \bibinfo {author} {\bibfnamefont
  {H.}~\bibnamefont {Lin}}, \bibinfo {author} {\bibfnamefont {T.-R.}\
  \bibnamefont {Chang}}, \bibinfo {author} {\bibfnamefont {L.}~\bibnamefont
  {Fu}}, \bibinfo {author} {\bibfnamefont {Q.}~\bibnamefont {Ma}}, \bibinfo
  {author} {\bibfnamefont {N.}~\bibnamefont {Ni}},\ and\ \bibinfo {author}
  {\bibfnamefont {S.-Y.}\ \bibnamefont {Xu}},\ }\bibfield  {title} {\bibinfo
  {title} {Layer {{Hall}} effect in a {2D} topological axion antiferromagnet},\
  }\href {https://doi.org/https://doi.org/10.1038/s41586-021-03679-w}
  {\bibfield  {journal} {\bibinfo  {journal} {Nature}\ }\textbf {\bibinfo
  {volume} {595}},\ \bibinfo {pages} {521} (\bibinfo {year}
  {2021})}\BibitemShut {NoStop}%
\bibitem [{\citenamefont {Gao}\ \emph {et~al.}(2023)\citenamefont {Gao},
  \citenamefont {Liu}, \citenamefont {Qiu}, \citenamefont {Ghosh},
  \citenamefont {V.~Trevisan}, \citenamefont {Onishi}, \citenamefont {Hu},
  \citenamefont {Qian}, \citenamefont {Tien}, \citenamefont {Chen} \emph
  {et~al.}}]{Gao.sci2023}%
  \BibitemOpen
  \bibfield  {author} {\bibinfo {author} {\bibfnamefont {A.}~\bibnamefont
  {Gao}}, \bibinfo {author} {\bibfnamefont {Y.-F.}\ \bibnamefont {Liu}},
  \bibinfo {author} {\bibfnamefont {J.-X.}\ \bibnamefont {Qiu}}, \bibinfo
  {author} {\bibfnamefont {B.}~\bibnamefont {Ghosh}}, \bibinfo {author}
  {\bibfnamefont {T.}~\bibnamefont {V.~Trevisan}}, \bibinfo {author}
  {\bibfnamefont {Y.}~\bibnamefont {Onishi}}, \bibinfo {author} {\bibfnamefont
  {C.}~\bibnamefont {Hu}}, \bibinfo {author} {\bibfnamefont {T.}~\bibnamefont
  {Qian}}, \bibinfo {author} {\bibfnamefont {H.-J.}\ \bibnamefont {Tien}},
  \bibinfo {author} {\bibfnamefont {S.-W.}\ \bibnamefont {Chen}}, \emph
  {et~al.},\ }\bibfield  {title} {\bibinfo {title} {Quantum metric nonlinear
  {Hall} effect in a topological antiferromagnetic heterostructure},\
  }\href {DOI: 10.1126/science.adf1506} {\bibfield  {journal} {\bibinfo  {journal} {Science}\ }\textbf
  {\bibinfo {volume} {381}},\ \bibinfo {pages} {181} (\bibinfo {year}
  {2023})}\BibitemShut {NoStop}%
\bibitem [{\citenamefont {Bosnar}\ \emph {et~al.}(2023)\citenamefont {Bosnar},
  \citenamefont {Vyazovskaya}, \citenamefont {Petrov}, \citenamefont
  {Chulkov},\ and\ \citenamefont {Otrokov}}]{Bosnar.npj2d2023}%
  \BibitemOpen
  \bibfield  {author} {\bibinfo {author} {\bibfnamefont {M.}~\bibnamefont
  {Bosnar}}, \bibinfo {author} {\bibfnamefont {A.~Y.}\ \bibnamefont
  {Vyazovskaya}}, \bibinfo {author} {\bibfnamefont {E.~K.}\ \bibnamefont
  {Petrov}}, \bibinfo {author} {\bibfnamefont {E.~V.}\ \bibnamefont
  {Chulkov}},\ and\ \bibinfo {author} {\bibfnamefont {M.~M.}\ \bibnamefont
  {Otrokov}},\ }\bibfield  {title} {\bibinfo {title} {{High {Chern} number van
  der {Waals} magnetic topological multilayers {MnBi}$_2${Te}$_4$/{hBN}}},\
  }\href {https://doi.org/https://doi.org/10.1038/s41699-023-00396-y}
  {\bibfield  {journal} {\bibinfo  {journal} {npj 2D Mater. Appl.}\ }\textbf
  {\bibinfo {volume} {7}},\ \bibinfo {pages} {33} (\bibinfo {year}
  {2023})}\BibitemShut {NoStop}%
\bibitem [{\citenamefont {Qiu}\ \emph {et~al.}(2023)\citenamefont {Qiu},
  \citenamefont {Tzschaschel}, \citenamefont {Ahn}, \citenamefont {Gao},
  \citenamefont {Li}, \citenamefont {Zhang}, \citenamefont {Ghosh},
  \citenamefont {Hu}, \citenamefont {Wang}, \citenamefont {Liu} \emph
  {et~al.}}]{Qiu.nmat2023}%
  \BibitemOpen
  \bibfield  {author} {\bibinfo {author} {\bibfnamefont {J.-X.}\ \bibnamefont
  {Qiu}}, \bibinfo {author} {\bibfnamefont {C.}~\bibnamefont {Tzschaschel}},
  \bibinfo {author} {\bibfnamefont {J.}~\bibnamefont {Ahn}}, \bibinfo {author}
  {\bibfnamefont {A.}~\bibnamefont {Gao}}, \bibinfo {author} {\bibfnamefont
  {H.}~\bibnamefont {Li}}, \bibinfo {author} {\bibfnamefont {X.-Y.}\
  \bibnamefont {Zhang}}, \bibinfo {author} {\bibfnamefont {B.}~\bibnamefont
  {Ghosh}}, \bibinfo {author} {\bibfnamefont {C.}~\bibnamefont {Hu}}, \bibinfo
  {author} {\bibfnamefont {Y.-X.}\ \bibnamefont {Wang}}, \bibinfo {author}
  {\bibfnamefont {Y.-F.}\ \bibnamefont {Liu}}, \emph {et~al.},\ }\bibfield
  {title} {\bibinfo {title} {Axion optical induction of antiferromagnetic
  order},\ }\href {https://doi.org/https://doi.org/10.1038/s41563-023-01493-5}
  {\bibfield  {journal} {\bibinfo  {journal} {Nat. Mater.}\ }\textbf {\bibinfo
  {volume} {22}},\ \bibinfo {pages} {583} (\bibinfo {year} {2023})}\BibitemShut
  {NoStop}%
\bibitem [{\citenamefont {Qiu}\ \emph {et~al.}(2025)\citenamefont {Qiu},
  \citenamefont {Ghosh}, \citenamefont {Sch{\"u}tte-Engel}, \citenamefont
  {Qian}, \citenamefont {Smith}, \citenamefont {Yao}, \citenamefont {Ahn},
  \citenamefont {Liu}, \citenamefont {Gao}, \citenamefont {Tzschaschel} \emph
  {et~al.}}]{Qiu.nat2025}%
  \BibitemOpen
  \bibfield  {author} {\bibinfo {author} {\bibfnamefont {J.-X.}\ \bibnamefont
  {Qiu}}, \bibinfo {author} {\bibfnamefont {B.}~\bibnamefont {Ghosh}}, \bibinfo
  {author} {\bibfnamefont {J.}~\bibnamefont {Sch{\"u}tte-Engel}}, \bibinfo
  {author} {\bibfnamefont {T.}~\bibnamefont {Qian}}, \bibinfo {author}
  {\bibfnamefont {M.}~\bibnamefont {Smith}}, \bibinfo {author} {\bibfnamefont
  {Y.-T.}\ \bibnamefont {Yao}}, \bibinfo {author} {\bibfnamefont
  {J.}~\bibnamefont {Ahn}}, \bibinfo {author} {\bibfnamefont {Y.-F.}\
  \bibnamefont {Liu}}, \bibinfo {author} {\bibfnamefont {A.}~\bibnamefont
  {Gao}}, \bibinfo {author} {\bibfnamefont {C.}~\bibnamefont {Tzschaschel}},
  \emph {et~al.},\ }\bibfield  {title} {\bibinfo {title} {Observation of the
  axion quasiparticle in {2D} {MnBi}${}_2${Te}${}_4$},\ }\href
  {https://doi.org/10.1038/s41586-025-08862-x} {\bibfield  {journal} {\bibinfo
  {journal} {Nature}\ }\textbf {\bibinfo {volume} {641}},\ \bibinfo {pages}
  {62} (\bibinfo {year} {2025})}\BibitemShut {NoStop}%
\bibitem [{\citenamefont {Lian}\ \emph {et~al.}(2025)\citenamefont {Lian},
  \citenamefont {Wang}, \citenamefont {Wang}, \citenamefont {Dong},
  \citenamefont {Feng}, \citenamefont {Dong}, \citenamefont {Ma}, \citenamefont
  {Yang}, \citenamefont {Xu}, \citenamefont {Li} \emph
  {et~al.}}]{Lian.nat2025}%
  \BibitemOpen
  \bibfield  {author} {\bibinfo {author} {\bibfnamefont {Z.}~\bibnamefont
  {Lian}}, \bibinfo {author} {\bibfnamefont {Y.}~\bibnamefont {Wang}}, \bibinfo
  {author} {\bibfnamefont {Y.}~\bibnamefont {Wang}}, \bibinfo {author}
  {\bibfnamefont {W.-H.}\ \bibnamefont {Dong}}, \bibinfo {author}
  {\bibfnamefont {Y.}~\bibnamefont {Feng}}, \bibinfo {author} {\bibfnamefont
  {Z.}~\bibnamefont {Dong}}, \bibinfo {author} {\bibfnamefont {M.}~\bibnamefont
  {Ma}}, \bibinfo {author} {\bibfnamefont {S.}~\bibnamefont {Yang}}, \bibinfo
  {author} {\bibfnamefont {L.}~\bibnamefont {Xu}}, \bibinfo {author}
  {\bibfnamefont {Y.}~\bibnamefont {Li}}, \emph {et~al.},\ }\bibfield  {title}
  {\bibinfo {title} {Antiferromagnetic quantum anomalous {Hall} effect under
  spin flips and flops},\ }\href
  {https://doi.org/https://doi.org/10.1038/s41586-025-08860-z} {\bibfield
  {journal} {\bibinfo  {journal} {Nature}\ }\textbf {\bibinfo {volume} {641}},\
  \bibinfo {pages} {70} (\bibinfo {year} {2025})}\BibitemShut {NoStop}%
\bibitem [{\citenamefont {Vyazovskaya}\ \emph {et~al.}(2025)\citenamefont
  {Vyazovskaya}, \citenamefont {Bosnar}, \citenamefont {Chulkov},\ and\
  \citenamefont {Otrokov}}]{Vyazovskaya.commsmat2025}%
  \BibitemOpen
  \bibfield  {author} {\bibinfo {author} {\bibfnamefont {A.~Y.}\ \bibnamefont
  {Vyazovskaya}}, \bibinfo {author} {\bibfnamefont {M.}~\bibnamefont {Bosnar}},
  \bibinfo {author} {\bibfnamefont {E.~V.}\ \bibnamefont {Chulkov}},\ and\
  \bibinfo {author} {\bibfnamefont {M.~M.}\ \bibnamefont {Otrokov}},\
  }\bibfield  {title} {\bibinfo {title} {Intrinsic magnetic topological
  insulators of the {MnBi}${}_2${Te}${}_4$ family},\ }\href
  {https://doi.org/https://doi.org/10.1038/s43246-025-00794-3} {\bibfield
  {journal} {\bibinfo  {journal} {Commun. Mater.}\ }\textbf {\bibinfo {volume}
  {6}},\ \bibinfo {pages} {88} (\bibinfo {year} {2025})}\BibitemShut {NoStop}%
\bibitem [{\citenamefont {Zhao}\ \emph {et~al.}(2021)\citenamefont {Zhao},
  \citenamefont {Zhou}, \citenamefont {Wang}, \citenamefont {Wang},
  \citenamefont {Song}, \citenamefont {Ovchinnikov}, \citenamefont {Yi},
  \citenamefont {Mei}, \citenamefont {Wang}, \citenamefont {Chan},
  \citenamefont {Liu}, \citenamefont {Xu},\ and\ \citenamefont
  {Chang}}]{Zhao.nl2021}%
  \BibitemOpen
  \bibfield  {author} {\bibinfo {author} {\bibfnamefont {Y.-F.}\ \bibnamefont
  {Zhao}}, \bibinfo {author} {\bibfnamefont {L.-J.}\ \bibnamefont {Zhou}},
  \bibinfo {author} {\bibfnamefont {F.}~\bibnamefont {Wang}}, \bibinfo {author}
  {\bibfnamefont {G.}~\bibnamefont {Wang}}, \bibinfo {author} {\bibfnamefont
  {T.}~\bibnamefont {Song}}, \bibinfo {author} {\bibfnamefont {D.}~\bibnamefont
  {Ovchinnikov}}, \bibinfo {author} {\bibfnamefont {H.}~\bibnamefont {Yi}},
  \bibinfo {author} {\bibfnamefont {R.}~\bibnamefont {Mei}}, \bibinfo {author}
  {\bibfnamefont {K.}~\bibnamefont {Wang}}, \bibinfo {author} {\bibfnamefont
  {M.~H.~W.}\ \bibnamefont {Chan}}, \bibinfo {author} {\bibfnamefont {C.-X.}\
  \bibnamefont {Liu}}, \bibinfo {author} {\bibfnamefont {X.}~\bibnamefont
  {Xu}},\ and\ \bibinfo {author} {\bibfnamefont {C.-Z.}\ \bibnamefont
  {Chang}},\ }\bibfield  {title} {\bibinfo {title} {Even–odd layer-dependent
  anomalous {Hall} effect in topological magnet {MnBi}${}_2${Te}${}_4$ thin
  films},\ }\href {https://doi.org/10.1021/acs.nanolett.1c02493} {\bibfield
  {journal} {\bibinfo  {journal} {Nano Lett.}\ }\textbf {\bibinfo {volume}
  {21}},\ \bibinfo {pages} {7691} (\bibinfo {year} {2021})}\BibitemShut
  {NoStop}%
\bibitem [{\citenamefont {Wang}\ \emph {et~al.}(2025)\citenamefont {Wang},
  \citenamefont {Fu}, \citenamefont {Wang}, \citenamefont {Lian}, \citenamefont
  {Yang}, \citenamefont {Li}, \citenamefont {Xu}, \citenamefont {Gao},
  \citenamefont {Yang}, \citenamefont {Wang} \emph {et~al.}}]{Wang.ncomms2025}%
  \BibitemOpen
  \bibfield  {author} {\bibinfo {author} {\bibfnamefont {Y.}~\bibnamefont
  {Wang}}, \bibinfo {author} {\bibfnamefont {B.}~\bibnamefont {Fu}}, \bibinfo
  {author} {\bibfnamefont {Y.}~\bibnamefont {Wang}}, \bibinfo {author}
  {\bibfnamefont {Z.}~\bibnamefont {Lian}}, \bibinfo {author} {\bibfnamefont
  {S.}~\bibnamefont {Yang}}, \bibinfo {author} {\bibfnamefont {Y.}~\bibnamefont
  {Li}}, \bibinfo {author} {\bibfnamefont {L.}~\bibnamefont {Xu}}, \bibinfo
  {author} {\bibfnamefont {Z.}~\bibnamefont {Gao}}, \bibinfo {author}
  {\bibfnamefont {X.}~\bibnamefont {Yang}}, \bibinfo {author} {\bibfnamefont
  {W.}~\bibnamefont {Wang}}, \emph {et~al.},\ }\bibfield  {title} {\bibinfo
  {title} {Towards the quantized anomalous {Hall} effect in {AlO}${}_x$-capped
  {MnBi}${}_2${Te}${}_4$},\ }\href
  {https://doi.org/https://doi.org/10.1038/s41467-025-57039-7} {\bibfield
  {journal} {\bibinfo  {journal} {Nat. Commun.}\ }\textbf {\bibinfo {volume}
  {16}},\ \bibinfo {pages} {1727} (\bibinfo {year} {2025})}\BibitemShut
  {NoStop}%
\bibitem [{\citenamefont {Zhang}\ \emph {et~al.}(2025)\citenamefont {Zhang},
  \citenamefont {Lu}, \citenamefont {Wang}, \citenamefont {Huang},
  \citenamefont {Zhang}, \citenamefont {Cao}, \citenamefont {Wang},
  \citenamefont {Zhou}, \citenamefont {Watanabe}, \citenamefont {Taniguchi}
  \emph {et~al.}}]{Zhang.ncomms2025}%
  \BibitemOpen
  \bibfield  {author} {\bibinfo {author} {\bibfnamefont {C.}~\bibnamefont
  {Zhang}}, \bibinfo {author} {\bibfnamefont {X.}~\bibnamefont {Lu}}, \bibinfo
  {author} {\bibfnamefont {N.}~\bibnamefont {Wang}}, \bibinfo {author}
  {\bibfnamefont {T.}~\bibnamefont {Huang}}, \bibinfo {author} {\bibfnamefont
  {H.}~\bibnamefont {Zhang}}, \bibinfo {author} {\bibfnamefont
  {N.}~\bibnamefont {Cao}}, \bibinfo {author} {\bibfnamefont {A.}~\bibnamefont
  {Wang}}, \bibinfo {author} {\bibfnamefont {X.}~\bibnamefont {Zhou}}, \bibinfo
  {author} {\bibfnamefont {K.}~\bibnamefont {Watanabe}}, \bibinfo {author}
  {\bibfnamefont {T.}~\bibnamefont {Taniguchi}}, \emph {et~al.},\ }\bibfield
  {title} {\bibinfo {title} {Zero-field chiral edge transport in an intrinsic
  magnetic topological insulator {MnBi}${}_2${Te}${}_4$},\ }\href
  {https://doi.org/https://doi.org/10.1038/s41467-025-59160-z} {\bibfield
  {journal} {\bibinfo  {journal} {Nat. Commun.}\ }\textbf {\bibinfo {volume}
  {16}},\ \bibinfo {pages} {5587} (\bibinfo {year} {2025})}\BibitemShut
  {NoStop}%
\bibitem [{\citenamefont {\v{Z}uti\'{c}}\ \emph {et~al.}(2004)\citenamefont
  {\v{Z}uti\'{c}}, \citenamefont {Fabian},\ and\ \citenamefont
  {Das~Sarma}}]{Zutic.rmp2004}%
  \BibitemOpen
  \bibfield  {author} {\bibinfo {author} {\bibfnamefont {I.}~\bibnamefont
  {\v{Z}uti\'{c}}}, \bibinfo {author} {\bibfnamefont {J.}~\bibnamefont
  {Fabian}},\ and\ \bibinfo {author} {\bibfnamefont {S.}~\bibnamefont
  {Das~Sarma}},\ }\bibfield  {title} {\bibinfo {title} {Spintronics:
  Fundamentals and applications},\ }\href
  {https://doi.org/10.1103/RevModPhys.76.323} {\bibfield  {journal} {\bibinfo
  {journal} {Rev. Mod. Phys.}\ }\textbf {\bibinfo {volume} {76}},\ \bibinfo
  {pages} {323} (\bibinfo {year} {2004})}\BibitemShut {NoStop}%

  \setcounter{lastbib}{\value{enumiv}} 
\end{thebibliography}
\end{document}